\documentclass[prb, twocolumn, floatfix, superscriptaddress, showpacs, longbibliography]{revtex4-2}

\pdfoutput=1						
\usepackage[charter]{mathdesign}	
\usepackage{amsmath}				

\usepackage{mathrsfs}					

\usepackage{mathtools}
\usepackage[]{stackengine}

\newlength\correct
\usepackage[pdftex]{graphicx}
\usepackage{microtype}
\usepackage{dsfont}
\usepackage{bm}\let\vec\bm
\usepackage{gensymb}
\usepackage{mdframed}
\usepackage[caption=false]{subfig}

\usepackage[svgnames]{xcolor}
\usepackage[
	colorlinks=True,linkcolor=DarkRed,citecolor=ForestGreen,urlcolor=MediumBlue,
	pdfstartview=FitH,bookmarks=False,pdfpagemode=UseNone
]{hyperref}

\usepackage{feynmp-auto}
\usepackage{natbib}
\usepackage{hyperref}
\usepackage{float}
\usepackage{cancel}

\usepackage[T1]{fontenc}
\usepackage{inputenc}
\usepackage{tcolorbox}
\usepackage{stackengine}

\definecolor{KIT-green}{RGB}{0, 150,130}
\definecolor{KIT-blue}{RGB}{70,100,170}

\makeatletter
\let\cat@comma@active\@empty
\makeatother

\begin{document}

\title{Correlation between phase stiffness and condensation energy across the non-Fermi to Fermi-liquid crossover in the Yukawa-Sachdev-Ye-Kitaev model on a lattice}

\author{D. Valentinis}
\affiliation{Institut für Quantenmaterialien und Technologien, Karlsruher Institut
für Technologie, 76131 Karlsruhe, Germany}
\affiliation{Institut für Theorie der Kondensierten Materie, Karlsruher Institut
für Technologie, 76131 Karlsruhe, Germany}
\author{G. A. Inkof}
\affiliation{Institut für Theorie der Kondensierten Materie, Karlsruher Institut
für Technologie, 76131 Karlsruhe, Germany}
\author{J. Schmalian}
\affiliation{Institut für Quantenmaterialien und Technologien, Karlsruher Institut
für Technologie, 76131 Karlsruhe, Germany}
\affiliation{Institut für Theorie der Kondensierten Materie, Karlsruher Institut
für Technologie, 76131 Karlsruhe, Germany}

\date{\today}

\begin{abstract}
We construct and analyze a lattice generalization of the Yukawa-Sachdev-Ye-Kitaev model, where spinful fermions experience on-site, random, all-to-all interactions with an Einstein bosonic mode, and random intersite coherent hopping. We obtain the exact self-consistent numerical solution of the model at mean-field level, and analytical approximations, for all values of fermion-boson coupling and hopping, under the spin-singlet ansatz and at particle-hole symmetry, both in the normal and superconducting states, thus tracing the entire phase diagram. In the normal state, the competition between hopping and coupling leads to crossovers between Fermi-liquid and non-Fermi liquid states, as reflected by the fermionic and bosonic spectral functions and the normal-state entropy. We  calculate the finite phase stiffness of the superconducting state through the equilibrium electromagnetic response. Furthermore, we study the critical temperature $T_c$, as well as the spectral functions, the quasiparticle weight, the gap, and the condensation energy in the superconducting state. At weak coupling, we retrieve a disordered generalization of Bardeen-Cooper-Schrieffer theory. At strong coupling, asymptotically $T_c$ saturates but the stiffness decreases, which suggests strong superconducting fluctuations. $T_c$ is maximum in the single-dot limit, while the stiffness peaks exactly at the crossover between non-Fermi liquid and Fermi-liquid phases. 
We discover that the quasiparticle weight, the stiffness, and the condensation energy, are all correlated as a function of coupling, reminiscent of the correlations observed in high-temperature cuprate superconductors. 

\end{abstract}

\maketitle

\section{Introduction}

\begin{figure*}[t]
\includegraphics[width=1.0\textwidth]{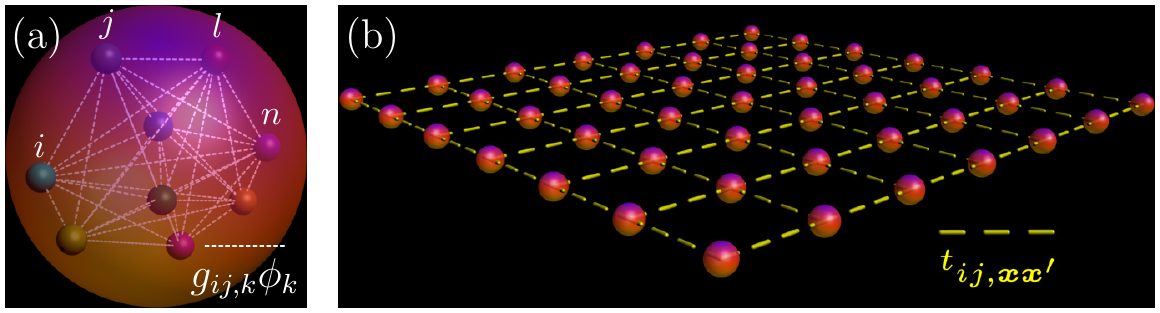}
\caption{\label{fig:Lattice3D} Schematic representation of a 2D lattice of Yukawa-SYK dots with all-to-all interactions. (a) All-to-all interactions (white dashed lines) between fermion flavors $\left\{i, j, l, n\right\}=\left\{1, \cdots \mathscr{N}\right\}$, mediated by a bosonic mode $\phi_{\vec{k}}$ with flavor $k=\left\{1, \cdots \mathscr{M}\right\}$ and with random couplings $g_{i j, k}$. (b) Construction of a 2D lattice where the building blocks are the Yukawa-SYK dots, connected by random coherent hopping $t_{i j, \vec{x} \vec{x}'}$ (dashed yellow lines) between nearest-neighbor sites.}
\end{figure*}

Unconventional superconductivity continues to represent a major challenge of contemporary condensed-matter research.
Primary examples, such as cuprates, are characterized by puzzling normal state behavior, in which the electronic spectrum  near so-called anti-nodal points in momentum space is structureless and devoid of sharp quasiparticle excitations 
\cite{Zaanen-2004,Valla-1999,Bruin-2013,Giraldo-2018,Naqib-2019,Mandal-2019,Michon-2019,Legros-2019,Grissonnanche-2021,Keimer-2015}. This is in contrast to the emergence of a well-defined coherence peak in the spectral function for $T<T_c$, which suggests that incoherent normal-state fermions are transformed, at least partially, into coherent Bogoliubov excitations of the superconducting condensate \cite{Dessau-1991,Campuzano-1996,Loeser-1997,Shen-1997,Fedorov-1999,Kaminski-2016,Orenstein-2000,Hashimoto-2014}.  The partial transformation is reflected in the fact that the spectral weight of  Bogoliubov  quasiparticles is small, in particular for underdoped systems, located not too far from the Mott-insulating parent state.

To rationalize this transformation from an incoherent normal towards a coherent superconducting state, it is desirable to design and analyze controlled theoretical approaches, that reproduce key  features of unconventional superconducting states stemming from normal-state incoherent electrons (the ``strange-metal phase'' \cite{Emery-1994,Hartnoll-2014b,Zaanen-2019}). A striking features is the observed correlation among the coherence-peak spectral weight, the superfluid stiffness, and the condensation energy in cuprates \cite{Shen-1997,Feng-2000}. 
In this work, we  analyze a solvable model that describes the FL to NFL crossover and that
reproduces this correlation between Bogoliubov quasiparticle weight, phase stiffness, and  condensation: the Yukawa-Sachdev-Ye-Kitaev model on a lattice, to be described in the following.

Our approach inserts itself into the general framework of quantum critical (QC) superconductors, regarded as a valuable route towards understanding pairing in NFL systems. There, the fermionic spectral function for single-particle excitations -- see Eq.\@ (\ref{eq:A_def}) for a formal definition -- assumes the power-law (``branch cut'' \cite{Zaanen-2021_preprint}) form $A_{\rm QC}(\omega)= A_0 \left|\omega\right|^{2 \Delta-1}$ as a function of frequency $\omega$, with constant $A_0$ and exponent $\Delta$. The limit of fully incoherent electrons is then $\Delta=1/2$, where $A_{\rm QC}(\omega)$ becomes frequency-independent. The physical origin of such power law is rooted in strong electronic interactions, which destroy Landau quasiparticles. These interactions are generated by the proximity to a quantum critical point (QCP), marking a zero-temperature phase transition tuned through an external parameter, e.g.\@, chemical doping or pressure \cite{Sachdev-2000,Sachdev-2000QPT,Zaanen-2004,Coleman-2005,Zaanen-2019,Hartnoll-2018holo}. The influence of the QCP on finite-temperature physics is manifested through an extremely short ``Planckian'' characteristic lifetime for single-particle excitations \cite{Zaanen-2004,Bruin-2013,Hartnoll-2022}, which in turn produces anomalous spectroscopic and transport properties like a linear-in-temperature resistivity \cite{Valla-1999,Damascelli-2003,Wang-2004,Hartnoll-2022}; such anomalous scalings have been observed in various families of materials, like cuprate superconductors \cite{Valla-1999,Bruin-2013,Giraldo-2018,Naqib-2019,Mandal-2019,Michon-2019,Legros-2019,Grissonnanche-2021}, graphene \cite{Sheehy-2007,Martin-2008,Crossno-2016,Gallagher-2019}, delafossites \cite{Moll-2016,Mackenzie-2017,Nandi-2018}, heavy fermions \cite{Lohneysen-1994,Trovarelli-2000,Schroder-2000,Si-2001,Stewart-2001,Paglione-2003,Tokiwa-2013,Grbic-2022} and WP$_2$ \cite{Gooth-2018,Jaoui-2018}. 
Concurrently, the same quantum fluctuations close to the QCP engender soft bosonic modes, which are capable of generating an effective attractive interaction between electrons and thus mediating Cooper-pair formation.  

If one investigates superconducting instabilities with the spectral function $A_{\rm QC}(\omega)$ and an instantaneous BCS-like pairing interaction $V_{\rm SC}(\omega)=-V_0 \Theta(\omega_0-\omega)$ \footnote{$\Theta(x)$ is the Heaviside step function}, which is negative and constant below a cutoff $\omega_0$, one concludes that a condensed phase would arise only above a threshold coupling to the bosonic mediator \cite{Balatsky-1993,Sudbo-1995,Muthukumar-1995,Yin-1996}, and not for all values of the coupling constant like in BCS theory. The inferred rarity of superconductivity would then be difficult to reconcile with its observed ubiquitousness in various classes of QC systems, such as heavy fermions \cite{Mathur-1998,Petrovic-2001,Sidorov-2002,Nakatsuji-2008,Knebel-2011} and pnictides \cite{Kasahara-2010,Bohmer-2014,Shibauchi-2014,Kuo-2016}. 
In reality, the self-consistently determined QC interaction, which takes into account the feedback of electrons on bosons, acquires the power-law form $V_{\rm SC}(\omega)=V_0 \left|\omega\right|^{1-4 \Delta}$ with the same exponent $\Delta$ as the spectral function: it is this singular behavior that amplifies the tendency to QC pairing, thus compensating for the weakened ability of NFL electrons to form Cooper pairs. Hence, the origin of NFL states and of Cooper pairing are intimately connected in such QC models, including our proposed one. In this sense, although the details of our specific model will be developed in the Sachdev-Ye-Kitaev (SYK) approach, ensuing superconducting properties like the gap equation will bear the same common structure of QC superconductors, i.e.\@, a ``generalized Cooper instability'' \cite{Bonesteel-1996,Son-1999,Abanov-2001a,Abanov-2001b,Chubukov-2005,She-2009,Moon-2010,Metlitski-2015,Roussev-2001,Raghu-2015}; hence, we expect our results to qualitatively hold for generic NFL, QC systems.

Within the recently developed theoretical advances in QC superconductivity, which allow for sign-problem-free quantum Monte Carlo simulations \cite{Berg-2012,Schattner-2016a,Schattner-2016b,Dumitrescu-2016,Lederer-2017,Li-2017,Wang-2017,Esterlis-2018b,Berg-2019}, an appealing formulation stems from the SYK picture. Initially formulated for Majorana fermions in 0+1 dimensions (0+1D) \cite{Maldacena-2016a,Maldacena-2016b,Kitaev2015a,Kitaev2015b,Bagrets-2017,Garcia-Garcia-2018,Chowdhury-2022}, and generalized to complex fermions to analyze spin glasses \cite{Sachdev-1993, Schmalian-2000,Georges-2000,Georges-2001,Westfahl-2003} and NFL normal phases \cite{Parcollet-1999,Song-2017,Gu-2020,Kim-2020,Esterlis-2021,Guo-2022}, the SYK paradigm was found to include superconducting ground states below a critical temperature $T_c$ \cite{Chowdhury-2020a,Salvati-2021,Lantagne-Hurtubise-2021,Patel-2018,Wang2020b,Wang-2021,Esterlis-2019,Hauck-2020,Classen-2022,Choi-2022,Li-2023}. 
In general, the SYK model describes $\mathscr{N}$ species (``flavors'') of fermions interacting through a random and infinitely ranged interaction \cite{Chowdhury-2022}. Such interaction might be fermionic (e.g.\@, a four-fermion term) or bosonic (e.g.\@, a Yukawa coupling to an Einstein boson) in nature. 
Averaging over the disordered configurations implied by randomness, one discovers a critical phase with a vanishing quasiparticle weight and a power-law spectral function at low temperatures and energies. The appeal of SYK formulations is manifold: they are exactly solvable in the $\mathscr{N}\rightarrow +\infty$ limit, in contrast to other approaches, yielding an artificially built but controlled example of QC strongly interacting electrons; they reproduce many experimentally observed aspects of NFL and strange-metal physics, like an extended regime where the resistivity is proportional to temperature; they are maximally chaotic, and therefore allow for exact studies of quantum chaos \cite{Patel-2017,Banerjee-2017,Kobrin-2021}; they also allow for an explicit gravity dual in an asymptotic anti–de Sitter (AdS) space AdS$_2$ \cite{Kitaev2015a,Kitaev2015b,Zaanen-2015holo,Maldacena-2016a,Maldacena-2016b,Kitaev-2018,Sarosi2017a,Nayak2018,Hartnoll-2018holo, Moitra2019,Sachdev2019,Inkof-2022,Schmalian2022}, thus contributing complementary insights into the ``holographic'' AdS/CFT correspondence between AdS gravity models and conformal field theories (CFTs) \cite{Maldacena-1999,Zaanen-2015holo,Hartnoll-2018holo,Franz-2018,Kruchkov-2020,Balm-2022_preprint}. 

The three fundamental assumptions of the SYK pictures may find physical grounding. The all-to-all  interactions resemble mean-field approaches that are local in space, i.e.\@, where the physical properties do not depend on spatial coordinates and fluctuations in space are neglected, like Dynamical Mean-Field Theory (DMFT) \cite{Georges-1996,Kotliar-2004}. 
Whether the SYK method discussed here can serve as a toy model for strongly interacting electrons is an open question. However, it is worthwhile pointing out a recent development in describing the phenomenology of the cuprates in terms of a theory with random electron-boson interactions \cite{Esterlis-2021,Guo-2022,Patel-2023_preprint}. Key results, like the linear-in-temperature resistivity, can obtained in a limit where the mean value of the coupling vanishes. Then the underlying many-body problem is almost identical to the one discussed in the present work. 
The large number $\mathscr{N}\rightarrow +\infty$ of flavors required to achieve an exact solution may be interpreted as an abundance of internal energy levels, or degrees of freedom, of a local quantum dot isolated from the environment. Fluctuations beyond the large-$\mathscr{N}$ limit can then included through numerical studies \cite{Fu-2016,Lunkin-2018,Sun-2018,Dai-2019,Wang-2020b,Kobrin-2021}, to check the validity of the large-$\mathscr{N}$ calculations. In this respect, another complementary technique is the AdS/CFT correspondence, which allows one to map a strongly interacting condensed-matter system to a weakly-interacting gravity theory; the holographic dual is another promising path to investigate fluctuations around the saddle-point solutions \cite{Davison-2017,Maldacena-2016a}, and non-equilibrium effects \cite{Grunwald-thesis-2022}. 
The randomness of the model may simulate real disorder, like in granular matter or nanoscopic flakes \cite{Chen-2018,Altland-2019a,Altland-2019b,Can-2019,Sahoo-2020}, or be understood as an effective description of a clean system with a rich spectrum of low-energy excitations (the flavors), over which we average to extract macroscopic properties. Notably, randomness can be self-generated in strongly interacting systems, such as frustrated magnets \cite{Schmalian-2000,Westfahl-2003} and in the spin-freezing region of multi-orbital Hubbard models \cite{Florens-2013,Werner-2008, Werner-2018,Tsuji-2019}: these speculations might provide a physical justification for effective SYK descriptions. 

To construct a minimal model of NFL superconductor in the SYK picture, we have first to complete a generalization, and secondly we have to assess whether the found low-temperature instability is really a superconducting ground state. 
The generalization involves: firstly, including a notion of space dimensions, which can be realized by employing SYK dots as building blocks and placing them on a lattice -- see Fig.\@ \ref{fig:Lattice3D}(b) for an illustrative depiction in 2D; secondly, investigating the emergence of a low-temperature electronic instability in such a lattice. 
The first step is technically implemented in a variety of ways through hopping parameters, which can be fixed, or random \cite{Beenakker-1997} as the SYK interactions: this protocol has been adopted for Majorana fermions with random hopping \cite{Salvati-2021} and four-body random interaction \cite{Gu-2017,Berkooz-2017,Bi-2017}, spinless complex fermions with random \cite{Song-2017,Can-2019} and non-random hopping \cite{Chowdhury-2018,Patel-2018,Zhang-2017,Haldar-2018,Garcia-Garcia-2021}, and spinful complex fermions with non-random hopping \cite{Chowdhury-2018,Cha-2020,Chowdhury-2020a}. 
Physically, non-random hopping parameters with translational invariance generate a band structure in reciprocal space of momenta $\hbar \vec{k}$, while random hopping preserves the local character of the SYK approach, as it leads to momentum-independent fermionic propagators after the disorder average. We will adopt the latter simpler assumption of randomness, which has the drawback of neglecting non-locality altogether, but it allows one to retrieve analytical results even in the condensed phase. 
Due to the lattice embedding, the complex creation and annihilation operators $ \hat{c}_{i,\sigma,\vec{x}}, \hat{c}_{i,\sigma,\vec{x}}^\dagger $ for fermions depend on a lattice site index $\vec{x}$ in addition to the flavor $i=\left\{1, \cdots \mathscr{N}\right\}$ and spin $\sigma=\left\{\uparrow, \downarrow \right\}$ indices. One realizes that, already in the normal state, the competition between on-site SYK interactions and two-body intersite hoppings determines crossovers between NFL phases (where interactions are dominant) and FL behavior (where fermions are itinerant). 

Secondly, to identify an instability towards a condensed phase, the development of anomalous terms of the kind $\mathscr{N}^{-1}\sum_i \hat{c}_{i,\sigma,\vec{x}}^\dagger \hat{c}_{i,\sigma',\vec{x}'}^\dagger$ -- see Sec.\@ \ref{Model} for formal definitions -- is regarded as a hallmark. This criterion on the disorder-averaged theory is in direct correspondence with the appearance of anomalous averages in non-disordered models, which constitute a superconducting order parameter. The associated gap equation allows one to determine the critical temperature $T_c$ where the anomalous term appears, as well as spectral and thermodynamic properties in the condensed phase. In the single-dot limit, a superconducting state of this kind has been retrieved for spinful fermions coupled by a negative Hubbard on-site term \cite{Li-2023}, two-body interactions \cite{Lantagne-Hurtubise-2021}, or through pair-hopping terms \cite{Wang-2020b}, and for fermions randomly coupled to an Einstein boson: the Yukawa-SYK model \cite{Esterlis-2019,Inkof-2022,Hauck-2020,Wang2020b,Wang-2020a,Classen-2022,Choi-2022}. We will adopt the latter route for on-site interactions, as developed in Sec.\@ \ref{Model}.
Recently, lattice calculations in the superconducting state also became available, for different pairing sources: collective chargeless excitations induced by the SYK lattice \cite{Salvati-2021}; additional correlations between the interaction matrix-elements of spinful fermions \cite{Chowdhury-2020a}; coupling of dots through two-body interactions that conserve charge \cite{Lantagne-Hurtubise-2021}; instantaneous attractive interaction between spinless fermions \cite{Patel-2018}; Hubbard on-site interaction and random hopping \cite{Li-2023}.

The embedding of SYK dots in a lattice is crucial to demonstrate that the found condensed phase is indeed superconducting. In fact, superconductivity is an electrodynamic phenomenon: a superconductor below $T_c$ is, first and foremost, a perfect diamagnet, which completely screens static magnetic fields from its bulk due to non-decaying circulating supercurrents, in accordance with the Mei{\ss}ner-Ochsenfeld effect \cite{Tinkham-1996int}. Thus, a notion of space must be included in the model to account for supercurrent circulation, and to calculate the associated response function in the presence of a magnetic field. In our case, such notion of space is introduced through intersite hopping. 
In turn, the response function is connected with the characteristic length scale within which the magnetic field is exponentially suppressed from the surface of the superconductor to its interior: this is the magnetic penetration depth $\lambda_L \propto 1/\sqrt{\rho_S}$ \cite{Tinkham-1959,Uemura-1991, Tinkham-1996int,Prozorov-2006}, which depends on the phase stiffness $\rho_S$ \cite{Lantagne-Hurtubise-2021} -- see Sec.\@ \ref{Stiffness} for technical details. Therefore, a true superconducting state involves a finite phase rigidity, or stiffness. One can interpret this phenomenon in the language of second-order phase transitions, as a spontaneous breaking of the global $U(1)$ symmetry with associated Goldstone bosons \cite{Altland-cm2010}.

The  question that we want to answer in this work is: does the condensed phase in a lattice of Yukawa-SYK dots possess a finite phase stiffness? Furthermore, in the light of the correlation between $\rho_S$, condensation energy, and relative weight of the superconducting coherence peak, observed in cuprates \cite{Shen-1997,Feng-2000}, do we retrieve a similar correlation in the Yukawa-SYK lattice model? 
The answer to both  questions is yes: in the following, we analyze the consequences of this analogy, comparing the spectral and thermodynamic properties of our superconductor. 
In synthesis, we generalize the Yukawa-SYK model to a lattice with random hopping parameters. Using the replica trick to perform the disorder average on the effective action, we solve the model exactly in the large-$\mathscr{N}$ limit at particle-hole symmetry, we construct the phase diagram, and we characterize the FL to NFL crossovers both numerically and analytically, in the normal and superconducting states.

The manuscript is organized as follows: Sec.\@ \ref{Summary} contains a summary of the main results of this paper. Our model is introduced in Sec.\@ \ref{Model}, where we derive the effective disorder-averaged action and the associated saddle-point equations, valid in the normal and superconducting states. The normal-state FL and NFL fixed points of our theory are identified in Sec.\@ \ref{Normal_fixed}, with reference to the fermionic and bosonic spectral functions obtained from the exact numerical solution of the saddle-point equations. Sec.\@ \ref{Normal_cross} hosts the analysis of the crossovers between the FL and NFL phases of our model, while Sec.\@ \ref{S_SYK} compares the derived crossover criteria with the normal-state entropy, as an exemplary application. The superconducting instability is first studied in Sec.\@ \ref{Tc_SYK}, where the critical temperature $T_c$ is identified by the self-consistent numerical solution of the linearized gap equation, and is compared with approximate analytical expressions in all distinct regimes. Sec.\@ \ref{Gap} discusses the full numerical solution and analytical approximations for the zero-temperature zero-energy gap, and the gap-to-$T_c$ ratio in the FL and NFL phases. The superconducting spectral functions of fermions and bosons are reported and analyzed in Sec.\@ \ref{Spectral_func_SC}, while Sec.\@ \ref{Z_SC} deals with the evolution of the dynamical quasiparticle weight with fermion-boson coupling in the superconducting state. Sec.\@ \ref{Stiffness} reports the derivation and the numerical results for the phase stiffness $\rho_S$ as a function of temperature and coupling, as well as asymptotic analytical formulae in the FL and strong-coupling regimes. The condensation energy on the lattice is computed in Sec.\@ \ref{Cond_en} and compared with the analogous evolution of the phase stiffness with coupling. 
Our conclusions, and perspectives for future developments of our theory, are summarized in Sec.\@ \ref{Disc}. Multiple appendices report technical details of our work: App.\@ \ref{App:dis_aver_action} reports the derivation of the disorder-averaged saddle-point action; App.\@ \ref{app:Numerics} describes the numerical methods employed to self-consistently solve the saddle-point equations on the imaginary and real axis; Apps.\@ \ref{Normal_prop} and \ref{SC_prop} contain the derivations of the approximate analytical results for various thermodynamic and spectral quantities, analyzed in the normal and superconducting states respectively; a separate App.\@ \ref{App:Tc} focuses on analytical approximations for the critical temperature; App.\@ \ref{App:Omega} shows the derivation of the thermodynamic grand potential in the normal and superconducting states. Finally, App.\@ \ref{Action_for_charge_fluctuations} contains the derivation of the action for charge fluctuations and of the electrodynamic kernel, used to calculate the phase stiffness. 

\begin{figure}[t]
\includegraphics[width=0.9\columnwidth]{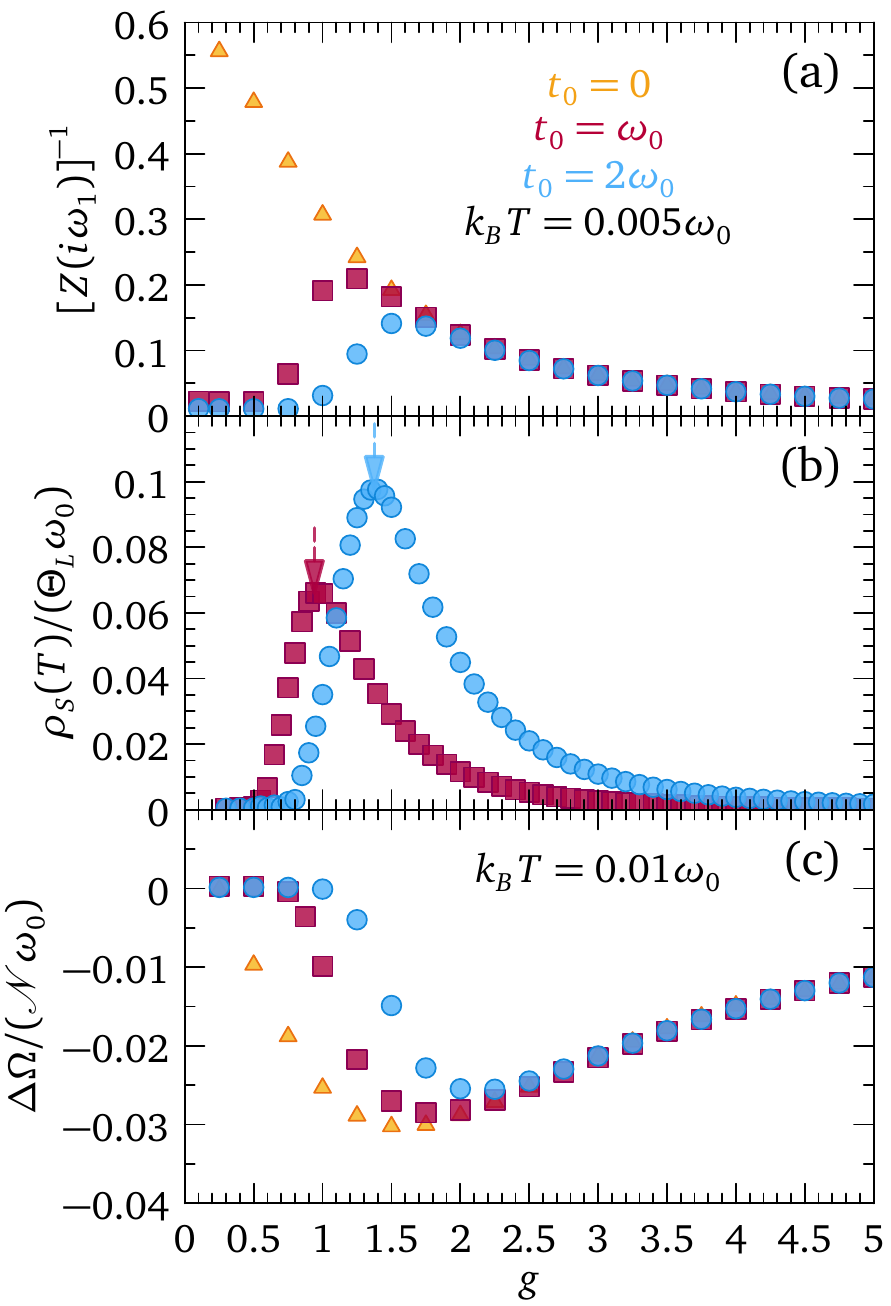}
\caption{\label{rhoS_Z_SC_DeltaOmega} Correlation between quasiparticle residue, phase stiffness and condensation energy as a function of fermion-boson coupling $g$, found in the Yukawa-SYK model on a lattice, i.e.\@, our model described by the Hamiltonian (\ref{eq:H_SYK_phon_hop}). All curves are calculated deep in the superconducting state, for different hopping parameters $t_0$. (a) Inverse of the dynamical weight $Z(i \omega_1)$ at the first Matsubara frequency, calculated on the imaginary axis; see Sec.\@ \ref{Z_SC}. (b) Energy scale $\rho_S(T)/\Theta_L$ corresponding to the phase stiffness $\rho_S(T)$; see Sec.\@ \ref{Stiffness}. (c) Condensation energy computed from the difference between the grand potentials in the normal and superconducting states; see Sec.\@ \ref{Cond_en}. }
\end{figure}

\section{Summary of main results}\label{Summary}

Our main result is that, in a lattice of coupled  Yukawa-SYK sites described by the Hamiltonian (\ref{eq:H_SYK_phon_hop}), that serves as a solvable model for superconductivity in a regime without quasiparticles,  interesting phenomenological correlations among the quasiparticle residue in the superconducting state, the phase stiffness, and the condensation energy emerge. In our model we vary the fermion-boson coupling $g$ and the hopping parameter $t_0$ to establish these correlations. They are, at least qualitatively, analogous to experimental observations in cuprate superconductors \cite{Shen-1997,Feng-2000}, where the tuning parameter was the carrier concentration. These findings are summarized in Fig.\@ \ref{rhoS_Z_SC_DeltaOmega}: the quasiparticle residue is obtained from the inverse of the dynamical weight $Z(i \omega_1)$ at the first Matsubara frequency $\omega_1$ -- see Sec.\@ \ref{Z_SC} and App.\@ \ref{App:Z}, and is shown in Fig.\@ \ref{rhoS_Z_SC_DeltaOmega}(a); the phase stiffness $\rho_S(T)$, converted into an energy scale as $\rho_S(T)/\Theta_L$, where $\Theta_L$ is a constant-- see Eq.\@ (\ref{eq:Theta_L}) in Sec.\@ \ref{Stiffness} -- that encodes the underlying lattice structure of the theory \cite{Emery-1994} and the number of fermion flavors, is displayed in Fig.\@ \ref{rhoS_Z_SC_DeltaOmega}(b); finally, Fig.\@ \ref{rhoS_Z_SC_DeltaOmega}(c) shows the condensation energy $\Delta \Omega/(\mathscr{N}  \omega_0)$ per fermion flavor, in units of the bare boson frequency $\omega_0$, evaluated from the difference between the thermodynamic grand potentials in the normal and superconducting states; see Sec.\@ \ref{Cond_en}. All curves are computed from the exact numerical solution of the saddle-point equations (\ref{eq:Eliashberg_eph:hop}) in the superconducting state, in the large-$\mathscr{N}$ limit, corresponding to the Hamiltonian (\ref{eq:H_SYK_phon_hop}).

The stiffness is maximal precisely at the FL/NFL crossover, which is marked by vertical arrows in Fig.\@ \ref{rhoS_Z_SC_DeltaOmega}(b) for a given hopping. 
Our analytical approximations in the weak-coupling regime, which corresponds to a disordered FL in the normal state \cite{Beenakker-1997}, indicate that $Z^{-1}(i \omega_1)$ and $\rho_S(T)$ are exponentially increasing functions of $g$, because in this regime the superconductor follows a disordered version of BCS theory, where the zero-temperature superconducting gap $\Delta_0 \propto e^{-1/\bar{\lambda}}$ with coupling constant $\bar{\lambda} \propto g^2/t_0$ -- see Eq.\@ (\ref{eq:lambda_bar}); at finite hopping, the same qualitative behavior is retrieved for the condensation energy $\Delta \Omega$. In the opposite, strong coupling limit, where fermions are fully incoherent and the normal state is a NFL, we show that the residue $Z^{-1}(i \omega_1) \propto g^{-2}$ and the stiffness $\rho_S(T) \propto Z^{-2}(i \omega_1) \propto g^{-4}$; correspondingly, $\Delta \Omega$ slowly decreases in magnitude \cite{Esterlis-2019}, mirroring the evolution of the previously analyzed quantities. 

The non-monotonic evolution of $\rho_S(T)$ with $g$ is contrasted by the monotonic one of the critical temperature $T_c$, which is computed from the linearized gap equation (\ref{eq:Phi_Z_tilde}) and analyzed in Sec.\@ \ref{Tc_SYK}: in the weak-coupling regime, the disordered FL pairs in accordance with the BCS formula $T_c \propto e^{-1/\bar{\lambda}}$ -- see Eq.\@ (\ref{eq:BCS_Tc_SYK2}) -- and thus  $\lim_{T\rightarrow 0}\rho_S(T)/(k_B T_c)\propto \Delta_0/(k_B T_c)\approx 0.567$, i.e.\@, the stiffness and the pairing temperature are proportional to each other; conversely, in the strong-coupling NFL regime, we have $\lim_{g \rightarrow +\infty}T_c\approx 0.112 \omega_0/k_B$, and so the characteristic energy scale $\rho_S(T)/\Theta_L \propto g^{-4}$ of the stiffness per fermion flavor is much smaller than $k_B T_c$: this effect is indicative of strong superconducting phase fluctuations.

The relation between the stiffness and the finite-frequency electromagnetic response of the Yukawa-SYK superconductor is further investigated in the companion paper Ref.\@ \onlinecite{short-paper}, where it is shown that the non-monotonicity of $\rho_S(T)$ with fermion-boson coupling is reflected into a different spectral weight removal from the low-energy optical conductivity, upon entering the superconducting state: this phenomenon offers a direct way to observe the NFL/FL crossovers in the condensed phase of the Yukawa-SYK model on a lattice.  
Moreover, the FL/NFL crossovers manifest themselves in qualitative differences between the spectral functions in the FL and NFL phases -- see Secs.\@ \ref{Normal_cross} and \ref{Spectral_func_SC}: in the normal state, the power-law fermionic spectral function and the soft boson excitations of the NFL regime leave the stage to the semicircular Wigner spectral function of the disordered FL \cite{Beenakker-1997}, with almost free bosons; in the superconducting state, the BCS-like gap of the FL regime transforms into a spectral function with multiple peak-dip-hump features in the strong-coupling regime, which are self-trapped states created by the pairing field \cite{Combescot-1994, Esterlis-2019}. Finally, the NFL/FL crossovers can be identified by the zero-temperature entropy in the large-$\mathscr{N}$ limit and in the normal state, which is finite in the NFL phase but is vanishing in the FL regime -- see Sec.\@ \ref{S_SYK}. 
The entire phase diagram of the model (\ref{eq:H_SYK_phon_hop}) is schematically drawn in Fig.\@ \ref{fig:phase_scheme}.  

\section{Model}\label{Model}

We consider the following model of electrons with all-to-all interactions, coupled to phonons: 
    \begin{align}\label{eq:H_SYK_phon_hop}
    \hat{H}&=-\sum_{i=1}^{\mathscr{N}} \sum_{\vec{x}} \sum_{\sigma} \mu \hat{c}_{i,\vec{x},\sigma}^\dagger \hat{c}_{i,\vec{x},\sigma} \notag \\ & +\frac{1}{2} \sum_{k=1}^{\mathscr{M}} \sum_{\vec{x}}\left[ \pi_{k \vec{x}}^2+\omega_0^2 \phi_{k \vec{x}}^2\right]
    \notag
    \\ &+ \sum_{\left\{i,j\right\}=1}^{\mathscr{N}}\sum_{\sigma}\sum_{k=1}^{\mathscr{M}} \sum_{\vec{x}} (g_{ij,k}+g_{ji,k}^{\star}) \hat{c}_{i,\vec{x},\sigma}^\dagger \hat{c}_{j,\vec{x},\sigma}\phi_{k \vec{x}}
    \notag\\ 
    &+ \sum_{\left\langle \vec{x},\vec{x}' \right\rangle} \sum_{\left\{i,j\right\}=1}^{\mathscr{N}} \sum_{\sigma,\sigma'} t_{ij, \vec{x} \vec{x}'} \hat{c}_{j,\vec{x}',\sigma}^\dagger \hat{c}_{i,\vec{x},\sigma},
    \end{align}
with $\hat{c}_{i,\vec{x},\sigma}^\dagger,  \hat{c}_{i,\vec{x},\sigma}$ fermionic operators on a lattice with site indexes $\vec{x}$ for spin $\sigma=\left\{\uparrow, \downarrow\right\}$, obeying usual anticommutation relations, and scalar bosonic degrees of freedom $\phi_{k \vec{x}}$ with canonical momentum $\pi_{k \vec{x}}$, such that $\phi_{k \vec{x}} \pi_{k'\vec{x}'}-\pi_{k' \vec{x}'}\phi_{k \vec{x}}=i \delta_{k k'}\delta_{\vec{x}\vec{x}'}$. The indices $\left\{i,j\right\}=\left\{1,\cdots \mathscr{N}\right\}$ run through the fermionic flavors, while the index $k=\left\{1,\cdots \mathscr{M}\right\}$ refers to the bosonic species. The all-to-all couplings $g_{ij,k}$ are assumed to be random real numbers obeying a Gaussian distribution with null average and variance $\bar{g}^2/(2 \mathscr{N}^2)$. Likewise, by assumption the hopping parameters $t_{ij, \vec{x} \vec{x}'}$ follow a Gaussian distribution with zero average and variance ${t_0^2}/(2 \mathscr{N})$, and they act only between nearest-neighbour sites $\left\langle \vec{x},\vec{x}' \right\rangle$; see also App.\@ \ref{App:dis_aver_action}. 
Formally, the theory (\ref{eq:H_SYK_phon_hop}) is a generalization of the electron-phonon model of SYK superconductivity \cite{Esterlis-2019} to a lattice of SYK fermions. We employ the replica trick to perform averages over the ``disorder'' determined by the randomness of the couplings $g_{ij,k}$ and of the hoppings $t_{ij, \vec{x} \vec{x}'}$, as described in App.\@ \ref{App:dis_aver_action}. This way, we deduce that the model \eqref{eq:H_SYK_phon_hop} is controlled by the following disorder-averaged effective action $\mathscr{S}=\sum_{\vec{x}} \mathscr{S}_{\vec{x}}$, where
\begin{widetext}
\begin{align}
		\frac{\mathscr{S}_{\vec{x}}}{{\mathscr{N}}}
		& = -  {\rm Tr}\log\left(\hat{G}_{0}^{-1}-\hat{\Sigma}_{\vec{x}}\right)+\frac{1}{2}{\rm Tr}\log\left(D_{0}^{-1}-\Pi_{\vec{x}}\right) -  2\int_{\tau\tau'} G_{\vec{x}}(\tau',\tau)\Sigma_{\vec{x}}(\tau,\tau')+\frac{1}{2}\int_{\tau\tau'} D_{\vec{x}}(\tau',\tau)\Pi_{\vec{x}}(\tau,\tau')
		\nonumber \\
		& -  \int_{\tau\tau'}\left[F_{\vec{x}}(\tau'\tau)\Phi_{\vec{x}}^{\dagger}(\tau,\tau')+F_{\vec{x}}^{\dagger}(\tau'\tau)\Phi_{\vec{x}}(\tau,\tau')\right]
		 + \frac{z t_{0}^{2}}{2}\int_{\tau\tau'}\left[G_{\vec{x}}(\tau,\tau')G_{\vec{x}'}(\tau',\tau)-F_{\vec{x}}^{\dagger}(\tau,\tau')F_{\vec{x}'}(\tau',\tau)\right]
		\nonumber \\
		& +  \bar{g}^{2}\int_{\tau\tau'}D_{\vec{x}}(\tau,\tau')\left[ G_{\vec{x}}(\tau,\tau')G_{\vec{x}}(\tau',\tau)-F_{\vec{x}}^{\dagger}(\tau,\tau')F_{\vec{x}}(\tau',\tau)\right] .
	\label{eq:effective_action}
\end{align}
\end{widetext}
Here $\int_{\tau\tau'}\equiv\int d\tau\int d\tau'$, $z$ is the coordination number (the number of nearest-neighbour sites) and we have introduced the cumulative bilocal fields
\begin{subequations}\label{eq:bilocal_fields_def}
\begin{equation}
		G_{\vec{x}}(\tau,\tau')  = \frac{1}{\mathscr{N}}\sum_{i=1}^\mathscr{N}\hat{c}_{i\vec{x}\sigma}^{\dagger}(\tau')\hat{c}_{i\vec{x}\sigma }(\tau),
\end{equation}
\begin{equation}
		F_{\vec{x}}(\tau,\tau') = \frac{1}{\mathscr{N}}\sum_{i=1}^\mathscr{N}\hat{c}_{i\vec{x}\downarrow}(\tau')\hat{c}_{i\vec{x}\uparrow }(\tau),
\end{equation}
\begin{equation}
		D_{\vec{x}}(\tau,\tau')  = \frac{1}{\mathscr{M}}\sum_{k=1}^\mathscr{M}\phi_{k\vec{x}}(\tau')\phi_{k\vec{x}}(\tau),
	\end{equation}
	\end{subequations}
where the normal fermionic propagator $G_{\vec{x}}(\tau,\tau')$ is diagonal in the spin index $\sigma$. The normal and anomalous fermionic self-energies, as well as the bosonic self-energy, are respectively given by $\Sigma_{\vec{x}}(\tau,\tau')$, $\Phi_{\vec{x}}(\tau,\tau')$ and $\Pi_{\vec{x}}(\tau,\tau')$. They are defined as the Lagrange multipliers for the identities which introduce the bilocal fields (\ref{eq:bilocal_fields_def}); see App.\@ \ref{App:dis_aver_action}.

In the following, we set $\mathscr{M}=\mathscr{N}$ for simplicity. Generalizations to a different number of fermion and boson flavors were reported for the single-dot limit \cite{Esterlis-2019,Wang-2020a, Wang-2021,Classen-2022}, and their lattice counterpart for the model (\ref{eq:H_SYK_phon_hop}) is left for future work. Notice that, after performing the disorder averages in the replica-symmetric ansatz, the bilocal fields are expected to acquire translational invariance in time and space, i.e.\@, the disorder realization is the same for any lattice site $\vec{x}$. For this reason, in the following we drop the index $\vec{x}$ when referring to the disorder-averaged theory. In the limit of large ${\mathscr{N}}$, the saddle-point equations coming from the action Eq.\@ \eqref{eq:effective_action} are given by
\begin{widetext}
\begin{subequations}\label{eq:Eliashberg_eph:hop}
        \begin{align}
        &\Sigma(i \omega_n)=\bar{g}^2 k_B T \sum_{m=-\infty}^{+\infty} D(i\Omega_m) G(i \omega_n-i\Omega_m) {+\frac{z {t_0^2}}{2} G(i \omega_n)}, \label{eq:Sigma}
        \\
        &
        D(i\Omega_n)=\frac{1}{\Omega_n^2+\omega_0^2-\Pi(i \Omega_n)},
        \label{eq:Phon}
        \\
        &
        \Pi(i \Omega_n)=-2 \bar{g}^2 k_B T \sum_{m=-\infty}^{+\infty}\left[ G(i \omega_m) G(i \omega_m+i \Omega_n)-F(i \omega_m) F(i \omega_m+i \Omega_n)\right],
        \label{eq:Polar}
        \\
        &\Phi(i \omega_n)=-\bar{g}^2 k_B T \sum_{m=-\infty}^{+\infty} F(i \omega_m) D(i \omega_n-i \omega_m){-\frac{z {t_0^2}}{2} F(i \omega_n)},
        \label{eq:Phi}
        \\
        &
        G(i \omega_n)=\frac{i \omega_n-\mu+\Sigma^{*}(i \omega_n)}{\left[i \omega_n+\mu-\Sigma(i\omega_n)\right]\left[i \omega_n-\mu+\Sigma^{*}(i\omega_n)\right]-\left|\Phi(i \omega_n)\right|^2},\label{eq:G2}
        \\
        & F(i \omega_n)=\frac{\Phi(i \omega_n)}{\left[i \omega_n+\mu-\Sigma(i\omega_n)\right]\left[i \omega_n-\mu+\Sigma^{*}(i\omega_n)\right]-\left|\Phi(i \omega_n)\right|^2}.
        \label{eq:F2}
        \end{align}
\label{saddle_point_equations}
\end{subequations}
\end{widetext}
Eqs.\@ (\ref{eq:G2}) and (\ref{eq:F2}) have the standard form of the normal and anomalous fermion propagators found in Eliashberg theory \cite{Berthod-2018,Marsiglio-2008,Marsiglio-2020}, here written for dispersionless fermions with chemical potential $\mu$. 
A crucial aspect of the saddle-point theory (\ref{eq:Eliashberg_eph:hop}) is that the boson themselves acquire the self-energy (\ref{eq:Polar}), due to their interaction with fermions. This is essential for the non-Fermi liquid physics entailed by our model: the fermion-boson interaction is at once responsible for the destruction of fermionic quasiparticles and the softening of the bosonic mode with bare frequency $\omega_0$. It is also responsible for the low-temperature instability towards a superconducting ground state, to be explored in the following. 
As a result of fermion-boson coupling, the natural frequency $\omega_0$ of the boson oscillator is renormalized according to
\begin{equation}\label{eq:omegar_gen}
\omega_r^2=\omega_0-\Pi(0),
\end{equation}
where $\Pi(0)$ is the static value of the boson self-energy (\ref{eq:Polar}).

The system (\ref{eq:Eliashberg_eph:hop}) realizes a self-consistent problem for the SYK fermions on a lattice, with random hoppings -- characterized by variance ${t_0^2}/(2 \mathscr{N})$ and coordination number $z$ -- and coupling to an Einstein phonon mode of natural frequency $\omega_0$, dressed by the fermion polarization bubble according to Eq.\@ (\ref{eq:Polar}). The coupling constant $\bar{g}$ has dimensions $\left[\mathrm{energy}\right]^{3/2}$ \cite{Esterlis-2019}. 
The equations are simplified by introducing 
\begin{equation}\label{eq:Sigma_Z}
\Sigma(i \omega_n)=i \omega_n \left[1-Z(i \omega_n)\right]
\end{equation} 
and 
\begin{equation}\label{eq:Phi_Delta}
\Phi(i\omega_n)= Z(i \omega_n) \Delta(i \omega_n),
\end{equation}
as in standard Eliashberg theory. 
Our main goals are to calculate the superconducting critical temperature $T_c$, thermodynamic properties such as the grand potential in the normal and superconducting states, and the phase stiffness of the system (\ref{eq:Eliashberg_eph:hop}). In the following, we will focus on the particle-hole symmetric case $\mu=0$. The main details of our numerical methods are reported in App.\@ \ref{app:Numerics}.

\section{Fixed points in the normal state}\label{Normal_fixed}

Let us first solve the coupled saddle-point equations (\ref{eq:Eliashberg_eph:hop}) in the normal state, that is, assuming that the anomalous propagator $F(i \omega_n)$ and the anomalous self-energy $\Phi(i \omega_n)$ both vanish. For couplings $g_{ij,k}$ taken from the Gaussian orthogonal ensemble (GOE), i.e.\@, real-valued and preserving time reversal symmetry, the solution below is valid in the normally conducting state at $T>T_c$; if the couplings are extracted from the Gaussian unitary ensemble (GUE), they are complex-valued and break time-reversal symmetry, so that there is no superconducting phase and the following represents the full solution of the model for any temperature \cite{Esterlis-2019,Hauck-2020}. 

The normal-state coupled equations on the imaginary axis are
\begin{subequations}\label{eq:saddle_point_NS}
    \begin{align}
    \Sigma(i \omega_n) 
    &=\bar{g}^2 k_B T \sum_{m=-\infty}^{+\infty} D(i\Omega_m) \frac{1}{i(\omega_n -\Omega_m) Z(i \omega_n-i \Omega_m)} \label{eq:Sigma_Z_hop}
    \notag\\\,\,&{+\frac{z {t_0^2}}{2} G(i \omega_n)},
    \\
    \Pi(i \Omega_n)&=-2 \bar{g}^2 k_B T \sum_{m=-\infty}^{+\infty} \frac{1}{i \omega_m Z(i \omega_m)} \notag\\ 
    &\qquad\qquad\times \frac{1}{i (\omega_m +\Omega_n) Z(i \omega_m+i \Omega_n)},
    \label{eq:Polar_Z}
    \end{align}
\end{subequations}
together with the Dyson equations 
    \begin{equation}\label{eq:G_NS}
    G^{-1}(i \omega_n)={i \omega_n-\Sigma(i\omega_n)}
    \end{equation}
and (\ref{eq:Phon}) for the fermion and boson propagators. The real-axis versions of Eqs.\@ (\ref{eq:saddle_point_NS}) formally stem from the analytic continuation $i \omega_n \rightarrow \omega+i 0^+$ of the Matsubara propagators, and they can be numerically solved by resorting to the spectral (Lehmann) representation; see App.\@ \ref{App:num_real_axis}. This representation allows to study the spectroscopic properties of the system, such as the fermionic spectral function
\begin{equation}\label{eq:A_def}
A(\omega)=-\frac{1}{\pi} \mathrm{Im}\left\{G^R(\omega)\right\},
\end{equation}
where $G^R(\omega)=\left. G(i \omega_n)\right|_{i \omega_n \rightarrow \omega+i 0^+}$ is the retared fermionic propagator \footnote{Furthermore, since the fermions are dispersionless, $A(\omega)$ from Eq.\@ (\ref{eq:A_def}) also counts how many states are available at the energy $\omega$, i.e.\@, it is equivalent to the single-particle density of states of the system}. 

In what follows we will compare the full numerical solution of Eqs.\@ (\ref{eq:saddle_point_NS}) on both the imaginary and real axis, to approximate analytical expressions that are valid in specific regimes of the model. 

\begin{figure}[t]
\includegraphics[width=0.95\columnwidth]{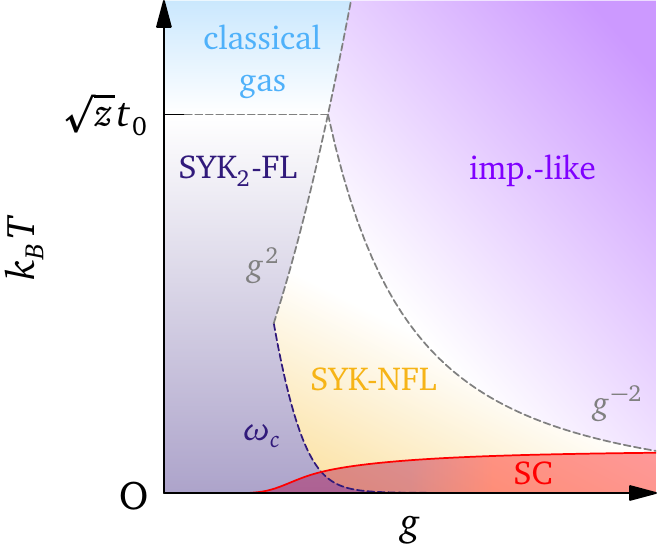}
\caption{\label{fig:phase_scheme} Large-$\mathscr{N}$ schematic phase diagram of the Yukawa-SYK model on a lattice, represented by the Hamiltonian (\ref{eq:H_SYK_phon_hop}), as a function of dimensionless fermion-boson coupling constant $g$ and at fixed hopping $t_0$. The classical-gas phase (light-blue shaded region) develops at small coupling, at temperatures $k_B T >\sqrt{z}t_0$, where $z$ is the coordination number. At lower temperatures, the system forms a disordered Fermi liquid, the SYK$_2$-FL phase dominated by hopping (dark-blue shaded region); the low-temperature boundary of this phase is determined by the crossover energy $\omega_c$, as calculated in Sec.\@ \ref{Normal_cross}. At the lowest temperatures and right to the crossover scale $\omega_c$, the system is a non-Fermi liquid labeled SYK-NFL (orange-shaded region). At strong coupling, for $k_B T \gtrapprox g^{-2}$, the impurity-like regime (purple-shaded region) appears. A low-temperature superconducting phase (red-shaded region) is delimited by $T_c$ (red curve), which is exponentially suppressed with decreasing coupling in the SYK$_2$-FL regime, while it reaches asymptotically $k_B T_c \approx 0.112 \omega_0$ in the impurity-like regime, where $\omega_0$ is the bare boson frequency \cite{Esterlis-2019}. At higher values of hopping, the FL/NFL crossover energy $\omega_c$ occurs at higher $g$, such that a direct crossover between the SYK$_2$-FL and impurity-like regimes occurs; see Fig.\@ 1(c) of Ref.\@ \onlinecite{short-paper}. The normal-state crossovers are analyzed in Sec.\@ \ref{Normal_fixed}.}
\end{figure}

There is a total of four fixed points for Eqs.\@ (\ref{eq:saddle_point_NS}): one is the trivial classical-gas noninteracting regime, other two give rise to non-Fermi liquids, and the remaining one represents a disordered Fermi liquid; the latter is the only new fixed point that arises due to nonzero hopping, while the other regimes are quantitatively affected by the lattice embedding of the SYK dots, but are qualitatively similar to their already studied counterparts in the single-dot limit \cite{Esterlis-2019}. The four fixed points are schematically depicted in Fig.\@ \ref{fig:phase_scheme}, together with the low-temperature superconducting phase. 

The classical-gas regime occurs at high temperatures with respect to coupling and hopping, that is, for $k_B T \gg \sqrt{z} t_0 \gg \omega_0 g$: it is characterized by approximately noninteracting fermions and bosons with propagators $G(i \omega_n) \approx 1/(i \omega_n)$ and $D(i \Omega_n) \approx 1/(\Omega_n^2 +\omega_0^2)$, respectively. At lower temperatures, such unstable free fermions flow towards the other fixed points of the model, the physics of which we now describe. Quantitative details on our analysis of the imaginary-axis equations of our model can be found in Appendix \ref{Normal_prop}. 

\subsection{Quantum critical non-Fermi liquid fixed point: NFL-SYK}\label{SYK-NFL_fixed_point}

When the fermion-boson coupling is dominant with respect to hopping, such that $g^2 \omega_0 \gg z t_0^2$, but is sufficiently small so that $g^2 < k_B T< g^{-2}$, the system enters the non-Fermi liquid quantum critical phase, which we label NFL-SYK. In these conditions, a nonzero $t_0$ only quantitatively affects the fermionic and bosonic properties, but the physics is qualitatively analogous to the single-dot limit. Thus, the properties of the NFL-SYK phase are analogous to the ones studied in Sec.\@ IIIa of Ref.\@ \onlinecite{Esterlis-2019}. For a self-contained explanation, let us recall such properties. 

\begin{figure*}[t]
\includegraphics[width=0.95\textwidth]{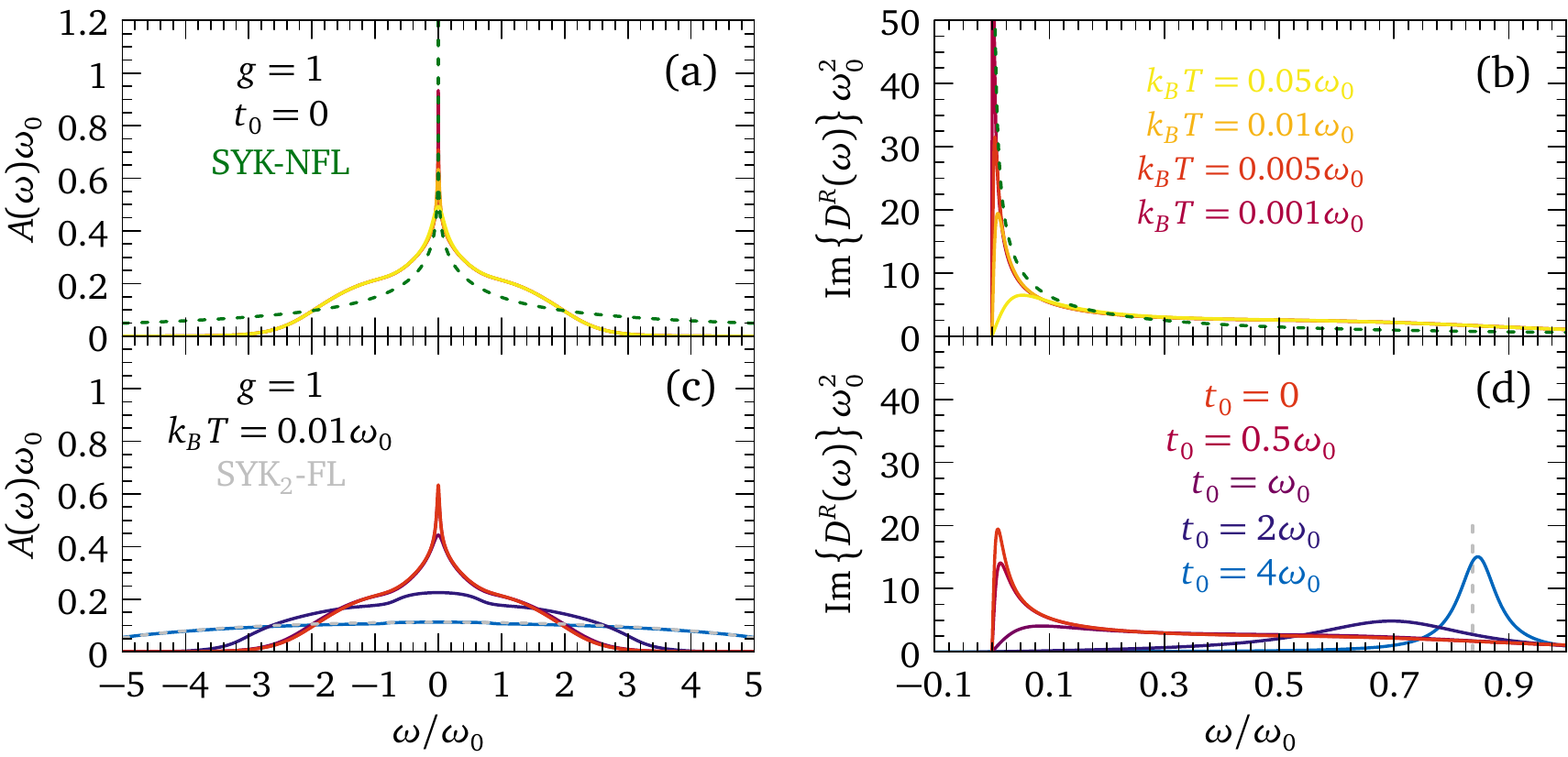}
\caption{\label{fig:spectr_NFL-SYK_g} Numerically exact spectral function and bosonic propagator in the normal state, showing the crossover from the NFL-SYK to the SYK$_2$-FL regimes as a function of hopping. (a) Fermionic spectral function in the single-dot limit ($t_0=0$) and for coupling $g=1$, as a function of energy $\omega$ and at temperature $k_B T/\omega_0=\left\{0.001,0.005,0.01,0.05\right\}$. The dashed green curve is the analytical low-energy SYK-NFL spectral function, given by Eq.\@ (\ref{eq:G_NFL_SYK}) for $g=1$. (b) Imaginary part of the bosonic propagator, for the same parameters as in panel (a). The dashed green curve corresponds to the low-energy analytical boson propagator (\ref{eq:D_NFL_SYK}) in SYK-NFL regime for $g=1$. (c) Fermionic spectral function at fixed temperature $k_B T/\omega_0=0.01$ and coupling $g=1$, as a function of energy $\omega$ and for different hoppings $t_0/\omega_0=\left\{0,0.5,1,2,4\right\}$. The dashed gray curve corresponds to the SYK$_2$-FL spectral function (\ref{eq:A_SYK2_analyt}), for $t_0/\omega_0=4$. (d) Imaginary part of the bosonic propagator, for the same parameters as in panel (c). The dashed vertical line is given by the approximate renormalized boson frequency (\ref{eq:omegar_0_T0}) in SYK$_2$-FL regime. }
\end{figure*}
The key findings of the NFL-SYK state are that the low-energy fermion and boson propagators follow a power-law (``branch cut'' \cite{Zaanen-2021_preprint}) form in frequency, characteristic of quantum critical systems: 
    \begin{equation}\label{eq:G_NFL_SYK}
    G^{-1}(i \omega_n)={i \omega_n \left(1+c_1 \left|\frac{g^2}{\omega_n}\right|^{2 \Delta}\right)},
    \end{equation}
    \begin{equation}\label{eq:D_NFL_SYK}
    D^{-1}(i \Omega_n)={\Omega_n^2 +\omega_r^2+c_3 \left|\frac{\Omega_n}{g^2}\right|^{4 \Delta-1}},
    \end{equation}
with the renormalized boson frequency
\begin{equation}\label{eq:omegar_NFL_SYK}
\omega_r^2=c_2 \left(\frac{k_B T}{g^2}\right)^{4 \Delta-1}. 
\end{equation}
The numerical coefficients in Eqs.\@ (\ref{eq:G_NFL_SYK})-(\ref{eq:omegar_NFL_SYK}) are $c_1\approx 1.154 700$, $c_2 \approx 0.561228$, and $c_3 \approx 0.709618$, while the exponent is $1/4 < \Delta < 1/2$ depending on the ratio $\mathscr{N}/\mathscr{M}$ between the number of fermion and boson flavors. For $\mathscr{N}=\mathscr{M}$, we have $\Delta \approx 0.420374134464041$ \cite{Esterlis-2019}. 
Through the Dyson equation (\ref{eq:G_NS}), the self-energy that corresponds to the Green's function (\ref{eq:G_NFL_SYK}) assumes the low-energy form \cite{Esterlis-2019}
\begin{equation}\label{eq:Sigma_NFL_SYK}
\Sigma(i \omega_n)=-i \mathrm{sign}(\omega_n) c_2 g^{4 \Delta} \left|\omega_n\right|^{1-2 \Delta},
\end{equation}
which differs from purely fermionic versions of SYK models only in the value of the exponent $\Delta$ \cite{Chowdhury-2022}. 

The propagators (\ref{eq:G_NFL_SYK}) and (\ref{eq:D_NFL_SYK}) imply that fermion-boson interaction destroys both fermionic and bosonic quasiparticles on the same footing. On one hand, the fermions acquire an SYK-like imaginary self-energy (\ref{eq:Sigma_NFL_SYK}) with anomalous exponent $\Delta$, while the boson dynamics is dominated by an anomalous Landau damping governed by the same exponent $\Delta$. 
That the system is quantum critical can be deduced from the renormalized frequency (\ref{eq:omegar_NFL_SYK}), which vanishes at $T\rightarrow 0$ for all values of $g$ and $\omega_0$. This is because the degeneracy stemming from the diverging charge susceptibility of bare fermions is lifted in the SYK-NFL state, thus providing the stability of the latter with respect to charge fluctuations. 

Fig.\@ \ref{fig:spectr_NFL-SYK_g}(a) shows the NFL-SYK spectral function $A(\omega)$ from Eq.\@ (\ref{eq:A_def}) at different temperatures $k_B T/\omega_0$, obtained from the exact numerical solution of Eqs.\@ (\ref{eq:saddle_point_NS}) on the real axis. We choose $g=1$ and $t_0=0$ (single-dot limit). 
The dashed green curve is given by the zero-temperature limit of Eq.\@ (\ref{eq:G_NFL_SYK}), where the Matsubara frequencies become the continuous variable $i \omega$. We clearly appreciate the power-law low-energy feature that develops at low temperature, a signature of the quantum critical phase. At the same time, the bosons are strongly renormalized with a progressive softening of their frequency $\omega_r$, as shown in Fig.\@ \ref{fig:spectr_NFL-SYK_g}(b) which displays the imaginary part of the retarded boson propagator $D^R(\omega)$ at different temperatures. The dashed green curve is given by Eq.\@ (\ref{eq:D_NFL_SYK}) at $T=0$. The softening is accompanied by the development of the power-law divergence (cut by temperature) signaling the critical dynamics of bosons. 

By increasing hopping such that the terms depending on $z t_0^2$ in Eqs.\@ (\ref{eq:saddle_point_NS}) are no longer negligible, the fermionic and bosonic dynamics progressively transform. Then, what is the new phase to which the system flows to? To answer this question, we need to study the large-hopping limit of Eqs.\@ (\ref{eq:saddle_point_NS}), as presented in the next section. 

\subsection{Disordered Fermi-liquid fixed point: SYK$_2$-FL}\label{SYK2-FL_fixed_point}

We now increase the hopping such that $g^2 \omega_0 \ll z t_0^2$ keeping the temperature in the window $g^2 < k_B T< g^{-2}$. Hence, the physics will no longer be dominated by intra-dot fermion-boson coupling, but by coherent hopping between nearest-neighbors sites on the lattice. The single-dot quantum critical dynamics of Sec.\@ \ref{SYK-NFL_fixed_point} then crosses over to another fixed point, which we baptize SYK$_2$-FL. The label ``SYK$_2$'' refers to the fact that the physics is dominated by the two-body term given by $t_0$ in the Hamiltonian (\ref{eq:H_SYK_phon_hop}), similarly to the SYK models SYK$_q$ with $q$-fermion interactions \cite{Davison-2017}, while ``FL'' indicates that it is a kind of Fermi liquid endowed with fermionic quasiparticles \cite{Chowdhury-2022}. While strictly speaking the hopping term is random \cite{Anderson-2010int}, such that the system is disordered, the many-body density of states given by the hopping-dependent terms in Eqs.\@ (\ref{eq:saddle_point_NS}) has a polynomial number of energy levels lying at low energy, which allows mapping the problem to a Fermi-liquid picture \cite{Chowdhury-2022}. Other distinctive features that such disordered phase shares with standard Fermi liquids include a DC resistivity $\rho_{DC} \propto T^2$ and a heat capacity $C_V \propto T$ \cite{Song-2017, Chowdhury-2022}. 
This purely fermionic limit of SYK$_2$ has been extensively studied in the condensed-matter and holographic contexts \cite[see, e.g.\@,][]{Beenakker-1997,Maldacena-2016c}. In our setting, we have the distinction that fermion-boson coupling, while weak compared to hopping in this regime, still has a non-negligible effect on the low-temperature physics of the model: notably, the weak-coupling transition to a superconducting state, that we will analyze in Sec.\@ \ref{Tc_SYK}. 

In the normal state, the saddle-point equations (\ref{eq:saddle_point_NS}) can be analytically decoupled and solved in the SYK$_2$-FL regime: they yield
\begin{align}\label{eq:G_SYK2_NS}
G(i \omega_n)&=\frac{i \omega_n -i \mathrm{sign}(\omega_n) \sqrt{(\omega_n)^2+ 2 z {t_0^2}}}{z t_0^2} \nonumber \\ & \equiv\frac{-2 i \mathrm{sign}(\omega_n)}{\left|\omega_n\right|+\sqrt{(\omega_n)^2+ 2 z {t_0^2}}}
\end{align}
and 
\begin{equation}\label{eq:Sigma_Dyson_SYK2}
\Sigma(i \omega_n)=\frac{z t_0^2}{2}G(i \omega_n)
\end{equation}
for the fermion Green's function and self-energy, respectively. Notice that the above solution is formally exact in the $g\rightarrow 0$ limit, while for finite coupling there will be small but finite corrections. Once continued to the real axis by the means of Eq.\@ (\ref{eq:A_def}), Eq.\@ (\ref{eq:G_SYK2_NS}) yields the Wigner semicircular (or more precisely, semielliptical) spectral function
\begin{equation}\label{eq:A_SYK2_analyt}
A(\omega)=\frac{1}{\pi} \frac{\sqrt{2 z t_0^2-\omega^2}}{z t_0^2} \Theta(2 z t_0^2-\omega^2),
\end{equation}
where $\Theta(x)$ is the Heaviside theta function. 
Fig.\@ \ref{fig:spectr_NFL-SYK_g}(c) shows how the crossover between the SYK-NFL phase and the SYK$_2$-FL regime affects $A(\omega)$. This evolution is obtained by numerically solving the full saddle-point problem (\ref{eq:saddle_point_NS}) on the real axis. We work at fixed temperature $T=0.01 \omega_0/k_B$ and coupling $g=1$, and we increase the hopping from zero -- see red curve, same as in panel (a) -- to $t_0=4 \omega_0$. We visually appreciate that the low-energy spectral function evolves from the power-law behavior of Eq.\@ (\ref{eq:G_NFL_SYK}) for $t_0=0$, to the SYK$_2$-FL Wigner spectral function (\ref{eq:A_SYK2_analyt}) shown by the dashed gray line. Thus, the fermions become more coherent and approach the energy distribution typical of the disordered Fermi liquid. Although quasiparticles are formally defined in this regime, their quasiparticle weight $Z=\left[1-\left.\partial \mathrm{Re}\left\{\Sigma^R(\omega)\right\}/\partial \omega\right|_{\omega=0}\right]^{-1}$ is still small compared to unity, and their lifetime as given by $1/\mathrm{Im}\left\{\Sigma^R(\omega) \right\}$ is still short compared to standard non-disordered Fermi liquids (see the broadening of the spectral function for $t_0=4 \omega_0$). 

\begin{figure*}[t]
\includegraphics[width=0.95\textwidth]{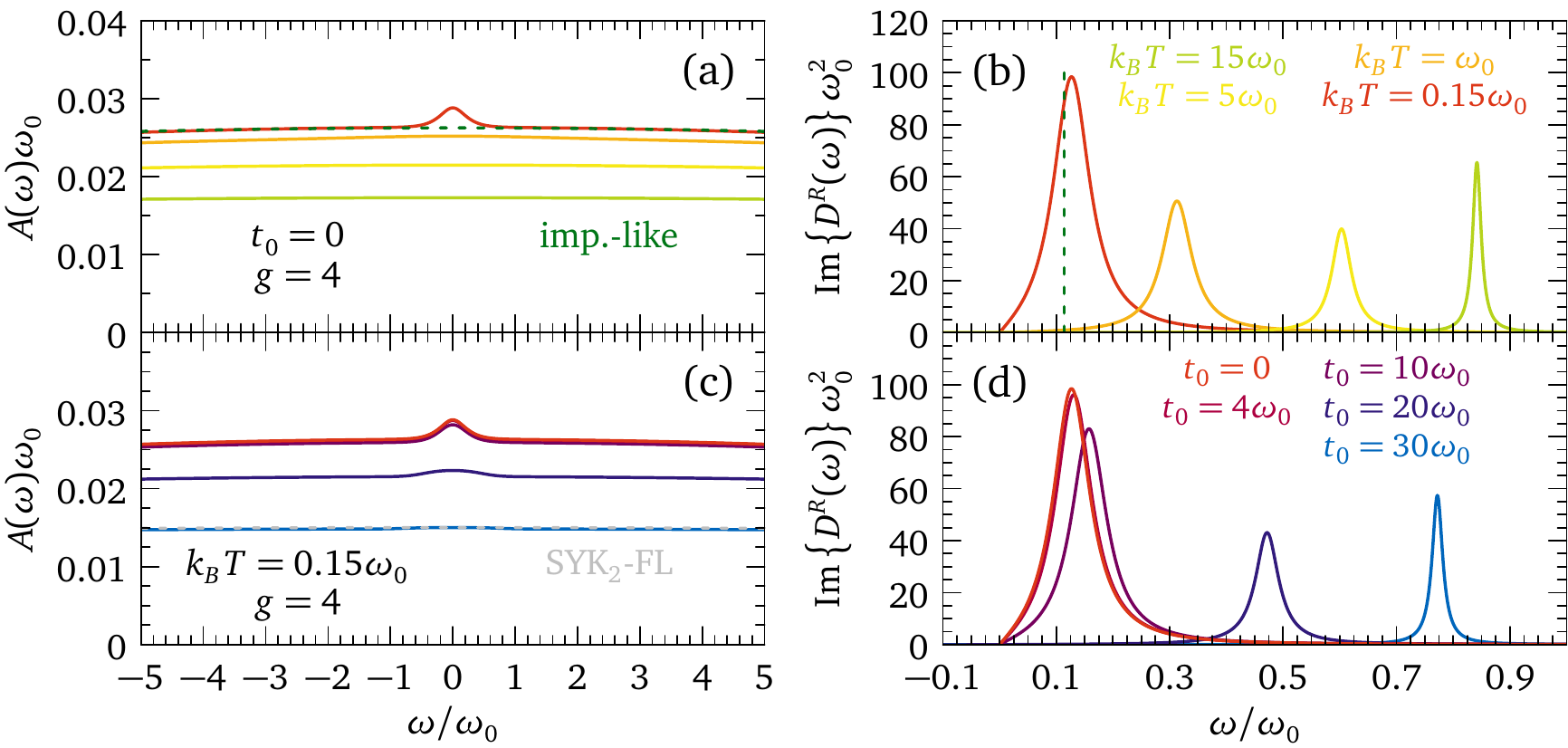}
\caption{\label{fig:spectr_imp_g} Numerically exact spectral function and bosonic propagator in the normal state, showing the crossover from the impurity-like to the SYK$_2$-FL regimes as a function of hopping. (a) Fermionic spectral function in the single-dot limit ($t_0=0$) and for coupling $g=4$, as a function of energy $\omega$ and at temperature $k_B T/\omega_0=\left\{0.15, 1, 5, 15\right\}$. The dashed green curve is $1.12$ times the analytical low-energy impurity-like spectral function, stemming from Eq.\@ (\ref{eq:A_imp_analyt}) for $g=4$. (b) Imaginary part of the bosonic propagator, for the same parameters as in panel (a). The dashed vertical green line is the approximate renormalized boson frequency (\ref{eq:omegar_imp}) in impurity-like regime. (c) Fermionic spectral function at fixed temperature $k_B T/\omega_0=0.15$ and coupling $g=4$, as a function of energy $\omega$ and for different hoppings $t_0/\omega_0=\left\{0,4,10,20,30\right\}$. The dashed gray curve corresponds to the SYK$_2$-FL spectral function (\ref{eq:A_SYK2_analyt}), for $t_0/\omega_0=30$. (d) Imaginary part of the bosonic propagator, for the same parameters as in panel (c). }
\end{figure*}
Notably, the effect of fermion-boson interaction on bosons themselves is not entirely negligible. This is illustrated in Fig.\@ \ref{fig:spectr_NFL-SYK_g}(d), which shows the imaginary part of the real-axis boson propagator $D^R(\omega)$ at temperature $T=0.01 \omega_0/k_B$ and coupling $g=1$, as a function of hopping. Increasing the latter, the strongly damped and softened peak for $t_0=0$ evolves into a well-defined excitation that moves towards $\omega_r \lessapprox \omega_0$: the bosons become progressively more free. Nevertheless, the residual broadening of the peak for $t_0=4 \omega_0$ and its renormalized natural frequency indicate that fermion-boson coupling still plays a role. For instance, we can demonstrate that the renormalized boson frequency is approximately given by
\begin{equation}\label{eq:omegar_0_T0}
\left[\omega_r(0)\right]^2\approx \omega_0^2-\frac{8 \sqrt{2} \bar{g}^2}{3 \pi \sqrt{z} t_0}
\end{equation}
at zero temperature in the SYK$_2$-FL regime; see Appendix \ref{SYK2_omegar}. Eq.\@ (\ref{eq:omegar_0_T0}) produces the dashed vertical gray line in Fig.\@ \ref{fig:spectr_NFL-SYK_g}(d). Notice that the $T=0$ value (\ref{eq:omegar_0_T0}) is finite, meaning that neither fermions nor bosons are critical in the Fermi-liquid regime.  

The crossover between the SYK-NFL and SYK$_2$-FL regimes analyzed so far departed from the hypothesis $g^2 < k_B T< g^{-2}$. However, in the single-dot limit we know that the impurity-like phase appears for $k_B T> g^{-2}$ \cite{Esterlis-2019}. Therefore, it is interesting to analyze how hopping affects the impurity-like fixed point as well. We perform such analysis in the next section. 

\subsection{Strong-coupling impurity-like fixed point}\label{Fixed_point_imp}

In the intermediate temperature window $g^{-2}<k_B T< g^2$ and for zero hopping, the system (\ref{eq:saddle_point_NS}) enters another non-Fermi liquid regime, labeled impurity-like and also analyzed in Ref.\@ \onlinecite{Esterlis-2019}. We recall the essential properties of such phase, before studying how the lattice embedding modifies the single-dot results. 
In the impurity-like regime, the characteristic energy scale of fermions 
\begin{equation}\label{eq:Omega_0}
\Omega_0=\frac{16 g^2 \omega_0}{3 \pi}
\end{equation}
is large with respect to $\omega_0$, so that fermions are ``cold'' and perceive bosons essentially as static. Therefore, the fermion self-energy has a similar form as the one for static impurities, while the fermionic propagator $G(i \omega_n)$ behaves as if fermions were quantum critical with exponent $\Delta=1/2$ -- see Eq.\@ (\ref{eq:G_NFL_SYK}). Formally, we have 
\begin{align}\label{eq:G_imp_NS}
G(i\omega_n)&=\frac{-2 i \mathrm{sign}(\omega_n)}{\left|\omega_n\right| +\sqrt{(\omega_n)^2+\Omega_0^2}} \nonumber \\ & \equiv\frac{i \omega_n- i \mathrm{sign}(\omega_n) \sqrt{(\omega_n)^2 +\Omega_0^2}}{\Omega_0^2/2}
\end{align}
for fermions and
\begin{equation}\label{eq:D_imp_NS}
D^{-1}(i \Omega_n)={\Omega_n^2 +\omega_r^2}
\end{equation}
for bosons, with the renormalized frequency
\begin{equation}\label{eq:omegar_imp}
\omega_r=\frac{3 \pi}{8 g} \sqrt{\omega_0 k_B T}. 
\end{equation}
The fermionic self-energy corresponding to Eq.\@ (\ref{eq:G_imp_NS}) is
\begin{align}\label{eq:Sigma_T_implike_NS}
\Sigma(i \omega_n) &=\frac{\Omega_0}{4} G(i \omega_n)\nonumber \\ & =\frac{i \omega_n}{2}\left(1-\sqrt{1+\frac{4 \left[8 g^2 \omega_0/(3 \pi)\right]^2}{\omega_n^2}}\right) \nonumber \\ & \approx -i \mathrm{sign}(\omega_n)\frac{8 \bar{g}^2}{3 \pi \omega_0^2},
\end{align}
where the last step is the leading small-frequency expansion, which is Eq.\@ (26) of Ref.\@ \onlinecite{Esterlis-2019}.

Notice the formal equivalence of Eqs.\@ (\ref{eq:G_SYK2_NS}), (\ref{eq:Sigma_Dyson_SYK2}) in SYK$_2$-FL regime and Eqs.\@ (\ref{eq:G_imp_NS}), (\ref{eq:Sigma_T_implike_NS}) for the impurity-like fixed point, upon the mapping $2 z t_0^2 \leftrightarrow \Omega_0^2$: static and random boson scattering acts in the same way as random hopping (which is also static by definition) on fermions. This feature will be crucial to derive approximate expressions for the fermion propagators and self-energies, which interpolate across the whole SYK$_2$-FL to impurity-like crossover and allow us to analyze the phase stiffness in the superconducting phase; see Sec.\@ \ref{Stiffness}. 
Although the propagators are formally equivalent in the SYK$_2$-FL and impurity-like phases, their nature is different. The SYK$_2$-FL is a Fermi liquid, while in the impurity-like regime we have to analyze multiple boson configurations, even for a given disorder configuration of the couplings $g_{i j,k}$ \cite{Esterlis-2019}: bosons and fermions strongly interact and influence each other, and the bosons are approximately static only as a consequence of their interaction with fermions; this hinders a mapping to a Fermi-liquid problem \cite{Esterlis-2019, Grunwald-thesis-2022}. 

While the fermion dynamics is similar in the Fermi-liquid and impurity-like regimes, the boson behavior is what differentiates most the two regimes. This is perhaps most evident from the respective renormalized frequencies (\ref{eq:omegar_0_T0}) and (\ref{eq:omegar_imp}), which are finite and vanishing at zero temperature respectively. Thus, in the impurity-like regime the bosons are sharp but soft excitations with a strongly renormalized frequency $\omega_r$. Such properties are visually illustrated in Fig.\@ \ref{fig:spectr_imp_g}(b), which shows the imaginary part of the retarded boson propagator at coupling $g=4$ and $t_0=0$, for different temperatures, from the exact numerical solution of the saddle-point problem (\ref{eq:saddle_point_NS}) on the real axis. Notice the progressive softening of the boson excitation peak, which remains well defined. The dashed green vertical line is given by Eq.\@ (\ref{eq:omegar_imp}) for temperature $T=0.15 \omega_0/k_B$. A similar softening dynamics has been pointed out in the context of magnetic precursors in cuprates \cite{Schmalian-1998, Schmalian-1999}. 

The fermionic spectral function from Eqs.\@ (\ref{eq:A_def}) and (\ref{eq:G_imp_NS}) is again a Wigner semicircle: 
\begin{equation}\label{eq:A_imp_analyt}
A(\omega)=\frac{2}{\pi} \frac{\sqrt{\omega_0^2-\omega^2}}{\Omega_0^2} \Theta(\Omega_0^2-\omega^2),
\end{equation}
which is the same as Eq.\@ (\ref{eq:A_SYK2_analyt}) upon the substitution $2 z t_0^2 \mapsto \Omega_0^2$. 
Fig.\@ \ref{fig:spectr_imp_g}(a) shows the numerically exact $A(\omega)$ from Eqs.\@ (\ref{eq:saddle_point_NS}) for $g=4$ and $t_0=0$, for different temperatures. The extremely broad spectral function signals incoherent fermions, and the dashed green curve stems from $1.12 G(i \omega)$, with $G(i \omega)$ the zero temperature version of Eq.\@ (\ref{eq:G_imp_NS}). 

Due to the mapping between the SYK$_2$-FL and impurity-like fermion propagators, we can expect a smooth crossover between the two regimes, characterized by a semicircular spectral function which changes in width and height but not in shape. This is exactly what happens, as shown by Fig.\@ \ref{fig:spectr_imp_g}(c), numerically calculated on the real axis for $g=4$, $T=0.15 \omega_0/k_B$, and for different hoppings. The blue curve for $t_0=30 \omega_0$ consistently agrees with the SYK$_2$-FL spectral function given by Eq.\@ (\ref{eq:A_SYK2_analyt}). 
Correspondingly, by increasing the hopping, the strongly renormalized bosons of the impurity-like regime stiffen and become more free, since the boson self-energy $\Pi(0)$ decreases due to the relatively weaker fermion-boson interaction. This is shown in Fig.\@ \ref{fig:spectr_imp_g}(d) by the full numerical solution for the imaginary part of the boson propagator, for the same parameters as in Fig.\@ \ref{fig:spectr_imp_g}(c). On the SYK$_2$-FL side of the crossover and close to zero temperature, the renormalized boson frequency still follows Eq.\@ (\ref{eq:omegar_0_T0}), which would give $\omega_r \approx 0.6 \omega_0$ for $t_0=30 \omega_0$. For the same hopping, the blue curve in Fig.\@ \ref{fig:spectr_imp_g}(d) displays a larger $\omega_r$, which is due to thermal effects since $T=0.15 \omega_0/k_B$ is still relatively larger than zero. We checked that further lowering the temperature makes the real-axis numerical solution agree with Eq.\@ (\ref{eq:omegar_0_T0}) in the SYK$_2$-FL regime.  

Notice that the crossover to the disordered Fermi-liquid regime occurs at much higher values of hopping for the impurity-like phase, compared to the SYK-NFL phase; this is seen by comparing Figs.\@ \ref{fig:spectr_imp_g}(c),(d) and \ref{fig:spectr_NFL-SYK_g}(c),(d). Thus, the question arises of where precisely the different crossovers occur at a given dimensionless ratio ${t_0^2}\omega_0/\bar{g}^2$. We give quantitative estimations of the specific crossover energies in the next section. 

\section{Normal-state crossovers}\label{Normal_cross}

The NFL/FL crossovers, described in Sec.\@ \ref{Normal_fixed} and summarized in the phase diagram sketched in Fig.\@ \ref{fig:phase_scheme}, can be quantitatively characterized as a function of the ratio between fermion-boson interaction and coherent lattice hopping. Our chosen criterion refers to the respective fermion self-energy $\Sigma(i \omega_n)$ in each regime. 

\begin{figure}[t]
\includegraphics[width=0.9\columnwidth]{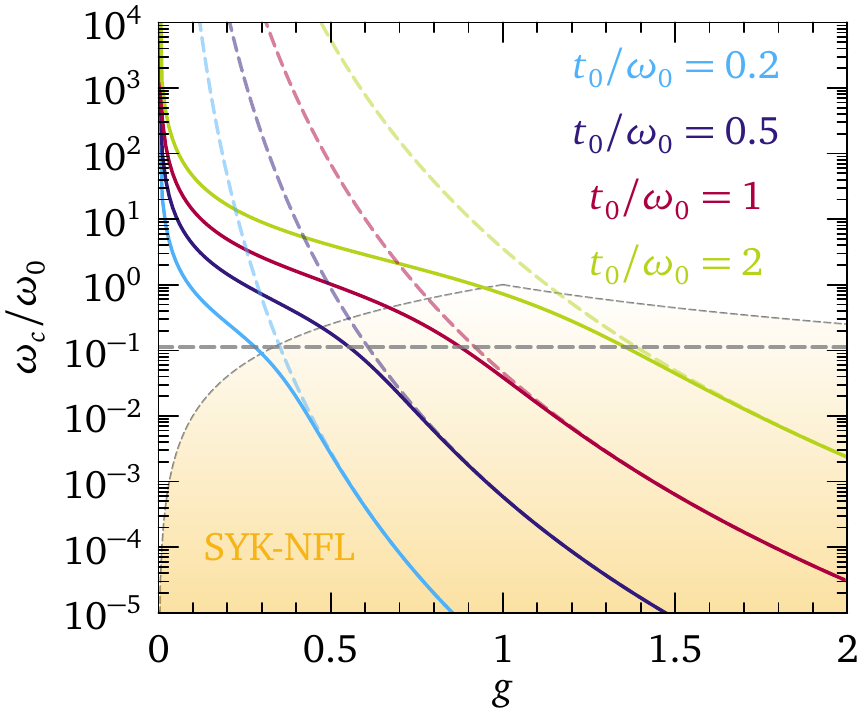}
\caption{\label{fig:omegac} Characteristic energy/temperature scale for the crossover between SYK-NFL fixed point and SYK$_2$-FL regime as a function of fermion-boson coupling $g$ for different hopping parameters $t_0$. The dashed gray line is the asymptotic large-coupling limit (\ref{eq:Tc_largeg}) for the superconducting critical temperature $T_c$, which is valid for $g \rightarrow +\infty$ both in the single-dot limit and on a lattice; see Sec.\@ \ref{Tc_imp_like}. The shaded orange area delimits the SYK-NFL region of the single-dot phase diagram, where the crossover estimation is valid. The analytical low-energy expansion (\ref{eq:omegac_analyt}) produces the dashed curves of the same color as the solid curves, the latter corresponding to the full numerical solution for $\omega_c$.}
\end{figure}

First, consider the crossover between the SYK-NFL and SYK$_2$-FL phases, described for the spectral functions in Sec.\@ \ref{SYK2-FL_fixed_point}. We can estimate the characteristic energy $\omega$ (or temperature $T$) scale at which such crossover occurs, as the energy $\omega$ at which the respective self-energies (\ref{eq:Sigma_NFL_SYK}) and (\ref{eq:Sigma_Dyson_SYK2}) of the two phases coincide. 
Actually, it is convenient to compare the corresponding real-valued dynamical quasiparticle residues through Eq.\@ (\ref{eq:Sigma_Z}), which yields
\begin{equation}\label{eq:NFL_Z}
Z_{\rm NFL}(i \omega_n)=1+\frac{ \mathrm{sign}(\omega_n)}{\omega_n} c_1 g^{4 \Delta} \left|\omega_n\right|^{1-2 \Delta},
\end{equation} 
for the SYK-NFL fixed point and
\begin{equation}\label{eq:SYK2_Z}
Z_{{\rm SYK}_2}(i \omega_n)=\frac{1}{2}+\frac{1}{2 \left|\omega_n\right|} \sqrt{(\omega_n)^2+2 z t_0^2}. 
\end{equation}
in the SYK$_2$-FL regime. 
A solution to $Z_{\rm NFL}(i \omega_n)=Z_{{\rm SYK}_2}(i \omega_n)$ can be obtained numerically at arbitrary $i \omega_n \equiv i \omega_c$. This gives the crossover scale $\omega_c$ depicted in Fig.\@ \ref{fig:omegac}. Schematically, for $\left\{k_B T, \omega\right\} <\omega_c$ the physics is dominated by hopping, while for $\left\{k_B T, \omega\right\} >\omega_c$ the NFL fixed point of single-dot dynamics prevails. 
At low energy, we can approximate Eq.\@ (\ref{eq:NFL_Z}) with $Z_{\rm NFL}(i \omega_n)\approx c_1/\left|\omega_n\right| g^{4 \Delta} \omega^{1-2 \Delta}$ and Eq.\@ (\ref{eq:SYK2_Z}) with $Z_{{\rm SYK}_2}(i \omega_n)\approx t_0 \sqrt{z}/(\sqrt{2} \left|\omega_n\right|)$, so that equating the last two expressions gives an analytical result for $\omega_n \equiv \omega_c$:
\begin{equation}\label{eq:omegac_analyt}
\omega_c=2^{-\frac{1}{2-4 \Delta}} \left(\frac{t_0 \sqrt{z} g^{-4 \Delta}}{c_1}\right)^{\frac{1}{1-2\Delta}}.
\end{equation}
Eq.\@ (\ref{eq:omegac_analyt}) yields the dashed lines in Fig.\@ \ref{fig:omegac}, for the corresponding values of $t_0$.
For $k_B T=0.01 \omega_0 \equiv \omega_c$ and $g=1$, Eq.\@ (\ref{eq:omegac_analyt}) predicts that the crossover occurs at $t_0/\omega_0 \approx 0.784$. This estimation qualitatively agrees with the real-axis calculations in Fig.\@ \ref{fig:spectr_NFL-SYK_g}(c), where we observe the first discernible differences with respect to the SYK-NFL form of the spectral function for $t_0/\omega_0=1$. Such evaluation of the crossover energy also agrees with the imaginary-axis numerics, as shown in Fig.\@ \ref{fig:Fig_G_hop_SYK} for the fermionic and bosonic propagators in Appendix \ref{Normal_prop}.

\begin{figure}[t]
\includegraphics[width=0.9\columnwidth]{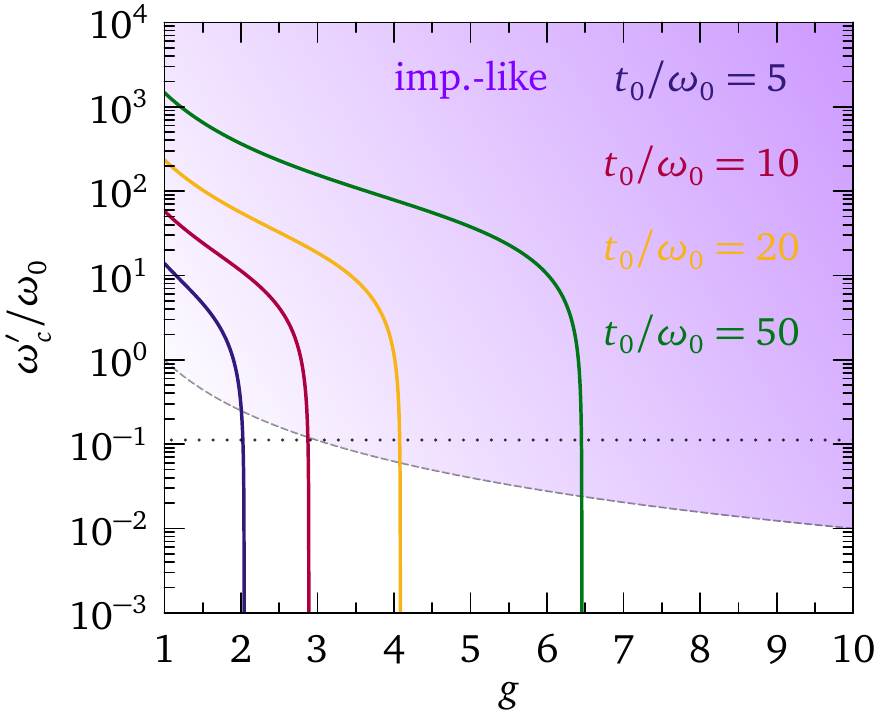}
\caption{\label{fig:omegacIL} Characteristic energy/temperature scale for the crossover between impurity-like fixed point and SYK$_2$-FL regime as a function of fermion-boson coupling $g$ for different hopping parameters $t_0$, in accordance with Eq.\@ (\ref{eq:omegac_IL_analyt}). The dotted black line is the asymptotic large-coupling limit (\ref{eq:Tc_largeg}) for $T_c$, which is valid for $g \rightarrow +\infty$ both in the single-dot limit and on a lattice; see Sec.\@ \ref{Tc_imp_like}. The shaded purple area delimits the impurity-like region of the single-dot phase diagram. }
\end{figure}
\begin{figure*}[t]
\includegraphics[width=0.8\textwidth]{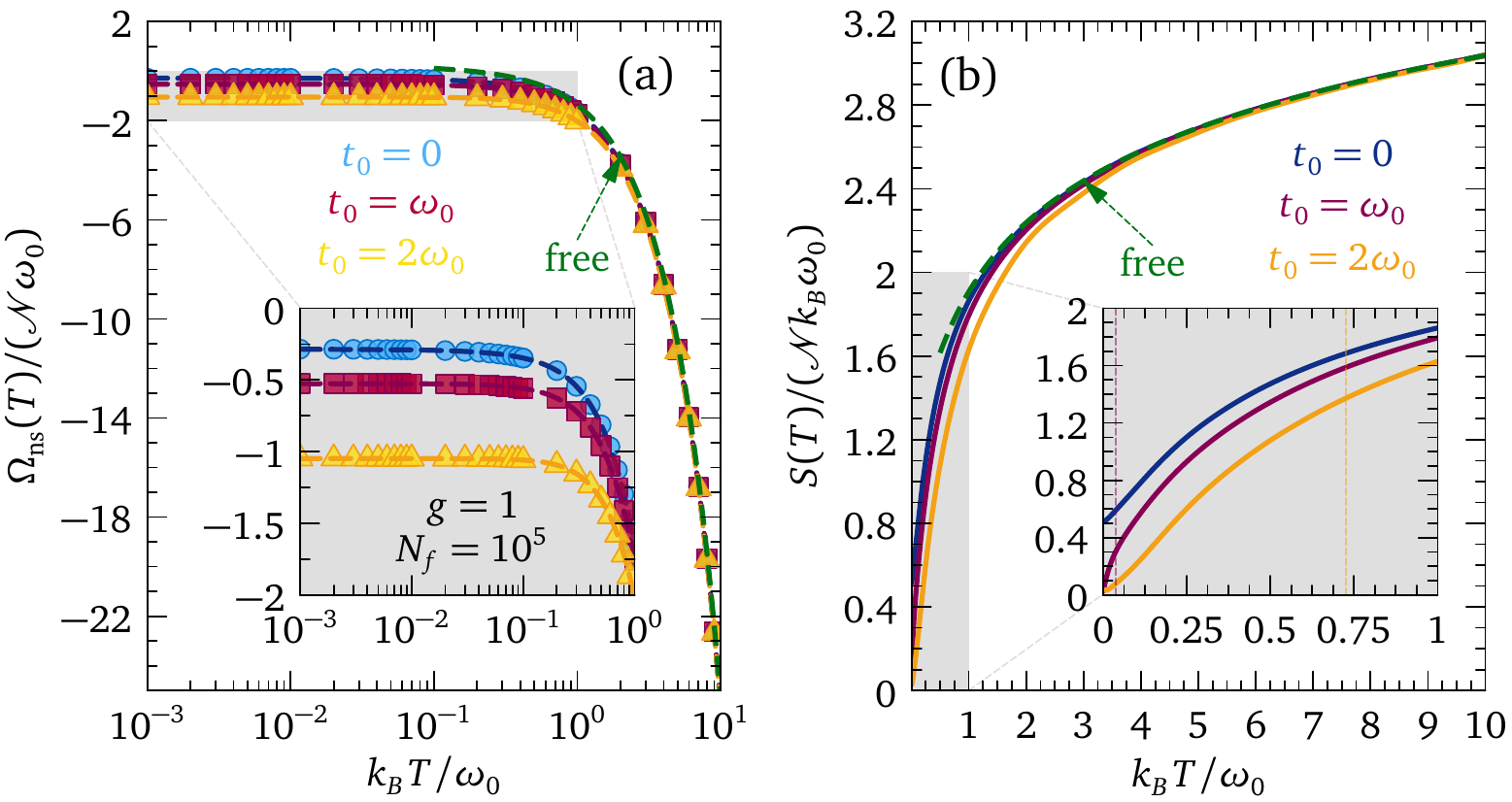}
\caption{\label{S_ph_NFL_SYK2} (a) Normal-state large-$\mathscr{N}$ grand potential $\Omega_{\rm ns}(T)$ of the Yukawa-SYK model on a lattice as a function of normalized temperature $k_B T/\omega_0$ for coupling constant $g=1$ and at fixed hopping $t_0/\omega_0=\left\{0,1,2\right\}$; data points are the results of the numerical solution of Eq.\@ (\ref{eq:Omega_SYK_ph}), with the saddle-point propagators stemming from Eqs.\@ (\ref{eq:saddle_point_NS}). The dashed green curve is the result for free fermions and bosons, corresponding to the last two terms in Eq.\@ (\ref{eq:Omega_SYK_ph}). The numerical calculations are performed with a number of Matsubara frequencies $N_f=10^5$. The inset zooms in on the low-temperature region; the dashed curves are smooth polynomial interpolations of the numerical data. (b) Normal-state entropy $S(T)$ as a function of normalized temperature $k_B T/\omega_0$, for the same parameters as in panel (a), calculated from the numerical derivative of the polynomial interpolations of $\Omega_{\rm ns}(T)$. The inset zooms in on the low-temperature regime, where we see the finite $T=0$ entropy in the single-dot limit ($t_0=0$).}
\end{figure*}
Next we analyze the crossover between the impurity-like and the SYK$_2$-FL regimes, which occurs at stronger values of coupling $g$ as discussed in Sec.\@ \ref{Fixed_point_imp}. The low-frequency expansion of the self-energy in the impurity-like regime is given by the last line of Eq.\@ (\ref{eq:Sigma_T_implike_NS}), and it is associated with the dynamical quasiparticle weight
\begin{equation}\label{eq:Z_imp}
Z_{\rm IL}(i \omega_n)=1+ \frac{8 \omega_0}{3 \pi \left|\omega_n\right|} g^2. 
\end{equation}
Eq.\@ (\ref{eq:Z_imp}) must be compared to the residue (\ref{eq:SYK2_Z}) in the SYK$_2$-FL regime. The solution to $Z_{\rm IL}(i \omega_n)=Z_{{\rm SYK}_2}(i \omega_n)$ is analytical at all frequencies, and reads (for $\omega_c'=\omega_n>0$)
\begin{equation}\label{eq:omegac_IL_analyt}
\omega_c'=\frac{9 \pi^2 z t_0^2-128 g^4 \omega_0^2}{48 \pi g^2 \omega_0}. 
\end{equation}
The curves in Fig.\@ \ref{fig:omegacIL} originates from Eq.\@ (\ref{eq:omegac_IL_analyt}), each corresponding to different values of $t_0$. Notice that Eq.\@ (\ref{eq:omegac_IL_analyt}) would predict a maximum coupling strength at which $\omega_c'=0$, i.e., above which the SYK$_2$-FL regime completely disappears; however, such coupling falls outside the regime where Eq.\@ (\ref{eq:omegac_IL_analyt}) is valid, since it occurs at energies $\omega <\omega_0^2/\bar{g}^2$. 
At $g=4$ and $k_B T= 0.15 \omega_0 \equiv \omega_c'$, Eq.\@ (\ref{eq:omegac_IL_analyt}) is satisfied for $t_0/\omega_0 \approx 19.312$. Such an estimation is in qualitative agreement with the spectral functions in Fig.\@ \ref{fig:spectr_imp_g}(c): the purple curve for $t_0=10 \omega_0$ is the one of lowest hopping for which deviations with respect to the impurity-like form of the propagator are noticeable. Eq.\@ (\ref{eq:omegac_IL_analyt}) is also consistent with the imaginary-axis numerical solution of the saddle-point equations (\ref{eq:saddle_point_NS}), as shown in Fig.\@ \ref{fig:Fig_G_hop_imp} for the fermion and boson Green's functions in Appendix \ref{Normal_prop}.

Having quantitatively characterized all normal-state crossovers on the saddle point, implied by Eqs.\@ (\ref{eq:saddle_point_NS}), a consequent task is identifying measurable observables of such crossovers. Of particular interest in this respect are thermodynamic quantities, that can all be calculated from the normal-state grand potential of the fermions coupled to bosons. For this reason, in the next section we tackle the problem of the saddle-point grand potential, and as an example of crossover-sensitive thermodynamic variable we calculate the entropy, both in the single-dot limit and on the lattice. 

\section{Normal-state thermodynamics: grand potential and entropy}\label{S_SYK}

To identify observable consequences of the crossovers analyzed in Sec.\@ \ref{Normal_cross} on the thermodynamic properties of the lattice, we compute the on-shell normal-state grand potential $\Omega_{\rm ns}$. As an exemplary application, we will focus on the entropy $S=S(T)$: this quantity is finite in the $T\rightarrow 0$ limit taken \emph{after} the $\mathscr{N}\rightarrow +\infty$ limit, for theories where the fermion propagator displays a branch-cut singularity, which include the Majorana and complex-fermion versions of the SYK model \cite{Parcollet-1999,Georges-2001,Sachdev-2015,Maldacena-2016a,Davison-2017,Song-2017,Kruchkov-2020,Gu-2020, Milekhin-2022_preprint}. We will shortly realize (see Fig.\@ \ref{S_ph_NFL_SYK2}) that this property persists in the zero-hopping limit $t_0=0$ of our Yukawa-SYK lattice, that is, for the NFL-SYK and impurity-like single-dot fixed points \cite{Esterlis-2019,Hauck-2020}. Instead, for any $t_0>0$ we will find $\lim_{T \rightarrow 0}S(T)=0$, so that the excess entropy is released on a lattice even in the large-$\mathscr{N}$ limit \cite{Song-2017}. 

The on-shell normal-state grand potential per fermion flavor is $\Omega/\mathscr{N}=-k_B T \ln \mathscr{Z}_{\rm sp}$, where $\mathscr{Z}_{sp}=e^{-\mathscr{S}_{\rm sp}}$ is the partition function linked to the saddle-point action $\mathscr{S}_{\rm sp}$; the latter corresponds to the disorder-averaged effective action (\ref{eq:effective_action}) where the bilocal fields have been substituted with their saddle-point expressions, yielded by the stationarity conditions (\ref{eq:saddle_point_NS}). In this section we focus on the normal-state grand potential $\Omega=\Omega_{\rm ns}$, where $F(i \omega_n)$ and $\Phi(i \omega_n)$ vanish. 
Aforementioned derivation, sketched in Appendix \ref{app:Omega_NS}, leads us to the following expression for generic chemical potential $\mu$: 
    \begin{align}\label{eq:Omega_SYK_ph}
    \frac{\Omega_{\rm ns}}{\mathscr{N}}&=k_B T \sum_{i \omega_n} \left\{2 \ln\left[\frac{G(i \omega_n)}{G_0(i \omega_n)}\right]-z\frac{{t_0^2}}{2}\left[G(i \omega_n)\right]^2\right\}
    \notag\\ 
    &+k_B T \sum_{i \Omega_m} \left\{-\frac{1}{2} \ln\left[\frac{D(i \Omega_m)}{D_0(i \Omega_m)}\right]+ D(i \Omega_m)\Pi(i \Omega_m)\right\} \notag\\ 
    &-2k_B T \ln\left[1+e^{\mu/(k_B T)}\right]
    \notag\\ 
    &+\frac{k_B T}{4} \left\{\frac{\omega_0}{k_B T} + 2 \ln[1 - e^{-\omega_0/(k_B T)}]\right\} .
    \end{align}
In general, the Matsubara sums in Eq.\@ (\ref{eq:Omega_SYK_ph}) must be evaluated numerically. To ease the convergence of these sums, the grand potential for free fermions and free bosons, given by the last two terms in Eq.\@ (\ref{eq:Omega_SYK_ph}), has been simultaneously added and subtracted in Eq.\@ (\ref{eq:Omega_SYK_ph}).

Fig.\@ \ref{S_ph_NFL_SYK2}(a) shows the evolution of the temperature-dependent grand potential across the SYK-NFL to SYK$_2$-FL crossover, for coupling $g=1$, $\mu=0$ (particle-hole symmetry), and hoppings $t_0/\omega_0=\left\{0,1,2\right\}$. The full numerical solution of Eqs.\@ (\ref{eq:Omega_SYK_ph}) and (\ref{eq:saddle_point_NS}) is marked by the colored data points. The dashed green curve labeled ``free'' shows the grand potential of free fermions and bosons, to which the numerical results consistently converge for large temperature with respect to hopping and coupling: this is the classical-gas fixed point of the model, where temperature dominates over all interactions and all particles are essentially free. 
We notice that temperatures of order $k_B T \gtrapprox 5 \omega_0$ are necessary, in order for the Yukawa-SYK results to approach the free-fermion limit. At first glance, this is surprising because fermions are already free at a temperature scale $k_B T/\omega_0 \gtrapprox g^2$ in the single-dot limit \cite{Esterlis-2019} and of $k_B T/\omega_0 \gtrapprox \mathrm{min}\left\{ g^2, \sqrt{z} t_0/\omega_0\right\}$ on the lattice; see Fig.\@ \ref{fig:phase_scheme}. However, the bosons are not completely free until temperatures of $k_B T \gtrapprox 10 \omega_0$ are reached, as shown by the imaginary-axis results for the SYK-NFL renormalized boson frequency in Fig.\@ \ref{fig:omrsq_t0_NFL}: this is the reason why the ``free'' limit of the theory is approached only above temperatures of order $10 \omega_0/k_B$: fermions are already essentially free at $T \sim \omega_0/k_B$, but bosons necessitate of temperatures about ten times larger to reach the independent-particle limit. 

The inset of Fig.\@ \ref{S_ph_NFL_SYK2}(a) zooms on the low-temperature regime, where we see that increasing hopping lowers $\Omega_{\rm ns}(T)$. The dashed colored curves superimposed on the numerical data points represent a smooth polynomial interpolation, which we partially differentiate with respect to temperature $T$ to find the entropy:
\begin{equation}\label{eq:S_def}
\frac{S}{\mathscr{N}}=-\frac{1}{\mathscr{N}}\frac{\partial \Omega_{\rm ns}}{\partial T}.
\end{equation}
The results of Eq.\@ (\ref{eq:S_def}) are displayed in Fig.\@ \ref{S_ph_NFL_SYK2}(b), where again the dashed green curve shows that the Yukawa-SYK entropy approaches the one of free fermions and bosons in the large-temperature limit. The most remarkable differences between curves at different hoppings appear in the low-temperature regime, marked by the gray shading and zoomed upon in the inset. There, we observe that the zero-temperature entropy in the single-dot SYK-NFL limit (blue curve) is finite, namely $S(0)/\mathscr{N} \approx 0.52$. This value differs from purely fermionic versions of the SYK model, where $S(0)/\mathscr{N} \approx 0.46$ \cite{Song-2017}. Such discrepancy can be traced back to the different anomalous exponent $\Delta$ of the SYK-NFL self-energy (\ref{eq:Sigma_NFL_SYK}) with respect to SYK$_q$ models. 
At any finite hopping, as expected \cite{Song-2017} the entropy vanishes (within numerical accuracy) at $T\rightarrow 0$. This feature happens because of the crossover between the SYK-NFL and SYK$_2$-FL regimes, which occurs here below a temperature scale $k_B T < \omega_c$ with $\omega_c$ the SYK-NFL/SYK$_2$-FL crossover energy analyzed in Sec.\@ \ref{Normal_cross}. The latter estimation gives the dashed red and gold vertical lines in the inset of Fig.\@ \ref{S_ph_NFL_SYK2}(b), below which the SYK$_2$-FL physics progressively takes over. There, we know that the fermionic entropy $S(0)/\mathscr{N} =0$ \cite{Chowdhury-2022}, and the same is realized here for our Yukawa-SYK lattice model. 

Here the results in Fig.\@ \ref{S_ph_NFL_SYK2} merely serve as a demonstration of the practical observability of the crossovers analyzed in Sec.\@ \ref{Normal_cross}. They provoke multiple interesting questions, for instance related to the dependence of the zero-temperature entropy on coupling $g$ in the single-dot limit, and its relation to the heat capacity $C_V$ in the normal and superconducting states of our model. We believe that these points deserve their own dedicated discussion, to which we defer such analysis. In what follows, we focus instead on the superconducting transition that appears when the coupling constants $\left\{g_{ij, k}\right\}$ are real-valued, taken from the GOE. We will then endeavor to investigate the properties of the superconducting state on the lattice, notably the phase stiffness which is the main focus of the present work.  

\section{Critical temperature}\label{Tc_SYK}

\subsection{Linearized saddle-point equations and numerical results}

\begin{figure*}[t]
\includegraphics[width=0.9\textwidth]{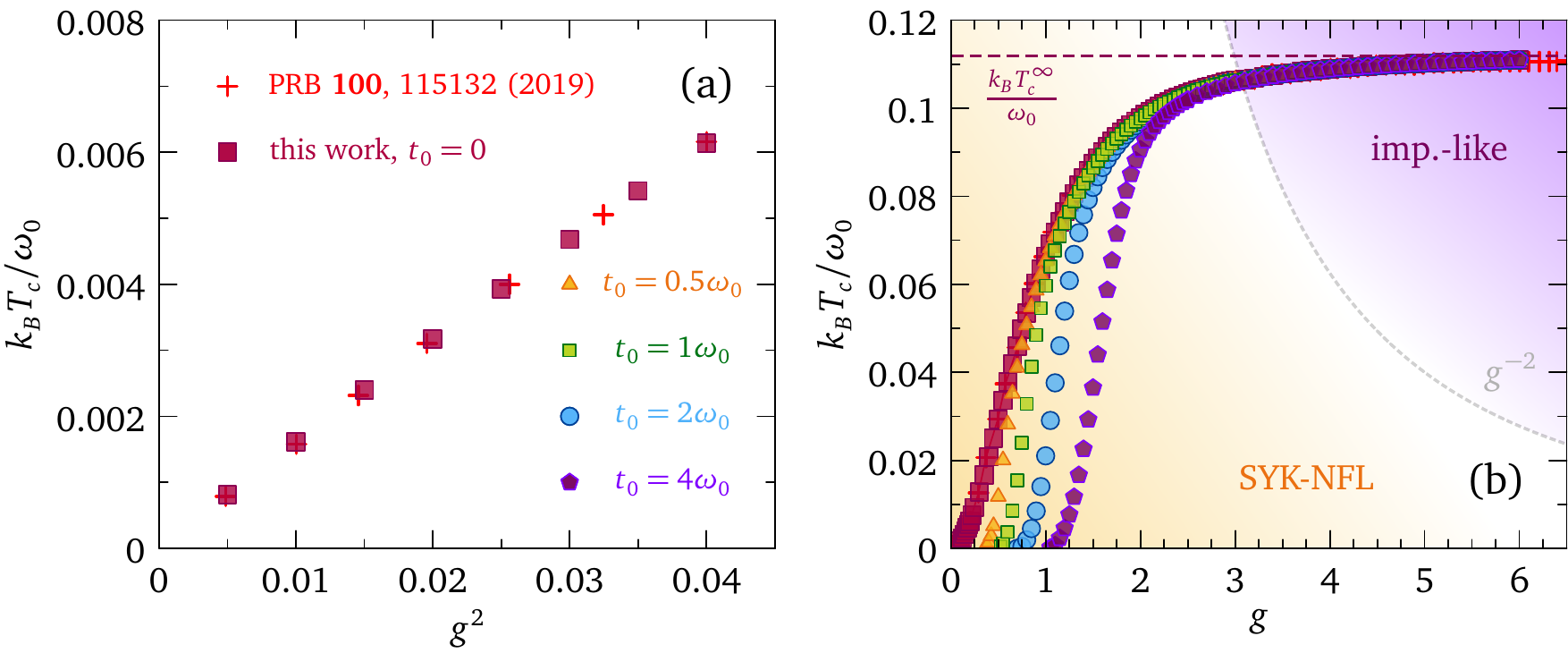}
\caption{\label{fig:Tc_SYK_t0} Superconducting critical temperature as a function of fermion-boson interaction $g$, at fixed hopping $t_0$, obtained from the self-consistent solution of the linearized saddle-point equations (\ref{eq:saddle_Tc_linearized}). (a) Comparison between the results of Ref.\@ \onlinecite{Esterlis-2019} in the single-dot limit ($t_0=0$) and our data for $k_B T_c/\omega_0$, as a function of $g^2$, showing the scaling $k_B T_c \approx 0.16 \bar{g}^2/(\omega_0)^2$. (b) Normalized critical temperature $k_B T_c/\omega_0$ as a function of $g$ for different hopping $t_0/\omega_0=\left\{ 0,1,2,4\right\}$. $T_c^\infty$, given by the dashed horizontal line, is the asymptotic strong-coupling limit (\ref{eq:Tc_largeg}) for $g \rightarrow +\infty$.}
\end{figure*}
To investigate the onset of superconducting pairing, we compute the corresponding critical temperature $T_c$ in our large-$\mathscr{N}$ theory. Let us emphasize that we define $T_c$ as the temperature at which the superconducting gap function $\Delta(i\omega_n)$ vanishes at all frequencies, i.e.\@, the \emph{gap-closing} temperature. The latter does not necessarily coincide at finite $\mathscr{N}$ with the temperature at which superconducting phase coherence sets in \cite{Emery-1994,Emery-1995,Keimer-2015}, and true long-range order is established. The distinction between $k_B T_c$ and the energy scale associated with the phase stiffness will be the main subject of Sec.\@ \ref{Stiffness}. 
To obtain $T_c$, we linearize eqs.\@ (\ref{eq:Sigma})-(\ref{eq:Polar}) in $\Phi(i \omega_n)$ neglecting all higher-order terms. This way we obtain \cite{Esterlis-2019} 
    \begin{subequations}\label{eq:saddle_Tc_linearized}
    \begin{align}
    \Sigma(i \omega_n) 
    &=\bar{g}^2 k_B T \sum_{m=-\infty}^{+\infty} D(i\Omega_m) \frac{1}{i(\omega_n -\Omega_m) Z(i \omega_n-i \Omega_m)} \label{eq:Sigma_Z_hop_lin}
    \notag\\\,\,&{+\frac{z t_0^2}{2} G(i \omega_n)},
    \\
    \Pi(i \Omega_n)&=-2 \bar{g}^2 k_B T \sum_{m=-\infty}^{+\infty} \frac{1}{i \omega_m Z(i \omega_m)} \notag\\ 
    &\qquad\qquad\qquad\qquad\times \frac{1}{i (\omega_m +\Omega_n) Z(i \omega_m+i \Omega_n)},
    \label{eq:Polar_Z_lin}
    \\
    \Phi(i \omega_n)&=\bar{g}^2 k_B T \sum_{m=-\infty}^{+\infty} \frac{D(i \omega_n-i \omega_m)}{\left[\omega_m Z(i \omega_m)\right]^2} \Phi(i \omega_m) \notag\\\,\,& {+\frac{z t_0^2}{2} \frac{\Phi(i \omega_n)}{\left[\omega_n Z(i \omega_n)\right]^2}}. 
    \label{eq:Phi_Z_lin}
    \end{align}
    \end{subequations}
The single-dot case ($t_0=0$) was analyzed in reference \cite{Esterlis-2019}. 
At finite $t_0$, the gap equation (\ref{eq:Phi_Z_lin}) can be equivalently reformulated in terms of a modified function 
\begin{equation}\label{eq:phi_tilde}
\tilde{\Phi}(i \omega_n)=\Phi(i \omega_n)\left\{1-\frac{z t_0^2}{2\left[\omega_n Z(i \omega_n)\right]^2}\right\}, 
\end{equation}
giving the linearized gap equation for $\tilde{\Phi}(i \omega_n)$
\begin{equation}\label{eq:Phi_Z_tilde}
\tilde{\Phi}(i \omega_n)=\bar{g}^2 k_B T \sum_{m=-\infty}^{+\infty} \frac{D(i \omega_n-i \omega_m)}{\left[\omega_m Z(i \omega_m)\right]^2-\frac{z t_0^2}{2}} \tilde{\Phi}(i \omega_m). 
\end{equation}
Eq.\@ (\ref{eq:Phi_Z_tilde}) reduces to the $\Phi(i \omega_n)$-linearized version of Eq.\@ (28) in ref.\@ \onlinecite{Esterlis-2019} for $t_0=0$ (isolated-dot limit). With finite hopping, Eq.\@ (\ref{eq:Phi_Z_tilde}) represents an eigenvalue problem for $\tilde{\Phi}(i \omega_n)$: at
$T < T_c$, the largest eigenvalue is greater than one, and consequently the self-consistent loop tends to increase the value of $\tilde{\Phi}(i \omega_n)$. Conversely, at $T > T_c$ the self-consistency drives $\tilde{\Phi}(i \omega_n)$ to zero and all eigenvalues of the problem (\ref{eq:Phi_Z_tilde}) are therefore smaller than one \cite{Berthod-2018}. At the critical temperature $T = T_c$, the largest eigenvalue is exactly one. This criterion is used to solve Eq.\@ (\ref{eq:Phi_Z_tilde}) exactly for $T_c=T_c(g, t_0)$, and further details on our numerical method are collected in Appendix \ref{App:numeric_Tc}. The ensuing results are displayed in Fig.\@ \ref{fig:Tc_SYK_t0} as a function of coupling $g$, with differently colored data points corresponding to different hopping $t_0$.  
Fig.\@ \ref{fig:Tc_SYK_t0}(a) shows that our results in the single-dot limit (red squares) are perfectly consistent with the results of Ref.\@ \onlinecite{Esterlis-2019} (red crosses). This indicates that $k_B T_c/\omega_0 \propto g^2$ at small $g \ll 1$ and $t_0=0$. Such quadratic dependence is qualitatively altered for nonzero coherent hopping, as illustrated in Fig.\@ \ref{fig:Tc_SYK_t0}(b). There, we observe that increasing $t_0$ leads to an exponential suppression of $T_c$ at small coupling, compared to the single-dot results. 
In fact, the detrimental effect of hopping on $T_c$ lends itself to an intuitive interpretation: since on-site fermion-boson coupling is at the origin of both quantum criticality and superconductivity for fermions in the single dot, a large ratio $z t_0^2/\bar{g}^{2/3}$ weakens quantum-critical superconductivity, because inter-site hopping competes with on-site pairing. In the limit $\sqrt{z}t_0/\bar{g}^{1/3}\rightarrow +\infty$, where coupling to bosons is negligible, Eq.\@ (\ref{eq:Phi_Z_tilde}) only has the trivial solution $\tilde{\Phi}(i \omega_n)=0$ so that fermions do not superconduct. Therefore, the critical temperature is much higher in the non-Fermi liquid SYK-NFL phase, than in the Fermi-liquid SYK$_2$-FL one, at fixed $g$. However, even in the SYK$_2$-FL phase, pairing does happen for an arbitrarily small $g$, which reminds us of the standard Cooper instability. Indeed, in Sec.\@ \ref{SYK_BCS_lattice} we will find that $T_c$ in this regime exactly follows the BCS formula \cite{Bardeen-1957a,Bardeen-1957b,Schrieffer-1963th}.  

On the other hand, in the impurity-like strong-coupling regime -- see purple-shaded area in Fig.\@ \ref{fig:Tc_SYK_t0}(b) -- $T_c$ is negligibly affected by hopping for values as large as $t_0=4 \omega_0$. Hence, the single-dot impurity-like results of Sec.\@ \ref{Fixed_point_imp} still apply on the lattice, which proves that the pairing transition in the impurity-like regime is more robust with respect to lattice embedding than Cooper-pair formation in the SYK-NFL state. 
Naturally, further increasing hopping would lead to the impurity-like/SYK$_2$-FL crossover as predicted by Eq.\@ (\ref{eq:omegac_IL_analyt}). 

The above arguments lead to the key conclusion that coherent hopping always reduces $T_c$ with respect to its single-dot value, at fixed coupling. Then, the maximum $T_c$ of the lattice is found for $t_0 \rightarrow 0$, which is the limit analyzed in Ref.\@ \onlinecite{Esterlis-2019}. 
This realization leads to questioning how the critical temperature is exponentially suppressed when $z t_0^2/\bar{g}^{2/3}$ is large, in other words, what is the nature of pairing in the Fermi-liquid phase. Another relevant aspect is whether hopping affects the limit $g \rightarrow +\infty$ of $T_c$. These questions are addressed in the next sections.

\subsection{Critical temperature at weak coupling: disordered BCS theory}\label{SYK_BCS_lattice}

In the single-dot limit $t_0=0$, since the only temperature scale in Eq.\@ (\ref{eq:gap_weak}) is $T_f/T$ \cite{Esterlis-2019}, the transition temperature $T_c$ is of the order of $T_f$ and is numerically found to be $k_B T_c \approx 0.16 \bar{g}^2/(\omega_0)^2$; see Appendix \ref{Tc_weak_dot}.  

The quadratic dependence of $T_c$ on coupling in the SYK-NFL phase is dramatically altered when we cross over to the SYK$_2$-FL regime, by increasing hopping. In this limit, fermion-boson coupling is relatively weak and the normal state is a disordered Fermi liquid, as analyzed in Sec.\@ \ref{SYK2-FL_fixed_point}. These situations are very similar to the standard Cooper problem in a wide-band non-disordered Fermi liquid, for which the BCS pairing instability is triggered. In fact, as analytically shown in Appendix \ref{Tc_weak_dot_BCS}, in our model we do retrieve the BCS formula for $T_c$
\begin{equation}\label{eq:BCS_Tc_SYK2}
T_c=\frac{2 e^\gamma}{\pi} \frac{\omega_0}{k_B} e^{-\frac{1}{\bar{\lambda}}},
\end{equation}
with $\gamma\approx 0.5772156649$ Euler-Mascheroni constant, and where the lattice coupling constant is
\begin{equation}\label{eq:lambda_bar}
\bar{\lambda}=\frac{\sqrt{2} g^2 \omega_0}{\pi \sqrt{z} t_0}.
\end{equation}
We can directly make sense of Eq.\@ (\ref{eq:lambda_bar}) by connecting it to BCS theory: by analogy with the latter, we can write $\bar{\lambda}=N(0) V_{\rm eff}$, where $N(0)\equiv A(0)=\sqrt{2}/(\pi \sqrt{z} t_0)$ is the ``density of states'' at energy $\omega=0$ -- see Eq.\@ (\ref{eq:A_SYK2_analyt}) -- and $V_{\rm eff}=g^2 \omega_0$ is the effective attractive interaction. 
That we retrieve the BCS pairing instability, usually derived in systems without disorder, in a model like (\ref{eq:H_SYK_phon_hop}) with fully random interactions might seem counterintuitive. However, the argument as to why $T_c$ is unaffected by disordered hoppings is similar to the Anderson theorem for static nonmagnetic impurities \cite{Anderson-1959,Abrikosov-1959a,Abrikosov-1959b,Abrikosov-1961,Potter-2011,Abrikosov-2012meth,Kang-2016}, which affect neither the superconducting transition temperature nor the zero-temperature gap. In the same way, in our model the SYK$_2$-FL phase is dominated by interactions that are static and nonmagnetic, i.e.\@, random coherent hopping between nearest-neighbors sites. Therefore, by a similar mechanism as for Anderson theorem, the weak-coupling regime of our theory is not affected by the disordered nature of the hopping, and our theory falls into the universal class of BCS models in such regime. As for $T_c$, we will later see that also the zero-temperature, zero-energy gap $\Delta_0=\lim_{T\rightarrow 0}\Delta(0)$ obeys the BCS formula, but the phase stiffness does not, as the latter is not protected by an Anderson-like argument.

In addition, although we retrieve the celebrated weak-coupling formula (\ref{eq:BCS_Tc_SYK2}) in our model, the origin of such expression is not the same as in standard BCS theory. To appreciate such a distinction, it is instructive to consider our pairing problem from a point of view which resembles the Thouless criterion for the pairing susceptibility \cite{Berthod-2018}. 
Employing the normal-state Dyson equation (\ref{eq:G_NS}) and Eq.\@ (\ref{eq:Sigma_Z}), and exploiting the even parity $Z(i \omega_n)=Z(-i \omega_n) \, \forall i \omega_n$, we have $\left[(\omega_n) Z(i \omega_n)\right]^{-2} \equiv G( i \omega_n) G(-i \omega_n)$, so that we can rewrite the linearized gap equation (\ref{eq:Polar_Z_lin}) as
\begin{align}\label{eq:gap_lin_GG}
    \Phi(i \omega_n)&=\bar{g}^2 k_B T \sum_{m=-\infty}^{+\infty} D(i \omega_n-i \omega_m) \nonumber \\ & \qquad\qquad\qquad\qquad\times G(i \omega_m)G(-i \omega_m) \Phi(i \omega_m)
    \notag\\
    &+\frac{z {t_0^2}}{2} \Phi(i \omega_n)G(i \omega_n)G(-i \omega_n). 
    \end{align}
The form (\ref{eq:gap_lin_GG}) highlights that there are two contributions to the anomalous self-energy: on-site scattering of fermions off bosons (first term) and inter-site coherent hopping (second term). We remark that the former term is essential to superconductivity, otherwise for $g=0$ the gap equation would not have nontrivial solutions, while the latter term is detrimental to pairing. 

Now, in the boson propagator (\ref{eq:Phon}) we neglect the $\Omega_n^2$ term and consider the static response, since we expect pairing at small energies; this approximation coincides with the one employed in the single-dot case \cite{Esterlis-2019}. For ${t_0^2}\omega_0/\bar{g}^2 \gg 1$, the fermionic propagator is approximately the SYK$_2$-FL one (\ref{eq:G_SYK2_NS}). Using this propagator, we realize that the many-body polarization bubble (\ref{eq:Polar_Z}) is of order $\bar{g}^2 k_B T/\left[z {t_0^2}(\omega_0)^2\right]^2 \ll 1$, so that we can neglect the bosonic self-energy in the current regime. Indeed, this is consistent with the numerical results for the normal-state bosonic propagator in the large-hopping regime. Hence, we have simply $D(i \Omega_n) \approx \omega_0^{-2}$. The gap equation (\ref{eq:gap_lin_GG}) becomes 
    \begin{align}
    \Phi(i \omega_n)&=\frac{\bar{g}^2 k_B T}{\omega_0^2} \sum_{m=-\infty}^{+\infty} G(i \omega_m)G(-i \omega_m) \Phi(i \omega_m)
    \notag\\&+\frac{z {t_0^2}}{2} \Phi(i \omega_n)G(i \omega_n)G(-i \omega_n). 
    \label{eq:Delta_weak_t0large}
    \end{align}
Further simplifying Eq.\@ (\ref{eq:Delta_weak_t0large}) with the help of Eq.\@ (\ref{eq:phi_tilde}) leads to
\begin{equation}\label{eq:lin_gap_GG}
\tilde{\Phi}(i \omega_n)=\bar{g}^2 k_B T/\omega_0^2 \sum_{m}\frac{P(i \omega_m)}{1-z t_0^2 P(\omega_m)/2} \tilde{\Phi}(i \omega_m),
\end{equation}
where $P(i \omega_m)=G(i \omega_m)G(-i \omega_m)$. If we now insert the SYK$_2$-FL expression (\ref{eq:G_SYK2_NS}) for $G(i \omega_n)$ into Eq.\@ (\ref{eq:lin_gap_GG}) and we expand the sum over $m$ for small $\omega_m$, we realize that the leading-order term scales as $1/\omega_m$, which yields the logarithmic infrared divergence generating the critical temperature (\ref{eq:BCS_Tc_SYK2}). Such instability would be absent if we still used the propagator (\ref{eq:G_SYK2_NS}) but we neglected the $t_0$-dependent term at the denominator of Eq.\@ (\ref{eq:lin_gap_GG}): this shows that the physical nature of the Cooper instability in the SYK$_2$-FL regime is the repeated coherent hopping of Cooper pairs, generated on one lattice site, onto adjacent sites; such process occurs with a diffusive dynamics, since $t_0$ follows a statistical distribution with zero mean. 
The order of magnitude of the critical temperature (\ref{eq:BCS_Tc_SYK2}) can also be estimated by considering Eq.\@ (\ref{eq:lin_gap_GG}) in the zero-temperature limit, where the sum over $m$ becomes an integral over $i \omega/(2 \pi)$, but cutting the integral extrema inferiorly by $k_B T_c$ and superiorly by $\omega_0$ \cite{Chubukov-2005}. Then, expanding the integrand in the low-energy limit and performing the integration over $\omega$, one retrieves $T_c \sim \omega_0 e^{-1/\bar{\lambda}}$, in agreement with Eq.\@ (\ref{eq:BCS_Tc_SYK2}). 

\subsection{Critical temperature at strong coupling: asymptotic Allen-Dynes formula}\label{Tc_imp_like}

Superconductivity at strong coupling is dominated by the intra-dot fermion-boson interaction, while hopping is negligible for $g^2/ \omega_0 \gg z t_0^2$ in the Hamiltonian (\ref{eq:H_SYK_phon_hop}): this is the impurity-like regime, analyzed in the normal state in Sec.\@ \ref{Fixed_point_imp}, where the single-dot physics described in Sec.\@ IVb of Ref.\@ \onlinecite{Esterlis-2019} holds. In essence, fermions are fully incoherent but are nevertheless able to precipitate in a partially coherent superconducting state of a strongly interacting Cooper-pair fluid. This coherence is preserved by a mechanism similar to Anderson's theorem \cite{Anderson-1959,Abrikosov-1959a,Abrikosov-1959b,Abrikosov-1961,Potter-2011,Abrikosov-2012meth,Kang-2016}, through which thermal fluctuations of static bosons, acting similarly to non-magnetic impurities \cite{Millis-1988,Abanov-2008}, deeply affect $Z(i \omega_n)$ and $\Phi(i \omega_n)$ -- see Eq.\@ (\ref{eq:G_imp_NS}) -- yet their effect cancels for $\Delta(i \omega_n)$ from Eq.\@ (\ref{eq:Phi_Delta}). Then, the latter pairing function is only influenced by the much weaker quantum fluctuations of bosons, which allows for the formation of time-reversal fermionic partners at $T=T_c$. 
\begin{figure*}[t]
\includegraphics[width=0.9\textwidth]{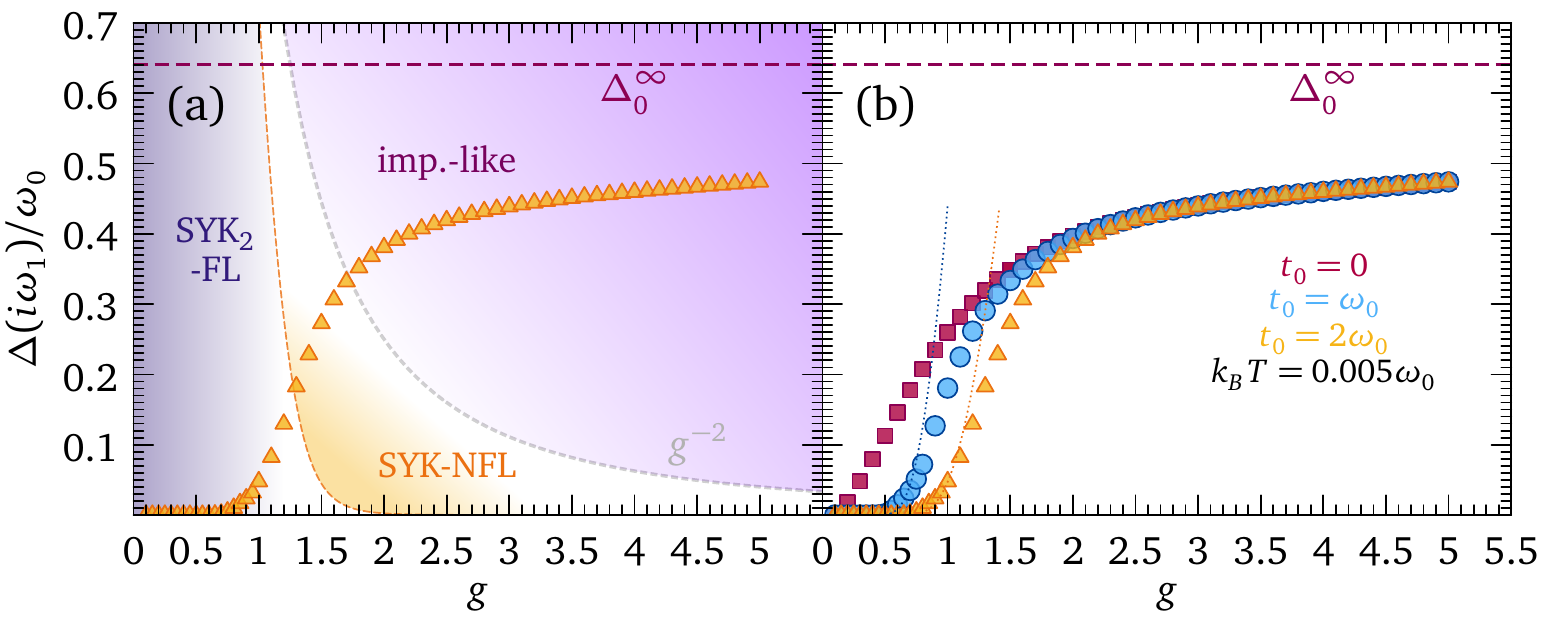}
\caption{\label{Delta_SYK_t0} Superconducting gap $\Delta(i \omega_1)/\omega_0$ at the first Matsubara frequency as a function of coupling $g$ for different hoppings $t_0$, at temperature $T=0.005 \omega_0/k_B$, stemming from the full numerical solution of the saddle-point equations (\ref{eq:Eliashberg_eph:hop}) in the superconducting state. (a) Results for $t_0/\omega_0=2$, showing the FL/NFL crossovers: the dashed blue curve is the SYK$_2$-FL/SYK-NFL crossover energy $\omega_c$, which is approximated at low energy by Eq.\@ (\ref{eq:omegac_analyt}), while the SYK-NFL/imp.-like crossover occurs for energies of the order of $g^{-2}$. $\Delta_0^\infty$ is the asymptotic zero-temperature zero-energy gap for $g \rightarrow+\infty$, i.e.\@, Eq.\@ (\ref{eq:imp_Delta0}). (b) Numerical results for $t_0/\omega_0=\left\{0,1,2\right\}$, and all other parameters equal to the ones of panel (a). The dashed curves show the analytical BCS formula (\ref{eq:SYK2-FL_FL_Delta0}) with the coupling constant (\ref{eq:lambda_bar}), for a given nonzero hopping. }
\end{figure*}

The maximum critical temperature can be analytically estimated in the limit $g \rightarrow +\infty$ and in the single-dot limit, using the fact that $k_B T_c \gg \omega_r$ \cite{Esterlis-2019} so that $\omega_r \approx 0$ in the gap equation. Such analysis, reported in Appendix \ref{app:Tc_implike_dot} for convenience, leads to the asymptotic value
\begin{equation}\label{eq:Tc_largeg}
T_c=0.111897 \omega_0/k_B. 
\end{equation}
Hence, in the large-coupling regime $T_c$ does not depend on $g$ for an isolated dot coupled to bosons. 

The asymptotic limit (\ref{eq:Tc_largeg}) continues to hold even with the lattice embedding of the Yukawa-SYK dots. This is confirmed both by the numerical results in Fig.\@ \ref{fig:Tc_SYK_t0}, and by the irrelevance of $t_0$ in the infinite-coupling limit where the impurity-like fixed point is stable \cite{Esterlis-2019}. This feature can be analytically confirmed, as proved in Appendix \ref{app:Tc_implike_lattice}: for $t_0>0$, the linearized gap equation is (\ref{eq:Phi_Z_tilde}) with $\tilde{\Phi}(i \omega_n)$ in accordance with Eq.\@ (\ref{eq:phi_tilde}); hopping also modifies the fermionic self-energy, as in Eq.\@ (\ref{eq:Sigma}), and the renormalized boson frequency, according to Eq.\@ (\ref{eq:omegar_gen}); yet, the modifications of the gap equation and of the fermion self-energy compensate each other in the $g \rightarrow +\infty$ limit, while we can still approximate $k_B T_c \gg \omega_r$ at finite hopping in the impurity-like regime, as the numerics confirm. Therefore, the maximum $T_c$ on the lattice, for a given coupling $g$, is still given by Eq.\@ (\ref{eq:Tc_largeg}). 
It is worth remarking that Eq.\@ (\ref{eq:Tc_largeg}) is also compatible with Allen-Dynes' solution $k_B T_c \approx 0.1827 \sqrt{\lambda} \omega_0$ of the Eliashberg equations, if we insert $\lambda=3/8$ for the coupling constant \cite{Allen-1975}. The applicability of this formula is usually confined to higher values of coupling, when the boson frequency to employ in the Eliashberg equations is the bare, unrenormalized one. However, such applicability is enabled by the extreme softening of the bosons for $g\gg1$ in our model, which lowers the required coupling to employ Allen-Dynes' result \cite{Esterlis-2019}. This argument will be shortly exploited to find an asymptotic strong-coupling value for the zero-temperature zero-energy gap. We then proceed in the next sections to explore the thermodynamic and spectroscopic properties deep into the superconducting state. 

\section{Zero-temperature superconducting gap}\label{Gap}

The energy gap in the single-particle spectrum is another characteristic spectroscopic figure of superconductivity, which is often invoked in the ratio with the critical temperature to assess the unconventionality of a superconducting state. In the case of our Yukawa-SYK model on the lattice, the pairing order parameter $\Delta(i\omega_n)=\Phi(i \omega_n)/Z(i\omega_n)$, which appears in the saddle-point equations (\ref{saddle_point_equations}), is the same as the single-particle gap in the spectral function (\ref{eq:A_def}), which appears in the superconducting state; see also Sec.\@ \ref{Spectral_func_SC}. This correspondence holds because the Yukawa interaction, which destroys fermionic quasiparticles, is also responsible for Cooper pairing. On the other hand, in similar models where pairing and non-Fermi liquidness are generated by distinct interactions, the order parameter and the spectral-function gap behave differently from each other \cite{Patel-2018}.
In our theory, the zero-temperature and zero-energy gap follows the BCS formula in the SYK$_2$-FL regime, it is quadratic in coupling $g$ in the SYK-NFL regime, while it asymptotically reaches a constant in the impurity-like regime for $g \rightarrow +\infty$. We proceed to discuss these results. 

In general, in our model the gap function $\Delta(i \omega_n)$ on the imaginary axis, or $\Delta(\omega)$ on the real axis, depends both on frequency and temperature \cite{Marsiglio-2008}. However, one can define a quantity that is independent from both aforementioned variables, as in 
\begin{equation}\label{eq:Delta0_def}
\Delta_0=\displaystyle \lim_{\substack{T \rightarrow 0 \\ \omega \rightarrow 0}} \Delta(\omega). 
\end{equation}
Technically, Eq.\@ (\ref{eq:Delta0_def}) refers to the real-axis gap function at zero temperature \cite{Marsiglio-2008}. However, one can also estimate $\Delta_0$ by extrapolation of the imaginary-axis solution $\Delta(i \omega_n)$ at very low temperature, down to $\omega=0$. 
$\Delta_0$ is the quantity that we will refer to, in discussing the underlying physics. 

In the SYK-NFL regime, reached when hopping is negligible, the gap $\Delta_0 \propto g^2$. This feature mirrors the analogous dependence of $k_B T_c \propto 0.156 g^2 \omega_0$ in this regime -- see Fig.\@ \ref{fig:Tc_SYK_t0}(a) and Appendix \ref{Tc_weak_dot} -- as the gap equation in both cases depends on the ratio $T/T_f$, with $2 \pi k_B T_f= c_1^{1/(2 \Delta)} \bar{g}^2/\omega_0^2\approx 0.1888 \bar{g}^2/\omega_0^2$ the crossover temperature between the free-fermion and SYK-NFL regimes for $t_0=0$ \cite{Esterlis-2019}. More precisely, from the numerical solution of Eqs.\@ (\ref{eq:Eliashberg_eph:hop}) we extract $\Delta_0 \approx \Delta(i\omega_1)$, approximated by the imaginary-axis gap at the first Matsubara frequency and at very low temperature $T=0.005 \omega_0/k_B$. As a matter of fact, the gap function at the first Matsubara frequency $\Delta(i \omega_1)$ is often regarded as an order parameter for superconducting calculations on the imaginary axis \cite{Marsiglio-2020}, and it has been shown to be of particular significance in quantum critical systems \cite{Wang-2016,Wu-2019}. This procedure yields $\Delta_0 \approx 0.5 g^2 \omega_0$ at small $g$. Therefore, in the SYK-NFL regime the gap-to-$T_c$ ratio results
\begin{equation}\label{eq:gaptoTc_SYK-NFL}
\frac{2 \Delta_0}{k_B T_c} \approx 6.414.
\end{equation}
The significant deviation of the ratio (\ref{eq:gaptoTc_SYK-NFL}) from the BCS value is another indication of the unconventional nature of pairing deep in the non-Fermi liquid state. 

Conversely, in the SYK$_2$-FL regime we can analytically show that the zero-temperature gap follows the BCS formula
\begin{equation}\label{eq:SYK2-FL_FL_Delta0}
\Delta_0=\frac{\omega_0}{\sinh(1/\bar{\lambda})} \approx 2 \omega_0 e^{-1/\bar{\lambda}} \, : \, \bar{\lambda} \ll 1. 
\end{equation}
Eq.\@ (\ref{eq:SYK2-FL_FL_Delta0}) is derived in Appendix \ref{App:gap0_SYK2-FL}, and it yields a good estimate for the zero-energy superconducting gap at very low temperatures: it is compared in Fig.\@ \ref{Delta_SYK_t0} with the gap $\Delta(i \omega_1)$ calculated on the imaginary axis at the first Matsubara frequency $\omega_1= \pi k_B T$, from the full numerical solution of the mean-field Eliashberg equations (\ref{eq:Eliashberg_eph:hop}). We should have $\Delta(i \omega_1)\approx \Delta_0$ for $T \rightarrow 0^+$, and indeed the evolution of the numerical $\Delta(i \omega_1)$ with $g$ at finite hopping is qualitatively consistent with Eq.\@ (\ref{eq:SYK2-FL_FL_Delta0}) in the SYK$_2$-FL regime. From the point of view of the gap-to-$T_c$ ratio, the SYK$_2$-FL state appears conventional, as it follows the prediction of the BCS model: 
\begin{equation}\label{eq:gaptoTc_SYK2_FL}
\frac{2 \Delta_0}{k_B T_c}= \frac{\pi}{e^\gamma}\approx 3.528.  
\end{equation}
As for $T_c$ discussed in Sec.\@ \ref{SYK_BCS_lattice}, the conventionality of Eq.\@ (\ref{eq:SYK2-FL_FL_Delta0}) is rationalized by invoking a version of Anderson's theorem, that confirms the insensitivity of the gap to disordered hoppings that act similarly to non-magnetic impurities \cite{Anderson-1959,Abrikosov-1959a,Abrikosov-1959b,Abrikosov-1961,Potter-2011,Abrikosov-2012meth,Kang-2016,Millis-1988,Abanov-2008}. However, although the ratio (\ref{eq:gaptoTc_SYK2_FL}) is conventional even in our fully disordered model, other thermodynamic quantities like the phase stiffness are not protected by an Anderson-like argument, and in fact they differ from the BCS prediction in the SYK$_2$-FL regime. We will return to this point in Sec.\@ \ref{Stiffness}.
\begin{figure*}[t]
\includegraphics[width=0.95\textwidth]{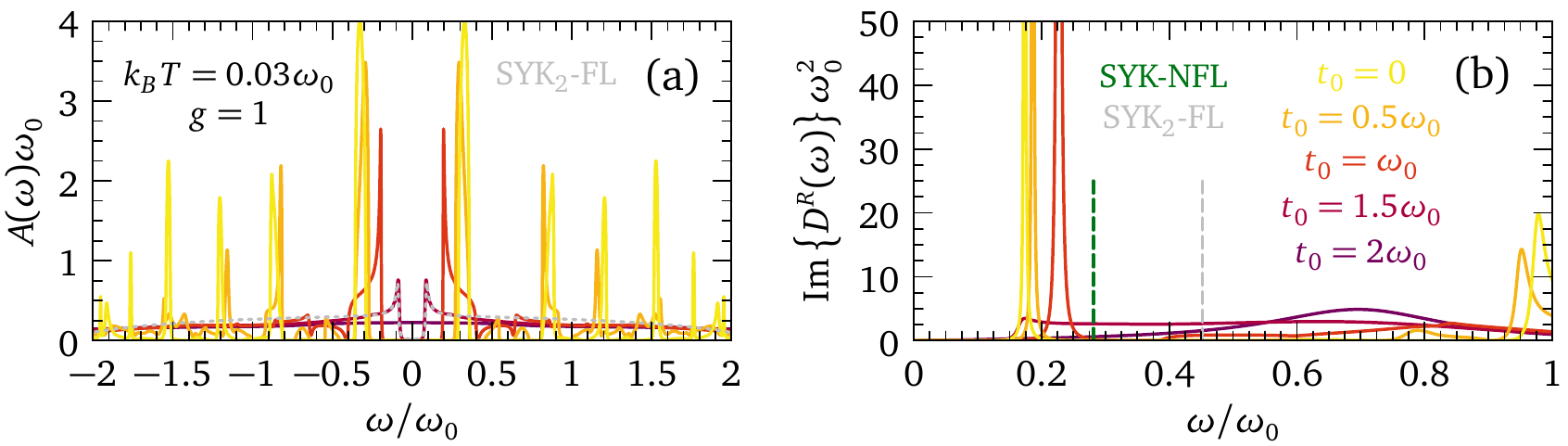}
\caption{\label{fig:spectr_NFL-SYK_g_SC} Numerically exact spectral function and bosonic propagator in the superconducting state, showing the crossover from the SYK-NFL to the SYK$_2$-FL regimes as a function of hopping. (a) Fermionic spectral function as a function of energy $\omega$, at temperature $k_B T/\omega_0=0.03$ and coupling $g=1$, and for different hoppings $t_0/\omega_0=\left\{0,0.5,1,1.5,2\right\}$. The dashed gray curve stems from the analytical approximation (\ref{eq:Sigma_SYK2_gapconst_real}) with $g=1$, $t_0/\omega_0=1.5$, and $\Delta \approx 0.089 \omega_0$. (b) Imaginary part of the bosonic propagator, for the same parameters as in panel (a). The dashed green vertical line is the zero-temperature numerical result for the renormalized boson frequency in SYK-NFL regime, and the dashed gray vertical line is the analytical approximation (\ref{eq:omegar_SC_SYK2_expl}) for the renormalized boson frequency in SYK$_2$-FL regime, using $\Delta \approx 0.089 \omega_0$, $g=1$, and $t_0=1.5 \omega_0$.}
\end{figure*}  

Finally, in the impurity-like regime $\Delta_0$ tends to a constant in the infinite-coupling limit, in the same way as $T_c$ does, as previously argued in Sec.\@ \ref{Tc_imp_like}. We find \cite{Esterlis-2019} 
\begin{equation}\label{eq:imp_Delta0}
\displaystyle\lim_{g \rightarrow +\infty}\Delta_0\approx 0.641, 
\end{equation}
which gives a highly unconventional gap-to-$T_c$ ratio of 
\begin{equation}\label{eq:gaptoTc_imp}
\frac{2 \Delta_0}{k_B T_c}\approx 11.456. 
\end{equation}
Such large values of $2 \Delta_0/(k_B T_c)$ are known to occur in the strong-coupling limit of Eliashberg theory for small phonon frequencies \cite{Carbotte-1990,Wu-2019b}. 

The gap $\Delta_0$ is the first excitation of the fermions in the superconducting state, so it should correspond to a spectroscopic gap in the single-particle density of states. Therefore, it is interesting to compare the results just obtained with the spectral functions calculated on the real axis, which is the subject of the next section. 

\section{Superconducting spectral functions}\label{Spectral_func_SC}

The spectral functions in the superconducting state for our Yukawa-SYK model (\ref{eq:H_SYK_phon_hop}) are obtained by solving the saddle-point equations (\ref{eq:Eliashberg_eph:hop}) on the real axis. While we relegate more technical aspects of our self-consistent method to Appendix \ref{App:num_real_axis}, here we comment on the physical aspects of the obtained solutions. In general, we expect a modification of the normal-state spectral functions discussed in Sec.\@ \ref{Normal_fixed}, due to the opening of a gap in the single-particle density of states. This modification should affect all regimes of the model, Fermi-liquid and non-Fermi liquid alike. 
In this section, we concentrate on the crossovers between the Fermi-liquid SYK$_2$-FL phase, and the non-Fermi liquid phases (SYK-NFL and impurity-like) as a function of hopping. A complementary discussion of the relation between the real-axis propagators and the optical conductivity can be found in the companion paper \cite{short-paper}. 

Fig.\@ \ref{fig:spectr_NFL-SYK_g_SC}(a) shows the exact fermionic spectral function (\ref{eq:A_def}) at $g=1$ and $T=0.03 \omega_0/k_B$, in the superconducting state, for a range of hopping spanning the crossover from the SYK-NFL to the SYK$_2$-FL regimes. For $t_0=0$ (single-dot limit), we notice a first sharp excitation corresponding to the spectroscopic gap, which indicates the development of coherent Bogoliubov quasiparticle excitations, even if the SYK-NFL normal state is incoherent \cite{Esterlis-2019}. The coherence peaks are followed by consecutive high-energy structures. As discussed in Ref.\@ \onlinecite{Esterlis-2019}, these shakeoff peaks are different from the polaronic states due to strong electron-phonon coupling found in wide-band Eliashberg theory \cite{Marsiglio-2008,Berthod-2018}, in that they correspond to self-trapping states of excited fermionic quasiparticles due to the pairing field generated by the other fermions \cite{Marsiglio-1991,Karakozov-1991,Combescot-1994}. Apart from the unconventional gap-to-$T_c$ ratio discussed in Sec.\@ \ref{Gap}, the shakeoff resonances are another indication of the unconventional character of the SYK-NFL paired state. They could be related to peak-dip-hump features observed in angle-resolved photoemission spectroscopy (ARPES) on cuprate superconductors \cite{Dessau-1991,Campuzano-1996,Loeser-1997,Shen-1997,Fedorov-1999,Kaminski-2016,Orenstein-2000,Hashimoto-2014}, and they could potentially be revealed through the AC Josephson effect \cite{Esterlis-2019}. Similar resonances in the superconducting spectral function are retrieved in a similar disordered model with on-site attractive Hubbard interactions, spin exchange, as well as single-particle and Cooper-pair hopping \cite{Li-2023}.  
As we increase hopping, we initially observe that the spectroscopic gap around zero energy understandably decreases, due to the detrimental effect of $t_0$ on the on-site fermion-boson coupling, but also the shakeoff peaks change and split into several structures -- see orange curve for $t_0=0.5 \omega_0$ in \ref{fig:spectr_NFL-SYK_g_SC}(a); the alteration in shape of the peaks makes sense, since coherent hopping competes with electron-boson coupling, thus affecting the self-trapping pairing field. Further increasing $t_0$, the shakeoff resonances progressively disappear altogether, which indicates the crossover from strong-coupling pairing in the non-Fermi liquid regime, to pairing in a disordered Fermi liquid with weak electron-boson coupling. In the SYK$_2$-FL regime, the spectral function is well approximated by employing the analytical expression
\begin{equation}\label{eq:Sigma_SYK2_gapconst_real}
G^R(\omega)=\frac{\omega+i 0^+- \sqrt{(\omega+i 0^+)^2+\frac{2 z t_0^2 (\omega+i 0^+)^2}{\Delta^2-(\omega+i 0^+)^2}}}{z t_0^2}
\end{equation}
for the retarded fermionic propagator. Eq.\@ (\ref{eq:Sigma_SYK2_gapconst_real}) yields the dashed gray curve in Fig.\@ \ref{fig:spectr_NFL-SYK_g_SC}(a) for $t_0=1.5 \omega_0$, which is in excellent agreement with the exact numerical output for the same hopping. Here the analytical propagator Eq.\@ (\ref{eq:Sigma_SYK2_gapconst_real}) has been complemented by $\Delta \approx \Delta_0 \approx 0.089 \omega_0$ stemming from the numerical solution on the imaginary axis, for $g=1$ and $t_0=1.5 \omega_0$; an artificial broadening $\omega \mapsto \omega+i 10^{-2.3}$ has also been introduced to simulate the temperature smearing of the coherence peaks. 
Further increasing hopping, the critical temperature of the system drops below the used value of temperature, in accordance with Sec.\@ \ref{SYK_BCS_lattice}, until the transition to the normal SYK$_2$-FL state is made; see purple curve for $t_0=2$ in Fig.\@ \ref{fig:spectr_NFL-SYK_g_SC}(a), for which the corresponding $T_c \approx 0.021 \omega_0/k_B <T$. Thus, in this limit we recover the normal-state results of Sec.\@ \ref{SYK2-FL_fixed_point}. 
\begin{figure*}[t]
\includegraphics[width=0.95\textwidth]{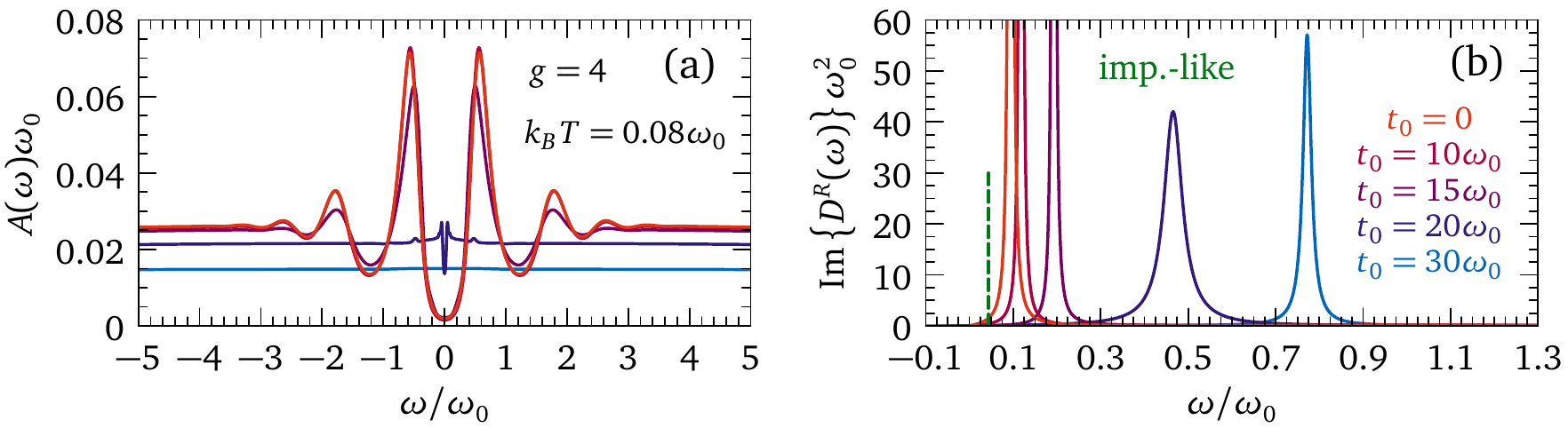}
\caption{\label{fig:spectr_imp_g_SC} Numerically exact spectral function and bosonic propagator in the superconducting state, showing the crossover from the impurity-like to the SYK$_2$-FL regimes as a function of hopping. (a) Fermionic spectral function as a function of energy $\omega$, at temperature $k_B T/\omega_0=0.08$ and coupling $g=4$, and for different hoppings $t_0/\omega_0=\left\{0,10, 15, 20, 30\right\}$. (b) Imaginary part of the bosonic propagator, for the same parameters as in panel (a). The dashed green vertical line is the zero-temperature analytical expression (\ref{eq:omegar_SC_imp}) for the renormalized boson frequency in the superconducting impurity-like phase.}
\end{figure*}

Fig.\@ \ref{fig:spectr_NFL-SYK_g_SC}(b) displays the exact imaginary part of the boson propagator, computed numerically on the real axis, for the same parameter as in Fig.\@ \ref{fig:spectr_NFL-SYK_g_SC}(a). For $t_0=0$, the renormalized boson frequency checked both with the real-axis and imaginary-axis codes is $\omega_r^{\rm SC}\approx 0.281 \omega_0$. This value corresponds to the dashed green line in Fig.\@ \ref{fig:spectr_NFL-SYK_g_SC}(b), but it does not correspond to the visible peaks in the $t_0=0$ yellow curve. Instead, the peaks seen there reflect the self-trapped states previously noticed in the fermionic spectral function. Thus, the bosons as well are affected by the self-trapping pair field. We can understand the discrepancy between $\omega_r^{\rm SC}$ and the lowest-energy peak in $\mathrm{Im}\left\{D^R(\omega)\right\}$ by referring to the Dyson equation (\ref{eq:Phon}) anlytically continued to $i \Omega_n \rightarrow \omega+i 0^+$: since at low energies the numerical boson self-energy schematically reads $\Pi(\omega)\approx \omega_0^2-(\omega_r^{\rm{SC}})^2+ A \omega^2$, the boson propagator becomes $D^R(\omega)\approx 1/\left[-(1+A)(\omega+i0^+)^2+(\omega_r^{\rm{SC}})^2\right]$. This means that, while the static part of the boson self-energy renormalizes the natural frequency to $\omega_r^{\rm{SC}}$, the first peak of the boson propagator is in fact shifted to $\omega \approx \omega_r^{\rm SC}/\sqrt{1+A} < \omega_r^{\rm SC}$. Higher-energy peaks in $\Pi(\omega)$ produce more resonances in $\mathrm{Im}\left\{D^R(\omega)\right\}$. 
Increasing hopping, the self-trapping structures move in energy, until they disappear on the weak-coupling Fermi-liquid side of the crossover, and a single broad boson peak appears for $t_0=2 \omega_0$ in the normal state when $T_c<T$; see corresponding curve in Fig.\@ \ref{fig:spectr_NFL-SYK_g}(b). One can qualitatively estimate the stiffening of the renormalized boson frequency in SYK$_2$-FL regime at $T=0$:
\begin{equation}\label{eq:omegar_SC_SYK2_expl}
(\omega_r^{SC})^2\approx \omega_0^2-\frac{8 \bar{g}^2}{6 \pi \omega_0 z t_0^2} \left(-3 \Delta^2 +  2 \sqrt{2 z} \omega_0 t_0\right).
\end{equation}
Notice that Eq.\@ (\ref{eq:omegar_SC_SYK2_expl}) consistently reduces to Eq.\@ (\ref{eq:omegar_0_T0}) for $\Delta=0$. It is derived in Appendix \ref{App:omegar_SC_SYK2}. Using $\Delta \approx \Delta_0 \approx 0.089 \omega_0$, $g=1$ and $t_0=1.5 \omega_0$ in Eq.\@ (\ref{eq:omegar_SC_SYK2_expl}), we obtain the dashed gray line in Fig.\@ \ref{fig:spectr_NFL-SYK_g_SC}(b), in fairly good agreement with the very broad peak in the numerical $\mathrm{Im}\left\{D^R(\omega)\right\}$ for $t_0=1.5 \omega_0$. 

We now investigate the crossover at higher $g$ between the impurity-like and SYK$_2$-FL regimes. Fig.\@ \ref{fig:spectr_imp_g_SC}(a) shows the evolution of the fermionic spectral function with hopping, across the impurity-like/SYK$_2$-FL crossover, for $g = 1$ and $k_B T=0.08 \omega_0$, and in the superconducting
state. The curve for $t_0=0$ is consistent with Fig.\@ 7 of Ref.\@ \onlinecite{Esterlis-2019}: the spectral density is nonzero inside the spectroscopic gap at this temperature, and at higher temperatures the gap is progressively filled without changing the position of the lowest-energy coherence peaks. Shakeoff features, similar to the one discussed for the SYK-NFL phase of Fig.\@ \ref{fig:spectr_NFL-SYK_g_SC}, develop at higher energies, and their shape is less sensitive to an increase in hopping in the impurity-like regime, compared to the SYK-NFL state. When $t_0 \gtrapprox g^2$, consistently with the crossover energy $\omega_c'$ in Sec.\@ \ref{Normal_cross}, the crossover to the SYK$_2$-FL phase occurs. The spectroscopic gap fills and decreases its width -- see blue curve for $t_0/\omega_0=20$ -- but some remnants of the shakeoff peaks are still visible, and the shape of the gap cannot be well approximated by the BCS-like piecewise-constant expression (\ref{eq:gap_const}). Therefore, the constant-gap approximation does not capture the spectral function oscillations and the general shape of the spectral function in the impurity-like to SYK$_2$-FL crossover. However, the same approximation will prove useful to understand the general trend of thermodynamic quantities like the phase stiffness, in the impurity-like regime; see Sec.\@ \ref{Stiffness}. 
Further increasing hopping, for $t_0/\omega_0=30$ we have $T>T_c$, and the system makes the transition to the normal state: the spectral function is the same as the corresponding normal-state curve for $t_0/\omega_0=30$ in Fig.\@ \ref{fig:spectr_imp_g}. 

Fig.\ref{fig:spectr_imp_g_SC}(b) reports the imaginary part of the bosonic propagator, for the same parameters as in Fig.\ref{fig:spectr_imp_g_SC}(a). When $g^2 \gg t_0^2$, the bosons are well defined but soft and essentially static excitations. The dashed green vertical line shows the analytical estimation
\begin{equation}\label{eq:omegar_SC_imp}
\omega_r^{SC}=\frac{\omega_0}{2}\left(\frac{3 \pi}{8}\right)^2 \frac{1}{g^2}, 
\end{equation}
for $g=4$, which is valid at $T=0$ in the impurity-like regime; see App.\@ \ref{Imp_omegar_SC} and Ref.\@ \onlinecite{Esterlis-2019}. The boson peak for $t_0=0$ tends to the frequency (\ref{eq:omegar_SC_imp}) in the vanishing temperature limit. Increasing hopping, $\omega_r^{SC}$ stiffens, in analogy to what observed in the SYK-NFL/SYK$_2$-FL crossover of Fig.\@ \ref{fig:spectr_NFL-SYK_g_SC}. The peak for $t_0/\omega_0=30$ corresponds to the normal-state result in Fig.\@ \ref{fig:spectr_imp_g}.

A prominent effect of hopping on the superconducting spectral functions in Figs.\@ \ref{fig:spectr_NFL-SYK_g_SC} and \ref{fig:spectr_imp_g_SC} is the modification of the height and shape of the coherence peaks at energies $\omega \approx \pm \Delta_0$, which signal the presence of coherent Bogoliubov quasiparticles in the paired state. To investigate the dependence of the coherent spectral weight on coupling and hopping, in the next section we focus specifically on the zero-energy quasiparticle weight. 

\section{Superconducting quasiparticle weight}\label{Z_SC}

The spectral functions analyzed in Sec.\@ \ref{Spectral_func_SC} highlight the presence of coherence peaks, which decrease in magnitude as the ratio $g^2 \omega_0/(\sqrt{z} t_0)$ decreases; see Figs.\@ \ref{fig:spectr_NFL-SYK_g_SC}(a) and \ref{fig:spectr_imp_g_SC}(a). Therefore, hopping considerably affects quasiparticles in the superconducting state. To further elucidate this phenomenon, here we calculate the quasiparticle weight $Z_{\rm qp}$ in the superconducting state as a function of fermion-boson coupling, in the different regimes of our model. 

To extract the quasiparticle residue $Z_{\rm qp}$, we use its definition in terms of the retarded self-energy \cite{Berthod-2018}: 
\begin{equation}\label{eq:Z_def}
\frac{1}{Z_{\rm qp}}=1-\left.\frac{\partial \mathrm{Re}\Sigma^R(\omega)}{\partial \omega}\right|_{\omega \rightarrow 0}. 
\end{equation}
The quantity (\ref{eq:Z_def}) can be directly extracted from the real-axis solution of the Eliashberg equations (\ref{eq:Eliashberg_eph:hop}). Moreover, an equivalent estimation can be attained using the imaginary-axis code in the $T \rightarrow 0^+$ limit, since we also have
\begin{equation}\label{eq:Z_ImSigma}
\frac{1}{Z_{\rm qp}}=\left.Z(i \omega)\right|_{\omega \rightarrow 0},
\end{equation}
where $Z(i \omega)$ is the zero-temperature limit of the dynamical weight function on the imaginary axis, defined from Eq.\@ (\ref{eq:Sigma_Z}). Since the convergence of the self-consistent loop (\ref{eq:Eliashberg_eph:hop}) at very low temperatures is less difficult on the imaginary axis than on the real axis, we employ Eq.\@ (\ref{eq:Z_ImSigma}) and approximate $\left.Z(i \omega)\right|_{\omega \rightarrow 0} \approx Z(i \omega_1)$ at the first Matsubara frequency $\omega_1=\pi k_B T$, at very low temperatures $T \ll T_c$. We checked the consistency of this approach with the definition (\ref{eq:Z_def}) using our real-axis code. The method (\ref{eq:Z_ImSigma}) is sufficiently accurate, since we are mainly interested in qualitative trends as a function of $g$ and $t_0$. 

\begin{figure}[t]
\includegraphics[width=0.95\columnwidth]{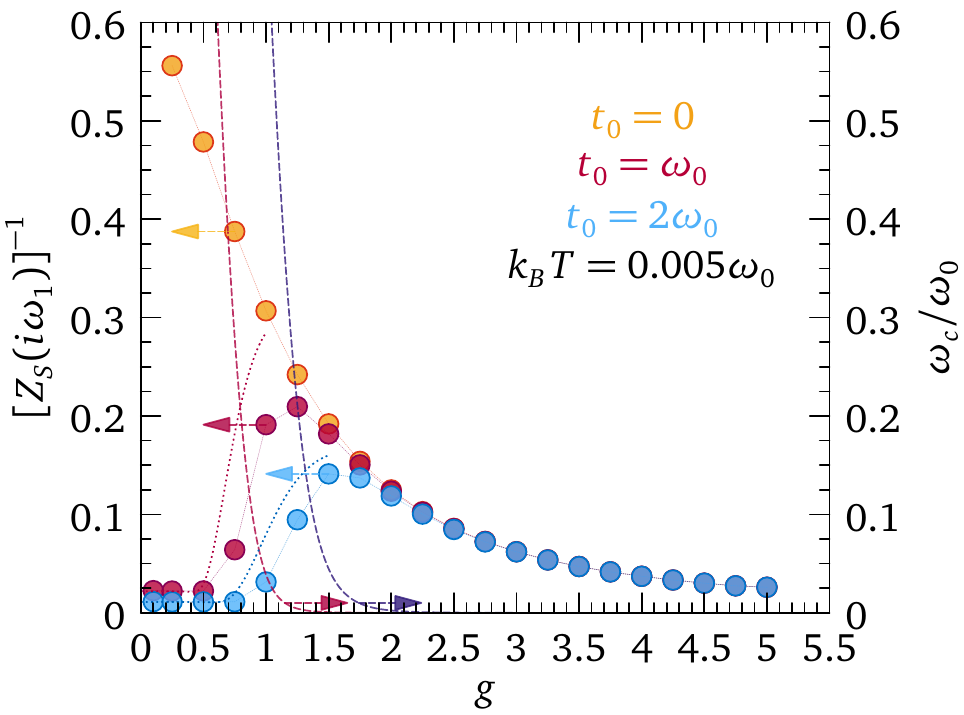}
\caption{\label{fig:Z_SC} Quasiparticle weight $Z_{\rm qp}$ deep into the superconducting state, estimated on the imaginary axis as $Z_{\rm qp}\approx \left[Z(i\omega_1)\right]^{-1}$, where $Z(i \omega)$ follows Eq.\@ (\ref{eq:Z_ImSigma}) and $\omega_1=\pi k_B T$. Data points correspond to the numerical solution of the saddle-point equations (\ref{eq:Eliashberg_eph:hop}), and their values are marked on the left $y$ axis, as indicated by arrows. The dashed curves correspond to the FL/NFL crossover energy $\omega_c$ at given hopping from Sec.\@ \ref{Normal_cross}, and their values are marked on the right $y$ axis, as signaled by arrows. The dotted lines correspond to the analytical approximation (\ref{eq:Z_T_implike}) with the gap approximated by Eq.\@ (\ref{eq:SYK2-FL_FL_Delta0}).}
\end{figure}

Fig.\@ \ref{fig:Z_SC} shows $1/Z(i\omega_1)$ stemming from the exact numerical solution of Eqs.\@ (\ref{eq:Eliashberg_eph:hop}), as a function of coupling $g$ and for different hoppings. In the single-dot limit (orange circles for $t_0=0$), the quasiparticle weight is a decreasing function of coupling. However, it remains finite at all couplings in the superconducting state. This finiteness is in contrast with the normal-state SYK-NFL expression (\ref{eq:NFL_Z}) stemming from Eqs.\@ (\ref{eq:Sigma_NFL_SYK}), (\ref{eq:Sigma_Z}), and (\ref{eq:Z_ImSigma}):
\begin{equation}\label{eq:Z_SYK-NFL_NS}
Z_{\rm qp}= \lim_{\omega \rightarrow 0}\left(\displaystyle  1+c_1 \left|\frac{g}{\omega}\right|^{2 \Delta}\right)^{-1}=0. 
\end{equation}
The normal-state result (\ref{eq:Z_SYK-NFL_NS}), which corresponds to the branch-cut divergence of the real-axis SYK-NFL fermion propagator in Fig.\@ \ref{fig:spectr_NFL-SYK_g}(a), shows the absence of quasiparticles in the non-Fermi liquid regime. In contrast, the superconducting phase generated from the SYK-NFL regime always has a degree of fermionic coherence signaled by a finite weight $Z_{\rm qp}$. 

The results at finite hopping in the SYK$_2$-FL and impurity-like regimes are best discussed together, due to the formal equivalence of the respective propagators and spectral functions, pointed out in Sec.\@ \ref{Fixed_point_imp} for the normal state. In the superconducting phase this equivalence persists, as derived in Appendix \ref{Z_imp}; see Eqs.\@ (\ref{eq:Z_SYK2_gapconst_expl}) and (\ref{eq:Z_T0_implike}). Therefore, we can write an imaginary-axis expression that interpolates between the Fermi-liquid and impurity-like regimes as follows:
\begin{equation}\label{eq:Z_T_implike}
Z(i \omega_n)\approx\frac{1}{2} \left[1+ \sqrt{1+\frac{\tilde{\Omega}^2}{\omega_n^2+\Delta^2}}\right],
\end{equation}
where
\begin{equation}\label{eq:tilde_Omega_interp}
\tilde{\Omega}=\sqrt{2 z {t_0^2}+\left(\frac{16 g^2 \omega_0}{3 \pi}\right)^2}. 
\end{equation}
Here $\Delta$ is again approximated by Eq.\@ (\ref{eq:gap_const}). 
We can exploit Eq.\@ (\ref{eq:Z_T_implike}) to qualitatively analyze the whole SYK$_2$-FL/impurity-like crossover. We will continue to employ this qualitative analysis in Sec.\@ \ref{Stiffness}, where we will discuss the condensate phase stiffness. Naturally, Eq.\@ (\ref{eq:Z_T_implike}) is most inaccurate in the SYK-NFL regime which is not captured by this expression. 
Using Eqs.\@ (\ref{eq:Z_T_implike}) and (\ref{eq:Z_ImSigma}), we obtain the superconducting quasiparticle weight 
\begin{equation}\label{eq:Z_imag_imp_SC_cross}
Z_{\rm qp}= \frac{2}{1+\sqrt{\tilde{\Omega}^2/\Delta^2}}. 
\end{equation}

In the superconducting SYK$_2$-FL regime, $\tilde{\Omega}^2\approx 2 z t_0^2$ is independent from $g$, while the gap follows the BCS-like expression (\ref{eq:SYK2-FL_FL_Delta0}) which has an inverse exponential dependence on $g^2/t_0$ through the coupling constant (\ref{eq:lambda_bar}).
Therefore, we have in Fermi-liquid regime
\begin{equation}\label{eq:Z_imag_SYK2_FL_SC_expl}
Z_{\rm qp}= \frac{2}{1+\sqrt{(2 z t_0^2)/(2 \omega_0 e^{-1/\bar{\lambda}})^2}}. 
\end{equation}
The exponential suppression of the coherent weight in the SYK$_2$-FL regime is consistent with the numerics in Fig.\@ \ref{fig:Z_SC}: the numerical data points are in qualitative agreement with Eq.\@ (\ref{eq:Z_T_implike}), which gives the dotted curves and where the gap is approximated by the SYK$_2$-FL expression (\ref{eq:SYK2-FL_FL_Delta0}). 
Increasing $g$, we cross over to the impurity-like regime, where the superconducting weight is well captured by the expression derived from Eqs.\@ (\ref{eq:G_implike_SC_T}), (\ref{eq:G_NS}), (\ref{eq:Sigma_Z}), and (\ref{eq:Z_ImSigma}):
\begin{align}\label{eq:Z_imag_imp_SC}
Z_{\rm qp}&=\left[ \displaystyle \lim_{\omega \rightarrow 0} \frac{1}{2}\left(1+\sqrt{1+\frac{\Omega_0^2}{\omega_n^2+\Delta^2}}\right)\right]^{-1} \notag \\ &= \frac{2}{1+\sqrt{\Omega_0^2/\Delta^2}}, 
\end{align}
where $\Omega_0$ follows Eq.\@ (\ref{eq:Omega_0}), and the gap $\Delta$ in Eq.\@ (\ref{eq:Z_imag_imp_SC}) stems from the frequency-independent approximation (\ref{eq:gap_const}). From Eqs.\@ (\ref{eq:Z_imag_imp_SC}) and (\ref{eq:Omega_0}) we see that $Z_{\rm qp} \propto g^{-2}$ at large $g$ \cite{Esterlis-2019}, since the gap is nearly independent on $g$ in this regime and it approaches the constant given by Eq.\@ (\ref{eq:imp_Delta0}). This evolution is indeed seen in Fig.\@ \ref{fig:Z_SC} at large coupling $g$, where the data points for finite hopping collapse on the single-dot result since hopping becomes negligible in this limit. 
Notice that even in the impurity-like regime the normal-state quasiparticle weight would be $Z=0$. This is derived from Eq.\@ (\ref{eq:Z_imag_imp_SC}) where the $\Delta \rightarrow 0$ limit is taken \emph{before} the $\omega \rightarrow 0$ limit, or equivalently from Eq.\@ (\ref{eq:Z_imp}). This vanishing reflects the incoherence of the impurity-like fermions, similarly to the SYK-NFL phase. Such incoherence is modified in the superconducting state, and a finite quasiparticle weight (\ref{eq:Z_imag_imp_SC}) appears. 

At finite hopping, the weight $Z_{\rm qp}\approx 1/Z(i \omega_1)$ is maximum around the SYK$_2$-FL/SYK-NFL crossover energy $\omega_c$, discussed in Sec.\@ \ref{Normal_cross} and marked by the dashed red and blue curves in Fig.\@ \ref{fig:Z_SC} for $t_0=\left\{1,2\right\} \omega_0$ (see Fig.\@ \ref{fig:omegac}). On the other hand, $Z_{\rm qp}$ is exponentially suppressed at small coupling because the gap is exponentially smaller in SYK$_2$-FL regime, while $Z_{\rm qp} \propto g^{-2}$ at large $g$ because of the quadratic dependence of $\Omega_0$ from Eq.\@ (\ref{eq:Omega_0}) on coupling. Such non-monotonic evolution with fermion-boson interaction is reminiscent of the behavior of the relative weight of the coherence peak in cuprate superconductors with respect to doping \cite{Shen-1997,Feng-2000}. 
Inspired by this analogy, the question arises of whether other thermodynamic quantities in the superconducting state share a similar non-monotonic dependence on coupling. In particular, one of the most crucial observables is the phase stiffness, which is the signature of a true superconducting ground state. Hence, the next section is devoted to the derivation and the analysis of the phase stiffness for our Yukawa-SYK model. 

\section{Phase stiffness}\label{Stiffness} 

A compelling question is whether the low-temperature condensed phase of our lattice model, so far analyzed in Secs.\@ \ref{Tc_SYK}-\ref{Z_SC}, possesses a finite phase stiffness, i.e.\@, whether such phase shows perfect diamagnetism and hence real superconductivity.  
From a thermodynamic standpoint, the phase stiffness of a Cooper-pair electronic condensate corresponds to the rigidity of the system, quantified by the second derivative of the grand potential $\Omega^{{\rm sc}}$ with respect to global shifts of the phase of the condensate wave function \cite{Fisher-1973,Taylor-2006,Taylor-2007}. 
However, since condensing electrons in a superconductor are electrically charged, phase fluctuations couple to charge and can thus be excited through external electromagnetic fields. Hence, the phase stiffness of a superconducting state also corresponds to the rigidity with respect to the penetration of a static magnetic field inside the superconducting specimen: the latter is a perfect diamagnet, i.e., it expels the applied field from its bulk, in accordance with Mei{\ss}ner-Ochsenfeld effect; the expulsion can be either complete (pure state in type-I superconductors below the critical field, or in type-II superconductors below the lower critical field), or partial (mixed state or Abrikosov lattice, between the lower and the upper critical fields in type-II superconductors) \cite{Tinkham-1996int}. 

In the following, we employ both aforementioned definitions of the phase stiffness, by first calculating the low-energy action, the propagator and the total energy difference for charge-coupled phase fluctuations that perturb the saddle-point solution of Eqs.\@ \eqref{saddle_point_equations} -- see Appendix \ref{Action_for_charge_fluctuations} -- then by coupling such phase fluctuations to a vector potential via Peierls substitution, and finally by computing the electrodynamic linear response at equilibrium. The latter is encoded in the electromagnetic kernel tensor $\mathcal{K}_{\alpha\beta}(\vec{q},\omega)$, which corresponds to the response function for a current density with respect to an external vector potential $\vec{A}(\vec{q},\omega)$: 
    \begin{align}
        \delta J_\alpha(\vec{q},\omega)& \equiv {J}_{\text{tot},\alpha}(\vec{q},\omega)-{J}_{\text{ext},\alpha}(\vec{q},\omega) \notag \\ & = \mathcal{K}_{\alpha\beta}(\vec{q},\omega)\, A_{\beta}(\vec{q},\omega).
        \label{definition_of_SC_kernel}
    \end{align}
Here $\delta \vec{J}(\vec{q},\omega)$ is the induced current density, which depends on frequency $\omega$ and wave vector $\vec{q}$. The latter is defined from the Fourier transform of the fluctuating lattice phase field $\phi(\vec{r},t)$ -- see App.\@ \ref{Action_for_charge_fluctuations}. The quantities ${J}_{\text{tot},\alpha}(\vec{q},\omega)$ and ${J}_{\text{ext},\alpha}(\vec{q},\omega)$ are the total and external currents, respectively.

\begin{figure}[t]
\includegraphics[width=0.9\columnwidth]{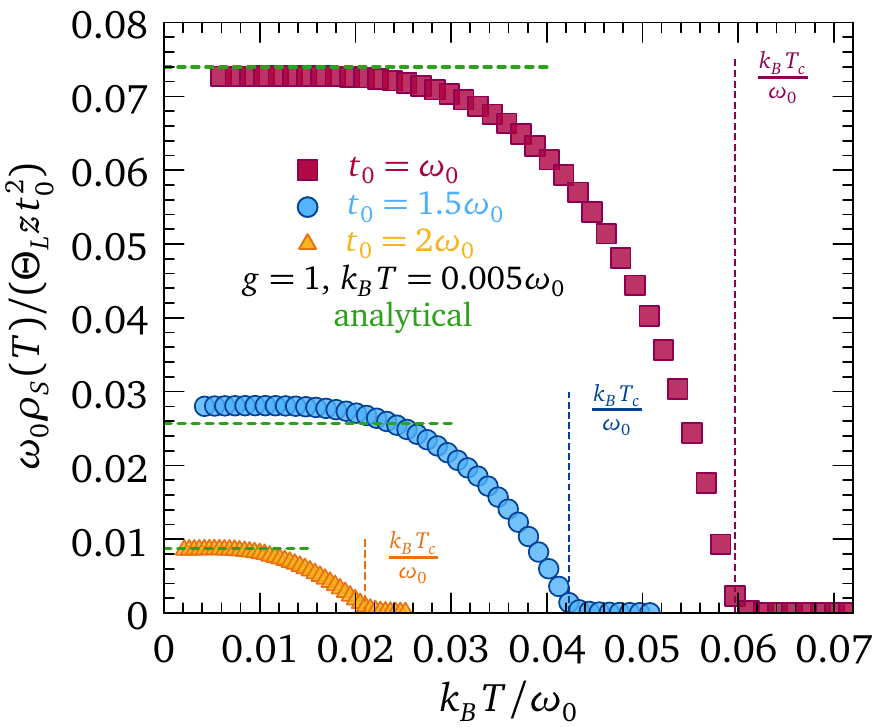}
\caption{\label{rhoS_SYK_T} Temperature dependence of the normalized phase stiffness $\omega_0\rho_S(T)/(\Theta_L z t_0^2)$, at coupling $g=1$, temperature $T=0.005 \omega_0/k_B$ and for different hopping $t_0$, numerically calculated from the saddle-point equations (\ref{eq:Eliashberg_eph:hop}) in the superconducting state. $\Theta_L$ obeys Eq.\@ (\ref{eq:Theta_L}). The normalized pairing temperature $k_B T_c/\omega_0$ is marked by dashed vertical lines for each hopping, and it corresponds to the numerical solution of the linearized gap equation, in accordance with Fig.\@ \ref{fig:Tc_SYK_t0}. The dashed horizontal green lines mark the analytical estimation of the zero-temperature stiffness from Eq.\@ (\ref{eq:rhoS_implike_cross_expl0}). }
\end{figure}

\begin{figure*}[t]
\includegraphics[width=0.8\textwidth]{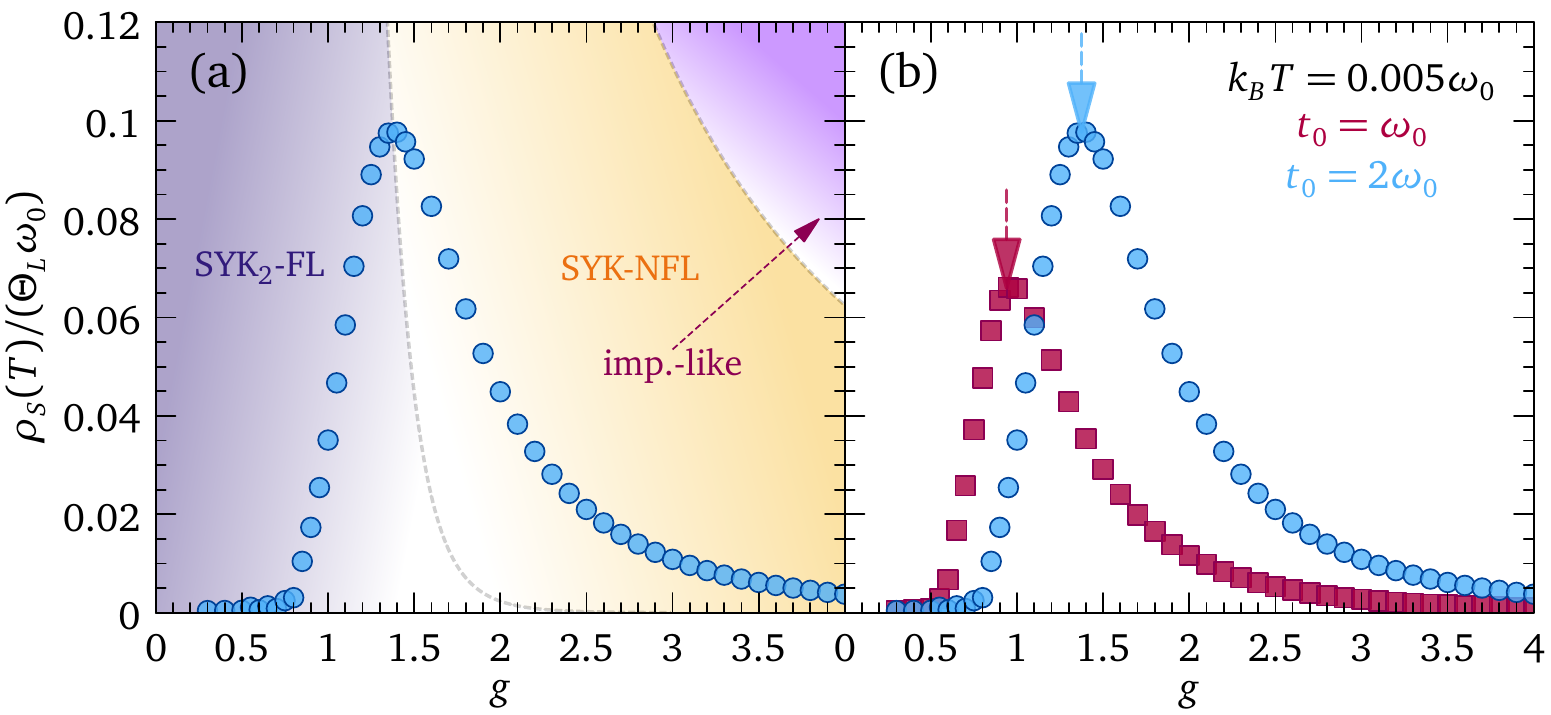}
\caption{\label{rhoS_SYK_g} Normalized superconducting stiffness $\rho_S(T)/(\Theta_L \omega_0)$ as a function of coupling $g$ for different hoppings $t_0$, where $\Theta_L$ obeys Eq.\@ (\ref{eq:Theta_L}). (a) Summary of regimes for the stiffness: Data points correspond to  the numerical solution of the saddle-point equations (\ref{eq:Eliashberg_eph:hop}) in the superconducting state, for temperature $T=0.005 \omega_0/k_B$ and hopping $t_0/\omega_0=2$. The maximum stiffness is found at the SYK$_2$-FL/SYK-NFL crossover energy $\omega_c$, marked by the dashed gray line and corresponding to the estimation in Sec.\@ \ref{Normal_cross}. (b) Comparison of results for different hoppings $t_0/\omega_0=\left\{1,2\right\}$, with all other parameters as in panel (a). The arrows mark the coupling $g$ at which the SYK$_2$-FL/SYK-NFL crossover occurs for a given hopping.}
\end{figure*}

Here we assume a static magnetic field $\vec{B}(\vec{r})$, to derive the Meissner effect. A more general derivation of the finite-frequency electromagnetic response leads to the optical conductivity reported in the companion paper \cite{short-paper}. 
In the static limit $\omega=0$, the intrinsic magnetic response is fully determined by the magnetization via $\delta\vec{J}(\vec{r})=\vec{\nabla}\times\vec{M}(\vec{r})$, where by definition $\vec{M}(\vec{r})=\vec{B}(\vec{r})-\mu_0 \vec{H}(\vec{r})$, with $\vec{B}(\vec{r})$ and $\vec{H}(\vec{r})$ magnetic flux density and magnetic field strength, respectively. $\mu_0=1.25663706212 \times 10^{-6}$ N/A$^2$ is the magnetic permeability of vacuum (in SI units). To make contact with Eq. \eqref{definition_of_SC_kernel}, we Fourier-transform the magnetization to the space of wave vectors $\vec{q}$ and frequencies $\omega$, and we employ the constitutive relation $\mu_0 \vec{H}(\vec{q},\omega)=\sum_{\alpha \beta} B_\beta(\vec{q},\omega) \hat{u}_\alpha/\mu_{\beta \alpha}(\vec{q},\omega)$, where $\hat{u}_\alpha$ is the unit vector in the direction $\alpha$, and $\underline{\underline{\mu}}(\vec{q},\omega)=\left\{\mu_{\alpha\beta}(\vec{q},\omega)\right\}$ constitutes the magnetic permeability tensor. For $\omega=0$, we thus obtain
\begin{equation}\label{eq:magnetiz_A}
\vec{M}(\vec{q},0) =i \vec{q} \times \left\{ \sum_{\alpha,\beta} \hat{u}_\alpha \left[\delta_{\beta \alpha}-\frac{1}{\mu_{\beta \alpha}}\right] B_\alpha(\vec{q},0)\right\}, 
\end{equation}
where by definition the vector potential satisfies $\vec{B}(\vec{q},\omega)=i \vec{q} \times \vec{A}(\vec{q},\omega)$. We then insert the latter relation into Eq.\@ (\ref{eq:magnetiz_A}), and we use the decomposition of vectors and tensors into longitudinal and transverse parts with respect to the wave vector $\vec{q}$ \cite{Dressel-2001,Robertson-1940}: $A_{\alpha}(\vec{q},\omega) = \frac{q_{\alpha}}{|\vec{q}|}A_{L}(\vec{q},\omega)+A_{T,\alpha}(\vec{q},\omega)$ and $\left[\mu_{\alpha\beta}(\vec{q},\omega)\right]^{-1}=\left[\mu_{L}(\vec{q},\omega)\right]^{-1}\frac{q_{\alpha}q_{\beta}}{|\vec{q}|^{2}}+\left[\mu_{T}(\vec{q},\omega)\right]^{-1}\big(\delta_{\alpha\beta}-\frac{q_{\alpha}q_{\beta}}{|\vec{q}|^{2}}\big)$. Tracing back to the induced current density $\delta J_\alpha(\vec{q},\omega)=\frac{q_{\alpha}}{|\vec{q}|}\delta J_{L,\alpha}(\vec{q},\omega)+\delta J_{T,\alpha}(\vec{q},\omega)$ via Eq.\@ (\ref{eq:magnetiz_A}), taking the transverse part, and going to the static limit, we finally achieve
\begin{equation}
        \delta J_{T,\alpha}(\vec{q},0) = \left[1-\frac{\mu_0}{\mu_T(\vec{q},0)}\right]\frac{q^{2}}{\mu_0}A_{T,\alpha}(\vec{q},0).\label{eq:current1}
    \end{equation}
Comparing to Eq. \eqref{definition_of_SC_kernel}, it then follows for the static transverse part of the kernel that
    \begin{equation}\label{eq:kernel_transverse_static}
    \mathcal{K}_T(\vec{q},0)=\left[1-\frac{\mu_0}{\mu_T(\vec{q},0)}\right]\frac{q^2}{\mu_0}.
    \end{equation}
The Meissner effect corresponds to a vanishing permeability, i.e.\@, $\mathcal{K}_{T}(\vec{q},0)$ has to vanish slower than $q^{2}$ for $q \rightarrow 0$. Let us analyze the macroscopic consequences of this effect from Maxwell's equations, and link them to the microscopically derived kernel (\ref{eq:kernel_transverse_static}). From Amp\`ere's law written in terms of the vector potential, we have $\left[q^2-\mu_0  \mathcal{K}_T(\vec{q},0) \right] \vec{J}_{T}(\vec{q},0)=\mu_0 \vec{J}_{T,\text{ext}}(\vec{q},0)$, where $\vec{J}_{T,\text{ext}}(\vec{q},0)$ is the transverse part of the external current density. Transforming back to the real space of coordinates, and using Eq.\@ (\ref{eq:current1}), we obtain for the magnetic field
    \begin{align}\label{eq:Ampere_law_gen}
    \vec{B}(\vec{r})& =\mu_0\int\frac{d\vec{q}}{(2\pi)^{3}}\frac{i\vec{q}\times\vec{J}_{T,\text{ext}}}{q^{2}-\mu_0 \mathcal{K}_{T}(\vec{q},0)}e^{i\vec{q}\cdot\vec{r}} \notag \\ & =\mu_0\int\frac{d\vec{q}}{(2\pi)^{3}}\frac{i\vec{q}\times\vec{J}_{T,\text{ext}}}{q^{2}+\lambda_L^2}e^{i\vec{q}\cdot\vec{r}}.
    \end{align}
The characteristic length scale for the decay of the magnetic field near the system surface is therefore given by 
    \begin{equation}\label{London_Stiffness_appendix}
    \lambda_L^{-2}=-\mu_0 \lim_{\vec{q}\to0}\mathcal{K}_{t}(\vec{q},\omega=0)=\mu_0 e^2 \rho_S,
    \end{equation}
where $e$ is the carrier electric charge, and $\rho_{S}$ denotes the superconducting phase stiffness. The latter is often written as $\rho_S=n_S/m^\ast$, where $n_S$ is the superfluid density and $m^\ast$ is the effective mass of the condensed particles. However, penetration depth experiments only probe $\rho_S$, and not its individual components $n_S$ and $m^\ast$ \cite{Tinkham-1959,Tinkham-1996int,Prozorov-2006}. 
Connecting the electromagnetic kernel $\mathcal{K}(\vec{q},0)$ to the low-energy action for phase fluctuations, as derived in App.\@ \ref{Action_for_charge_fluctuations}, we arrive at an explicit expression of the phase stiffness in terms of the anomalous propagator $F^{\dagger}(\tau)$ \footnote{The connection between the kernel $\lim_{q \rightarrow 0}\mathcal{K}(\vec{q},0)$ and the macroscopic magnetic response (\ref{eq:Ampere_law_gen}) can be realized by utilizing either the longitudinal or the transverse part of the kernel (\ref{eq:Kernel_micro}): in fact, in the long-wavelength limit $q \rightarrow 0$, the distinction between transverse and longitudinal waves with respect to $q$ disappears, as can also be explicitly checked from our expression (\ref{eq:Kernel_micro}) of the electromagnetic kernel in the vanishing-momentum limit.}. In SI units, we have
	\begin{equation}\label{SC_stiffness}
	\rho_{S}=\frac{\mathscr{N} z t_{0}^{2}}{a \hbar^2 }\int d\tau F^{\dagger}(\tau)F(-\tau),
	\end{equation}
	where $a$ is the typical microscopic distance over which the phase field varies in space (i.e.\@, the lattice constant of our system), $z$ is the coordination number, and $\hbar$ is the reduced Planck's constant. One can verify that Eq.\@ (\ref{SC_stiffness}) has the units of a number density divided by a mass, consistently with the general definition of the phase stiffness. As expected $\rho_s=0$ in the normal state, i.e. for $F(\tau)=F^{\dagger}(\tau)=0$.
The stiffness (\ref{SC_stiffness}) can be written through a Matsubara transform as
\begin{subequations}
\begin{equation}\label{eq:stiff_SYK_lattice}
\rho_S(T) =\Theta_L z t_0^2 \, k_B T\sum_{i \omega_n} F^\dagger(i \omega_n) F(i \omega_n),
\end{equation}
\begin{equation}\label{eq:Theta_L}
\Theta_L=\frac{\mathscr{N}}{a \hbar^2},
\end{equation}
\end{subequations}
which is amenable to numerical computation. Specifically, Eq.\@ (\ref{eq:stiff_SYK_lattice}) is employed together with the exact numerical solution for the anomalous propagator, stemming from the saddle-point equations (\ref{eq:Eliashberg_eph:hop}), to obtain the data points reported in Figs.\@ \ref{rhoS_SYK_T} and \ref{rhoS_SYK_g}. Furthermore, notice that $E_\rho=\rho_S/\Theta_L$ yields an energy scale: this is the characteristic energy scale (or temperature scale $T_\rho=E_\rho/k_B$) \emph{per fermion flavor} below which long-range phase coherence of the order parameter is established. 
Fig.\@ \ref{rhoS_SYK_T} shows the temperature dependence of $\rho_S(T)/\Theta_L$ for $g=1$ and different hopping. All curves are normalized by $z t_0^2/\omega_0$ for a better visual comparison, so that $\omega_0\rho_S(T)/(\Theta_L z t_0^2)$ is a dimensionless quantity. The stiffness reaches a finite constant value for $T \rightarrow 0$. This is a fundamental result, as it demonstrates that the Yukawa-SYK model (\ref{eq:H_SYK_phon_hop}), once embedded in a lattice, has a \emph{true} superconducting ground state at $T<T_c$ characterized by a phase rigidity of the order parameter and the associated diamagnetic supercurrents. Then, the condensed phases of Refs.\@ \onlinecite{Esterlis-2019, Hauck-2020, Wang-2021,Classen-2022} can be regarded as a specific limit of a lattice model for $t_0 \rightarrow 0$, where the phase stiffness per flavor is vanishingly small. 

\begin{figure*}[t]
\includegraphics[width=0.8\textwidth]{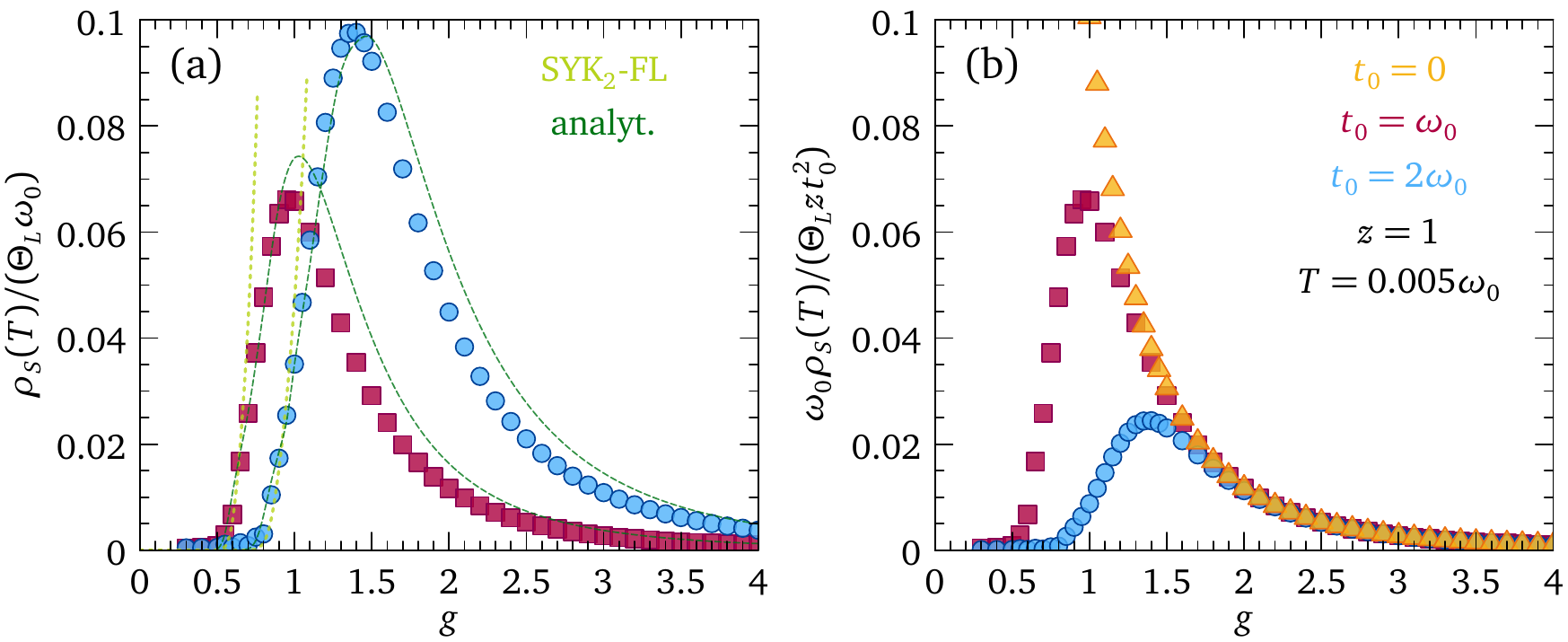}
\caption{\label{rhoS_SYK_g_analys} Analysis of the phase stiffness as a function of coupling $g$, at temperature $T=0.005 \omega_0/k_B$ and for different hopping $t_0$, where the data points are numerically calculated from the saddle-point equations (\ref{eq:Eliashberg_eph:hop}) in the superconducting state. (a) Normalized stiffness $\rho_S(T)/(\Theta_L \omega_0)$ as a function of coupling $g$, for hoppings $t_0/\omega_0=\left\{1,2\right\}$. The dashed light-green curves show the analytical estimation (\ref{eq:SYK2_FL_stiff_T0}) in SYK$_2$-FL regime. The dark green curves are the analytical approximation (\ref{eq:rhoS_implike_cross_expl0}), which interpolates between the SYK$_2$-FL and impurity-like regimes, and where the gap $\Delta_0\approx \Delta(i \omega_1)$ is calculated at each $g$ from the self-consistent imaginary-axis solution at the first Matsubara frequency; see Fig.\@ \ref{Delta_SYK_t0}. (b) Example of scaling collapse of $\omega_0\rho_S(T)/(\Theta_L \omega_0 z t_0^2) \propto g^{-4}$ as a function of $g$ in the strong-coupling limit: the curves for $t_0=\left\{0,1,2\right\}$ all collapse to the single-dot result at large $g$. All other parameters are the same as in panel (a).}
\end{figure*}

The analysis of the stiffness in SYK$_2$-FL and impurity-like regimes is best performed simultaneously, as for the quasiparticle weight (\ref{eq:Z_imag_imp_SC_cross}), to qualitatively describe the whole crossover. We report the associated derivations in Appendices \ref{rhoS_SYK2-FL} and \ref{rhoS_imp}, while here we discuss the final result: 
\begin{equation}\label{eq:rhoS_implike_cross_expl2}
\rho_S(T)= \frac{2 \Theta_L z t_0^2}{\tilde{\Omega}^2}\Delta\left[1-2 f_{FD}(\Delta)\right], 
\end{equation}
where $\tilde{\Omega}$ obeys Eq.\@ (\ref{eq:tilde_Omega_interp}).
Notice that at $T=0$ Eq.\@ (\ref{eq:rhoS_implike_cross_expl2}) implies
\begin{equation}\label{eq:rhoS_implike_cross_expl0}
\rho_S(0)= \frac{2 \Theta_L z t_0^2}{\tilde{\Omega}^2}\Delta. 
\end{equation}
The result (\ref{eq:rhoS_implike_cross_expl0}) is in contrast with the conventional expression of BCS theory in a clean system, where the stiffness $\rho_S(0)=n/m$ with $n$ the entire available electron density and $m$ the carrier mass \cite{Tinkham-1996int,Leggett-1998,Leggett-2006quantum,Altland-cm2010}. On the contrary, Eq.\@ (\ref{eq:rhoS_implike_cross_expl0}) is consistent with the expression for the stiffness in the dirty limit: for disordered superconductors with impurity scattering rate $\tau^{-1}$ it holds in the limit $\Delta \ll \tau^{-1} \ll E_{\rm F}$, with $E_F$ Fermi energy, that the ground-state stiffness is reduced to $\rho_S/(n/m)=\pi \Delta \tau$, therefore it scales with the superconducting gap $\Delta$ \cite{Abrikosov-2012meth}. 
Eq.\@ (\ref{eq:rhoS_implike_cross_expl2}), together with the numerical gap $\Delta \approx \Delta(i \omega_1)$ from Fig.\@ \ref{Delta_SYK_t0}(b), yields the dashed dark-green curves in Fig.\@ \ref{rhoS_SYK_g_analys}(a), which are in good agreement with the numerical data points for all couplings $g$; the latter data points are the same as in Fig.\@ \ref{rhoS_SYK_g}(b).
In SYK$_2$-FL regime $\tilde{\Omega} \mapsto \sqrt{2 z} t_0$, and we have
\begin{equation}\label{eq:SYK2_FL_stiff}
\rho_S(T)=\Theta_L \Delta \left[1-2 f_{FD}(\Delta)\right].
\end{equation}
Therefore, the stiffness is proportional to the superconducting gap in the regime $\Delta/(k_B T) \gg 1$. Indeed, at $T=0$ using Eq.\@ (\ref{eq:SYK2-FL_FL_Delta0}) we obtain 
\begin{equation}\label{eq:SYK2_FL_stiff_T0}
\rho_S(0)=\Theta_L \Delta_0 = \Theta_L 2 \omega_0 e^{-\frac{1}{\bar{\lambda}}}, 
\end{equation}
with coupling constant (\ref{eq:lambda_bar}), which shows that $\rho_S(0)$ is exponentially suppressed as $g$ is decreased in SYK$_2$-FL regime. Such exponential suppression is shared by the quasiparticle weight (\ref{eq:Z_imag_SYK2_FL_SC_expl}).
Let us remark that the result (\ref{eq:SYK2_FL_stiff_T0}) is consistent with the dirty limit of BCS theory: there, the stiffness is affected by nonmagnetic disorder, but the pairing temperature and the zero-temperature gap are not. This difference is reflected by our findings for $T_c$ and $\Delta_0$ in SYK$_2$-FL regime -- see Eqs.\@ (\ref{eq:BCS_Tc_SYK2}) and (\ref{eq:SYK2-FL_FL_Delta0}) -- which are in agreement with the respective BCS formulae. 

The result (\ref{eq:SYK2_FL_stiff_T0}) corresponds to the dashed light-green curves in Fig.\@ \ref{rhoS_SYK_g_analys}(a), and it is there compared with the same numerical data points reported in Fig.\@ \ref{rhoS_SYK_g}(b). Comparing Eq.\@ (\ref{eq:SYK2_FL_stiff_T0}) with Eqs.\@ (\ref{eq:SYK2-FL_FL_Delta0}) and (\ref{eq:BCS_Tc_SYK2}), we deduce that the ratio between the energy scale given by the zero-temperature stiffness, and the mean-field transition temperature $k_B T_c$, is a universal constant in SYK$_2$-FL regime:
\begin{equation}\label{eq:Tc_stif_universal_SYK2}
\frac{\rho_S(T)}{\Theta_L k_B T_c}=\frac{\Delta_0}{k_B T_c}=\frac{\pi}{e^{\gamma}}=0.5669329586555488,
\end{equation}
where $\gamma$ is again the Euler-Mascheroni constant. A constant ratio of the order of Eq.\@ (\ref{eq:Tc_stif_universal_SYK2}) between $k_B T_c$ and stiffness is indeed retrieved in the SYK$_2$-FL regime, from the full numerical solution of the Eliashberg equations. This is shown in Fig.\@ 3 of the companion paper Ref.\@ \onlinecite{short-paper}, and there further discussed. 

In the impurity-like regime $\tilde{\Omega} \mapsto \Omega_0$, and Eq.\@ (\ref{eq:rhoS_implike_cross_expl2}) gives
\begin{align}\label{eq:rhoS_implike_expl}
\rho_S(T) & = \frac{\Theta_L z t_0^2}{2 \left[8 g^2 \omega_0/(3 \pi )\right]^2}\Delta\left[1-2 f_{FD}(\Delta)\right]\notag \\ &=\frac{\Theta_L}{2} z t_0^2 \left(\frac{3 \pi}{8} \right)^2 \frac{1}{\omega_0^2 g^4} \Delta\left[1-2 f_{FD}(\Delta)\right]. 
\end{align}
In particular, at $T=0$ we have
\begin{equation}\label{eq:rhoS_implike_expl_T0}
\rho_S(0)=\frac{\Theta_L}{2} z t_0^2 \left(\frac{3 \pi}{8} \right)^2 \frac{1}{\omega_0^2 g^4} \Delta_0. 
\end{equation}
In the impurity-like regime, the gap asymptotically reaches a constant in the infinite-interaction limit, in accordance with Eq.\@ (\ref{eq:imp_Delta0}). 
Therefore, Eqs.\@ (\ref{eq:rhoS_implike_expl_T0}) and (\ref{eq:imp_Delta0}) tell us that the stiffness goes like $\rho_S(0) \propto \bar{g}^{-4}$ at large interaction $g \gg 1$. We can also make sense of this conclusion by realizing that $Z(i \omega_n) \propto \bar{g}^{2}$ in the impurity-like regime \cite{Esterlis-2019}, and the stiffness contains $\left[Z(i \omega_n)\right]^{-2} \propto \bar{g}^{-4}$ as seen from Eq.\@ (\ref{eq:stiff_SYK_lattice}). 
Therefore, Eq.\@ (\ref{eq:rhoS_implike_expl_T0}) predicts that the normalized quantity $\rho_S(0)/(\Theta_L z t_0^2)$ is independent from hopping in impurity-like regime: this is exactly what we see in Fig.\@ \ref{rhoS_SYK_g_analys}(b), where the numerical data points for $t_0=\left\{1,2\right\}\omega_0$ (red squares and light-blue circles) collapse onto the $t_0=0$ single-dot results (golden triangles) at large $g$. 

In summary, we find that the low-temperature phase stiffness is controlled by the dependence of $Z(i \omega_n)$ and $\Delta_0$ on coupling: in the Fermi-liquid SYK$_2$-FL regime, $\rho_S(0) \propto \Delta_0$ is exponentially diminished as we decrease $g$, because $\Delta_0$ follows the BCS formula; in the impurity-like regime, $\rho_S(0)\propto \sum_{i \omega_n} \left[Z(i \omega_n)\right]^{-2} \propto g^{-4}$, because the gap $\Delta_0$ is approximately constant and the imaginary-axis weight $Z(i \omega_n) \propto g^{2}$. 
In the intermediate coupling regime, the stiffness reaches a maximum. In fact, this maximum occurs precisely at the SYK$_2$-FL/SYK-NFL crossover energy $\omega_c$, previously found in Sec.\@ \ref{Normal_cross} and marked by the red and blue arrows in Fig.\@ \ref{rhoS_SYK_g}(b) for $t_0=\left\{1,2\right\} \omega_0$. 
This is the second fundamental result of this section: the phase stiffness is maximum at the crossover between the SYK-NFL phase (non-Fermi liquid) and the SYK$_2$-FL regime (disordered Fermi liquid). Such non-monotonic evolution is reminiscent of the doping evolution of the superfluid density extracted from muon relaxation experiments in cuprate high-temperature superconductors \cite{Shen-1997,Feng-2000}. 
In the same experimental works \cite{Shen-1997,Feng-2000}, a correlation between the superfluid density and the condensation energy, extracted from calorimetric measurements, was discovered. In the same spirit, in the next section we conclude our investigation of the superconducting properties of our model (\ref{eq:H_SYK_phon_hop}) with a discussion of the grand potential and the condensation energy in the paired phase. 

\section{Condensation energy}\label{Cond_en}

\begin{figure*}[t]
\includegraphics[width=1.0\textwidth]{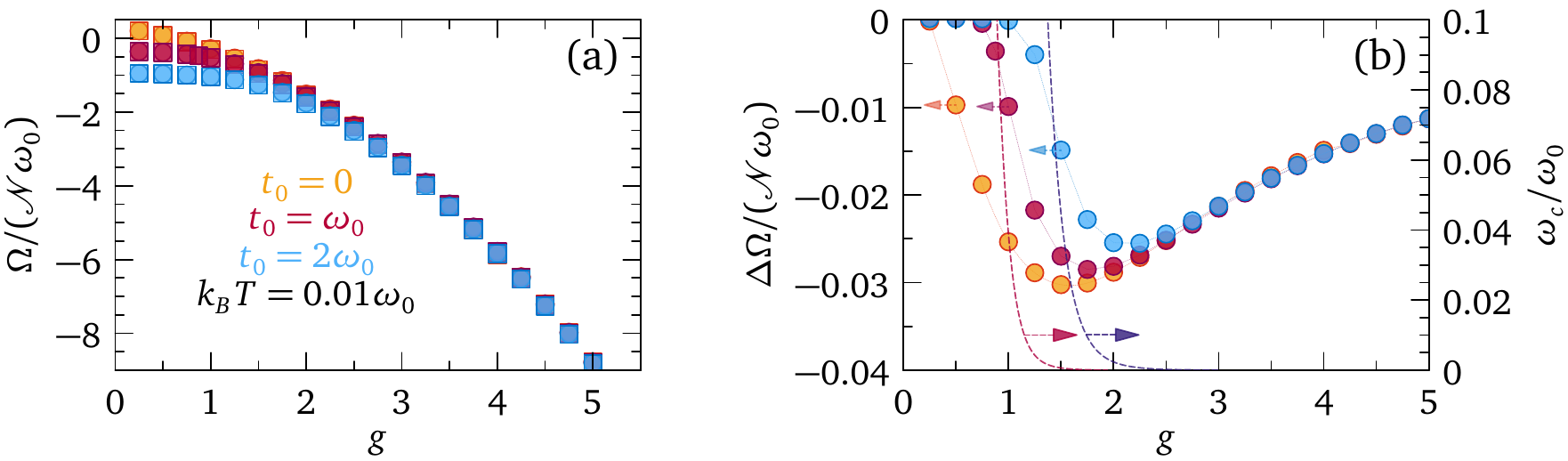}
\caption{\label{DeltaOmega_g} Grand potential and condensation energy as a function of coupling $g$ for different hoppings $t_0$, numerically computed from Eqs.\@ (\ref{eq:Omega_ph_SC}) and (\ref{eq:Omega_SYK_ph}). (a) Normalized grand potential $\Omega/(\mathscr{N}\omega_0)$ in the normal state (circles) and in the superconducting state (squares) at temperature $k_B T=0.01 \omega_0$, for hopping $t_0/\omega_0=\left\{0,1,2\right\}$. (b) Condensation energy $\Delta \Omega/(\mathscr{N}\omega_0)$, evaluated from the difference between the normal-state and superconducting-state grand potentials in panel (a), with values reported on the left $y$ axis, as indicated by arrows. The dashed lines are the SYK$_2$-FL/SYK-NFL crossover energies $\omega_c/\omega_0$ at a given hopping, according to Sec.\@ \ref{Normal_cross}, with values shown on the right $y$ axis, as indicated by arrows.}
\end{figure*}
The superconducting transition in the Yukawa-SYK model entails a lowering of the total energy of the system, expressed by the grand potential $\Omega$, with respect to the normal-state grand potential (\ref{eq:Omega_SYK_ph}). In a strict sense is the condensation energy  an ill-defined concept, as the rigorous solution of a statistical mechanics problem for a superconductor knows nothing about the behavior of the unstable normal state. The exception may be a normal state stabilized by a magnetic field, which may however dramatically alter the system on its own right. On the other hand, the condensation energy can be well-defined and is physically insightful within mean-field theory. Thus, keeping the formal caveat in mind we will, in what follows, analyze the energy gain of superconducting solutions if compared to the corresponding normal-state case.
In the single-dot limit, such condensation energy was analyzed in Ref.\@ \onlinecite{Esterlis-2019}. To calculate the same quantity on the lattice, we have to derive an expression for the grand potential $\Omega^{{\rm sc}}$ for $T<T_c$. We report the technical derivation in Appendix \ref{app:Omega_SC}, while the final result is quoted here below: 
\begin{eqnarray}\label{eq:Omega_ph_SC}
\frac{\Omega^{{\rm sc}}}{\mathscr{N}} & = & k_B T\sum_{i \omega_n}\left\{2\log\left[\frac{\sqrt{G^{2}(i\omega_n)-F^{\dagger}(i\omega_n)F(i\omega_n)}}{G_{0}(i\omega_n)}\right] \right.\nonumber \\ &-&\left.\frac{zt_{0}^{2}}{2}\left[G^{2}(i\omega_{n})-F^{\dagger}(i\omega_{n})F(i\omega_{n})\right]\right\}\nonumber \\  &-&k_B T\sum_{i \Omega_n}\left\{\frac{1}{2}\log\left[\frac{D(i\Omega_{n})}{D_{0}(i\Omega_{n})}\right]+D(i\Omega_{n})\Pi(i\Omega_{n}) \right\} \nonumber \\
 & - & 2 k_B T\log\left[1+e^{\mu/(k_B T)}\right]-\frac{\omega_{0}}{4} \nonumber \\ &-& \frac{k_B T}{2}\log\left[1-e^{-\omega_{0}/(k_B T)}\right].
\end{eqnarray}
As in the normal-state case, the grand potentials of free fermions and bosons have been simultaneously added and subtracted in Eq.\@ (\ref{eq:Omega_ph_SC}), to improve the numerical convergence of the Matsubara sums. 
Fig.\@ \ref{DeltaOmega_g}(a) shows the grand potential in the superconducting state (colored squares) and in the normal state (colored circles), as a function of coupling and at $k_B T =0.01 \omega_0$, for different hoppings. The data points are numerically calculated from Eqs.\@ (\ref{eq:Omega_ph_SC}) and (\ref{eq:Omega_SYK_ph}) respectively, and using the saddle-point propagators self-consistently evaluated from the saddle-point equations (\ref{eq:Eliashberg_eph:hop}).
We see that the difference between the normal-state and the condensed-state results is very small on the scale of the grand potentials themselves. 

Subtracting the normal-state result (\ref{eq:Omega_SYK_ph}) from the superconducting grand potential (\ref{eq:Omega_ph_SC}) gives the condensation energy:
\begin{equation}\label{eq:DeltaOmega_YSYK}
\frac{\Delta \Omega}{\mathscr{N}}=\frac{\Omega^{{\rm sc}}}{\mathscr{N}}-\frac{\Omega^{{\rm ns}}}{\mathscr{N}}.
\end{equation}
Eq.\@ (\ref{eq:DeltaOmega_YSYK}) is employed together with the grand potentials in Fig.\@ \ref{DeltaOmega_g}(a), to obtain the numerical results shown in Fig.\@ \ref{DeltaOmega_g}(b). We observe that the almost linear evolution $\Delta \Omega \propto g$ at small $g$, in the single-dot case (golden circles) and in the SYK-NFL phase, is replaced by an exponential suppression of the condensation energy in the SYK$_2$-FL regime (red and light-blue circles), for finite hopping. Indeed, $\Delta \Omega$ starts to significantly increase in magnitude only for couplings above the SYK$_2$-FL/SYK-NFL crossover, signaled by $\omega_c$; see Sec.\@ \ref{Normal_cross}. The crossover energy $\omega_c$ is marked by the dashed red and blue curves for hopping $t_0=\left\{1,2\right\}\omega_0$, respectively. 
Therefore, the non-Fermi liquid phase is characterized by a significantly higher condensation energy than the disordered Fermi-liquid regime; such phenomenon might bear observable consequences for the specific-heat jump at the superconducting transition in non-Fermi liquid superconductors \cite{Fisher-1988,Loram-2001,Movshovich-2001}.

On the other hand, the data points in the impurity-like regime at finite hopping are superimposed to the single-dot results, since $z t_0^2 \ll g^2 \omega_0^2$ and hopping is an irrelevant perturbation. Then, we find the same slow decrease of $\Delta \Omega$ with increasing $g$ as found in the single-dot limit \cite{Esterlis-2019}. 

Notice that the exponential-like suppression of $\Delta \Omega$ in the SYK$_2$-FL phase, as well as the decrease with $g$ in the impurity-like state, remarkably correlate with the coupling evolution of the quasiparticle weight and of the stiffness, as summarized in Fig.\@ \ref{rhoS_Z_SC_DeltaOmega}. This is the fundamental result of the present section: in our Yukawa-SYK superconductor on a lattice, the superfluid stiffness, the condensation energy, and the quasiparticle weight are all correlated with each other and share similar behaviors as functions of fermion-boson coupling. Given the analogy with the experimental findings on cuprate superconductors \cite{Shen-1997,Feng-2000}, the qualitative trends proposed in the present model may transcend the specific aspects of SYK physics, and be generic for non-Fermi liquid superconductors. We discuss these speculations, together with future useful extensions of our lattice model, in the conclusive remarks.

\section{Conclusions and perspectives}\label{Disc}

In summary, we formulated and solved a model for spinful fermions with multiple flavors, interacting at local sites through all-to-all random couplings to many Einstein phonon modes and through a random coherent single-particle hopping between nearest-neighbor sites. This forms  a lattice of Yukawa-SYK dots. Both couplings and hoppings follow a distribution function with zero mean and finite variance. Our mean-field equations on the saddle point of the disorder-averaged action (\ref{eq:effective_action}) are exact in the limit of a large number of fermion and boson flavors; they assume a diagonal structure in replica space for the disordered interactions, an equal number of fermions and boson, and particle-hole symmetry (zero chemical potential). They are equivalent to self-consistent Eliashberg equations, where fermion-boson interaction is at once responsible for the destruction of fermionic quasiparticles (NFL state), the softening of the renormalized boson frequency to criticality, and a superconducting state at low temperatures if every disorder realization of the couplings preserves time-reversal symmetry. 

In the normal state, the competition between coupling and hopping leads to crossovers between FL and NFL phases. In addition to the high-temperature classical-gas phase, the low-temperature SYK-NFL phase, and the intermediate-temperature impurity-like phase, all identified and studied in the single-dot limit in Refs.\@ \onlinecite{Esterlis-2019,Hauck-2020,Classen-2022}, we find crossovers to a disordered Fermi liquid, the SYK$_2$-FL state, when hoppings prevail over on-site coupling. 
The characteristic crossover energies, estimated in Sec.\@ \ref{Normal_cross}, are decreasing functions of coupling at fixed hopping, and they correspond to observable changes in the fermionic and bosonic spectral functions: in the SYK-NFL phase, the fermion and boson propagators follow a power-law form governed by the exponent $\Delta$, and by increasing hopping this evolves into the semicircular Wigner spectral function of SYK$_2$-FL fermions with bandwidth controlled by hopping, with almost free bosons and weak fermion-boson coupling. The crossover from the impurity-like regime is more subtle, as here the fermions are also broadened into a Wigner spectral function controlled by coupling, but bosons behave like static impurities, with a sharp but soft renormalized frequency. This frequency stiffens towards the SYK$_2$-FL regime. 
The crossovers leave qualitative differences in the normal-state entropy, which is finite at $T=0$ in the single-dot limit of negligible hopping (SYK-NFL and imp.-like phases), and vanishes for any finite hopping. 
Superconductivity emerges at low temperature in all regimes. The critical temperature saturates to a fraction of the bare boson frequency $\omega_0$, $k_B T_c \sim 0.11 \omega_0$ \cite{Esterlis-2018a, Chowdhury-2020eff}, in the strong-coupling imp.-like regime where fermions are fully incoherent \cite{Esterlis-2019,Hauck-2020}. Conversely, in the weak-coupling SYK$_2$-FL phase we retrieve a disordered version of BCS theory, where $T_c$ and the gap at zero temperature and energy follow the BCS formulae with a coupling constant $\bar{\lambda} \propto g^2\omega_0/t_0$ that depends on both coupling and hopping. The tell-tale signature of the crossovers in the superconducting spectral functions is the alteration and eventual suppression of peak-dip-hump features, in the SYK-NFL and imp.-like states, with increasing hopping towards the SYK$_2$-FL regime. Such resonances are self-trapping bound states of the interacting Cooper-pair fluid \cite{Esterlis-2019}.

Our most essential result is that the superconducting state of our model has a finite phase stiffness. The latter is calculated through the electromagnetic linear response function to an external vector potential, which couples to phase fluctuations of the order parameter away from the saddle-point through electric charge. At vanishing temperature and weak coupling (SYK$_2$-FL regime), the stiffness follows the exponential evolution of the gap with coupling $\bar{\lambda}$, so that the ratio between $T_c$ and the stiffness corresponds to the ``universal'' BCS ratio $\pi/e^{\gamma}\approx 0.567$. On the contrary, at strong coupling (imp.-like phase) the gap saturates, but the stiffness decreases as the inverse fourth power of coupling due to the decrease of the quasiparticle weight. 
Remarkably, we find a correlation in the evolution with coupling of the stiffness, the quasiparticle weight, and the condensation energy. Their magnitudes all peak at the crossover between NFL and FL behavior. Such correlation is reminiscent of the experimental measurements of the same quantities in cuprate superconductors \cite{Shen-1997,Feng-2000}. 
More generally, our results demonstrate that the low-temperature condensed phase of the Yukawa-SYK model is indeed superconducting, as it exhibits perfect diamagnetism according to the Mei{\ss}ner effect. Therefore, our calculations offer a suitable platform to compare the behavior of the stiffness within similar SYK-derived models, as well as in different classes of models for quantum-critical superconductors \cite{Chubukov-2005,Aji-2007,She-2009,Lederer-2015,Wang-2016,Fernandes-2016,Lederer-2017,Caprara-2017,Huang-2019,Chubukov-2020,Phillips-2020,Chowdhury-2020eff,Caprara-2022}.
For the Yukawa-SYK model in particular, the single-dot limit corresponds to a superconductor with negligible interdot hopping and vanishingly small stiffness per flavor, which represents a special case of our lattice embedding. 

The difference between a saturating mean-field $T_c$ and a decreasing stiffness at strong coupling, as well as a decreasing condensation energy in the same regime, suggest strong superconducting phase fluctuations in the system \cite{Emery-1994,Esterlis-2019}. If we assume that phase fluctuations are the driving cause of the disappearance of superconducting coherence, the phase stiffness obtained in our theory
would suggest that the fluctuation-corrected transition temperature $T_\Theta$ vanishes as $T_\Theta \sim g^{-4}$ for $g \rightarrow +\infty$.
However, performing a quantitative analysis of superconducting phase fluctuations requires to include $1/\mathscr{N}$ corrections with respect to our saddle-point results, which hold for  $\mathscr{N}\rightarrow+\infty$. The effect of such corrections has recently been studied in an SYK-type system which superconducts due to an attractive local Hubbard interaction: the $1/\mathscr{N}$ terms corresponding to phase fluctuations lead to a pseudogap behavior \cite{Wang-2020b}. A similar phenomenon, that has been pointed out in various microscopic models for quantum critical superconductors \cite{Chubukov-2005,Chowdhury-2020a}, could also be revealed at order $1/\mathscr{N}$ in our model. In this respect, the mean-field $T_c$ calculated in this work would not be the real superconducting temperature where zero resistivity appears, but rather a pseudogap temperature. A future analysis of such pseudogap phase is in order, in the light of the analogous phenomenon found in strange metals \cite{Keimer-2015}. 

Since the stiffness is connected to the electrodynamic response, it is  interesting to investigate whether the evolution of the stiffness with coupling and hopping leaves distinctive traces in the long-wavelength optical conductivity $\sigma(\omega)$ of our lattice model in the superconducting state as a function of frequency $\omega$. This is the subject of a companion paper \cite{short-paper}, in which we demonstrate that the removed spectral weight for energies $\omega<2 \Delta_{S}$, where $\Delta_S$ is the spectroscopic gap, is a direct signature of the different behavior of the stiffness in the SYK$_2$-FL, SYK-NFL and imp.-like regimes. Moreover, the Cooper-pair bound states, appearing in the spectral functions of the SYK-NFL and imp.-like phases, are reflected in resonances of $\sigma(\omega)$, which disappear in the crossover towards the SYK$_2$-FL regime. Then, it is crucial to compare our results with optical spectroscopy measurements in strongly correlated superconductors such as cuprates, which show a Planckian-like power-law behavior of the conductivity in energy and temperature \cite{Collins-1989,Orenstein-1990,vanderMarel-2003,vanHeumen-2022,Michon-2022_preprint}. 

Numerous future developments and generalizations of our theory could provide further connections with experiments and with complementary theoretical perspectives. 

A thermodynamic quantity which is equally found to correlate with the stiffness in cuprate superconductors is the heat-capacity jump at $T_c$ \cite{Shen-1997,Feng-2000}. This quantity can be computed within our formalism, from the second derivative of the condensation energy at $T=T_c$, and if the correlation with the stiffness persists in our model, the latter would see its position strengthen as a suitable effective toy model for quantum critical superconductors and strange metals. Nevertheless, other ingredients related to strange-metal behavior, such as Mott-insulator physics \cite{Lee-2006}, are not included in our model. In this respect, it would be interesting to quantitatively compare predictions for the superfluid density evolution stemming from Mott-based pictures with our SYK-based approach.

Concerning the still unexplored parameter space, it would be interesting to extend the lattice calculations, in particular for $T_c$, the gap, and the phase stiffness, at finite chemical potential $\mu\neq 0$. A thorough study of this kind in the single-dot limit became recently available \cite{Classen-2022}, and could serve as a springboard reference to assess whether our found correlation among stiffness, quasiparticle weight and condensation energy persists as a function of $\mu$. Indeed, valence transitions were found by varying $\mu$ in the Yukawa-SYK dot \cite{Wang-2020a,Wang-2021} and in purely fermionic versions of the SYK model \cite{Azeyanagi-2018, Ferrari-2019,Sorokhaibam-2020,Smit-2021}. These phenomena could serve as an ideal playground for comparison of the thermodynamic and optical observables of the Yukawa-SYK lattice model with experiments on one hand, and with numerical methods such as Functional Renormalization Group (FRG) \cite{Smit-2021} and DMFT \cite{Werner-2008,Werner-2018} on the other hand. 
Even more precisely, a self-consistent calculation of the chemical potential \cite{Vandermarel-1990,Valentinis-2016a,Valentinis-2016b,Valentinis-2017}, $T_c$, and the stiffness at fixed density $\mathscr{Q}\in\left[0,1\right]$ per fermion flavor \cite{Wang-2020a} would allow one to explore the phase diagram as a function of $\mathscr{Q}$, which is analogous to chemical doping; this exploration would permit a direct comparison of our theoretical results with  the observed experimental correlations among thermodynamic and spectroscopic data in cuprates \cite{Shen-1997,Feng-2000}. 
Another insightful additional variable to tune is the pair-breaking parameter $\alpha \in\left(0,1\right)$, which permits a continuous interpolation between complex and real fermion-boson coupling constants, that violate or preserve time-reversal symmetry, respectively \cite{Hauck-2020,Inkof-2022,Inkof-thesis-2021}. Specifically as a function of hopping, how the crossover between FL and NFL physics affects the QCP, where $T_c(\alpha) \rightarrow 0$ with BKT-like evolution, remains an open question. 
The numerical results for thermodynamic and dynamical observables as a function of $\alpha$ also offer a direct comparison with the AdS/CFT correspondence: recently, a holographic dual of the Yukawa-SYK model in the single-dot limit was explicitly constructed, and shown to be equivalent to the Eliashberg formulation of the same model in the regime $T_c(\alpha) \rightarrow 0$ \cite{Inkof-2022}. The comparison of quantities like the pairing susceptibility, computed from the numerical saddle-point solution and from holography, could extend the newly found proof of the correspondence between the dual theories \cite{Inkof-2022}.

Another crucial generalization of our formalism involves the extension to anisotropic pairing symmetries, for instance through coupling with a spin-1 boson $\phi_{\vec{k}}$ \cite{Esterlis-2019}. Anisotropic pairing is also relevant for dispersive fermions, for instance Dirac fermions which could model quasiparticles in flat-band systems like graphene \cite{Cao-2018b,Cao-2018a,Yankowitz-2019,Balents-2020} at the charge neutrality point, or nodal quasiparticles in cuprate superconductors \cite{Peng-2022}.  

The foundational assumptions on which our results rest also necessitate further clarification, in particular the replica-diagonal ansatz for disorder-averages assumed in the large-$\mathscr{N}$ limit \cite{Wang-2019}. For instance, in a bosonic variant of the SYK model it was shown that anharmonic boson-boson interactions are able to break the replica symmetry and give rise to a glassy phase \cite{Tulipman-2020,Tulipman-2021}. 

Finally, since our results are valid on the saddle point and at equilibrium, nonequilibrium numerical studies of the Yukawa-SYK model are crucial to assess the boundaries in parameter space where our results apply, and viceversa where new physics arises, e.g.\@, for the electrodynamic response and the pairing susceptibility \cite{Kennes-2017,Lunkin-2022a,Lunkin-2022b,Kennes-2022_preprint,Grunwald-thesis-2022}. 

All in all, our lattice theory offers an exactly solvable and analytically controlled scenario to study the interplay between NFL and FL phases, the thermodynamics, and the electromagnetic response in quantum-critical superconducting systems. It represents a platform based on the Eliashberg formalism, to incorporate fluctuations beyond the mean-field level and to investigate the properties of unconventional superconductors. 

\section{Acknowledgments}

D.\@ V.\@ acknowledges insightful discussions with C.\@ Berthod, J.\@ Zaanen, L.\@ Benfatto, B.\@ Gout\'{e}raux, R.\@ Riva, N.\@ Stegani, F.\@ Lombardi, S.\@ Caprara, R.\@ Arpaia, G.\@ Ghiringhelli, D.\@ van der Marel, A.\@ Chubukov, C.\@ Bernhard, P.\@ Werner, M.\@ M\"{u}ller, R.\@ Morin, V.\@ Stangier, D.\@ Hauck, G.\@ Mazza, D.\@ Kennes, L.\@ Grunwald, R.\@ Ojaj\"{a}rvi, R.\@ Willa. D.\@ V.\@ and J.\@ S.\@ also acknowledge fruitful discussions with S.\@ Sachdev, I.\@ Esterlis, C.\@ Li, A.\@ Patel, and H.\@ Guo. 
D.\@ V.\@ acknowledges partial support by the Swiss National Science Foundation (SNSF) through the SNSF Early Postdoc.Mobility Grant P2GEP2\_181450, and by the Mafalda cluster of the University of Geneva, on which some of the calculations were performed. G.\@ A.\@ I.\@ and J.\@ S.\@ were supported by the Deutsche Forschungsgemeinschaft (DFG, German Research Foundation) - TRR 288 - 422213477 Elasto-Q-Mat (project A07).  

\appendix

\begin{widetext}
\section{Disorder-averaged effective action for the SYK model}\label{App:dis_aver_action}
In this appendix we derive the disorder-averaged effective action of the SYK model, Eq.\@ \eqref{eq:H_SYK_phon_hop}. The partition function
of the $m$-replicated system is given by 
\begin{align}
\overline{\mathscr{Z}^{m}}=\int\mathcal{D}\Psi e^{-\sum_{a}\mathscr{S}_{0}[\Psi]}\int dg_{ij,k}\varrho(g_{ij,k})e^{-\sum_{a}\mathscr{S}_{g}[\Psi]}\int dt_{ij,\vec{x}\vec{x}'}\varrho(t_{ij,\vec{x}\vec{x}'})e^{-\sum_{a}\mathscr{S}_{t}[\Psi]},\label{partition_function_disorder}
\end{align}
where $\Psi=\{\hat{c}_{i\sigma a\vec{x}},\hat{c}_{i\sigma a\vec{x}}^{\dagger},\phi_{ka\vec{x}}\}$
denotes the fields of the model and $a=\left\{1 \cdots m\right\}$ is the replica index.
The coupling constants $g$ and $t$ are drawn from the Gaussian orthogonal ensemble, with probability distribution functions $\varrho(g)$ and
$\varrho(t)$ defined in terms of the second moments $\overline{g_{ijk}g_{i'j'k'}}=\frac{\bar{g}^{2}}{2\mathscr{N}^{2}}\delta_{kk'}\left(\delta_{ii'}\delta_{jj'}+\delta_{ij'}\delta_{ji'}\right)$ and $\overline{t_{ij\vec{x}\vec{x}'}t_{i'j'yy'}}=\frac{t_{0}^{2}}{2\mathscr{N}}\left(\delta_{ii'}\delta_{jj'}\delta_{\vec{x}\vec{y}}\delta_{\vec{x}'\vec{y}'}+\delta_{ij'}\delta_{ji'}\delta_{\vec{x}\vec{y}'}\delta_{\vec{x}'\vec{y}}\right)$.
The imaginary-time action is given by \begin{subequations} 
\begin{eqnarray}
\mathscr{S}_{0} & = & \sum_{i\sigma\vec{x}}\int d\tau\,\hat{c}_{i\sigma a\vec{x}}^{\dagger}(\tau)(\partial_{\tau}-\mu)\hat{c}_{i\sigma a\vec{x}}(\tau)+\frac{1}{2}\sum_{ia\vec{x}}\int d\tau\phi_{ia\vec{x}}(\tau)(-\partial_{\tau}^{2}+m_{0})\phi_{ia\vec{x}}(\tau),\\
\mathscr{S}_{g} & = & \sum_{\sigma\vec{x};ijk}\int d\tau\,\left[g_{ij,k}+g_{ji,k}^{\star}\right]\hat{c}_{i\sigma a\vec{x}}^{\dagger}(\tau)\hat{c}_{j\sigma a\vec{x}}(\tau)\phi_{ka\vec{x}}(\tau),\\
\mathscr{S}_{t} & = & \sum_{\substack{\langle\vec{x},\vec{x}'\rangle\\
ij\sigma
}
}\int d\tau\,t_{ij,\vec{x}\vec{x}'}\hat{c}_{j\sigma\vec{x}'a}^{\dagger}(\tau)\hat{c}_{i\sigma\vec{x}a}(\tau),
\end{eqnarray}
\end{subequations} where angled brackets denote the sum over nearest-neighbor sites. Performing the Gaussian integration over the coupling
constants $g$ and $t$ -- see Sec.\@ 6.1 and Apps.\@ C1-C3 of Ref.\@ \onlinecite{Inkof-thesis-2021} -- yields the disorder-averaged effective action 
\begin{eqnarray}
\mathscr{S}^{{\rm eff}} & = & \sum_{i\sigma a\vec{x}}\int d\tau \hat{c}_{i\sigma a\vec{x}}^{\dagger}(\tau)(\partial_{\tau}-\mu)\hat{c}_{i\sigma a\vec{x}}(\tau)+\frac{1}{2}\sum_{ia\vec{x}}\int d\tau\phi_{ia\vec{x}}(\tau)(-\partial_{\tau}^{2}+m_{0})\phi_{ia\vec{x}}(\tau)\nonumber \\
 & - & \frac{\bar{g}^{2}}{2\mathscr{N}}\sum_{ijk}\sum_{ab\sigma\sigma'\vec{x}}\int_{\tau\tau'}\left[\hat{c}_{j\sigma a\vec{x}}^{\dagger}(\tau)\hat{c}_{j\sigma'b\vec{x}}^{\dagger}(\tau')\,\hat{c}_{i\sigma'b\vec{x}}(\tau')\hat{c}_{i\sigma a\vec{x}}(\tau)-\hat{c}_{j\sigma a\vec{x}}^{\dagger}(\tau)\hat{c}_{j\sigma'b\vec{x}}(\tau')\,\hat{c}_{i\sigma'b\vec{x}}^{\dagger}(\tau')\hat{c}_{i\sigma a\vec{x}}(\tau)\right]\phi_{ka\vec{x}}(\tau)\phi_{kb\vec{x}}(\tau')\nonumber \\
 & - & \frac{t_{0}^{2}}{4\mathscr{N}}\sum_{ij}\sum_{\substack{\langle\vec{x},\vec{x}'\rangle\\
ab\sigma\sigma'
}
}\int_{\tau\tau'}\,\,\left[\hat{c}_{j\sigma\vec{x}'a}^{\dagger}(\tau)\hat{c}_{i\sigma\vec{x}a}(\tau)\hat{c}_{j\sigma'\vec{x}'b}^{\dagger}(\tau')\hat{c}_{i\sigma'\vec{x}b}(\tau')+\hat{c}_{j\sigma\vec{x}'a}^{\dagger}(\tau)\hat{c}_{i\sigma'\vec{x}b}(\tau')\hat{c}_{i\sigma\vec{x}a}^{\dagger}(\tau)\hat{c}_{j\sigma'\vec{x}'b}(\tau')\right],
\end{eqnarray}
where $\int_{\tau\tau'}\equiv\int d\tau\int d\tau'$.
We now introduce the following collective bi-local fields \begin{subequations}
\begin{align}
 & G_{\sigma\sigma',\vec{x}}^{ab}(\tau,\tau')=\frac{1}{\mathscr{N}}\sum_{i}\hat{c}_{i\sigma'b\vec{x}}^{\dagger}(\tau')\hat{c}_{i\sigma a\vec{x}}(\tau),\quad F_{\sigma\sigma',\vec{x}}^{ab}(\tau,\tau')=\frac{1}{\mathscr{N}}\sum_{i}\hat{c}_{i\sigma'b\vec{x}}(\tau')\hat{c}_{i\sigma a\vec{x}}(\tau),\\
 & D_{\vec{x}}^{ab}(\tau,\tau')=\frac{1}{\mathscr{N}}\sum_{k}\phi_{kb\vec{x}}(\tau')\phi_{ka\vec{x}}(\tau).
\end{align}
\end{subequations} In order to insert such variables into the action we make use of the following identity 
\begin{eqnarray}
1 & = & \int\mathcal{D}G\prod_{\substack{ab\tau\tau'\\
\sigma\sigma'\vec{x}
}
}\delta\left[\mathscr{N}G_{\sigma'\sigma,\vec{x}}^{ba}(\tau',\tau)-\sum_{i}\hat{c}_{i\sigma a\vec{x}}^{\dagger}(\tau)\hat{c}_{i\sigma'b\vec{x}}(\tau')\right]=\int\mathcal{D}G\mathcal{D}\Sigma e^{\left[\mathscr{N}G_{\sigma\sigma',\vec{x}}^{ba}(\tau',\tau)-\sum_{i}\hat{c}_{i\sigma a\vec{x}}^{\dagger}(\tau)\hat{c}_{i\sigma'b\vec{x}}(\tau')\right]\,\Sigma_{\sigma\sigma',\vec{x}}^{ab}(\tau,\tau')},\nonumber \\
\end{eqnarray}
where in the second equality we have introduced the Lagrange-multiplier $\Sigma$
of the field $G$ and the sum over repetated indices is kept implicit.
Alternatively, one can introduce such fields through a Hubbard-Stratonovich
transformation -- see Refs.\@ \onlinecite{Sachdev-2015,Inkof-thesis-2021}. Similarly, for the other fields we use: \begin{subequations} 
\begin{eqnarray}
1 & = & \int\mathcal{D}F^{\dagger}\mathcal{D}\Phi e^{\frac{1}{2}\left[\mathscr{N}F_{\sigma'\sigma,\vec{x}}^{\dagger ba}(\tau',\tau)-\sum_{i}\hat{c}_{i\sigma a\vec{x}}^{\dagger}(\tau)\,c^{\dagger}{}_{i\sigma'b\vec{x}}(\tau')\right]\,\Phi_{\sigma\sigma',\vec{x}}^{ab}(\tau,\tau')},\\
1 & = & \int\mathcal{D}F\mathcal{D}\Phi^{\dagger}e^{\frac{1}{2}\left[\mathscr{N}F_{\sigma'\sigma,\vec{x}}^{ba}(\tau',\tau)-\sum_{i}\hat{c}_{i\sigma a\vec{x}}(\tau)\,\hat{c}_{i\sigma'b\vec{x}}(\tau')\right]\,\Phi_{\sigma\sigma',\vec{x}}^{\dagger ab}(\tau,\tau')},\\
1 & = & \int\mathcal{D}D\mathcal{D}\Pi e^{\frac{1}{2}\left[\mathscr{N}D_{\vec{x}}^{ba}(\tau',\tau)-\sum_{i}\phi_{ia\vec{x}}(\tau)\,\phi_{ib\vec{x}}(\tau')\right]\,\Pi_{\vec{x}}^{ab}(\tau,\tau')}.
\end{eqnarray}
\end{subequations} The partition function Eq.\eqref{partition_function_disorder}
then becomes $\overline{\mathscr{Z}^{m}}=\int\mathcal{D}G\mathcal{D}\Sigma\mathcal{D}F\mathcal{D}\Phi\mathcal{D}F^{\dagger}\mathcal{D}\Phi^{\dagger}\mathcal{D}c\mathcal{D}c^{\dagger}\mathcal{D}\phi\,e^{-\mathscr{S}^{{\rm eff}}}$
with action
\begin{eqnarray}
\mathscr{S}^{{\rm eff}} & = & \sum_{iab\sigma\sigma'\vec{x}}\int_{\tau\tau'}\,\hat{c}_{i\sigma a\vec{x}}^{\dagger}(\tau)\,[(\partial_{\tau}-\mu)\delta_{ab}\delta_{\sigma\sigma'}\delta(\tau-\tau')+\Sigma_{\sigma\sigma',\vec{x}}^{ab}(\tau,\tau')]\,\hat{c}_{i\sigma'b\vec{x}}(\tau')\nonumber \\
 & + & \frac{1}{2}\sum_{iab\vec{x}}\int_{\tau\tau'}\,\phi_{ia\vec{x}}(\tau)[(-\partial_{\tau}^{2}+m_{0})\delta_{ab}\delta(\tau-\tau')-\Pi^{ab}(\tau,\tau')]\phi_{ib\vec{x}}(\tau')\nonumber \\
 & + & \frac{1}{2}\sum_{iab\sigma\sigma'\vec{x}}\int_{\tau\tau'}[\hat{c}_{i\sigma a\vec{x}}^{\dagger}(\tau)\,\Phi_{\sigma\sigma',\vec{x}}^{ab}(\tau,\tau')\hat{c}_{i\sigma'b\vec{x}}^{\dagger}(\tau')+\hat{c}_{i\sigma a\vec{x}}(\tau)\,\Phi_{\sigma\sigma',\vec{x}}^{\dagger ab}(\tau,\tau')\,\hat{c}_{i\sigma'b\vec{x}}(\tau')]\nonumber \\
 & - & \mathscr{N}\sum_{ab\sigma\sigma'\vec{x}}\int_{\tau\tau'} G_{\sigma'\sigma,\vec{x}}^{ba}(\tau',\tau)\Sigma_{\sigma\sigma',\vec{x}}^{ab}(\tau,\tau')+\frac{\mathscr{N}}{2}\int_{\tau\tau'} D_{\vec{x}}^{ba}(\tau',\tau)\Pi_{\vec{x}}^{ab}(\tau,\tau')\nonumber \\
 & - & \frac{\mathscr{N}}{2}\sum_{ab\sigma\sigma'\vec{x}}\int_{\tau\tau'}\left[F_{\sigma'\sigma,\vec{x}}^{ba}(\tau'\tau)\Phi_{\sigma\sigma',\vec{x}}^{\dagger ab}(\tau,\tau')+F_{\sigma'\sigma,\vec{x}}^{\dagger ba}(\tau'\tau)\Phi_{\sigma\sigma',\vec{x}}^{ab}(\tau,\tau')\right]\nonumber \\
 & + & \frac{\bar{g}^{2}}{2}\mathscr{N}\sum_{ab\sigma\sigma'\vec{x}}\int_{\tau\tau'}\left[G_{\sigma'\sigma,\vec{x}}^{ba}(\tau,\tau')G_{\sigma\sigma',\vec{x}}^{ab}(\tau',\tau)-F_{\sigma'\sigma,\vec{x}}^{\dagger ba}(\tau,\tau')F_{\sigma\sigma',\vec{x}}^{ab}(\tau',\tau)\right]D_{\vec{x}}^{ab}(\tau,\tau')\nonumber \\
 & + & \frac{t_{0}^{2}}{4}\mathscr{N}\sum_{\substack{\langle\vec{x},\vec{x}'\rangle\\
ab\sigma\sigma'\vec{x}
}
}\int_{\tau\tau'}\left[G_{\sigma'\sigma,\vec{x}'}^{ba}(\tau,\tau')G_{\sigma\sigma',\vec{x}}^{ab}(\tau',\tau)-F_{\sigma'\sigma,\vec{x}'}^{\dagger ba}(\tau,\tau')F_{\sigma\sigma',\vec{x}}^{ab}(\tau',\tau)\right].
\end{eqnarray}
The fermionic part can be reorganized in the Nambu representation
as: 
\begin{equation}
\mathscr{S}_{{\rm ferm}}=-\frac{1}{2}\sum_{iab}\int_{\tau\tau'}\,\,\bar{\psi}_{ia}^{\dagger}(\tau)\left[\bar{G}_{0,ab}^{-1}(\tau,\tau')-\bar{\Sigma}_{ab}(\tau,\tau')\right]\bar{\psi}_{ib}(\tau'),\label{preGaussianGrassmann}
\end{equation}
with Nambu-spinor $\bar{\psi}_{ia}(\tau)=(\hat{c}_{i\uparrow a}(\tau),\hat{c}_{i\downarrow a}(\tau),\hat{c}_{i\uparrow a}^{\dagger}(\tau),\hat{c}_{i\downarrow a}^{\dagger}(\tau))^{T}.$
Bars denote $4\times4$ matrices in Nambu space: 
\begin{align}
\bar{G}_{0}^{-1}(\tau,\tau')=\begin{pmatrix}\hat{G}_{0}^{-1}(\tau,\tau') & 0\\
0 & -\hat{\tilde{G}}_{0}^{-1}(\tau',\tau)
\end{pmatrix},\quad\bar{\Sigma}(\tau,\tau')=\begin{pmatrix}{\hat{\Sigma}}(\tau,\tau') & \hat{\Phi}(\tau,\tau')\\
\hat{\Phi}^{\dagger}(\tau,\tau') & -\hat{\Sigma}(\tau',\tau)
\end{pmatrix}.\label{Nambu_self_energies}
\end{align}
Hats denote $2\times2$ matrices in the spin subspace. The bare propagator
is given by $\hat{G}_{0}^{-1}(\tau,\tau')=-(\partial_{\tau}-\mu)\delta(\tau-\tau')\,\hat{\sigma}_{0},$
$\hat{\tilde{G}}_{0}^{-1}(\tau,\tau')=(\partial_{\tau}+\mu)\delta(\tau-\tau')\,\hat{\sigma}_{0}$,
with $\hat{\sigma}_{0}$ the identity matrix in the 2 dimensional
spin subspace. We now perform the Grassmann integral over the Nambu-fermions
$\bar{\psi}$ and the Gaussian one over the bosons $\phi$, and obtain:
\begin{eqnarray}
\frac{\mathscr{S}^{{\rm eff}}}{\mathscr{N}} & = & -\frac{1}{2}{\rm Tr}\log\left(\bar{G}_{0}^{-1}-\bar{\Sigma}\right)+\frac{1}{2}{\rm Tr}\log\left(D_{0}^{-1}-\Pi\right)
\notag\\
 & - & \sum_{ab\sigma\sigma'\vec{x}}\int_{\tau\tau'} G_{\sigma'\sigma,\vec{x}}^{ba}(\tau',\tau)\Sigma_{\sigma\sigma',\vec{x}}^{ab}(\tau,\tau')+\frac{1}{2}\int_{\tau\tau'} D_{\vec{x}}^{ba}(\tau',\tau)\Pi_{\vec{x}}^{ab}(\tau,\tau')
 \notag\\
 & - & \frac{1}{2}\sum_{ab\sigma\sigma'\vec{x}}\int_{\tau\tau'}\left[F_{\sigma'\sigma,\vec{x}}^{ba}(\tau'\tau)\Phi_{\sigma\sigma',\vec{x}}^{\dagger ab}(\tau,\tau')+F_{\sigma'\sigma,\vec{x}}^{\dagger ba}(\tau'\tau)\Phi_{\sigma\sigma',\vec{x}}^{ab}(\tau,\tau')\right]
 \notag\\
 & + & \frac{\bar{g}^{2}}{2}\sum_{ab\sigma\sigma'\vec{x}}\int_{\tau\tau'}\left[G_{\sigma'\sigma,\vec{x}}^{ba}(\tau,\tau')G_{\sigma\sigma',\vec{x}}^{ab}(\tau',\tau)-F_{\sigma'\sigma,\vec{x}}^{\dagger ba}(\tau,\tau')F_{\sigma\sigma',\vec{x}}^{ab}(\tau',\tau)\right]D_{\vec{x}}^{ab}(\tau,\tau')
 \notag\\
 & + & \frac{t_{0}^{2}}{4}\sum_{\substack{\langle\vec{x},\vec{x}'\rangle\\
ab\sigma\sigma'\vec{x}
}
}\int_{\tau\tau'}\left[G_{\sigma'\sigma,\vec{x}'}^{ba}(\tau,\tau')G_{\sigma\sigma',\vec{x}}^{ab}(\tau',\tau)-F_{\sigma'\sigma,\vec{x}'}^{\dagger ba}(\tau,\tau')F_{\sigma\sigma',\vec{x}}^{ab}(\tau',\tau)\right].
\end{eqnarray}
In addition, we work within the spin-singlet and replica-diagonal ansatz 
\begin{eqnarray}
{G}_{\sigma\sigma',\vec{x}}^{ab}(\tau,\tau')=G_{\vec{x}}(\tau,\tau')\hat{\sigma}_{0}\delta_{ab},\,\,\,\, &  & \quad{\Sigma}_{\sigma\sigma',\vec{x}}^{ab}(\tau,\tau')=\Sigma_{\vec{x}}(\tau,\tau')\hat{\sigma}_{0}\delta_{ab},\nonumber \\
{F}_{\sigma\sigma',\vec{x}}^{ab}(\tau,\tau')=F_{\vec{x}}(\tau,\tau')\,i\hat{\sigma}_{2}\delta_{ab},\,\,\, &  & \quad{\Phi}_{\sigma\sigma',\vec{x}}^{ab}(\tau,\tau')=\Phi_{\vec{x}}(\tau,\tau')\,i\hat{\sigma}_{2}\delta_{ab},\nonumber \\
{F}_{\sigma\sigma',\vec{x}}^{\dagger ab}(\tau,\tau')=-F_{\vec{x}}^{\dagger}(\tau,\tau')\,i\hat{\sigma}_{2}\delta_{ab}, &  & \quad{\Phi^{\dagger}}_{\sigma\sigma',\vec{x}}^{ab}(\tau,\tau')=-\Phi_{\vec{x}}^{\dagger}(\tau,\tau')\,i\hat{\sigma}_{2}\delta_{ab},\label{singlet_pairing}
\end{eqnarray}
with Pauli matrices $\hat{\sigma}_{i}$. Noting that ${\rm Tr}{\rm log}(\bar{G}_{0}^{-1}-\bar{\Sigma})=2{\rm Tr}{\rm log}(\hat{G}_{0}^{-1}-\hat{\Sigma})$,
we finally get Eq.\@ \eqref{eq:effective_action} of the main text.
\end{widetext}

\section{Numerical methods}\label{app:Numerics}

This Appendix sketches the numerical algorithms employed throughout this work to exactly solve the large-$\mathscr{N}$ saddle-point Eliashberg equations (\ref{eq:Eliashberg_eph:hop}) on the imaginary axis, in the normal and superconducting states. 
The propagators for fermions and bosons $G(i \omega_n)$, $F(i \omega_n)$, and $D(i \Omega_n)$, and their associtated self-energies $\Sigma(i \omega_n)$, $\Phi(i\omega_n)$, and $\Pi(i \Omega_n)$, are written in terms of fermionic and bosonic Matsubara frequencies, $\omega_n$ and $\Omega_n$ respectively. These quantities can be transformed to imaginary time $\tau \in \left[0, \beta\right]$ with $\beta=(k_B T)^{-1}$, which is especially useful for convolutions like the ones contained in Eqs.\@ (\ref{eq:Sigma}) or (\ref{eq:Polar}), since in the $\tau$-representation such convolutions become simple products between the convolved functions. 

\subsection{Matsubara Fourier transforms on the imaginary axis}

Numerically, we discretize the imaginary-time interval in equal steps according to $\tau_l=l/(2 N_f k_B T)$ with $l \in \left[0, 2N_f-1\right]$, such that the discretized Matsubara frequencies are $\omega_n=(2 n+1) \pi k_B T$ and $\Omega_n=2n \pi k_B T$, with $n \in \left[-N_f,N_f-1\right]$. $N_f$ is the high-frequency cutoff, that must be chosen to lie well into the ultraviolet regime of the theory where the propagators have already decayed from their low-frequency behavior. Then, the propagators and self-energies become finite discrete lists of values. After algebraic manipulations (circshifts) of these lists, the FFT protocol lends itself to the built-in implementation provided by the optimized \verb|Fourier[]| and \verb|InverseFourier[]| functions of Mathematica. Such implementation is similar to many other self-consistent loops used to solve Shwinger-Dyson saddle-point equations of SYK-like models \cite{Azeyanagi-2018,Patel-2018,Ferrari-2019,Patel-2019,Wang-2020b,Sorokhaibam-2020,Sachdev-1993,Maldacena-2016a,Song-2017,Smit-2021,Grunwald-thesis-2022}. 

\subsection{Self-consistent loops for the saddle-point equations on the imaginary axis}\label{Saddle_selfcons_num}

At the first iteration $j=0$, our self-consistent loop to solve Eqs.\@ (\ref{eq:Eliashberg_eph:hop}) starts with a guess on $\Sigma(i \omega_n)$, which for $g^2<k_B T <1/g^2$ is assumed to be given by the NFL-SYK low-energy solution (\ref{eq:G_NFL_SYK}), while for $k_B T> 1/g^2$ we adopt the impurity-like solution (\ref{eq:G_imp_NS}). We find these guesses to be sufficiently accurate that the loop converges both in the normal and superconducting states, although one could refine the algorithm to assume more accurate low-energy solutions in the superconducting state, such as Eq.\@ (\ref{eq:G_implike_SC_T}) for the fermionic propagator in the SYK$_2$-FL and impurity-like regimes. 
At the end of the current iteration $j>0$, the list of the normal self-energy $\Sigma_{j}(i \omega_n)$ is updated with a weighted sum of the solution $\bar{\Sigma}_j(i \omega_n)$ of Eq.\@ (\ref{eq:Sigma}) and the solution $\Sigma_{j-1}(i \omega_n)$ at the previous iteration $j-1$, according to
\begin{equation}\label{eq:Sigma_mix}
\Sigma_{j}(i \omega_n)=\alpha_s \bar{\Sigma}_j(i \omega_n)+ (1-\alpha_s) \Sigma_{j-1}(i \omega_n).
\end{equation}
The mixing factor $\alpha_s \in \left(0, 1\right)$ helps to stabilize the convergence of the self-consistent loop \cite{Grunwald-thesis-2022}. We heuristically choose $\alpha_s \in \left(0.004, 0.1\right)$ depending on the values of parameters like $g^2/(z t_0^2 \omega_0)$. In practice, we find that smaller values of $\alpha_s$ are required in the superconducting state, and especially at strong coupling (impurity-like regime). 
The error between the current iteration $j>0$ and the previous one $j-1$ is monitored by the sum over the normal self-energy $\epsilon_\Sigma=\sum_{i \omega_n} \left| \Sigma_{j}(i \omega_n)-\Sigma_{j-1}(i \omega_n)\right|$. Convergence is reached when $\epsilon_\Sigma$ falls below a user-imposed threshold. 
In the superconducting state, the gap function $\Delta_j(i \omega_n)$ is also weighted and summed to its value at the previous iteration: 
\begin{equation}\label{eq:Delta_mix}
\Delta_{j}(i \omega_n)=\alpha_d \Delta_j(i \omega_n)+ (1-\alpha_d) \Delta_{j-1}(i \omega_n),
\end{equation}
where $\alpha_d \in \left(0.004, 0.5\right)$ is another heuristically chosen weighting factor, which may be different from $\alpha_s$. We empirically find that the lowest values of $\alpha_d$ are required in the strong coupling, impurity-like regime. 

\subsection{Self-consistent calculation of the critical temperature}\label{App:numeric_Tc}

In order to find the superconducting critical temperature $T_c$, we solve the linearized gap equation (\ref{eq:Phi_Z_tilde}). Here we define $T_c$ at the mean-field, \emph{gap-closing} temperature at which $\Delta(i \omega_n) \rightarrow 0 \, \forall \omega_n$. As described in Sec.\@ \ref{Stiffness}, the superconducting transition temperature is further decreased by phase fluctuations at finite $\mathscr{N}$. 
Finding $T_c$ requires us to know the saddle-point normal-state converged lists for $G(i \omega_n)$, $\Sigma(i \omega_n)$, $D(i \Omega_n)$, and $\Pi(i \Omega_n)$, that we obtain through the self-consistent loop previously sketched in Sec.\@ \ref{Saddle_selfcons_num}. Indeed, after linearization of the Eliashberg equations, $G(i \omega_n)$ does not depend on $\Phi(i \omega_n)$, so it coincides with the normal-state solution. 
Then, the gap equation (\ref{eq:Phi_Z_tilde}) has the structure of an infinite-dimensional eigenvalue problem for the anomalous self-energy $\tilde{\Phi}(i \omega_n)$, which can be written schematically as $\tilde{\Phi}(i \omega_n)=\sum_{m} A_{nm} \tilde{\Phi}(i \omega_m)$, where $\underline{\underline{A}}=\left\{A_{nm}\right\}$ is the kernel matrix which depends on $D(i \Omega_n)$, $Z(i \omega_n)$, and temperature $T$. The dimensionality of $\underline{\underline{A}}$ is truncated to $2 N_f \times 2 N_f$ due to the high-frequency cutoff of the Matsubara propagators. At $T<T_c$, the largest eigenvalue of $\underline{\underline{A}}$ is greater than one, and consequently the self-consistent loop tends to increase the value of $\tilde{\Phi}(i \omega_m)$. Conversely, at $T>T_c$ the self-consistency drives $\tilde{\Phi}(i \omega_m)$ to zero and all eigenvalues of $\underline{\underline{A}}$ are therefore smaller than one. At $T=T_c$, the largest eigenvalue of $\underline{\underline{A}}$ is unitary \cite{Berthod-2018,Valentinis-2016a,Grunwald-thesis-2022}. Monitoring the $T$-dependent eigenvalues of $\underline{\underline{A}}$ is implemented by a root-finding loop on temperature, which allows us to find $T_c$. The accuracy on the final $T_c$ value is set to $dT=10^{-7}$. 
\begin{figure*}[t]
\includegraphics[width=0.8\textwidth]{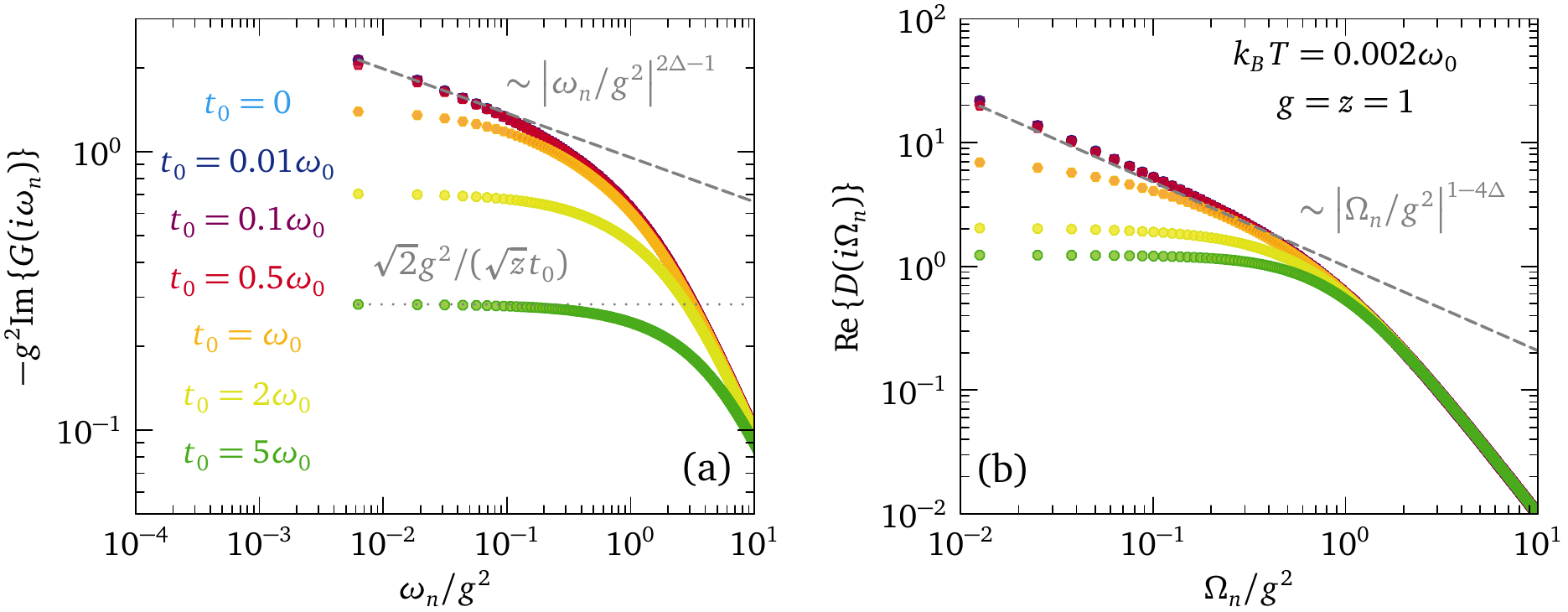}
\caption{\label{fig:Fig_G_hop_SYK} (a) Numerical solution for the fermionic propagator on the imaginary axis, showing the crossover from the SYK-NFL phase to the SYK$_2$-FL regime, at temperature $T=0.002 \omega_0/k_B$, coupling $g=1$, coordination number $z=1$, and for different hoppings $t_0$. The dashed and dotted gray lines are given by the low-energy expansions of Eqs.\@ (\ref{eq:G_NFL_SYK}) and (\ref{eq:G_SYK2_NS}), respectively. (b) Numerical solution for the bosonic propagator, for the same parameters as in panel (a). The dashed gray line stems from Eq.\@ (\ref{eq:D_NFL_SYK}).}
\end{figure*}

\subsection{Estimation of the quasiparticle residue}\label{App:Z}

To extract the quasiparticle residue $Z_{\rm qp}$, we use its definition (\ref{eq:Z_def}) on the real axis and its relation to the imaginary-axis quantity $Z(i \omega_n)$.  
From the Kramers-Kronig relation we have
\begin{equation}\label{eq:Sigma_KK}
\mathrm{Re}\Sigma^R(\omega)=-\frac{1}{\pi}\mathscr{P}\int_{-\infty}^{+\infty} d \omega' \frac{\mathrm{Im}\Sigma^R(\omega')}{\omega-\omega'},
\end{equation}
which in the $\omega \rightarrow 0$ limit yields ($\mathscr{P}$ stands for the principal part of the integral)
\begin{align}\label{eq:Sigma_KK_der0}
\left.\frac{\partial \mathrm{Re}\Sigma^R(\omega)}{\partial \omega}\right|_{\omega \rightarrow 0}= \mathscr{P}\int_{-\infty}^{+\infty} \frac{d \omega'}{\pi} \frac{\mathrm{Im}\Sigma^R(\omega')}{(\omega')^2}. 
\end{align}
On the other hand, from the spectral representation
\begin{equation}\label{eq:Sigma_spectral}
\Sigma(i \omega_n)=\int_{-\infty}^{+\infty} \frac{d \epsilon}{\pi} \frac{\mathrm{Im}\Sigma^R(\epsilon)}{\epsilon-i \omega_n}.
\end{equation}
In the $T\rightarrow0$ limit, and taking the imaginary part of Eq.\@ (\ref{eq:Sigma_spectral}) we have
\begin{equation}\label{eq:Sigma_spectral_Im0}
\mathrm{Im}\Sigma(i \omega)=\int_{-\infty}^{+\infty}  \frac{d \epsilon}{\pi} \frac{\mathrm{Im}\Sigma^R(\epsilon)\omega}{\epsilon^2+\omega^2}.
\end{equation}
Combining equations (\ref{eq:Sigma_spectral_Im0}) and (\ref{eq:Sigma_KK_der0}), we have 
\begin{equation}\label{eq:Z_ImSigma2}
\frac{1}{Z_{\rm qp}}=1-\left.\frac{\mathrm{Im}\Sigma(i \omega)}{\omega}\right|_{\omega \rightarrow 0}.
\end{equation}
Using the relation (\ref{eq:Sigma_Z}) valid for our model on the imaginary axis, we finally achieve Eq.\@ (\ref{eq:Z_ImSigma}).
Therefore, the real-axis zero-energy quasiparticle weight can be estimated through Eq.\@ (\ref{eq:Z_ImSigma}) by the inverse of the imaginary axis quantity $Z(i \omega)$, in the zero-temperature limit. In practice, we use the imaginary-axis code described in Sec.\@ \ref{Saddle_selfcons_num} at temperatures $T \ll T_c$ to approximate Eq.\@ (\ref{eq:Z_ImSigma}). As a consistency check, we also evaluate $Z_{\rm qp}$ for selected couplings and hoppings through the definition (\ref{eq:Z_def}) using the real-axis code described in Sec.\@ \ref{App:num_real_axis}, which allows us to go to the zero-energy limit but still requires finite temperatures. 
\begin{figure*}[t]
\includegraphics[width=0.8\textwidth]{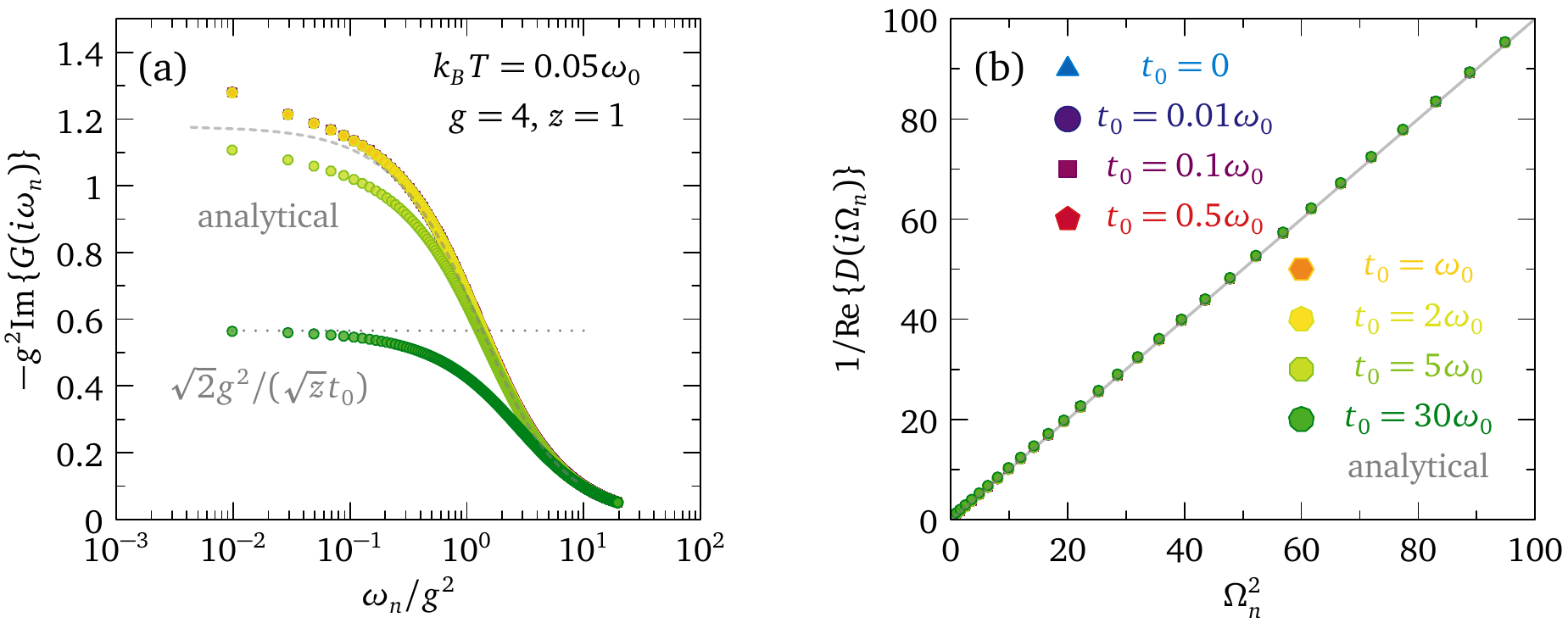}
\caption{\label{fig:Fig_G_hop_imp} (a) Numerical solution for the fermionic propagator on the imaginary axis, showing the crossover from the impurity-like phase to the SYK$_2$-FL regime, at temperature $T=0.05 \omega_0/k_B$, coupling $g=4$, coordination number $z=1$, and for different hoppings $t_0$. The dashed curve and dotted gray line are given by Eq.\@ (\ref{eq:G_imp_NS}) and by the low-energy expansion of Eq.\@ (\ref{eq:G_SYK2_NS}), respectively. (b) Numerical solution for the bosonic propagator, for the same parameters as in panel (a). The dashed gray line is generated by Eqs.\@ (\ref{eq:D_imp_NS}) and (\ref{eq:omegar_imp}).}
\end{figure*}

\subsection{Saddle-point equations on the real axis}\label{App:num_real_axis}

Our numerical solution of the saddle-point equations (\ref{eq:Eliashberg_eph:hop}) on the real axis is based on the spectral (Lehmann) representation of the fermionic and bosonic propagators \cite{Berthod-2018}, which can be implemented through Laplace transforms \cite{Schmalian-1996}. In essence, if we have to perform a convolution on the imaginary axis of the form 
\begin{equation}\label{eq:conv_imag}
\alpha(i \Omega_n)=-k_B T \sum_{m=-\infty}^{+\infty} g_1(i \omega_m+i\Omega_n)g_2(i \omega_m),
\end{equation}
on the real axis it follows that
\begin{align}\label{eq:conv_real}
\alpha(\omega)&=\int_{-\infty}^{+\infty} \frac{d \epsilon}{\pi} f_{FD}(\epsilon)\left[ g_1(\omega+\epsilon)\mathrm{Im}\left\{g_2(\epsilon)\right\} \right. \notag \\ & \left. +\mathrm{Im}\left\{g_1(\epsilon)\right\}g_2^\ast(\epsilon-\omega)\right].
\end{align}
Defining the Laplace transform $\alpha(t)$ as
\begin{equation}\label{eq:Laplace}
\alpha(\omega)=\int_0^{+\infty} dt \alpha(t) e^{i \omega t},
\end{equation}
Eq.\@ (\ref{eq:conv_real}) translates as 
\begin{equation}\label{eq:conv_Laplace}
\alpha(t)=i (2\pi)^2 \left[\rho_1(t) a_2^\ast(t)-\rho_2^\ast(t) a_1(t)\right],
\end{equation}
where 
\begin{equation}\label{eq:rhoi_t}
\rho_i(t)=-\int_{-\infty}^{+\infty}\frac{d \epsilon}{2 \pi}\frac{\mathrm{Im}\left\{g_i(\epsilon)\right\}}{\pi}e^{-i \epsilon t}
\end{equation}
and
\begin{equation}\label{eq:ai_t}
a_i(t)=-\int_{-\infty}^{+\infty}\frac{d \epsilon}{2 \pi}f_{FD}(\epsilon)\frac{\mathrm{Im}\left\{g_i(\epsilon)\right\}}{\pi}e^{-i \epsilon t}.
\end{equation}
Employing Eq.\@ (\ref{eq:conv_Laplace}), one can transform convolutions of the form (\ref{eq:conv_real}) on the real axis into simpler products between the functions (\ref{eq:rhoi_t}) and (\ref{eq:ai_t}). Then, an inverse Laplace transform of the resulting $\alpha(t)$ yields $\alpha(\omega)$ as a function of real frequency $\omega$. 
We apply the above approach to the fermionic self-energy (\ref{eq:Sigma}), the anomalous self-energy (\ref{eq:Phi}), and the boson self-energy (\ref{eq:Polar}), written on the real axis by the means of the spectral representation (\ref{eq:conv_real}). The respective Laplace transforms $\Sigma^R(t)$, $\Phi^R(t)$, and $\Pi^R(t)$, where the superscript $R$ denotes the retarded quantities, are complemented by the Dyson equations
\begin{equation}\label{eq:G_real}
G^R(\omega)=\frac{\omega+i \gamma-\Sigma^R(\omega)}{\left[\omega+i \gamma-\Sigma^R(\omega)\right]^2-\left[\Phi^R(\omega)\right]^2},
\end{equation}
\begin{equation}\label{eq:F_real}
F^R(\omega)=\frac{\Phi^R(\omega)}{\left[\omega+i \gamma-\Sigma^R(\omega)\right]^2-\left[\Phi^R(\omega)\right]^2},
\end{equation}
and
\begin{equation}\label{eq:D_real}
D^R(\omega)=\frac{1}{\omega_0^2-(\omega+i \gamma)^2-\Pi^R(\omega)},
\end{equation}
where $\gamma=i 0^+$, and iterated self-consistently starting from either a BCS-like solution of the form \cite{Bruus-2004mb}
\begin{subequations}\label{eq:G_F_BCS}
\begin{equation}
G^R(\omega)=\frac{\omega+i \gamma}{(\omega+i \gamma)^2-\Delta^2},
\end{equation}
\begin{equation}
F^R(\omega)=\frac{\Delta}{(\omega+i \gamma)^2-\Delta^2},
\end{equation}
\end{subequations}
or from the results of a previously converged self-consistent loop. In practice, we employ a finite damping $\gamma=10^{-5}\omega_0$. The fermionic self-energy $\Sigma(\omega)$ is weighted at each self-consistent iteration with the result of the previous iteration, similarly to Eq.\@ (\ref{eq:Sigma_mix}), with a mixing factor $\alpha_s \in \left(0, 1\right)$. 
The error between the current iteration $j>0$ and the previous one $j-1$ is monitored by the sum over the normal self-energy $\epsilon_\Sigma=\sum_{\omega} \left| \Sigma^R_{j}(\omega)-\Sigma_{j-1}^R(\omega)\right|$: when $\epsilon_\Sigma$ decreases below a user-imposed value, convergence is reached. 

\section{Derivation of the normal-state results}\label{Normal_prop}

\subsection{Fermion and boson propagators on the imaginary axis}

In this section, for completeness we report the graphs of the normal-state fermion $G(i \omega_n)$ and boson $D(i \Omega_n)$ propagators on the imaginary axis. These propagators stem from the self-consistent solution of the saddle-point equations (\ref{eq:saddle_point_NS}), obtained with the methods described in Appendix \ref{Saddle_selfcons_num}. Notice that, on the imaginary axis, $G(i \omega_n)$ and $D(i \Omega_n)$ are purely imaginary and purely real, respectively, as deduced from Eqs.\@ (\ref{eq:saddle_point_NS}) and (\ref{eq:Sigma_Z}). 
\begin{figure*}[t]
\includegraphics[width=0.8\textwidth]{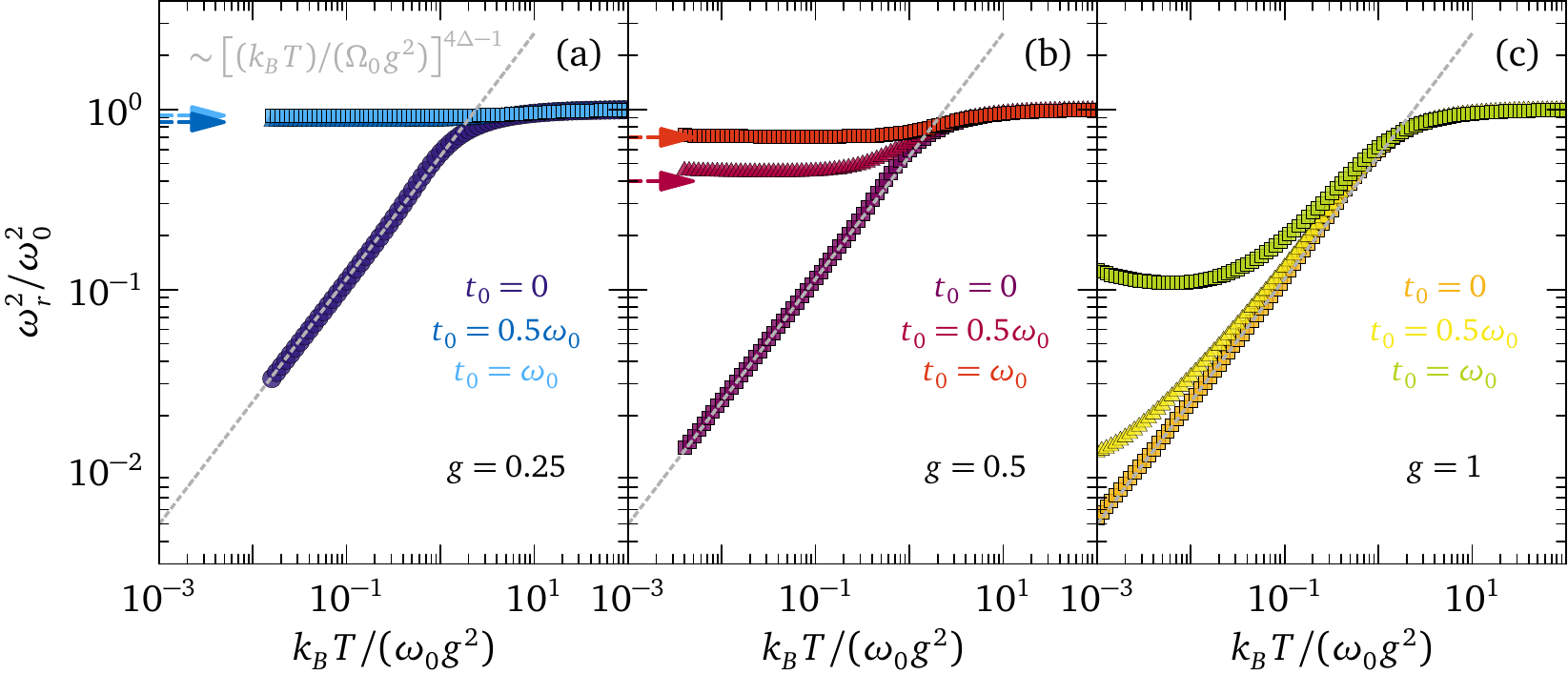}
\caption{\label{fig:omrsq_t0_NFL} Normal-state squared renormalized boson frequency $\omega_r^2/\omega_0^2$ as a function of normalized temperature $k_B T/(\omega_0 g^2)$ and fermion hopping parameter $t_0$, calculated from the full numerical solution of the Yukawa-SYK Eliashberg equations in accordance with Eq.\@ (\ref{eq:omegar_gen}). The dashed gray line is the power-law solution $\omega_r^2/\omega_0^2 \sim \left[(k_B T)/(\omega_0 g^2)\right]^{4 \Delta-1}$ found in the SYK-NFL regime of the single Yukawa-SYK dot \cite{Esterlis-2019}. Panels (a), (b), and (c) show the numerical results for $g=\left\{0.25, 0.5, 1\right\}$, respectively, for different hopping parameters $t_0/\omega_0=\left\{0, 0.5,1 \right\}$. Arrows indicate the analytical estimation (\ref{eq:omegar_0_T0}) of the zero-temperature value for $\omega_r^2/\omega_0^2$ in the SYK$_2$-FL regime, for the same parameters as the corresponding numerical data points of the same color.}
\end{figure*}
Fig.\@ \ref{fig:Fig_G_hop_SYK}(a) shows the results for $-g^2\mathrm{Im}\left\{G(i \omega_n)\right\}$ across the SYK-NFL/SYK$_2$-FL crossover, at temperature $T=0.002 \omega_0/k_B$, coupling $g=1$, coordination number $z=1$, and for different hoppings $t_0$. At the lowest values of $t_0$, the fermionic propagator displays the power-law dependence (\ref{eq:G_NFL_SYK}) at small energies, which is characteristic of the SYK-NFL regime, and is indicated by the dashed gray line. The crossover to the disordered FL regime is realized as higher hoppings, with visible differences in the propagator appearing for $t_0\approx \omega_0$, consistently with the crossover energy $\omega_c$ estimated in Sec.\@ \ref{Normal_cross} and with the spectral functions on the real axis in Fig.\@ \ref{fig:spectr_NFL-SYK_g}(c). In the SYK$_2$-FL regime, the low-energy part of the propagator tends to the constant $\sqrt{2}/(\sqrt{z} t_0)$ (the additional factor of $g^2$ is due to the normalization of the $y$ axis in the figure), as deduced from Eq.\@ (\ref{eq:G_SYK2_NS}) at small $\omega_n$; this behavior is shown by the dotted gray line. 
Fig.\@ \ref{fig:Fig_G_hop_SYK}(b) reports the results for $D(i \Omega_n)$, for the same parameters as in panel (a). In the SYK-NFL regime, i.e.\@, at low hopping, we recognize the power-law scaling of Eq.\@ (\ref{eq:D_NFL_SYK}), which is highlighted by the dashed gray line. Such scaling evolves into almost free bosons in the high-hopping SYK$_2$-FL regime, since the boson self-energy becomes negligible when hopping dominates over fermion-boson coupling; this stiffening of the renormalized boson frequency towards $\omega_r \approx \omega_0$ is consistent with the analogous trend of the imaginary part of the real-axis boson propagator, shown in Fig.\@ \ref{fig:spectr_NFL-SYK_g}(d). 
\begin{figure}[t]
\includegraphics[width=0.9\columnwidth]{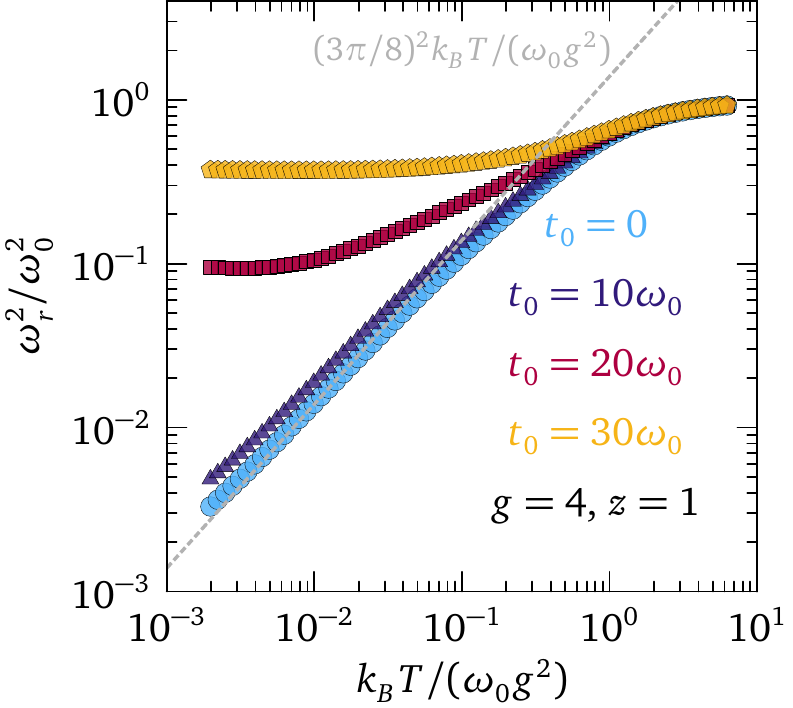}
\caption{\label{fig:omrsq_t0_imp} Normal-state squared renormalized boson frequency $\omega_r^2/\omega_0^2$ as a function of normalized temperature $k_B T/(\omega_0 g^2)$, for coupling $g=4$ and for different hoppings $t_0$. Data points are calculated from the full numerical solution of the Yukawa-SYK Eliashberg equations Eq.\@ (\ref{eq:omegar_gen}). The dashed gray line is the power-law solution (\ref{eq:omegar_imp}) found in the impurity-like regime of the single Yukawa-SYK dot \cite{Esterlis-2019}.}
\end{figure}

The results for $-g^2\mathrm{Im}\left\{G(i \omega_n)\right\}$ across the impurity-like/SYK$_2$-FL crossover are shown in Fig.\@ \ref{fig:Fig_G_hop_imp}(a), at temperature $T=0.05 \omega_0/k_B$, coupling $g=4$, coordination number $z=1$, and for different hoppings. At low hopping, the curves follow the analytical result (\ref{eq:G_imp_NS}), proper of the impurity-like regime, and indicated by the dashed gray curve. Increasing hopping, we cross over to the SYK$_2$-FL regime, and the low-energy part of the propagator approaches $\sqrt{2}/(\sqrt{z} t_0)$, as marked by the dotted gray line. The crossover of fermions to the disordered FL regime occurs at $t_0 \approx 19 \omega_0$ at $T=0.05 k_B/\omega_0$, consistently with the criterion (\ref{eq:omegac_IL_analyt}) in Sec.\@ \ref{Normal_cross}, and with the real-axis spectral functions shown in Fig.\@ \ref{fig:spectr_imp_g}(c). 
Fig.\@ \ref{fig:Fig_G_hop_imp}(b) shows the boson propagator, which is consistent with the impurity-like solution given by Eqs.\@ (\ref{eq:D_imp_NS}) and (\ref{eq:omegar_imp}) at low hoppings. At the highest value $t_0=30 \omega_0$, there are modifications which appear at low frequency, due to the stiffening of the renormalized boson frequency. These features are most easily seen by looking at the analogous real-axis calculations in Fig.\@ \ref{fig:spectr_imp_g}(d).

\subsection{Normal-state renormalized boson frequency}

The renormalized boson frequency is given by Eq.\@ (\ref{eq:omegar_gen}), while the boson self-energy in the normal state follows Eq.\@ (\ref{eq:Polar_Z}). We first present the full numerical results stemming from Eqs.\@ (\ref{eq:saddle_point_NS}), and then report the derivations of analytical approximations. 
Fig.\@ \ref{fig:omrsq_t0_NFL} shows $\omega_r^2/\omega_0^2$ as a function of normalized temperature $k_B T/(\omega_0 g^2)$, for different hopping parameters and different couplings. In the single-dot limit $t_0=0$, all curves follow the power-law scaling (\ref{eq:omegar_NFL_SYK}) found in SYK-NFL regime. Increasing hopping at fixed coupling, we cross over to the SYK$_2$-FL regime and $\omega_r^2$ stiffens because the static boson self-energy is negligible for $g^2 \omega_0 \ll z t_0^2$, as seen from Eq.\@ (\ref{eq:omegar_0_T0}). The latter analytical expression, valid at $T=0$ and derived in App.\@ \ref{SYK2_omegar}, yields the values indicated by arrows in Fig.\@ \ref{fig:omrsq_t0_NFL}. 
Fig.\@ \ref{fig:omrsq_t0_imp} displays $\omega_r^2/\omega_0^2$ as a function of normalized temperature $k_B T/(\omega_0 g^2)$, for $g=4$ and different hoppings. The single-dot curve $t_0=0$ is consistent with Eq.\@ (\ref{eq:omegar_imp}), characteristic of the impurity-like regime: the bosons are critical at zero temperature. As in Fig.\@ \ref{fig:omrsq_t0_NFL}, the crossover to the SYK$_2$-FL regime is realized by increasing hopping, which makes $\omega_r^2$ stiffen: the bosons are no longer critical in the disordered FL phase.

We now derive analytical expressions for $\omega_r$ at $T=0$ in the SYK$_2$-FL and impurity-like regimes. 

\subsubsection{Fermi-liquid SYK$_2$-FL regime}\label{SYK2_omegar}

In the SYK$_2$-FL regime, for $\bar{g}^2/\omega_0^2 \ll z t_0^2$, the dynamical quasiparticle weight is approximately given by Eq.\@ (\ref{eq:SYK2_Z}), i.e.\@,
\begin{equation}\label{eq:Z_SYK2_NS}
Z(i \omega_n)=\frac{1}{2} \left(1 + \sqrt{1+\frac{2 z t_0^2}{\omega_n^2}}\right) 
\end{equation} 
in the normal state, a result which follows from Eq.\@ (\ref{eq:G_SYK2_NS}) and the normal-state Dyson equation (\ref{eq:G_NS}). 
Let us first calculate the normal-state polarization bubble at zero boson frequency and at $T=0$. Using Eqs.\@ (\ref{eq:Polar_Z}) and (\ref{eq:Z_SYK2_NS}), we have
\begin{multline}\label{eq:Pi0_NS_SYK2}
\Pi(0)=2 \bar{g}^2 k_B T \sum_{m=-\infty}^{+\infty} \frac{1}{\omega_m^2} \frac{1}{\left[Z(i \omega_m)\right]^2} \\ \approx 2 \bar{g}^2 k_B T \sum_{m=-\infty}^{+\infty} \frac{1}{\omega_m^2} \frac{4}{\left[1+\sqrt{1+\frac{2 z t_0^2}{\omega_m^2}}\right]^2}\\ =8 \bar{g}^2 k_B T \sum_{m=-\infty}^{+\infty} \frac{1}{\left[\omega_m+\sqrt{\omega_m^2+ 2 z t_0^2}\right]^2}.
\end{multline}
At $T=0$, the sums over Matsubara frequencies become integrals, so that
\begin{align}\label{eq:Pi0_NS_SYK2_T0}
\Pi(0)&=8 \bar{g}^2 \int_{-\infty}^{+\infty} \frac{d \omega}{2 \pi} \frac{1}{\left(\omega+\sqrt{\omega^2+ 2 z t_0^2}\right)^2} \notag \\ &=\frac{8 \sqrt{2} \bar{g}^2}{3 \pi \sqrt{z} t_0}.
\end{align}
The renormalized boson frequency thus has an analytical expression at $T=0$, which stems from Eqs.\@ (\ref{eq:omegar_gen}) and (\ref{eq:Pi0_NS_SYK2_T0}): it is Eq.\@ (\ref{eq:omegar_0_T0}). 
Of course Eq.\@ (\ref{eq:omegar_0_T0}) is valid only as long as $\omega_r(0) > 0$, which means at sufficiently low interaction $\bar{g}^2/\omega_0^2 \ll 3 \pi \sqrt{z} t_0/(8 \sqrt{2})$. The estimation (\ref{eq:omegar_0_T0}) favorably compares with the full numerical solution of the Eliashberg equations in the SYK$_2$-FL regime, as shown by the arrows in Fig.\@ \ref{fig:omrsq_t0_NFL}.

\subsubsection{Strong-coupling impurity-like regime}\label{Imp_omegar}

For the dynamical weight $Z(i \omega_n)$ in the impurity-like regime we can employ the parametrization (\ref{eq:Z_T_implike_NS}) that interpolates between the SYK$_2$-FL and impurity-like regions of the phase diagram: 
\begin{equation}\label{eq:Z_T_implike_NS_Omegatilde}
Z(i \omega_n)\approx\frac{1}{2} \left[1+ \sqrt{1+\frac{\tilde{\Omega}^2}{(\omega_n)^2}}\right],
\end{equation}
where $\tilde{\Omega}$ follows Eq.\@ (\ref{eq:tilde_Omega_interp}): it is a characteristic frequency which interpolates between the SYK$_2$-FL and impurity-like regimes. 

Using Eqs.\@ (\ref{eq:Polar_Z}) and (\ref{eq:Z_T_implike_NS_Omegatilde}), we achieve
\begin{multline}\label{eq:Pi0_NS_implike}
\Pi(0)=2 \bar{g}^2 k_B T \sum_{m=-\infty}^{+\infty} \frac{1}{\omega_m^2} \frac{1}{\left[Z(i \omega_m)\right]^2} \\ \approx 2 \bar{g}^2 k_B T \sum_{m=-\infty}^{+\infty} \frac{1}{\omega_m^2} \frac{4}{\left[1+\sqrt{1+\frac{\tilde{\Omega}^2}{\omega_m^2}}\right]^2}\\ =8 \bar{g}^2 k_B T \sum_{m=-\infty}^{+\infty} \frac{1}{\omega_m^2} \frac{1}{\left[1+\sqrt{1+ \tilde{\Omega}^2/\omega_m^2}\right]^2}.
\end{multline}
At $T=0$, the sum over Matsubara frequencies becomes an integral, therefore
\begin{align}\label{eq:Pi0_NS_implike_T0}
\Pi(0)&=8 \bar{g}^2 \int_{-\infty}^{+\infty} \frac{d \omega}{2 \pi} \frac{1}{\omega^2}\frac{1}{\left(1+\sqrt{1+ \tilde{\Omega}^2/\omega^2}\right)^2} \notag \\ &=\frac{16 \bar{g}^2}{3 \pi \tilde{\Omega}}.
\end{align}
The renormalized boson frequency stems from Eqs.\@ (\ref{eq:omegar_gen}) and (\ref{eq:Pi0_NS_implike_T0}): 
\begin{align}\label{eq:omegar_0_T0_imp}
\left[\omega_r(0)\right]^2 &\approx \omega_0^2-\frac{16 \bar{g}^2}{3 \pi \tilde{\Omega}}\notag \\ &=\omega_0^2-\frac{16 \bar{g}^2}{3 \pi} \left[ 2 z t_0^2+4 \left(\frac{8 g^2 \omega_0}{3 \pi}\right)^2 \right]^{-\frac{1}{2}}.
\end{align}
In the SYK$_2$-FL limit $\bar{g}^2/\omega_0^2 \ll \sqrt{z} t_0$, we retrieve Eq.\@ (\ref{eq:omegar_0_T0}) from the just obtained result (\ref{eq:omegar_0_T0_imp}). On the other hand, in the impurity-like regime $\bar{g}^2/\omega_0^2 \gg \sqrt{z} t_0$, neglecting hopping altogether yields
\begin{equation}\label{eq:omegar_0_T0_imp_t00}
\left[\omega_r(0)\right]^2\approx \omega_0^2-\frac{16 \bar{g}^2}{3 \pi} \frac{ 3 \pi \omega_0^2}{16 \bar{g}^2}=0,
\end{equation}
which is consistent with Eq.\@ (\ref{eq:omegar_imp}): the bosons are fully massless at zero temperature in the limit of zero hopping. 

\section{Derivation of the superconducting-state results}\label{SC_prop}

In this Appendix, we show that analytical results are reachable for various quantities in the superconducting state, in both the Fermi-liquid and impurity-like regimes. The general strategy in the SYK$_2$-FL regime involves approximating $D(i \Omega_n)$ with the propagator for free bosons, 
\begin{equation}\label{eq:D_free}
D(i \Omega_n) \approx \frac{1}{\omega_0^2},
\end{equation}
since the boson self-energy is negligible in such regime, where $g^2 \omega_0 \ll z t_0^2$. Furthermore, in both the SYK$_2$-FL and impurity-like regimes, a useful approximation for the gap function $\Delta(i\omega_n)$ is the piecewise-constant 
\begin{equation}\label{eq:gap_const}
\Delta(i\omega_n)= \begin{cases} \Delta \, : \,  \left|\omega_n\right| \leq \omega_0,  \\ 0\, : \,  \left|\omega_n\right| > \omega_0.\end{cases}
\end{equation}
In the impurity-like regime, the rather crude approximation (\ref{eq:gap_const}) neglects all strong-coupling retardation effects on the gap $\Delta(i \omega_n)$, which manifest themselves in the oscillations observed in the superconducting spectral functions of the interacting Cooper-pair fluid; see Fig.\@ \ref{fig:spectr_imp_g_SC}. Nevertheless, Eq.\@ (\ref{eq:gap_const}) is sufficient to achieve qualitatively valid expressions for thermodynamic quantities even in the impurity-like regime, knowing the zero-energy gap $\Delta_0$ as a function of coupling and hopping. One example of such quantities is the phase stiffness, analyzed in Sec.\@ \ref{Stiffness}. 
Given the above assumptions, we proceed with the analysis of the superconducting state in the SYK$_2$-FL and impurity-like regimes. 

\subsection{Weak-coupling disordered Fermi liquid: the SYK$_2$-FL regime}

\subsubsection{Dynamical quasiparticle weight}\label{App:Z_SYK2-FL}

In the SYK$_2$-FL regime, we can assume the bosons to obey Eq.\@ (\ref{eq:D_free}), i.e.\@, for a negligible boson self-energy $\Pi(i \Omega_n)\approx 0$ and in the static limit, as verified with a full numerical solution of the Eliashberg equations for $\bar{g}^2/\omega_0^2 \ll z t_0^2$. Using Eqs.\@ (\ref{eq:Sigma}) and (\ref{eq:Sigma_Z}) in this regime, where we can neglect the term in $\Sigma(i \omega_n)$ that depends on fermion-boson coupling, we find
\begin{equation}\label{eq:Z_SYK2-SC}
Z(i \omega_n)\approx 1 +\frac{z t_0^2}{2} \frac{1}{Z(i \omega_n) \left\{\omega_m^2+\left[\Delta(i \omega_m)\right]^2\right\}}. 
\end{equation} 
We now make the assumption (\ref{eq:gap_const}) of a constant gap up to $\omega_0$, which is numerically confirmed by the full self-consistent solution of the lattice-SYK Eliashberg equations. Then, at low energies $\left|i \omega_n\right| \lessapprox \omega_0$, we can write
\begin{equation}\label{eq:Z_SYK2_gapconst}
Z(i \omega_n)\approx 1 +\frac{z t_0^2}{2} \frac{1}{Z(i \omega_n) \left(\omega_m^2+\Delta^2\right)}. 
\end{equation} 
The solution to the quadratic equation (\ref{eq:Z_SYK2_gapconst}) is 
\begin{equation}\label{eq:Z_SYK2_gapconst_expl_pm}
Z(i \omega_n)=\frac{1}{2} \left(1 \pm \sqrt{1+\frac{2 z t_0^2}{\Delta^2+\omega_n^2}}\right),
\end{equation} 
where we have to take the positive root in order for $Z(\omega)$ to have physically sound asymptotics, as in the normal state. We thus conclude that in the superconducting state, for $\bar{g}^2/\omega_0^2 \ll z t_0^2$,
\begin{equation}\label{eq:Z_SYK2_gapconst_expl}
Z(i \omega_n)=\frac{1}{2} \left(1 + \sqrt{1+\frac{2 z t_0^2}{\Delta^2+\omega_n^2}}\right). 
\end{equation} 
In the normal state, $\Delta \rightarrow 0$ and we retrieve the quasiparticle weight of the SYK$_2$-FL model, stemming from Eq.\@ (\ref{eq:G_SYK2_NS}): it is given by Eq.\@ (\ref{eq:Z_SYK2_NS}).

\subsubsection{Fermion propagators and self-energies in the superconducting state}

Using Eq.\@ (\ref{eq:Sigma_Z}), we can calculate the fermion self-energy from the previously obtained dynamical quasiparticle weight (\ref{eq:Z_SYK2_gapconst_expl}) in the superconducting state: 
\begin{align}\label{eq:Sigma_SYK2_gapconst}
\Sigma(i \omega)&=\frac{1}{2} \left[i \omega_n -i \mathrm{sign}(\omega_n) \right. \nonumber \\ & \times \left. \sqrt{\omega_n^2+\frac{ 2 z t_0^2 \omega_n^2}{\omega_n^2 +\Delta^2}}\right].
\end{align}
In the normal state, i.e.\@, for $\Delta\rightarrow 0$, Eq.\@ (\ref{eq:Sigma_SYK2_gapconst}) correctly yields the SYK$_2$-FL self-energy stemming from Eqs.\@ (\ref{eq:G_SYK2_NS}) and (\ref{eq:Sigma_Dyson_SYK2}).  
The fermion propagator results from the normal-state Dyson equation (\ref{eq:Sigma_Dyson_SYK2}), which is also approximately valid in the SYK$_2$-FL superconducting regime, since fermion-boson coupling is negligible with respect to hopping in such regime. We thus achieve 
\begin{equation}\label{eq:G_SYK2_gapconst}
G(i \omega_n)=\frac{1}{z t_0^2} \left[i \omega_n -i \mathrm{sign}(\omega_n) \sqrt{\omega_n^2+\frac{ 2 z t_0^2 \omega_n^2}{\omega_n^2 +\Delta^2}}\right]. 
\end{equation}
In the normal state we consequently obtain the SYK$_2$-FL propagator (\ref{eq:G_SYK2_NS}).
Performing the analytic continuation $i \omega_n \rightarrow \omega+i 0^+$ on Eq.\@ (\ref{eq:G_SYK2_gapconst}), we arrive at the (retarded) fermion propagator on the real axis: Eq.\@ (\ref{eq:Sigma_SYK2_gapconst_real}).
In the superconducting phase, the anomalous propagator $F(i \omega_n)$ appears as well, in accordance with Eq.\@ (\ref{eq:F2}). Combining the latter with Eq.\@ (\ref{eq:Phi_Delta}), we obtain 
\begin{equation}\label{eq:F_Z_Delta}
F(i \omega_n)=-\frac{ \Delta(i \omega_n)}{Z(i \omega_n)\left\{\omega_n^2+\left[\Delta(i \omega_n)\right]^2\right\}}.
\end{equation}
Making the assumption (\ref{eq:gap_const}) of an energy-independent gap in the SYK$_2$-FL state, and using the previously determined dynamical weight (\ref{eq:Z_SYK2_gapconst_expl}), we arrive at
\begin{equation}\label{eq:F_Z_Delta_SYK2}
F(i \omega_n)=-\frac{2\Delta}{\left[\omega_n^2+\Delta^2\right]\left[1+\sqrt{1+\frac{2 z {t_0^2}}{\omega_n^2+\Delta}^2}\right]}.
\end{equation}
The real-axis retarded anomalous propagator stems from setting $i \omega_n \rightarrow \omega+i 0^+$ in Eq.\@ (\ref{eq:F_Z_Delta_SYK2}):
\begin{align}\label{eq:F_Z_Delta_SYK2_real}
F^R(\omega)&=-\frac{2 \Delta}{\Delta^2-(\omega+i 0^+)^2}\notag \\ & \times \frac{1}{1+ \sqrt{1+\frac{2 z t_0^2}{\Delta^2-(\omega+i 0^+)^2}}}.
\end{align}

\subsubsection{Superconducting-state renormalized boson frequency}\label{App:omegar_SC_SYK2}

In the superconducting state, the boson frequency (\ref{eq:omegar_gen}) is altered with respect to the normal-state result (\ref{eq:omegar_0_T0}), due to the opening of the gap $\Delta(i \omega_n)$ in the spectral function. The boson self-energy follows Eq.\@ (\ref{eq:Polar}), which can be recast in terms of the dynamical quasiparticle weight $Z(i\omega_n)$ by the means of Eqs.\@ (\ref{eq:Sigma_Z}) and (\ref{eq:G_NS}):
\begin{multline}\label{eq:Pi_def_SC}
\Pi(i \Omega_n)=-2 \bar{g}^2 k_B T \sum_{m=-\infty}^{+\infty}  \sum_{m=-\infty}^{+\infty} \left[ \frac{1}{i \omega_m Z(i \omega_m)}\right. \\ \left. \times \frac{1}{(i \omega_m+i \Omega_n) Z(i \omega_m+i \Omega_n)} \right. \\ \left. -F(i \omega_m) F(i \omega_m+i \Omega_n)\right].
\end{multline}
We now approximate the dynamical weight $Z(i \omega_n)$ by Eq.\@ (\ref{eq:Z_SYK2_gapconst_expl}) in the superconducting state, with the piecewise-constant gap (\ref{eq:gap_const}). One can verify both numerically and analytically that the $Z(i \omega_n)$-dependent term in Eq.\@ (\ref{eq:Pi_def_SC}) dominates over the $F(i \omega_n)$-dependent term for $\Delta \lessapprox 1$. Therefore, in the following we neglect the $F(i \omega_n)$-dependent term in Eq.\@ (\ref{eq:Pi_def_SC}). Using Eqs.\@ (\ref{eq:Pi_def_SC}) and (\ref{eq:Z_SYK2_gapconst_expl}), and going to the $T=0$ limit, we are left with
\begin{multline}\label{eq:Pi0_SC_SYK2}
\Pi(0)=2 \bar{g}^2 \left\{ \int_{-\infty}^{-\omega_0} \frac{d \omega}{2\pi} \frac{1}{\omega^2} \frac{4}{\left[1+\sqrt{1+\frac{2 z t_0^2}{\omega^2}}\right]^2} \right. \\ \left. +\int_{\omega_0}^{+\infty} \frac{d \omega}{2\pi} \frac{1}{\omega^2} \frac{4}{\left[1+\sqrt{1+\frac{2 z t_0^2}{\omega^2}}\right]^2}\right. \\  \left. + \int_{-\omega_0}^{\omega_0} \frac{d \omega}{2\pi} \frac{1}{\omega^2} \frac{4}{\left[1+\sqrt{1+\frac{2 z t_0^2}{\omega^2+\Delta^2}}\right]^2} \right\}. 
\end{multline}
The three integrations in Eq.\@ (\ref{eq:Pi0_SC_SYK2}) can be performed analytically, with the latter one yielding a rather lengthy expression in terms of imcomplete elliptic integrals of the second kind. Expanding the explicit result of Eq.\@ (\ref{eq:Pi0_SC_SYK2}) at leading order for $z t_0^2 \rightarrow +\infty$, we achieve
\begin{equation}\label{eq:Pi0_SC_SYK2_expl}
\Pi(0)\approx \frac{8 \bar{g}^2}{6 \pi \omega_0 z t_0^2} \left(-3 \Delta^2 +  2 \sqrt{2} \omega_0 \sqrt{z} t_0\right).
\end{equation}
Notice that the normal-state limit of Eq.\@ (\ref{eq:Pi0_SC_SYK2_expl}), i.e.\@, $\Delta \rightarrow 0$, consistently yields Eq.\@ (\ref{eq:Pi0_NS_SYK2_T0}). Using the definition (\ref{eq:omegar_gen}) and Eq.\@ (\ref{eq:Pi0_SC_SYK2_expl}), the renormalized boson frequency becomes Eq.\@ (\ref{eq:omegar_SC_SYK2_expl}).

\subsubsection{Gap equation}

The gap equation stems from Eq.\@ (\ref{eq:Phi}) for the anomalous self-energy $\Phi(i \omega_n)$.
Here we assume to be in the SYK$_2$-FL regime, where $\bar{g}^2/\omega_0^2 \ll z {t_0^2}$. For the boson propagator, we neglect the boson self-energy $\Pi(i\Omega_n) \approx 0 \, \forall i \omega_n$ altogether, which nevertheless yields a satisfactory agreement of the ensuing analytical expressions, for the transition temperature and the zero-temperature gap at vanishing energy, with the full numerical solutions of the model. Hence, we assume Eq.\@ (\ref{eq:D_free}) in the present regime.
Using Eqs.\@ (\ref{eq:Sigma_Z}) and (\ref{eq:Phi_Delta}), together with Eq.\@ (\ref{eq:D_free}), the gap equation (\ref{eq:Phi}) becomes
\begin{align}\label{eq:Delta_Z_anyT}
Z(i\omega_n)\Delta(i \omega_n)&	\left\{1-\frac{z t_0^2}{2} \right. \notag \\ & \left. \times \frac{1}{\left[Z(i \omega_n)\right]^2 \left\{ \omega_n^2+\left[\Delta(i \omega_n)\right]^2\right\}} \right\} \notag \\ &=\frac{\bar{g}^2 k_B T}{\omega_0} \sum_{m=-\infty}^{+\infty}Z(i \omega_m) \Delta(i \omega_m) \notag \\ & \times \frac{1}{\left[Z(i \omega_m)\right]^2 \left\{ \omega_m^2+\left[\Delta(i \omega_m)\right]^2 \right\}}. 
\end{align}  
Further approximating the gap function with Eq.\@ (\ref{eq:gap_const}), and using the SYK$_2$-FL result (\ref{eq:Z_SYK2_gapconst_expl}) for the dynamical weight in the superconducting state, the left-hand side of Eq.\@ (\ref{eq:Delta_Z_anyT}) simplifies to $\Delta$, and we are left with
\begin{align}\label{eq:Delta_Z_SYK2_anyT_2}
\Delta &=\frac{\bar{g}^2 k_B T}{\omega_0} \sum_{n=-\infty}^{+\infty} 2 \Delta \notag \\ & \times \frac{1}{\left(1+\sqrt{1+\frac{2 z t_0^2}{\omega_n^2+\Delta^2}} \right) \left( \omega_n^2+\Delta^2 \right)}. 
\end{align}
In the large-hopping regime, we can approximate $1+\sqrt{1+\left(2 z t_0^2\right)/\left(\omega_n^2+\Delta^2\right)} \approx \sqrt{2 z} t_0/\sqrt{\omega_n^2+\Delta^2}$, so that
\begin{equation}\label{eq:Delta_Z_SYK2_anyT_3}
1=\frac{\sqrt{2} \bar{g}^2 k_B T}{ \omega_0 \sqrt{z} t_0} \sum_{n=-\infty}^{+\infty} \frac{1}{ \sqrt{ \omega_n^2+\Delta^2}}. 
\end{equation}
Eq.\@ (\ref{eq:Delta_Z_SYK2_anyT_3}) has now assumed the form of a BCS-like gap equation \cite{Bardeen-1957a,Bardeen-1957b,Schrieffer-1963th}, which we can employ to find the critical temperature and the zero-temperature zero-energy gap in Fermi-liquid regime. 

\subsubsection{Zero-temperature gap: BCS formula}\label{App:gap0_SYK2-FL}

In the $T=0$ limit, where the Matsubara sums become continuous integrals, the gap equation (\ref{eq:Delta_Z_SYK2_anyT_3}) in SYK$_2$-FL regime translates as
\begin{equation}\label{eq:Delta_Z_SYK2_T0}
1=\frac{\sqrt{2} \bar{g}^2}{\pi \omega_0 \sqrt{z} t_0} \int_{-\infty}^{+\infty} \frac{d \omega}{2} \frac{1}{ \sqrt{ \omega^2+\Delta^2}}. 
\end{equation} 
We use a ultraviolet cutoff $\omega_0$ for the integral in Eq.\@ (\ref{eq:Delta_Z_SYK2_T0}), since pairing occurs at much lower energies than the bare boson energy $\omega_0$; this feature is completely analogous to BCS theory \cite{Bardeen-1957a,Bardeen-1957b,Schrieffer-1963th}. Then we have
\begin{equation}\label{eq:Delta_Z_SYK2_T0_2}
\frac{1}{\bar{\lambda}}= \int_{-\omega_0}^{+\omega_0} \frac{d \omega}{2} \frac{1}{ \sqrt{\omega^2+\Delta^2}}=\mathrm{arcsinh}\left(\frac{\omega_0}{\Delta}\right), 
\end{equation} 
where we retrieve the superconducting coupling constant (\ref{eq:lambda_bar}). Eq.\@ (\ref{eq:Delta_Z_SYK2_T0_2}) is the BCS formula (\ref{eq:SYK2-FL_FL_Delta0}).

\subsubsection{Superfluid phase stiffness}\label{rhoS_SYK2-FL}

Approximating the gap to a constant with Eq.\@ (\ref{eq:gap_const}), and using the superconducting dynamical weight (\ref{eq:Z_SYK2_gapconst_expl}) in SYK$_2$-FL regime, together with Eq.\@ (\ref{eq:Phi_Delta}), Eq.\@ (\ref{eq:stiff_SYK_lattice}) becomes
\begin{align}\label{eq:stiff_SYK_lattice_2}
\rho_S(T) &=\Theta_L z t_0^2 k_B T \sum_{i \omega_n} \frac{4 \Delta^2}{\left[Z(i \omega_n)\right]^2 \left(\omega_n^2+\Delta^2\right)^2} \notag \\ &=\Theta_L z t_0^2 k_B T \sum_{i \omega_n} \frac{1}{\left(\omega_n^2+\Delta^2\right)^2}\notag \\ & \times \frac{4 \Delta^2}{\left(1+\sqrt{1+\frac{2 z t_0^2}{\Delta^2+\omega_n^2}}\right)^2 }. 
\end{align}
For $z t_0^2 \gg \left| \omega_n^2+\Delta^2\right|$, we can expand as
\begin{equation}\label{eq:stiff_SYK_lattice_3}
\rho_S(T) \approx 2 \Theta_L \Delta^2  k_B T k_B T \sum_{i \omega_n} \frac{1}{\omega_n^2 +\Delta^2}. 
\end{equation}
Using the Matsubara sum
\begin{equation}\label{eq:Mats_SYK_stiff}
k_B T \sum_{i \omega_n} \frac{1}{\omega_n^2+\mathscr{E}^2}=\frac{1}{2 \mathscr{E}} \left[1-2 f_{FD}(\mathscr{E})\right]
\end{equation}
with $\mathscr{E} \equiv \Delta$ in Eq.\@ (\ref{eq:stiff_SYK_lattice_3}), we finally achieve Eq.\@ (\ref{eq:SYK2_FL_stiff}).

\subsection{Strong-coupling non-Fermi liquid: the impurity-like regime}

\subsubsection{Dynamical quasiparticle weight}\label{Z_imp}

To analyze the dynamical quasiparticle weight in the impurity-like regime $\bar{g}^2/\omega_0 \gg z t_0^2$, we start again from the normal self-energy (\ref{eq:Sigma}) written in terms of $Z(i \omega_n)$, using Eq.\@ (\ref{eq:Sigma_Z}). We now employ the Eliashberg equation (\ref{eq:G2}) for $G(i \omega_n)$, and we approximate the boson propagator with its impurity-like expression in the superconducting state \cite{Esterlis-2019}
\begin{equation}\label{eq:D_implike}
D(i \omega_n-i \omega_m)\approx \frac{1}{(\omega_n-\omega_m)^2+(\omega_r^{SC})^2},
\end{equation}
with the finite renormalized boson frequency (\ref{eq:omegar_SC_imp}). The latter appears as an in-line equation before Eq.\@ (40) in Ref.\@ \onlinecite{Esterlis-2019}.
We also use Eq.\@ (\ref{eq:Phi_Delta}) for $\Delta(i \omega_n)$, with the result
\begin{align}\label{eq:Eliashberg_Z_T_implike}
Z(i \omega_n)&=1+\frac{\bar{g}^2 k_B T}{\omega_0^2} \sum_{m=-\infty}^{+\infty} \frac{1}{(\omega_n-\omega_m)^2+(\omega_r^{SC})^2} \frac{\omega_m}{\omega_n} \notag \\ & \times \frac{1}{Z(i \omega_n) \left\{\omega_m^2+\left[\Delta(i \omega_m)\right]^2\right\}} \notag \\ & +\frac{z t_0^2}{2} \frac{1}{Z(i \omega_n) \left\{\omega_m^2+\left[\Delta(i \omega_m)\right]^2\right\}}. 
\end{align} 
In the impurity-like regime, the $t_0$-dependent term in Eq.\@ (\ref{eq:Eliashberg_Z_T_implike}) is negligible with respect to the fermion-boson term that depends on $\bar{g}^2$. However, as we will see in the following, we can obtain an analytical result for $Z(i \omega_n)$ even at finite hopping, i.e.\@, retaining the $t_0$-dependent part of Eq.\@ (\ref{eq:Eliashberg_Z_T_implike}). While formally valid for $\bar{g}^2/\Omega_0 \gg z {t_0^2}$, the resulting expression -- see Eqs.\@ (\ref{eq:Z_T0_implike}) and (\ref{eq:Z_T_implike}), at zero and finite temperature respectively -- allows us to smoothly interpolate between the SYK$_2$-FL and impurity-like regimes, which occurs because the two regimes essentially share the same physics: fermions randomly interacting with static scattering centers.
At $T=0$, we convert the Matsubara sum in Eq.\@ (\ref{eq:Eliashberg_Z_T_implike}) into an integral and we achieve
\begin{align}\label{eq:Eliashberg_Z_T_implike_T0}
Z(i \omega)&=1+\frac{\bar{g}^2}{\omega_0^2} \int_{-\infty}^{+\infty} \frac{d \omega'}{2 \pi} \frac{1}{(\omega-\omega')^2+(\omega_r^{SC})^2} \frac{\omega'}{\omega} \notag \\ & \frac{1}{Z(i \omega') \left\{(\omega')^2+\left[\Delta(i \omega')\right]^2\right\}} \notag \\ &+\frac{z t_0^2}{2} \frac{1}{Z(i \omega) \left\{\omega^2+\left[\Delta(i \omega)\right]^2\right\}}. 
\end{align} 
Now, in the impurity-like regime, we have $\omega_r^{SC}\ll k_B T$. Then, the term depending on $\omega_r^{SC}$ in Eq.\@ (\ref{eq:Eliashberg_Z_T_implike_T0}) behaves like a Dirac delta function \cite{Esterlis-2019}, 
\begin{equation}\label{eq:omegar_T0}
\lim_{\omega_r^{SC}\rightarrow 0} \frac{1}{(\omega-\omega')^2+(\omega_r^{SC})^2} \approx \frac{\pi}{\omega_r^{SC}} \delta (\omega-\omega').
\end{equation}
Inserting Eq.\@ (\ref{eq:omegar_T0}) into the quasiparticle weight (\ref{eq:Eliashberg_Z_T_implike_T0}) yields
\begin{multline}\label{eq:Eliashberg_Z_T_implike_T0_2}
Z(i \omega)=1+\frac{\bar{g}^2}{\omega_0^2 \omega_r^{SC}} \int_{-\infty}^{+\infty} \frac{d \omega'}{2 \pi} \pi \delta(\omega-\omega') \\ \times \frac{1}{Z(i \omega') \left\{(\omega')^2+\left[\Delta(i \omega')\right]^2\right\}} \\  +\frac{z t_0^2}{2} \frac{1}{Z(i \omega) \left\{\omega^2+\left[\Delta(i \omega)\right]^2\right\}}\\= 1+\left[\left(\frac{8 g^2 \omega_0}{3 \pi}\right)^2+\frac{z {t_0^2}}{2} \right] \\ \times \frac{1}{Z(i \omega) \left\{\omega^2 +\left[\Delta(i \omega)\right]^2\right\}},
\end{multline} 
where we used $g^2=\bar{g}^2/\omega_0^3$.
Therefore, Eq.\@ (\ref{eq:Eliashberg_Z_T_implike_T0_2}) gives the solutions 
\begin{equation}\label{eq:Z_T0_implike_pm}
Z(i \omega)=\frac{1}{2} \left[1\pm \sqrt{1+\frac{\tilde{\Omega}^2}{\omega^2+\left[\Delta( i\omega)\right]^2}}\right]. 
\end{equation}
To comply with the normal-state result, we choose the positive root of Eq.\@ (\ref{eq:Z_T0_implike_pm}).
Actually, one can numerically confirm that the gap $\Delta(i \omega)$ can be approximated by a constant at low energies, as in Eq.\@ (\ref{eq:gap_const}), also in the impurity-like regime. Therefore, we can write
\begin{align}\label{eq:Z_T0_implike}
Z(i \omega)&=\frac{1}{2} \left[1+ \sqrt{1+\frac{\tilde{\Omega}^2}{\omega^2+\Delta^2}}\right] \notag \\ & \approx \frac{1}{2}\left(1+\sqrt{1+\frac{ \left[16 g^2 \omega_0/(3 \pi)\right]^2}{\omega^2 +\Delta^2}}\right),  
\end{align}
where the last step is valid for $\bar{g}^2/\omega_0 \gg z t_0^2$.

At low but nonzero temperatures, we can substitute $i \omega \mapsto i \omega_n$ in Eq.\@ (\ref{eq:Z_T0_implike}), which gives Eq.\@ (\ref{eq:Z_T_implike}). 

The normal-state limit of Eq.\@ (\ref{eq:Z_T_implike}) is obtained by taking $\Delta\rightarrow 0$, with the result
\begin{equation}\label{eq:Z_T_implike_NS}
Z(i \omega_n)\approx\frac{1}{2} \left[1+ \sqrt{1+\frac{\tilde{\Omega}^2}{\omega_n^2}}\right].
\end{equation}
At low frequencies $\left|\omega_n\right| \rightarrow 0^+$, we have
\begin{equation}\label{eq:Z_T_implike_NS_omn0}
Z(i \omega_n)\approx \frac{\tilde{\Omega}}{2\left|\omega_n\right|} \approx \frac{8 g^2 \omega_0/(3 \pi)}{\left|\omega_n\right|},
\end{equation}
The last step of Eq.\@ (\ref{eq:Z_T_implike_NS_omn0}), again valid for $\bar{g}^2/\omega_0 \gg z t_0^2$, is consistent with Eq.\@ (26) of Ref.\@ \onlinecite{Esterlis-2019}, through Eq.\@ (\ref{eq:Sigma_Z}). 

\subsubsection{Fermion propagators and self-energies in the superconducting state}

In the superconducting state, in general from Eqs.\@ (\ref{eq:Z_T0_implike}) and (\ref{eq:Sigma_Z}) we have
\begin{equation}\label{eq:Z_implike_SC_T0}
\Sigma(i \omega)=\frac{i \omega}{2} \left(1- \sqrt{1+\frac{\tilde{\Omega}}{\omega^2+\Delta^2}}\right)
\end{equation}
at $T=0$. Instead, at finite temperature we can employ
\begin{equation}\label{eq:Z_implike_SC_T}
\Sigma(i \omega_n)=\frac{i \omega_n}{2} \left(1- \sqrt{1+\frac{\tilde{\Omega}}{\omega_n^2+\Delta^2}}\right). 
\end{equation}
The fermion propagator which corresponds to the self-energy (\ref{eq:Z_implike_SC_T}) is
\begin{equation}\label{eq:G_implike_SC_T}
G(i \omega_n)=\frac{2}{i \omega_n+i \mathrm{sign}(\omega_n) \sqrt{\omega_n^2+\frac{\omega_n^2}{\omega_n^2+\Delta^2} \tilde{\Omega}^2}},
\end{equation}
which reduces to Eq.\@ (\ref{eq:G_imp_NS}) in the normal state, where $\Delta=0$.  
Performing the analytic continuation $i \omega_n \rightarrow \omega+i 0^+$ on Eq.\@ (\ref{eq:G_implike_SC_T}), we obtain the (retarded) fermion propagator on the real axis:  
\begin{equation}\label{eq:G_implike_SC_T_real}
G^R(\omega)=\frac{2}{(\omega+i 0^+)\left[1+ \sqrt{1+\frac{\tilde{\Omega}^2}{\Delta^2-(\omega+i 0^+)^2}}\right]}.
\end{equation}
In addition, in the superconducting state the anomalous propagator (\ref{eq:F2}) emerges. Performing the same steps leading to Eq.\@ (\ref{eq:F_Z_Delta_SYK2}), but now substituting the impurity-like dynamical weight (\ref{eq:Z_T_implike}) in Eq.\@ (\ref{eq:F2}), we obtain
\begin{equation}\label{eq:F_Z_Delta_imp}
F(i \omega_n)=-\frac{2 \Delta}{\left(\omega_n^2+\Delta^2\right)\left(1+\sqrt{1+\frac{\tilde{\Omega}^2}{\omega_n^2+\Delta^2}}\right)},
\end{equation}
where $\tilde{\Omega}$ satisfies Eq.\@ (\ref{eq:tilde_Omega_interp}).
The retarded anomalous propagator is obtained from the analytic continuation of Eq.\@ (\ref{eq:F_Z_Delta_imp}), and it reads
\begin{equation}\label{eq:F_Z_Delta_imp_real}
F^R(\omega)=-\frac{2 \Delta}{\left[\Delta^2-(\omega+i 0^+)^2\right]\left[1+ \sqrt{1+\frac{\tilde{\Omega}^2}{\Delta^2-(\omega+i 0^+)^2}}\right]}.
\end{equation}
Notice that, in the limit $\bar{g}^2/\omega_0 \ll z t_0^2$, we directly retrieve the SYK$_2$-FL result (\ref{eq:F_Z_Delta_SYK2_real}) from Eq.\@ (\ref{eq:F_Z_Delta_imp_real}).

\subsubsection{Superconducting-state renormalized boson frequency}\label{Imp_omegar_SC}

In the impurity-like superconducting state, the renormalized boson frequency does not vanish at $T=0$ as would be predicted by the normal-state result (\ref{eq:omegar_0_T0_imp_t00}), due to the gapped nature of the bosons at $T<T_c$ \cite{Esterlis-2019}. Instead, at $k_B T \ll \omega_0 g^2$ we have Eq.\@ (\ref{eq:omegar_SC_imp}). The latter can be obtained from the high-energy behaviour of the spectral function in the superconducting state \cite{Esterlis-2019}. More precisely, we exploit the approximate Dirac-delta form (\ref{eq:omegar_T0}) of the bosonic propagator (\ref{eq:D_implike}) in the impurity-like regime, and we insert it in Eq.\@ (\ref{eq:Sigma}) for the fermionic self-energy in the $T=0$ limit. Due to the Dirac delta sampling only the $\Omega=0$ component of the bosonic propagator $D(i\Omega)$, the integral over bosonic energy $\Omega$ yields 
\begin{equation}\label{eq:Sigma_G_omegarSC_imp}
\Sigma(i \omega)=\frac{\bar{g}^2}{2\pi} \frac{\pi}{\omega_r^{SC}(0)}G(i\omega)+\frac{z t_0^2}{2} G(i\omega). 
\end{equation}
Assuming $\Sigma(\omega)\approx -1/G(i\omega)$ in Eq.\@ (\ref{eq:Sigma_G_omegarSC_imp}) leaves a relation between $\omega_r^{SC}$ and $G(i \omega)$. Using $G(i\omega)=1/\left[i \omega Z(i\omega)\right]$ and the high-energy leading-order expansion of the dynamical weight (\ref{eq:Z_T0_implike}), which gives
\begin{equation}\label{eq:Z_imp_largeomega_SC}
 \left|\omega\right|Z(i\omega)\approx \frac{\tilde{\Omega}}{2}
\end{equation}
even in the superconducting state (the gap does not contribute at leading order for $\omega \rightarrow +\infty$), we finally achieve
\begin{align}\label{eq:G_omegarSC_imp2}
\omega_r^{SC}(0) & \approx -\frac{\bar{g}^2 \left[G(i\omega)\right]^2}{2+z t_0^2 \left[G(i\omega)\right]^2} \notag \\ &= \frac{\omega_0}{2} \left(\frac{3 \pi}{8 g}\right)^2.
\end{align}
Eq.\@ (\ref{eq:G_omegarSC_imp2}) is consistent with Eq.\@ (\ref{eq:omegar_SC_imp}). Notice that Eq.\@ (\ref{eq:G_omegarSC_imp2}) has been derived in the impurity-like regime, but in the presence of small hopping $z t_0^2 \ll \bar{g}^2/\omega_0$. This means that small hopping perturbations with respect to fermion-boson coupling are not expected to significantly alter the renormalized boson frequency. This is indeed observed in Fig.\@ \ref{fig:spectr_NFL-SYK_g_SC}(b). 

\subsubsection{Gap equation}

We consider again Eq.\@ (\ref{eq:Phi}) for the anomalous self-energy, with the anomalous propagator (\ref{eq:F2}), in the impurity-like regime where $\bar{g}^2/\omega_0^2 \gg z t_0^2$. The boson propagator can be approximated by Eq.\@ (\ref{eq:D_implike}). Then, using Eqs.\@ (\ref{eq:Sigma_Z}) and (\ref{eq:Phi_Delta}), and approximating the gap function with Eq.\@ (\ref{eq:gap_const}), we translate Eq.\@ (\ref{eq:Phi}) into
\begin{align}\label{eq:Delta_Z_implike_anyT}
& Z(i\omega_n)\left\{1-\frac{z t_0^2}{2} \frac{1}{\left[Z(i \omega_n)\right]^2 \left( \omega_n^2+\Delta^2\right)} \right\} \notag \\ &=\bar{g}^2 k_B T \sum_{m=-\infty}^{+\infty} \frac{1}{(\omega_n-\omega_m)^2+(\omega_r^{SC})^2} \notag \\ &\times\frac{1}{Z(i \omega_m) \left( \omega_m^2+\Delta^2 \right)}. 
\end{align}

\subsubsection{Zero-temperature gap: asymptotic strong-coupling limit}

Let us consider the gap equation (\ref{eq:Delta_Z_implike_anyT}) at $T=0$. Then, the Matsubara sum becomes an integral and
\begin{align}\label{eq:Delta_Z_implike_T0}
&Z(i\omega)	\left\{1-\frac{z t_0^2}{2} \frac{1}{\left[Z(i \omega)\right]^2 \left(\omega^2+\Delta^2\right)} \right\} \notag \\ &=\bar{g}^2 \int_{-\infty}^{+\infty} \frac{d \omega'}{2 \pi} \frac{1}{(\omega-\omega')^2+(\omega_r^{SC})^2} \notag \\ & \times \frac{1}{Z(i \omega) \left( \omega^2+\Delta^2 \right)}. 
\end{align}
Let us analyze Eq.\@ (\ref{eq:Delta_Z_implike_T0}) in the single-dot limit $t_0 \rightarrow 0$. In such limit, Eq.\@ (\ref{eq:Delta_Z_implike_T0}) for the dispersionless Yukawa-SYK model can be formally brought into the same form as the Eliashberg equation for a wide-band system interacting with an Einstein phonon of frequency $\Omega_E$ in the asymptotic strong-coupling limit \cite{Carbotte-1990} -- see also the discussion after Eq.\@ (32) in Ref.\@ \onlinecite{Esterlis-2019}. We can then borrow the asymptotic results for the gap and the quasiparticle weight from strong-coupling Eliashberg theory, as follows. 
Labeling $\Delta_0$ as in Eq.\@ (\ref{eq:Delta0_def}), we employ Eq.\@ (\ref{eq:Z_T_implike}) and for the quasiparticle weight, in the very large-coupling limit $g \rightarrow +\infty$, we obtain explicitly
\begin{equation}\label{eq:Z_quasistat_implike_T0}
Z(\omega)\approx \frac{8 g^2 \omega_0}{(3 \pi) \sqrt{\omega^2 +\Delta^2}}.
\end{equation}
On the other hand, the dynamical quasiparticle weight satisfies Eq.\@ (\ref{eq:Eliashberg_Z_T_implike_T0}), where we can neglect the $t_0$-dependent term deep in the impurity-like regime. Hence, inserting the result (\ref{eq:Z_quasistat_implike_T0}) into Eq.\@ (\ref{eq:Eliashberg_Z_T_implike_T0}), we have
\begin{multline}\label{eq:Z_implike_T0_gap_simpl}
Z(i \omega)=1+ \frac{g^2 \omega_0}{2 \pi \omega} \int_{-\infty}^{+\infty} d \omega' \frac{1}{(\omega-\omega')^2+(\omega_r^{SC})^2} \\ \times \frac{\omega'}{Z(i \omega) \left( \omega^2+\Delta^2 \right)} \\ = 1+\frac{3 \pi \omega_0^2}{8} \frac{1}{2 \pi \omega} \int_{-\infty}^{+\infty} d \omega' \frac{1}{(\omega-\omega')^2+(\omega_r^{SC})^2} \\ \times \frac{\omega'}{\sqrt{ \omega^2+\Delta^2}}.
\end{multline}
Eq.\@ (\ref{eq:Z_implike_T0_gap_simpl}) yields an expression for the weight $\left. Z(\omega)\right|_{\omega \approx 0}$ \cite{Carbotte-1990}:
\begin{align}\label{eq:Z_implike_T0_gap_simpl_2}
\left. Z(\omega)\right|_{\omega \approx 0}& =1+ 2 \frac{3 \omega_0^2}{16 \omega_r^{SC}} \frac{d}{d \omega_r^{SC}} \left[(\omega_r^{SC})^2 \mathscr{J}(\omega_r^{SC})\right] \notag \\ & +\frac{3 \omega_0^2}{16} \frac{1}{2 \Delta_0^2},
\end{align}
where
\begin{equation}\label{eq:J_omegar_SC}
\mathscr{J}(\omega_r^{SC})=\int_0^{2 \Delta_0} \frac{d \omega'}{(\omega')^2+(\omega_r^{SC})^2	} \frac{1}{\sqrt{(\omega')^2+\Delta_0^2}}.
\end{equation}
In the same way, using Eq.\@ (\ref{eq:J_omegar_SC}) the gap equation (\ref{eq:Delta_Z_implike_T0}) for $t_0=0$ reduces to
\begin{equation}\label{eq:Delta_Z_implike_T0_2}
\left. Z(\omega)\right|_{\omega \approx 0}=2 \frac{3 \omega_0^2}{16} 	\mathscr{J}(\omega_r^{SC}). 
\end{equation}
Therefore, we now have Eqs.\@ (\ref{eq:Z_implike_T0_gap_simpl_2}) and (\ref{eq:Delta_Z_implike_T0_2}) for the quasistatic weight, which constitute a closed problem for the gap $\Delta_0$. To obtain the latter analytically, we use the series expansion of Eq.\@ (\ref{eq:J_omegar_SC}), 
\begin{equation}\label{eq:J_omegar_SC_expand}
\mathscr{J}(\omega_r^{SC})\approx \frac{1}{\omega_r^{SC} \Delta_0}\left[\frac{\pi}{2}-\frac{\sqrt{5}}{2}\frac{\omega_r^{SC}}{\omega_0}-\frac{\sqrt{5}}{24}\left(\frac{\omega_r^{SC}}{\omega_0}\right)^3\right],
\end{equation}
into Eqs.\@ (\ref{eq:Z_implike_T0_gap_simpl_2}) and (\ref{eq:Delta_Z_implike_T0_2}). Solving the latter system for $\Delta_0$, we finally achieve
\begin{equation}\label{eq:Delta_T0_gap_analyt}
\Delta_0\approx \sqrt{\frac{-1+2 \sqrt{5}}{2}} \sqrt{\frac{3}{16}} \omega_0 \approx 0.5705372431149527 \omega_0. 
\end{equation}
An analogous calculation for $T_c$ in the asymptotic limit $g \rightarrow +\infty$ would lead to \cite{Carbotte-1990}
\begin{equation}\label{eq:Tc_ginf_Carbotte}
k_B T_c =\frac{1}{2\pi} \sqrt{\frac{3}{8}} \omega_0 \approx 0.09746210015420952 \omega_0,
\end{equation}
so that the gap to $T_c$ ratio would be 
\begin{equation}\label{eq:gaptoTc_ginf}
\frac{2 \Delta_0}{k_B T_c} \approx 11.708,
\end{equation}
in good agreement with Eq.\@ (\ref{eq:gaptoTc_imp}) obtained from the numerical solution of the Eliashberg equations (\ref{eq:Eliashberg_eph:hop}), and with Eq.\@ (38) in Ref.\@ \onlinecite{Esterlis-2019}. Actually, the approximated analytical expressions (\ref{eq:Delta_T0_gap_analyt}) and (\ref{eq:Tc_ginf_Carbotte}) underestimate the gap and $T_c$ by a factor $C^{-1} \approx 1/1.1481028752388993$, so that more accurate results are
\begin{equation}\label{eq_Delta_T0_gap_analy}
\Delta_0\approx \sqrt{\frac{-1+2 \sqrt{5}}{2}} \sqrt{\frac{3}{16}} C \omega_0 \approx 0.6550354492511521 \omega_0 
\end{equation}
and
\begin{equation}\label{eq:Tc_ginf_Carbotte_corr}
k_B T_{\rm c} =\frac{1}{2\pi} \sqrt{\frac{3}{8}} C \omega_0 \approx 0.111897 \omega_0.
\end{equation}
Eq.\@ (\ref{eq_Delta_T0_gap_analy}) and (\ref{eq:Tc_ginf_Carbotte_corr}) are in good agreement with the gap value quoted above Eq.\@ (38) and with Eq.\@ (30) in Ref.\@ \onlinecite{Esterlis-2019}. Notice that, even with the correction $C$, the gap to $T_{\rm c}$ ratio still has the value (\ref{eq:gaptoTc_ginf}).

\subsubsection{Superfluid phase stiffness}\label{rhoS_imp}

Recognizing the formal equivalence between the results for the dynamical weight $Z( i\omega_n)$ in the SYK$_2$-FL and impurity-like regimes, in accordance with Eqs.\@ (\ref{eq:Z_SYK2_gapconst_expl}) and (\ref{eq:Z_T_implike}), we can readily calculate the phase stiffness in the impurity-like regime as well, employing the results of Sec.\@ \ref{rhoS_SYK2-FL}. Starting from Eq.\@ (\ref{eq:stiff_SYK_lattice}), the derivation is exactly the same upon mapping $2 z {t_0^2} \rightarrow \tilde{\Omega}^2$ in $Z( i \omega_n)$, in accordance with Eq.\@ (\ref{eq:tilde_Omega_interp}). We then have
\begin{equation}\label{eq:rhoS_implike_cross}
\rho_S = \Theta_L z t_0^2 k_B T \sum_{i \omega_n} \frac{ 4 \Delta^2}{\left(1+\sqrt{1+\frac{ \tilde{\Omega}^2}{\omega_n^2 +\Delta^2}}\right)^2 \left(\omega_n^2+\Delta^2\right)^2}.
\end{equation}
In the regime where $\tilde{\Omega}^2 \gg \omega_n^2+\Delta^2$, we can write
\begin{equation}\label{eq:rhoS_implike_cross2}
\rho_S =\frac{4 \Theta_L \Delta^2 z t_0^2}{\tilde{\Omega}^2} k_B T \sum_{i \omega_n} \frac{1}{\omega_n^2+\Delta^2}.
\end{equation}
Using Eq.\@ (\ref{eq:Mats_SYK_stiff}) for the Matsubara sum, we are left with Eq.\@ (\ref{eq:rhoS_implike_cross_expl2}). 

\section{Analysis of the critical temperature}\label{App:Tc}

\subsection{Weak-coupling single-dot limit: the NFL-SYK regime}\label{Tc_weak_dot}

To analyze the weak-coupling limit in the single dot, we start from the linearized gap equation (\ref{eq:Phi_Z_lin}). For the SYK-NFL fixed point, the fermionic and bosonic low-energy propagators are given by Eqs.\@ (\ref{eq:G_NFL_SYK}) and (\ref{eq:D_NFL_SYK}), respectively, with the renormalized boson frequency (\ref{eq:omegar_NFL_SYK}). 

With Eqs.\@ (\ref{eq:G_NFL_SYK})-(\ref{eq:D_NFL_SYK}), neglecting the term $\Omega_n^2$ in Eq.\@ (\ref{eq:D_NFL_SYK}) at weak coupling \cite{Esterlis-2019}, and using the relation (\ref{eq:Phi_Delta}), the linearized gap equation (\ref{eq:Phi_Z_lin}) becomes
\begin{align}
\Delta(i \omega_n)&=a_0 \sum_{m=-\infty}^{+\infty}(T_f/T)^{2 \Delta} \mathrm{sign}(\omega_m) \times \nonumber \\ & \frac{1}{(T_f/T)^{-2 \Delta}\left|m+1/2\right|+\left|m+1/2\right|^{1-2 \Delta}}\notag\\
&\qquad \times\left[\frac{\Delta(i \omega_m)}{i \omega_m}-\frac{\Delta(i \omega_n)}{i \omega_n}\right] \frac{1}{m_0+\left|n-m\right|^{4 \Delta-1}},
\label{eq:gap_weak}
\end{align}
where $m_0=c_2/\left[c_3(2\pi)^{4\Delta-1}\right]$, $a_0=1/\left[2 \pi (c_1)^2 c_2\right]$, and $2 \pi k_B T_f= c_1^{1/(2 \Delta)} \bar{g}^2/\omega_0^2\approx 0.1888 \bar{g}^2/\omega_0^2$. Since the only temperature scale in Eq.\@ (\ref{eq:gap_weak}) is $T_f/T$, the transition temperature $T_c$ is of the order of $T_f$ and is numerically found to be $k_B T_c \approx 0.16 \bar{g}^2/\omega_0^2$ \cite{Esterlis-2019}. 
\begin{figure}[ht]
\includegraphics[width=0.9\columnwidth]{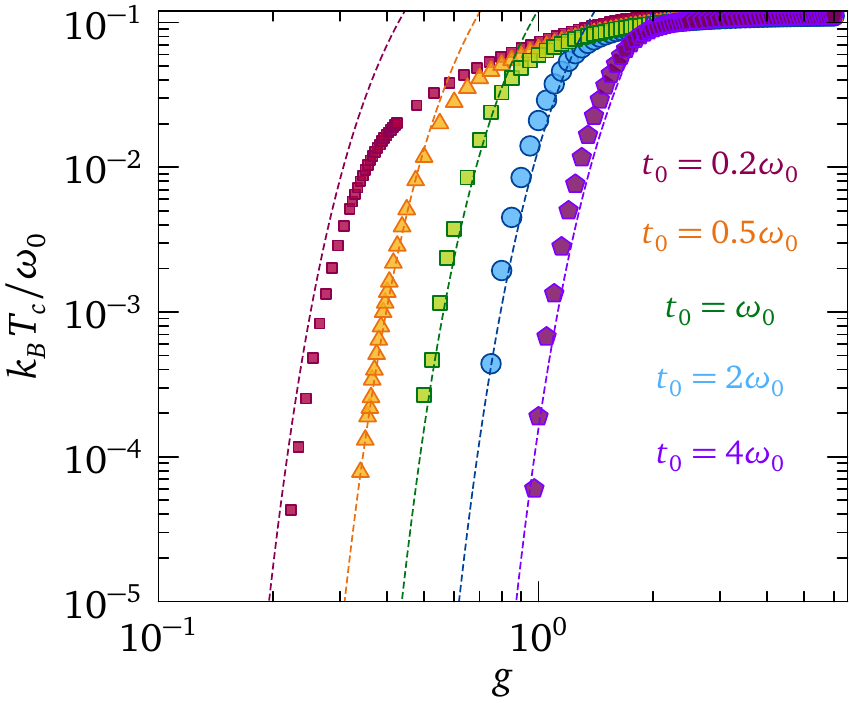}
\caption{\label{fig:Tclow}
Comparison between the numerical solution of the linearized Eliashberg equations (\ref{eq:saddle_Tc_linearized}) for the Yukawa-SYK model on a lattice (point markers) and the BCS-like formula (\ref{eq:BCS_Tc_SYK2}) with the coupling constant (\ref{eq:lambda_bar}) (dashed lines) as a function of dimensionless interaction $g$, for different hopping parameters $t_0$. 
}
\end{figure}

\subsection{Weak-coupling BCS formula from a disordered Fermi liquid: the SYK$_2$-FL regime}\label{Tc_weak_dot_BCS}

Let us consider the SYK$_2$-FL regime at $T=T_c$. Then the gap equation (\ref{eq:Delta_Z_SYK2_anyT_3}) for $\Delta\rightarrow 0$ becomes 
\begin{equation}\label{eq:Tc_SYK2}
1=\frac{\sqrt{2} \bar{g}^2 k_B T}{ \omega_0 \sqrt{z} t_0} \sum_{m=-\infty}^{+\infty} \frac{1}{\left|\omega_n\right|}. 
\end{equation}
The result (\ref{eq:Tc_SYK2}) is also consistent with the series expansion of the summation term in Eq.\@ (\ref{eq:lin_gap_GG}) at leading order in small $\omega_m$. 
At this point, we can exploit the analogy with standard BCS theory, and write Eq.\@ (\ref{eq:Tc_SYK2}) in analogy with its BCS counterpart \cite{Bardeen-1957a,Bardeen-1957b,Schrieffer-1963th}, using the result
\begin{equation}\label{eq:BCS_Tc_sum}
\sum_{m=-\infty}^{+\infty} \frac{1}{\left|\omega_n\right|} \approx \ln \left(\frac{\bar{\Omega}}{k_B T} \frac{2 e^{\gamma}}{\pi}\right), 
\end{equation}
where $\bar{\Omega}$ is an appropriate UV cutoff energy and $\gamma\approx0.5772156649$ is the Euler-Mascheroni constant. The approximation (\ref{eq:BCS_Tc_sum}) is valid for $\bar{\Omega}\gg k_B T$. The proper cutoff energy of the Yukawa-SYK problem is $\bar{\Omega}=\omega_0$, so in the SYK$_2$-FL regime we use Eqs.\@ (\ref{eq:Tc_SYK2}) and (\ref{eq:BCS_Tc_sum}) to achieve
\begin{equation}\label{eq:Tc_SYK2_step2}
\frac{1}{\bar{\lambda}}=\ln\left(\frac{\omega_0}{k_B T} \frac{2 e^{\gamma}}{\pi}\right),
\end{equation}
which is an effective BCS equation for the critical temperature in the Fermi-liquid regime, with the coupling constant  (\ref{eq:lambda_bar}). Inverting Eq.\@ (\ref{eq:Tc_SYK2_step2}), we find the critical temperature (\ref{eq:BCS_Tc_SYK2}). 
 
We find good agreement between Eq.\@ (\ref{eq:BCS_Tc_SYK2}) and the numerics for $T_c$, employing the full-fledged linearized Eliashberg equations (\ref{eq:saddle_Tc_linearized}), as shown Fig.\@ \ref{fig:Tclow}. 

\subsection{Strong-coupling single-dot limit: the impurity-like regime}\label{app:Tc_implike_dot}

For completeness, we start by reviewing the strong-coupling analysis in the single-dot limit, i.e.\@, $t_0=0$, reported in Ref.\@ \onlinecite{Esterlis-2019}. We will exploit this derivation to deduce the same infinite-coupling limit on the lattice. Using the functions (\ref{eq:Sigma_Z}) and (\ref{eq:Phi_Delta}), the gap equation (\ref{eq:Phi}) can be equivalently rewritten as
\begin{align}\label{eq:gap_Delta}
&Z(i \omega_n)\Delta(i\omega_n)\nonumber \\ &=\bar{g}^2 k_B T \sum_{m} \frac{D(\omega_n-\omega_m) \Delta(i\omega_m)}{Z(i \omega_m) \left\{(\omega_m)^2+\left[\Delta(i \omega_m)\right]^2\right\}}.
\end{align}
On the other hand, the self-energy equation (\ref{eq:Sigma}) for $t_0=0$ translates as
\begin{align}\label{eq:Z_Delta}
Z(i \omega_n)&=1+\bar{g}^2 k_B T \sum_m \frac{D(\omega_n-\omega_m)}{Z(i \omega_m) \left\{(\omega_m)^2+\left[\Delta(i \omega_m)\right]^2\right\}} \nonumber \\ & \times \frac{\omega_m}{\omega_n}.
\end{align}
Combining Eq.\@ (\ref{eq:gap_Delta}) and (\ref{eq:Z_Delta}) gives \cite{Esterlis-2019,Hauck-2020}
\begin{align}
\Delta(i\omega_n) &=\bar{g}^2 k_B T  \sum_{m} \frac{D(\omega_n-\omega_m)}{Z(i \omega_m) \sqrt{(\omega_m)^2+\left[\Delta(i \omega_m)\right]^2}} \nonumber \\ & \times \frac{1}{\sqrt{(\omega_m)^2+\left[\Delta(i \omega_m)\right]^2}}\left[\Delta(i\omega_m)-\frac{\omega_m}{\omega_n} \Delta(i\omega_n)\right].
\label{eq:Delta_dot_rearr}
\end{align}
The linearized version of Eq.\@ (\ref{eq:Delta_dot_rearr}) is 
\begin{align}\label{eq:Delta_dot_lin}
\Delta(i\omega_n)&=\bar{g}^2 k_B T  \sum_{m} \frac{D(\omega_n-\omega_m)}{Z(i \omega_m) (\omega_m)^2} \nonumber \\ & \times \left[\Delta(i\omega_m)-\frac{\omega_m}{\omega_n} \Delta(i\omega_n)\right].
\end{align}
Using the normal-state boson propagator (\ref{eq:D_imp_NS}) and the low-frequency strong-coupling quasiparticle residue $\left|\omega_n\right| Z(i \omega_n) \approx 8 \bar{g}^2/\left(3 \pi \omega_0^2\right)$, which results from Eqs.\@ (\ref{eq:Sigma_T_implike_NS}) and (\ref{eq:Sigma_Z}), we obtain
\begin{widetext}
\begin{equation}\label{eq:Delta_dot_lin2}
\Delta(i\omega_n)=\frac{3 \pi \omega_0^2 k_B T}{8} \sum_{m} \frac{1}{(\omega_n-\omega_m)^2+(\omega_r)^2}\left[\frac{\Delta(i\omega_m)}{\omega_m}-\frac{\Delta(i\omega_n)}{\omega_n} \right] \mathrm{sign}(\omega_m).
\end{equation}
Since $k_B T \gg \omega_r$ in the strong-coupling limit, and since the zeroth Matsubara frequency does not participate to pairing \cite{Esterlis-2019}, we set $\omega_r=0$ in Eq.\@ (\ref{eq:Delta_dot_lin2}). Hence, as the fermionic Matsubara frequencies are $\omega_n=(2n+1)\pi k_B T$, Eq.\@ (\ref{eq:Delta_dot_lin2}) becomes
\begin{equation}\label{eq:Delta_dot_lin3}
\Delta(i\omega_n)=\frac{3 \omega_0^2}{8 (\pi k_B T)^2}  \sum_{m \neq n} \frac{1}{\left[2n+1-(2m+1)\right]^2}\left[\frac{\Delta(i\omega_m)}{2 m+1}-\frac{\Delta(i\omega_n)}{2n+1} \right] \mathrm{sign}\left(m+\frac{1}{2}\right).
\end{equation}
Now, we can rewrite
\begin{multline}\label{eq:sum_nm}
\sum_{m \neq n} \frac{1}{\left[2n+1-(2m+1)\right]^2}\left[\frac{\Delta(i\omega_m)}{\Delta(i\omega_n)}\frac{1}{2 m+1}-\frac{1}{2n+1} \right] \mathrm{sign}\left(m+\frac{1}{2}\right)\\=\sum_{m \neq n} \frac{1}{\left[2n+1-(2m+1)\right]^2}\left[\frac{\Delta(i\omega_m)}{\Delta(i\omega_n)}\frac{1}{\left|2 m+1 \right|}-\frac{\mathrm{sign}\left(m+\frac{1}{2}\right)}{2n+1} \right]  \\= \sum_{m \neq n} \frac{1}{\left[2n+1-(2m+1)\right]^2}\left[\frac{r_{mn}}{\left|2 m+1 \right|}-\frac{r_{nn}}{2n+1} \mathrm{sign}\left(m+\frac{1}{2}\right) \right],
\end{multline}
where $r_{mn}=\Delta(i\omega_m)/\Delta(i\omega_n)$. Hence, Eq.\@ (\ref{eq:Delta_dot_lin3}) is an eigenvalue problem for the ratios $r_{mn}$. Additionally, in Eq.\@ (\ref{eq:Delta_dot_lin3}) we have
\begin{multline}\label{eq:sum_1}
-\sum_{m \neq n} \frac{1}{\left[2n+1-(2m+1)\right]^2}\frac{\mathrm{sign}\left(m+\frac{1}{2}\right)}{2n+1}= \\ \frac{1}{2n+1} \left\{ \sum_{m=-\infty}^{-1} \frac{1}{\left[2n+1-(2m+1)\right]^2}-\sum_{m=0}^{n-1} \frac{1}{\left[2n+1-(2m+1)\right]^2}-\sum_{m=n+1}^{+\infty} \frac{1}{\left[2n+1-(2m+1)\right]^2} \right\}= \\  \frac{1}{2 n +1} \left[\frac{\psi(1+n)}{4}-\frac{\pi^2}{24} +\frac{\psi(1+n)}{4} -\frac{\pi^2}{24} \right]=\left[-\frac{\pi^2}{12}+\frac{\psi(1+n)}{2}\right] \frac{1}{2 n +1},
\end{multline}
where $\psi(z)=\lim_{M\rightarrow +\infty}\left[ \ln M-\sum_{n=0}^M 1/(n+z)\right]$ is the digamma function. Therefore
\begin{equation}\label{eq:Delta_dot_lin4}
1= \alpha  \left\{\underbrace{\sum_{m \neq n}  \frac{1}{\left[2n+1-(2m+1)\right]^2}\frac{r_{mn}}{\left|2 m+1\right|}+\left[-\frac{\pi^2}{12}+\frac{\psi(1+n)}{2}\right] \frac{r_{nn}}{2 n +1}}_{\sum_{n,m}\Lambda_{nm} r_{mn}} \right\},
\end{equation}
\end{widetext}
with $\alpha=3 \omega_0^2/[8 (\pi k_B T)^2]$ and $r_{nn}=1 \, \forall n$.
The strategy to solve Eq.\@ (\ref{eq:Delta_dot_lin4}) is to diagonalize the matrix $\underline{\underline{\Lambda}}$ formed by the coefficients of the ratios $r_{mn}$: at $T=T_c$, the largest eigenvalue of $\alpha \underline{\underline{\Lambda}}$ has to be unitary. Of course, this is the same criterion we adopt to numerically find $T_c$ for the full linearized problem (\ref{eq:Phi_Z_tilde}), as described in Appendix \ref{App:numeric_Tc}. Without loss of generality, we take $0\leq n<+\infty$ and $0\leq m<+\infty$, and the diagonal matrix elements can be written as
\begin{equation}
\Lambda_{nn}=\frac{1}{2 n+1}\frac{1}{2^2(2n+1)^2} +\left[-\frac{\pi^2}{12}+\frac{\psi(1+n)}{2}\right] \frac{1}{2 n +1},
\end{equation}
where the first term comes from $m=-n-1$ in Eq.\@ (\ref{eq:Delta_dot_lin4}), when we rewrite the sum $\sum_{m\neq n}$ with $-\infty<m<\infty$ as a sum $\sum_{m\neq n}$ with $m>0$ exploiting the even parity of $\left|2 m+1\right|$ and of $r_{n m}$ under the exchange $m \mapsto -n-1$. The off-diagonal matrix elements are
\begin{align}
\Lambda_{nm}&=\frac{1}{2 n+1}\left\{\frac{1}{[2n+1-(2m+1)]^2} \right. \nonumber \\ & \left. +\frac{1}{[2n+1+(2m+1)]^2}\right\},
\end{align}
where the first term comes from $m>0$ and the second term from $m<0$. The largest eigenvalue of the matrix $\underline{\underline{\Lambda}}$ is 
\begin{equation}\label{eq:lambda_max}
\lambda_{\Lambda}=0.32953505303295694.
\end{equation}
Therefore, the $T_c$ equation amounts to 
\begin{equation}\label{eq:Tc_alpha}
1=\alpha_c \lambda_{\Lambda}, \, \alpha_c=\frac{3 (\omega_0)^2}{8 (\pi k_B T)^2} .
\end{equation}
Eq.\@ (\ref{eq:Tc_alpha}) is solved for $T_c>0$, to finally give with reasonable accuracy
\begin{align}\label{eq:Tc_largeg_precise}
k_B T_c &=\frac{\sqrt{\alpha_c}}{\pi}\sqrt{\frac{3}{8}}=0.18272624777228844 \sqrt{\frac{3}{8}} \omega_0 \nonumber \\ &= 0.11189651741386951 \omega_0, 
\end{align}
which is Eq.\@ (\ref{eq:Tc_largeg}). 
Hence, in the strong-coupling regime $T_c$ does not depend on $\bar{g}^2$ for an isolated dot coupled to an Einstein boson. Notice that Eq.\@ (\ref{eq:Tc_largeg}) is in satisfactory agreement with the analytical estimation (\ref{eq:Tc_ginf_Carbotte_corr}). 

\subsection{Impurity-like regime on the lattice}\label{app:Tc_implike_lattice}

The estimation of the asymptotic limit for $T_c$ on the lattice closely follows the developments in Appendix \ref{app:Tc_implike_dot}, modulo the alterations to the fermion self-energy, the gap equation, and the renormalized boson frequency due to hopping. The gap equation becomes Eq.\@ (\ref{eq:Phi_Z_lin}), while the normal part of the self-energy follows Eq.\@ (\ref{eq:Sigma}). However, combining Eqs.\@ (\ref{eq:Phi_Z_lin}) and (\ref{eq:Sigma}), together with Eqs.\@ (\ref{eq:Sigma_Z}) and (\ref{eq:Phi_Delta}), we realize that the hopping-dependent part disappears from the resulting gap equation. 
Then, using the low-frequency quasiparticle residue $\left|\omega_n\right| Z(i \omega_n) \approx \sqrt{ z t_0^2/2+\left[ 8 \bar{g}^2/\left(3 \pi \omega_0^2\right)\right]^2}$, which results from Eqs.\@ (\ref{eq:G_implike_SC_T}) and (\ref{eq:Sigma_Z}), neglecting the $t_0$-dependent part of $Z(i \omega_n)$ in the $g \rightarrow +\infty$ limit, we retrieve the single-dot Eq.\@ (\ref{eq:Delta_dot_lin2}). One can confirm that even on the lattice, in the impurity-like regime we can approximate $\omega_r \approx 0$, which yields again Eq.\@ (\ref{eq:Delta_dot_lin3}). 
Therefore, the limit $\lim_{g \rightarrow +\infty}T_c$ on the lattice is the same as the one in the single dot, namely, Eq.\@ (\ref{eq:Tc_largeg}). 

\section{Derivation of the lattice grand potential}\label{App:Omega}

\subsection{Normal-state grand potential}\label{app:Omega_NS}

The grand potential in the normal state stems from the disorder-averaged Yukawa-SYK action (\ref{eq:effective_action}), where the anomalous propagator $F(i\omega_n)$ and self-energy $\Phi(i \omega_n)$ are both null. Assuming translation invariance in time and space after the disorder average, the effective action per lattice site $\vec{x}$ in imaginary time is given by
    \begin{align}\label{effective_action_after_spin_singlet-1}
    \frac{\mathscr{S}}{\mathscr{N}}= & -\text{Tr}\log\left\{\left(\hat{G}_{0}^{-1}-\hat{\Sigma}\right)\right\}
    \nonumber \\
    & +\frac{1}{2}\text{Tr}\log\text{\ensuremath{\left\{\left(D_{0}^{-1}-\Pi\right)\right\} } } 
    \nonumber \\
     & -2\int_{\tau_{12}\tau_{21}} G(\tau_{21})\Sigma(\tau_{12}) 
     \nonumber \\ & 
     +\frac{1}{2}\int_{\tau_{12}\tau_{21}} D(\tau_{21})\Pi(\tau_{12})
     \nonumber \\
      & +\bar{g}^{2}\int_{\tau_{12}\tau_{21}} G(\tau_{12})G(\tau_{21})D(\tau_{12})
     \nonumber \\ 
     & +\frac{z t_{0}^{2}}{2}\int_{\tau_{12}\tau_{21}} G(\tau_{12})G(\tau_{21}).
    \end{align}
Here $\hat{\Sigma}(i\epsilon_{n})={\rm diag}\left\{\Sigma(i\epsilon_{n}),-\Sigma(-i\epsilon_{n})\right\}$, while $\hat{G}_{0}^{-1}(\tau,\tau')=-(\partial_{\tau}-\mu)\delta(\tau-\tau')\,\hat{\sigma}_{0}$ -- see also Appendix \ref{App:dis_aver_action}.

On the normal-state saddle point the fields are homogeneous in time and space, and are given by
    \begin{subequations}\label{eq:saddle_point_YSYK_tau}
    \begin{align}
        \label{eq:saddle_point_YSYK_tau_G}
        &\hat{G}^{-1}(i\omega_{n})=\hat{G}_{0}^{-1}(i\epsilon_{n})-\hat{\Sigma}(i\epsilon_{n}),
        \\
        \label{eq:saddle_point_YSYK_tau_D}
        &D^{-1}(i\Omega_{n})=D_{0}^{-1}(i\Omega_{n})-\Pi(i\Omega_{n}),
        \\
        \label{eq:saddle_point_YSYK_tau_Sigma}
        &\Sigma(\tau)=\bar{g}^{2}G(\tau)D(\tau)+\frac{z t_{0}^{2}}{2}G(\tau),
        \\
        \label{eq:saddle_point_YSYK_tau_Pi}
        &\Pi(\tau)=-2\bar{g}^{2}G(\tau)G(-\tau).
    \end{align}
    \end{subequations}
We now go on-shell and insert Eqs.\@ (\ref{eq:saddle_point_YSYK_tau}) into the action (\ref{effective_action_after_spin_singlet-1}). Focusing on the integrands in the last two lines of the effective action, and using Eqs.\@ (\ref{eq:saddle_point_YSYK_tau_Sigma}) and (\ref{eq:saddle_point_YSYK_tau_Pi}), we obtain
    \begin{align}
    & -2G(\tau_{21})\Sigma(\tau_{12})+\frac{1}{2}D(\tau_{21})\Pi(\tau_{12}) \nonumber \\ & \qquad+\bar{g}^{2}G(\tau_{12})G(\tau_{21})D(\tau_{12})+\frac{z t_{0}^{2}}{2}G(\tau_{12})G(\tau_{21})\nonumber \\ & \qquad\qquad\equiv D(\tau_{21})\Pi(\tau_{12})  - \frac{z t_{0}^{2}}{2}G(\tau_{12})G(\tau_{21}).
    \end{align}
The trace-log term can be written in Matsubara space as
		\begin{eqnarray}\label{eq:Mats_trace_log}
-\text{Tr}\log(\hat{G}^{-1}) & = & \sum_{i \omega_n}\text{Tr}_{{\rm N}}\log\left\{ \hat{G}(i\omega_{n})/\beta\right\},
\end{eqnarray}
where $\text{Tr}_{{\rm N}}$ denotes the trace over the Nambu subspace. 
Using also the Dyson equations (\ref{eq:saddle_point_YSYK_tau_G}) and (\ref{eq:saddle_point_YSYK_tau_D}), we end up with
    \begin{align}\label{eq:normal_case}
    \frac{\mathscr{S}^{{\rm sp}}}{\mathscr{N}}= & \sum_{n}\text{Tr}_{{\rm N}}\log\left\{ \hat{G}(i\omega_{n})/\beta\right\} -\frac{1}{2}\sum_{n}\text{\ensuremath{\log\left\{  D(\Omega_{n})/\beta^{2}\right\} } }\nonumber \\
    + & \int_{\tau_{12}\tau_{21}} D(\tau_{21})\Pi_{x}(\tau_{12}) -\frac{z t_{0}^{2}}{2}\int_{\tau_{12}\tau_{21}} G(\tau_{12})G(\tau_{21}).
    \end{align}
The fermionic trace-log term can be simplified by using the explicit expression of the Nambu propagator: 
    \begin{align}
    & \text{Tr}_{{\rm N}}\log\left\{ \hat{G}(i\omega_{n})/\beta\right\} \nonumber \\  & \qquad =\log{\rm det_{N}}\left(\begin{array}{cc}
    G(i\omega_{n})/\beta & 0\\
    0 & -G(-i\omega_{n})/\beta
    \end{array}\right) \nonumber \\ & \qquad\qquad= 2\log[G(i\omega_{n})/\beta],
    \end{align}
where we have used the odd parity of normal-state fermionic propagator with respect to $\omega_n$. We Matsubara-transform the terms depending on $\tau_{12}$ in Eq.\@ (\ref{eq:normal_case}), then we add and subtract the grand potential density (per flavor $\mathscr{N}$) for free fermions and bosons in equilibrium:
\begin{subequations}\label{Altland_identity-1-1}
\begin{align}
\frac{\beta \Omega_{c}^{(0)}}{\mathscr{N}} & = -\ln(1+e^{\beta \mu}) \nonumber \\ &=\sum_{i \omega_n}\log\left[G_{0}(i\omega_{n})/\beta\right],
\end{align}
\begin{align}
\frac{\beta \Omega_{b}^{(0)}}{\mathscr{N}} & = \frac{\beta \omega_{0}}{2}+\log(1-e^{-\beta \omega_{0}}) \nonumber \\ &=\sum_{i \Omega_n}\log\left[D_{0}(i\Omega_{n})/\beta^{2}\right],
\end{align}
\end{subequations}   
with bare propagators $G_{0}^{-1}(i\omega_{n})=i\omega_{n}+\mu$ and $D_{0}^{-1}(i\Omega_{n})=\Omega_{n}^{2}+\omega_{0}^{2}$. Inserting Eqs.\@ (\ref{Altland_identity-1-1}) in the Matsubara-transformed version of the action (\ref{eq:normal_case}) gives Eq.\@ (\ref{eq:Omega_SYK_ph}). 

\subsection{Superconducting-state grand potential}\label{app:Omega_SC}

To calculate the grand potential in the superconducting state, we have to consider the full Yukawa-SYK action (\ref{eq:effective_action}), including the anomalous Green's function and self-energy. Working within the spin-singlet ansatz, and assuming translation invariance in time and space, the effective action per lattice site $\vec{x}$ is given by
    \begin{align}\label{eq:saddle_SC_tau}
    \frac{\mathscr{S}}{\mathscr{N}}= & -\text{Tr}\log\left(\hat{G}_{0}^{-1}-\hat{\Sigma}\right)+\frac{1}{2}\text{Tr}\log\left({D}_{0}^{-1}-{\Pi}\right)\nonumber \\
     & -2\int_{\tau_{12}\tau_{21}} G(\tau_{21})\Sigma(\tau_{12}) +\frac{1}{2}\int_{\tau_{12}\tau_{21}} D(\tau_{21})\Pi(\tau_{12})\nonumber \\
     & -\int_{\tau_{12}\tau_{21}}\left[F(\tau_{21})\Phi^{\dagger}(\tau_{12}) +F^{\dagger}(\tau_{21})\Phi(\tau_{12})\right]\nonumber \\
     & +\bar{g}^{2}\int_{\tau_{12}\tau_{21}}\left[G(\tau_{12})G(\tau_{21}) -F^{\dagger}(\tau_{12})F(\tau_{21})\right]D(\tau_{12})\nonumber \\
     & +\frac{zt_{0}^{2}}{2}\int_{\tau_{12}\tau_{21}}\left[G(\tau_{12})G(\tau_{21}) -F^{\dagger}(\tau_{12})F(\tau_{21})\right].
    \end{align}
The saddle-point equations (\ref{eq:Eliashberg_eph:hop}), entirely written in imaginary time, are
\begin{subequations}\label{eq:phonon_SE}
\begin{eqnarray}
\hat{G}(\tau) & = & \left[\hat{G}_{0}^{-1}(\tau)-\hat{\Sigma}(\tau)\right]^{-1}, \\
D(\tau)&=&\left[{D}_{0}^{-1}(\tau)-{\Pi}(\tau)\right]^{-1},\\
\Sigma(\tau) & = & \bar{g}^{2}G(\tau)D(\tau)+\frac{zt_{0}^{2}}{2}G(\tau), \\
\Phi(\tau)&=&-\bar{g}^{2}F(\tau)D(\tau)-\frac{zt_{0}^{2}}{2}F(\tau),\\
\Pi(\tau) & = & -2\bar{g}^{2}\left[G(\tau)G(-\tau)-F^{\dagger}(\tau)F(-\tau)\right].
\end{eqnarray}
\end{subequations}
Introducting the saddle-point equations (\ref{eq:phonon_SE}) in Eq.\@ (\ref{eq:saddle_SC_tau}), we find the on-shell action
\begin{eqnarray}\label{eq:S_on_shell_SC_tau}
\frac{\mathscr{S}^{{\rm sp}}}{\mathscr{N}} & = & -\text{Tr}\log(\hat{G}^{-1})+\frac{1}{2}\text{Tr}\log(D^{-1})\nonumber \\ & - &2\bar{g}^{2}\int_{\tau_{12}\tau_{21}}\left[G(\tau_{21})G(\tau_{12})-F^{\dagger}(\tau_{21})F(\tau_{12})\right]D(\tau_{12})\nonumber \\
 & - & \frac{zt_{0}^{2}}{2}\int_{\tau_{12}\tau_{21}}\left[G(\tau_{12})G(\tau_{21})-F^{\dagger}(\tau_{12})F(\tau_{21})\right],
\end{eqnarray}
where we have used the notation $\tau_{12}=\tau_{1}-\tau_{2}.$ 
Now, if we further use Eq.\@ (\ref{eq:phonon_SE}) we can rewrite the second
line in terms of the boson self-energy:
\begin{eqnarray}
\frac{\mathscr{S}^{{\rm sp}}}{\mathscr{N}} & = & -\text{Tr}\log(\hat{G}^{-1})+\frac{1}{2}\text{Tr}\log(D^{-1}) \\ &+&\int_{\tau_{12}\tau_{21}}\Pi(\tau_{21})D(\tau_{12})\nonumber \\
 & - & \frac{zt_{0}^{2}}{2}\int_{\tau_{12}\tau_{21}}\left[G(\tau_{12})G(\tau_{21})-F^{\dagger}(\tau_{12})F(\tau_{21})\right].\nonumber
\end{eqnarray}
Let us manipulate the fermionic trace-log term. In analogy with the normal-state calculation, we go to Matsubara space and employ Eq.\@ (\ref{eq:Mats_trace_log}), where now the trace is performed in Nambu subspace. We recall that, in our conventions, $\hat{G}$ is a $2\times2$ matrix. We have
    \begin{align}\label{eq:trace_log_SC_manip}
    &\log\left[\frac{1}{\beta^{2}}\text{det}_{{\rm N}}\left\{ \hat{G}(i\omega_{n})\right\} \right] \nonumber \\  & \qquad= \log\left[\frac{1}{\beta^{2}}\text{det}_{{\rm N}}\left(\begin{array}{cc}
    G(i\omega_{n}) & F(i\omega_{n})\\
    F^{\dagger}(i\omega_{n}) & G(i\omega_{n})
    \end{array}\right)\right]\nonumber \\
     & \qquad\qquad= \log\left\{\frac{1}{\beta^{2}}\left[G^{2}(i\omega_{n})-F^{\dagger}(i\omega_{n})F(i\omega_{n})\right]\right\}.
    \end{align}
Using Eq.\@ (\ref{eq:trace_log_SC_manip}) in the on-shell action (\ref{eq:S_on_shell_SC_tau}), we obtain
    \begin{eqnarray}\label{eq:S_on_shell_SC_tau_2}
    \frac{\mathscr{S}^{{\rm sp}}}{\mathscr{N}} & = & \sum_{i \omega_n}\log\left[\frac{1}{\beta^{2}}(G^{2}(i\omega_{n})-F^{\dagger}(i\omega_{n})F(i\omega_{n}))\right] \nonumber \\ &-&\frac{1}{2}\sum_{i \Omega_n}\text{\ensuremath{\log\left\{  D(i\Omega_{n})/\beta^{2}\right\} } }+ \sum_{i \Omega_n}D(i\Omega_{n})\Pi(i\Omega_{n}) \nonumber \\ &-&\frac{zt_{0}^{2}}{2}\sum_{i \omega_n}\left[G^{2}(i\omega_{n})-F^{\dagger}(i\omega_{n})F(i\omega_{n})\right].
    \end{eqnarray}
Setting $F(i \omega_n)=F^\dagger(i \omega_n)=0$ in Eq.\@ (\ref{eq:S_on_shell_SC_tau_2}), we retrieve Eq.\@ (\ref{eq:Omega_SYK_ph}), consistently with the normal-state calculations. 
For numerical reasons we normalize the ${\rm log}$ terms as follows: 
    \begin{eqnarray}\label{eq:S_on_shell_SC_omn}
    \frac{\mathscr{S}^{{\rm sp}}}{{\mathscr{N}}_{t}} & = & \sum_{i \omega_n}\left\{\log\left[\frac{G^{2}(i\omega_{n})-F^{\dagger}(i\omega_{n})F(i\omega_{n})}{G_{0}^{2}(i\omega_{n})}\right] \right.  \\ &-&\frac{1}{2}\sum_{i \Omega_n} \log\left[\frac{D(i\Omega_{n})}{D_{0}(i\Omega_{n})}\right]+\sum_{i \Omega_n}D(i\Omega_{n})\Pi(i\Omega_{n}) \nonumber \\  &-& \left.\frac{zt_{0}^{2}}{2}\sum_{i \omega_n}\left[G^{2}(i\omega_{n})-F^{\dagger}(i\omega_{n})F(i\omega_{n})\right]\right\}\nonumber \\
     & + & 2\sum_{i \omega_n}\log\left[G_{0}(i\omega_{n})/\beta\right] \nonumber -\frac{1}{2}\sum_{i\Omega_n}\text{\ensuremath{\log\left[D_{0}(i\Omega_{n})/\beta^{2}\right]}},
    \end{eqnarray}
with the bare propagators $G_{0}^{-1}(i\omega_{n})$ and $D_{0}^{-1}(i\Omega_{n})$ defined after Eq.\@ (\ref{Altland_identity-1-1}). Using Eq.\@ \eqref{Altland_identity-1-1}, $\Omega^{{\rm sc}}/\mathscr{N}=\mathscr{S}^{\text{sp}}/(\beta{\mathscr{N}})$ and Eq.\@ (\ref{eq:S_on_shell_SC_omn}), we finally find Eq.\@ (\ref{eq:Omega_ph_SC}). 

\section{Action for charge fluctuations and superconducting kernel}
\label{Action_for_charge_fluctuations}
In this Appendix we analyze phase fluctuations around the solution to the stationary equations \eqref{saddle_point_equations}.
At low energies, a generic solution can be easily generated from the time and space translation-invariant one $\left\{\hat{G}_{{\rm s}}(\tau),\hat{\Sigma}_{{\rm s}}(\tau)\right\}$ through a $U(1)$ transformation \cite{Inkof-thesis-2021}:
    \begin{align}
    \hat{G}_{\vec{x}}\left(\tau_{1},\tau_{2}\right) & = & e^{i\varphi_{\vec{x}}(\tau_{1})\hat{\sigma}_{3}}\hat{G}_{{\rm s}}\left(\tau_{1}-\tau_{2}\right)e^{-i\varphi_{\vec{x}}(\tau_{2})\hat{\sigma}_{3}}, \notag\\
    \hat{\Sigma}_{\vec{x}}\left(\tau_{1},\tau_{2}\right) & = & e^{i\varphi_{\vec{x}}(\tau_{1})\hat{\sigma}_{3}}\hat{\Sigma}_{{\rm s}}\left(\tau_{1}-\tau_{2}\right)e^{-i\varphi_{\vec{x}}(\tau_{2})\hat{\sigma}_{3}}.
    \label{eq:u1transf-1}
    \end{align}
Hats denote $2\times2$ matrices in Nambu space, and $\varphi_{\vec{x}}(\tau)$ is a $U(1)$ phase field conjugated to charge fluctuations of the system \cite{Song-2017,Davison-2017}. 
Away from the IR limit, the equations are no longer invariant under the map Eq.\@ \eqref{eq:u1transf-1} and the $U(1)$ symmetry is broken. The associated Goldstone
mode is precisely given by the field $\varphi_{\vec{x}}(\tau)$, and in the following we derive the associated low-energy effective action. \\
Let us insert Eq.\@ \eqref{eq:u1transf-1} into the disorder-averaged SYK action \eqref{eq:effective_action}, and focus on the phase degrees of freedom only. The terms which are not invariant under the $U(1)$ transformation are the trace-log and hopping ones. This leads to the following expression:
     \begin{widetext}
        \begin{eqnarray}\label{eq:phase_mode_action}
        &\frac{\mathscr{S}[\varphi]}{\mathscr{N}} & =  -\sum_{\vec{x}}{\rm Tr}{\rm log}\left\{[(\mu-i\partial_{\tau}\varphi)\hat{\sigma}_{3}-\hat{\sigma}_{0}\partial_{\tau}]\delta(\tau_{1}-\tau_{2})-\hat{\Sigma}_{\vec{x}}(\tau_{1},\tau_{2})\right\} \\
         & + & \frac{t_{0}^{2}}{2}\sum_{\substack{\left\langle\vec{x},\vec{x}'\right\rangle}
        }\int_{\tau_{1}\tau_{2}}\left[G_{{\rm s}}(\tau_{1}-\tau_{2})G_{{\rm s}}(\tau_{2}-\tau_{1})\,e^{i\left[\partial_{\vec{x}}\varphi(\tau_{1})-\partial_{\vec{x}}\varphi(\tau_{2})\right]\cdot\vec{a}}-F_{{\rm s}}^{\dagger}(\tau_{1}-\tau_{2})F_{{\rm s}}(\tau_{2}-\tau_{1})\,e^{-i\left[\partial_{\vec{x}}\varphi(\tau_{1})+\partial_{\vec{x}}\varphi(\tau_{2})\right]\cdot\vec{a}}\right]
        \nonumber
        \end{eqnarray}
    \end{widetext}
with $\hat{\sigma}_{0}$ and $\hat{\sigma}_{3}$ Pauli matrices. Angled brackets denote the sum over nearest-neighbor sites on the lattice. Space gradients are defined via 
\begin{eqnarray}
\vec{a}\cdot\partial_{\vec{x}}\varphi(\tau) & = & \varphi_{\vec{x}}(\tau)-\varphi_{\vec{x}'}(\tau),
\end{eqnarray}
where $\vec{a}=\vec{x}-\vec{x}'$ gives the spacing between nearest-neighbor sites, which we have assumed to be the same for any pairs of sites.
Time gradients enter the action by shifting the chemical potential as $\mu\mapsto\mu-i\partial_{\tau}\varphi$. This is consequence of a gauge transformation of the fermion fields $\hat{c}_{i\vec{x}\uparrow}(\tau)\mapsto e^{-i\varphi_{\vec{x}}(\tau)}\hat{c}_{i\vec{x}\uparrow}(\tau),\,\,\hat{c}_{i\vec{x}\downarrow}^{\dagger}(\tau)\mapsto e^{i\varphi_{\vec{x}}(\tau)}\hat{c}_{i\vec{x}\downarrow}^{\dagger}(\tau)$ that eliminates phase factors from the self-energy term \cite{Inkof-thesis-2021}.
\\
 Expanding the action Eq.\@ \eqref{eq:phase_mode_action} to quadratic order in phase gradients, and ignoring contributions independent of $\varphi$, yields the Gaussian action:
    \begin{align}\label{eq:gaussian_action}
    \frac{\mathscr{S}[\varphi]}{{\mathscr{N}}}&=-\frac{k_B T}{a^3}\sum_{\vec{p}, i\Omega_{n}}\varphi_{-\vec{p}}(i \Omega_n) \\ & \qquad\qquad\quad\times \left[\Omega^2_{n}\bar{\Pi}(i\Omega_{n})+\epsilon(p)\Lambda(i\Omega_{n})\right]\varphi_{\vec{p}}(-i \Omega_n),     
	 \notag	\end{align}
which is the lowest-order effective theory for phase fluctuations.
Here we have used the double Fourier transform of the phase field to the space of momenta $\vec{p}$ and bosonic Matsubara frequencies $\Omega_n$: 
\begin{equation}\label{eq:phase_field_Fourier}
\varphi_{\vec{x}}(\tau)=\frac{k_B T}{a^3}\sum_{\vec{p},i\Omega_{n}}\varphi_{\vec{p}}(i \Omega_{n}) e^{i(\vec{p}\cdot\vec{x}-\Omega_{n}\tau)},
\end{equation}
where $a^3$ is the volume of a unit cell, which enters into the definition of the spatial Fourier transform $\vec{x} \mapsto \vec{p}$. The dispersion
relation of the phase fluctuations in the action (\ref{eq:gaussian_action}) is given by $\epsilon(p)=\frac{1}{2}\sum_{\left\langle x, x'\right\rangle}(\vec{p}\cdot\vec{a})^{2}$. In the case of isotropic Bravais lattices with spacing $|\vec{a}|=a$, the dispersion is simply $\epsilon(p)=z \, a^2 p^2/2$, where $p=|\vec{p}|$ and $z$ is the coordination number. The latter form of the dispersion $\epsilon(p)$ will be assumed from now on. Here $\bar{\Pi}(i\Omega_{n})$ is the Matsubara transform of $\bar{\Pi}(\tau)=\Pi_{G}(\tau)-\Pi_{F}(\tau)$, with standard and anomalous bubbles $\Pi_{G}(\tau)=G(\tau)G(-\tau)$
and $\Pi_{F}(\tau)=F^{\dagger}(\tau)F(-\tau)$. Moreover, $2\Lambda(i\Omega_{n})=t_{0}^{2}\left[\Pi_{G}(0)-\Pi_{G}(i\Omega_{n})-\Pi_{F}(0)-\Pi_{F}(i\Omega_{n})\right]$. 

In the following we analyze the effects of an external electromagnetic field and compute the superconducting kernel \eqref{definition_of_SC_kernel}. In order to introduce a vector potential $\vec{A}$, we perform a Peierls substitution \cite{Peierls-1933,Li-2020} of the fermions
\begin{equation}\label{eq:Peierls_subst}
\hat{c}_{i\vec{x}\sigma}(\tau)\mapsto \hat{c}_{i\vec{x}\sigma}(\tau)e^{-i e\int_{\vec{x}_{0}}^{\vec{x}}d\vec{y}\cdot\vec{A}(\vec{y},\tau)},
\end{equation}
with $e$ the fermion electric charge and $\vec{x}_{0}$ a reference location. 

Only the non-local-in-$\vec{x}$ hopping term of the SYK action Eq.\@ \eqref{eq:effective_action} is affected by the transformation (\ref{eq:Peierls_subst}). As a consequence, space gradients in Eq.\@ \eqref{eq:phase_mode_action} are shifted as $\vec{a}\cdot\partial_{\vec{x}}\varphi(\tau)\mapsto\vec{a}\cdot\partial_{\vec{x}}\varphi(\tau)-e \int_{\vec{x}'}^{\vec{x}}d\vec{y}\cdot\vec{A}(\vec{y},\tau)$.
Furthermore, for slowly varying-in-space vector fields we can approximate  $\int_{\vec{x}'}^{\vec{x}}d\vec{y}\cdot\vec{A}(\vec{y},\tau)\approx\vec{a}\cdot\vec{A}(\vec{x},\tau)$.
The effective theory in the presence of an external vector potential then simply follows from Eq.\@ \eqref{eq:gaussian_action} via the minimal substitution
$i\vec{p}\,\varphi_{\vec{p}}(i\Omega_{n})\mapsto i\vec{p}\,\varphi_{\vec{p}}(i\Omega_{n})-a \, e \,\vec{A}_{\vec{p}}(i\Omega_{n})$.
The change in the action with respect to its unperturbed counterpart (\ref{eq:phase_mode_action}), due to the presence of the vector potential, is given by 
\begin{widetext}
\begin{equation}\label{eq:quadratic_action_after_Peierls}
\frac{\Delta \mathscr{S}[\varphi,\vec{A}]}{{\mathscr{N}}} = \frac{1}{2}\frac{k_B T}{a^3} \sum_{\vec{p},i \Omega_n}\sum_{\alpha,\beta} A_{\alpha}(\vec{p},-i\Omega_{n})m_{\alpha\beta}(\vec{p},i \Omega_n)A_{\beta}(-\vec{p},i\Omega_{n})-\frac{k_B T}{a^3} \sum_{\vec{p},i \Omega_n}\sum_{\alpha}\,j_{\alpha}(\vec{p},-i\Omega_{n})A_{\alpha}(-\vec{p},i\Omega_{n}),
\end{equation}
\end{widetext}
where $\alpha$ and $\beta$ label the spatial directions. The mass and source terms are given by 
    \begin{subequations}\label{eq:mass_current_micro}
    \begin{equation}
    m_{\alpha\beta}(\vec{p},i \Omega_n) = -{\mathscr{N}}e^{2} z a^2 \Lambda(i\Omega_{n})\delta_{\alpha\beta}
		\end{equation}
		\begin{equation}\label{eq:kernel_inT}
    j_{\alpha}(\vec{p},-i\Omega_{n}) = -i{\mathscr{N}}e z a^2 \Lambda(i\Omega_{n})\varphi_{\vec{p}}(-i \Omega_n),
    \end{equation}
    \end{subequations}
where we have used $\Lambda(-i\omega_{n})=\Lambda(i\omega_{n})$, which enters into the propagator for phase fluctuations that we defined in Eq.\@ (\ref{eq:gaussian_action}).
The partition function, including the effect of the vector potential $\vec{A}$, is $\mathscr{Z}=\int D \varphi e^{-(\mathscr{S}[\varphi]+\Delta \mathscr{S}[\varphi,\vec{A}])}$. The functional derivative of the partition function with respect to the vector potential gives the total current $\vec{J}_{\text{tot}}(\vec{q},\omega)$ of the system \cite{Altland-cm2010,Inkof-thesis-2021}, which comprises the external current as well as the induced internal current, in accordance with Eq.\@ (\ref{definition_of_SC_kernel}) upon the analytic continuation $i \Omega_n \rightarrow \omega+i 0^+$ \cite{Schrieffer-1963th, Dressel-2001}. Formally, $a^3/(k_B T) J_{\text{tot},\alpha}(\vec{p},-i \Omega_n)=\delta \log \mathscr{Z}/\delta A_\alpha(-\vec{p},i\Omega_{n})=\left.\left\langle j_\alpha(\vec{p},-i \Omega_n) \right\rangle\right|_{\vec{A}}-m_{\alpha \beta}(\vec{p},i \Omega_n)A_\alpha(\vec{p},-i \Omega_n)$. Here the brackets $\left\langle (\cdots) \right\rangle_{\vec{A}}=\mathscr{Z}^{-1} \int D \varphi (\cdots) e^{-(\mathscr{S}[\varphi]+\Delta \mathscr{S}[\varphi,\vec{A}])}$. 
Then, the electromagnetic kernel is defined as the ratio between the total current $J_{\text{tot},\alpha}(\vec{p},-i \Omega_n)$ and the imposed vector potential $A_\beta(\vec{p},-i\Omega_{n})$, evaluated in the vanishing vector-potential limit -- see Eq.\@ \eqref{definition_of_SC_kernel}. This additional derivative yields two terms: one involves the product between two one-point functions $\left\langle j_\alpha(-\vec{p}, i\Omega_n)\right\rangle_{\vec{A}}\left\langle j_\alpha(\vec{p}, -i\Omega_n)\right\rangle_{\vec{A}}$, which is assumed to be null in the absence of the external source (in the unperturbed system, with $\vec{A}=0$); the second term is the correlator $\left\langle j_\alpha(\vec{p}, -i\Omega_n) j_\beta(-\vec{p}, i\Omega_n)\right\rangle_{\vec{A}}$. Employing Eq.\eqref{eq:kernel_inT}, we thus obtain for the kernel
    \begin{align}\label{eq:Kernel_micro}
    \mathcal{K}_{\alpha\beta}(\vec{p},i\Omega_{n})& =\frac{1}{k_B T}\frac{\delta^{2}\log{\cal Z}}{\delta A_{\beta}(\vec{p},-i\Omega_{n})\delta A_{\alpha}(-\vec{p},i\Omega_{n})}\Big\rvert_{\vec{A}=0}\nonumber \\
     & =\frac{\mathscr{N} e^2 \, z \Lambda(i\Omega_{n})}{a} \notag \\ & \times \left[ \delta_{\alpha\beta}+\frac{z a^2 \Lambda(i\Omega_{n})p_\alpha p_\beta}{\Omega_{n}^{2}\bar{\Pi}(i\Omega_{n})+z a^2 p^2\Lambda(i\Omega_{n})/2}\right].
    \end{align}
We immediately see that (minus) the static, zero-momentum limit of the kernel (\ref{eq:Kernel_micro}) corresponds to the phase stiffness of the system, consistently with Eq.\@ (\ref{London_Stiffness_appendix}): it is proportional to $-\Lambda(0)$, which can be equivalently obtained from the phase action (\ref{eq:gaussian_action}), written in real space of coordinates, by deriving the logarithm of $\mathscr{Z}_{\varphi}=\int D \varphi e^{-\mathscr{S}[\varphi]}$ twice with respect to spatial gradients $\nabla \varphi(\vec{r},\tau)$ of the phase. Hence, the long-wavelength static electromagnetic response function probes the rigidity of the system, i.e.\@ the second derivative of the grand potential, with respect to global phase twists \cite{Fisher-1973,Taylor-2006,Taylor-2007,Altland-cm2010}. 
As such, the stiffness has units $\left[\mathscr{E} L^{2-d}\right]$, where $\mathscr{E}$ is an energy, $L$ is a length, and $d$ is the system dimensionality. 
We rapidly comment on the modifications to the various quantities in this section, when we employ SI units instead: in this case, in the Peierls-substitution formula (\ref{eq:Peierls_subst}) we have $i e \mapsto i e/\hbar$, with $\hbar$ reduced Planck's constant; this constant propagates to Eqs.\@ (\ref{eq:mass_current_micro}), where $m_{\alpha\beta}(\vec{p},i \Omega_n) \mapsto m_{\alpha\beta}(\vec{p},i \Omega_n)/\hbar^2$ and $J_{\alpha}(\vec{p},-i\Omega_{n}) \mapsto J_{\alpha}(\vec{p},-i\Omega_{n})/\hbar^2$, and finally to the kernel $\mathcal{K}_{\alpha\beta}(\vec{p},i\Omega_{n}) \mapsto \mathcal{K}_{\alpha\beta}(\vec{p},i\Omega_{n})/\hbar^2$. Going to the $i\Omega_n=0$ and $q \rightarrow 0$ limit of Eq.\@ (\ref{eq:Kernel_micro}), and employing Eq.\@ (\ref{London_Stiffness_appendix}), finally yields Eq.\@ (\ref{SC_stiffness}) for the superconducting phase stiffness. 

The electromagnetic kernel (\ref{eq:Kernel_micro}) allows us to calculate the conductivity tensor of the Yukawa-SYK lattice through \cite{Schrieffer-1963th, Dressel-2001,Altland-cm2010}
\begin{equation}\label{eq:kernel_cond}
\sigma_{\alpha \beta}(\vec{p},\omega)=-i \frac{\mathcal{K}_{\alpha\beta}(\vec{p},\omega)}{\omega+i 0^+},
\end{equation}
where we have analytically continued the kernel to real frequencies as $\mathcal{K}_{\alpha\beta}(\vec{p},\omega)=\lim_{i \Omega_n \rightarrow \omega+i 0^+} \mathcal{K}_{\alpha\beta}(\vec{p},i\Omega_n)$. Taking the longitudinal part $\mathcal{K}_{L}(\vec{p},\omega)$ of the kernel -- see discussion after Eq.\@ (\ref{eq:magnetiz_A}) -- and going to the $p\rightarrow 0$ limit \cite{Inkof-thesis-2021}, we obtain the zero-momentum conductivity $\sigma(\omega) \equiv -i \lim_{p \rightarrow 0} \mathcal{K}_{L}(\vec{p},\omega)/(\omega+i 0^+)$. At $T>T_c$ (normal state) the conductivity only depends on the retarded normal propagator $G^R(\omega)$ through $\Lambda(\omega)$, while for $T<T_c$ (superconducting state) there is an additional contribution to the conductivity stemming from $F^R(\omega)$; see also Appendix \ref{App:num_real_axis}. An analysis of the conductivity in the normal and superconducting states is reported in the companion paper \cite{short-paper}. 


\begin{thebibliography}{243}%
\makeatletter
\providecommand \@ifxundefined [1]{%
 \@ifx{#1\undefined}
}%
\providecommand \@ifnum [1]{%
 \ifnum #1\expandafter \@firstoftwo
 \else \expandafter \@secondoftwo
 \fi
}%
\providecommand \@ifx [1]{%
 \ifx #1\expandafter \@firstoftwo
 \else \expandafter \@secondoftwo
 \fi
}%
\providecommand \natexlab [1]{#1}%
\providecommand \enquote  [1]{``#1''}%
\providecommand \bibnamefont  [1]{#1}%
\providecommand \bibfnamefont [1]{#1}%
\providecommand \citenamefont [1]{#1}%
\providecommand \href@noop [0]{\@secondoftwo}%
\providecommand \href [0]{\begingroup \@sanitize@url \@href}%
\providecommand \@href[1]{\@@startlink{#1}\@@href}%
\providecommand \@@href[1]{\endgroup#1\@@endlink}%
\providecommand \@sanitize@url [0]{\catcode `\\12\catcode `\$12\catcode
  `\&12\catcode `\#12\catcode `\^12\catcode `\_12\catcode `\%12\relax}%
\providecommand \@@startlink[1]{}%
\providecommand \@@endlink[0]{}%
\providecommand \url  [0]{\begingroup\@sanitize@url \@url }%
\providecommand \@url [1]{\endgroup\@href {#1}{\urlprefix }}%
\providecommand \urlprefix  [0]{URL }%
\providecommand \Eprint [0]{\href }%
\providecommand \doibase [0]{https://doi.org/}%
\providecommand \selectlanguage [0]{\@gobble}%
\providecommand \bibinfo  [0]{\@secondoftwo}%
\providecommand \bibfield  [0]{\@secondoftwo}%
\providecommand \translation [1]{[#1]}%
\providecommand \BibitemOpen [0]{}%
\providecommand \bibitemStop [0]{}%
\providecommand \bibitemNoStop [0]{.\EOS\space}%
\providecommand \EOS [0]{\spacefactor3000\relax}%
\providecommand \BibitemShut  [1]{\csname bibitem#1\endcsname}%
\let\auto@bib@innerbib\@empty
\bibitem [{\citenamefont {Zaanen}(2004)}]{Zaanen-2004}%
  \BibitemOpen
  \bibfield  {author} {\bibinfo {author} {\bibfnamefont {J.}~\bibnamefont
  {Zaanen}},\ }\bibfield  {title} {\bibinfo {title} {Superconductivity: Why the
  temperature is high},\ }\href
  {http://www.nature.com/nature/journal/v430/n6999/full/430512a.html}
  {\bibfield  {journal} {\bibinfo  {journal} {Nature}\ }\textbf {\bibinfo
  {volume} {430}},\ \bibinfo {pages} {512} (\bibinfo {year}
  {2004})}\BibitemShut {NoStop}%
\bibitem [{\citenamefont {Valla}\ \emph {et~al.}(1999)\citenamefont {Valla},
  \citenamefont {Fedorov}, \citenamefont {Johnson}, \citenamefont {Wells},
  \citenamefont {Hulbert}, \citenamefont {Li}, \citenamefont {Gu},\ and\
  \citenamefont {Koshizuka}}]{Valla-1999}%
  \BibitemOpen
  \bibfield  {author} {\bibinfo {author} {\bibfnamefont {T.}~\bibnamefont
  {Valla}}, \bibinfo {author} {\bibfnamefont {A.}~\bibnamefont {Fedorov}},
  \bibinfo {author} {\bibfnamefont {P.}~\bibnamefont {Johnson}}, \bibinfo
  {author} {\bibfnamefont {B.}~\bibnamefont {Wells}}, \bibinfo {author}
  {\bibfnamefont {S.}~\bibnamefont {Hulbert}}, \bibinfo {author} {\bibfnamefont
  {Q.}~\bibnamefont {Li}}, \bibinfo {author} {\bibfnamefont {G.}~\bibnamefont
  {Gu}},\ and\ \bibinfo {author} {\bibfnamefont {N.}~\bibnamefont
  {Koshizuka}},\ }\bibfield  {title} {\bibinfo {title} {Evidence for quantum
  critical behavior in the optimally doped cuprate
  {Bi$_2$Sr$_2$CaCu$_2$O}$_{8+\delta}$},\ }\href
  {https://science.sciencemag.org/content/285/5436/2110} {\bibfield  {journal}
  {\bibinfo  {journal} {Science}\ }\textbf {\bibinfo {volume} {285}},\ \bibinfo
  {pages} {2110} (\bibinfo {year} {1999})}\BibitemShut {NoStop}%
\bibitem [{\citenamefont {Bruin}\ \emph {et~al.}(2013)\citenamefont {Bruin},
  \citenamefont {Sakai}, \citenamefont {Perry},\ and\ \citenamefont
  {Mackenzie}}]{Bruin-2013}%
  \BibitemOpen
  \bibfield  {author} {\bibinfo {author} {\bibfnamefont {J.}~\bibnamefont
  {Bruin}}, \bibinfo {author} {\bibfnamefont {H.}~\bibnamefont {Sakai}},
  \bibinfo {author} {\bibfnamefont {R.}~\bibnamefont {Perry}},\ and\ \bibinfo
  {author} {\bibfnamefont {A.~P.}\ \bibnamefont {Mackenzie}},\ }\bibfield
  {title} {\bibinfo {title} {Similarity of scattering rates in metals showing
  {T}-linear resistivity},\ }\href {https://doi.org/10.1126/sciadv.abi8481}
  {\bibfield  {journal} {\bibinfo  {journal} {Science}\ }\textbf {\bibinfo
  {volume} {339}},\ \bibinfo {pages} {804} (\bibinfo {year}
  {2013})}\BibitemShut {NoStop}%
\bibitem [{\citenamefont {Giraldo-Gallo}\ \emph {et~al.}(2018)\citenamefont
  {Giraldo-Gallo}, \citenamefont {Galvis}, \citenamefont {Stegen},
  \citenamefont {Modic}, \citenamefont {Balakirev}, \citenamefont {Betts},
  \citenamefont {Lian}, \citenamefont {Moir}, \citenamefont {Riggs},
  \citenamefont {Wu}, \citenamefont {Bollinger}, \citenamefont {He},
  \citenamefont {Bo\v{z}ovi\'{c}}, \citenamefont {Ramshaw}, \citenamefont
  {McDonald}, \citenamefont {Boebinger},\ and\ \citenamefont
  {Shekhter}}]{Giraldo-2018}%
  \BibitemOpen
  \bibfield  {author} {\bibinfo {author} {\bibfnamefont {P.}~\bibnamefont
  {Giraldo-Gallo}}, \bibinfo {author} {\bibfnamefont {J.}~\bibnamefont
  {Galvis}}, \bibinfo {author} {\bibfnamefont {Z.}~\bibnamefont {Stegen}},
  \bibinfo {author} {\bibfnamefont {K.~A.}\ \bibnamefont {Modic}}, \bibinfo
  {author} {\bibfnamefont {F.}~\bibnamefont {Balakirev}}, \bibinfo {author}
  {\bibfnamefont {J.}~\bibnamefont {Betts}}, \bibinfo {author} {\bibfnamefont
  {X.}~\bibnamefont {Lian}}, \bibinfo {author} {\bibfnamefont {C.}~\bibnamefont
  {Moir}}, \bibinfo {author} {\bibfnamefont {S.}~\bibnamefont {Riggs}},
  \bibinfo {author} {\bibfnamefont {J.}~\bibnamefont {Wu}}, \bibinfo {author}
  {\bibfnamefont {A.~T.}\ \bibnamefont {Bollinger}}, \bibinfo {author}
  {\bibfnamefont {X.}~\bibnamefont {He}}, \bibinfo {author} {\bibfnamefont
  {I.}~\bibnamefont {Bo\v{z}ovi\'{c}}}, \bibinfo {author} {\bibfnamefont
  {B.~J.}\ \bibnamefont {Ramshaw}}, \bibinfo {author} {\bibfnamefont {R.~D.}\
  \bibnamefont {McDonald}}, \bibinfo {author} {\bibfnamefont {G.~S.}\
  \bibnamefont {Boebinger}},\ and\ \bibinfo {author} {\bibfnamefont
  {A.}~\bibnamefont {Shekhter}},\ }\bibfield  {title} {\bibinfo {title}
  {Scale-invariant magnetoresistance in a cuprate superconductor},\ }\href
  {https://science.sciencemag.org/content/361/6401/479} {\bibfield  {journal}
  {\bibinfo  {journal} {Science}\ }\textbf {\bibinfo {volume} {361}},\ \bibinfo
  {pages} {479} (\bibinfo {year} {2018})}\BibitemShut {NoStop}%
\bibitem [{\citenamefont {Naqib}\ and\ \citenamefont
  {Islam}(2019)}]{Naqib-2019}%
  \BibitemOpen
  \bibfield  {author} {\bibinfo {author} {\bibfnamefont {S.~H.}\ \bibnamefont
  {Naqib}}\ and\ \bibinfo {author} {\bibfnamefont {R.~S.}\ \bibnamefont
  {Islam}},\ }\bibfield  {title} {\bibinfo {title} {Possible quantum critical
  behavior revealed by the critical current density of hole doped high-{T$_c$}
  cuprates in comparison to heavy fermion superconductors},\ }\href
  {https://www.nature.com/articles/s41598-019-51467-4} {\bibfield  {journal}
  {\bibinfo  {journal} {Sci. Rep.}\ }\textbf {\bibinfo {volume} {9}},\ \bibinfo
  {pages} {1} (\bibinfo {year} {2019})}\BibitemShut {NoStop}%
\bibitem [{\citenamefont {Mandal}\ \emph {et~al.}(2019)\citenamefont {Mandal},
  \citenamefont {Sarkar},\ and\ \citenamefont {Greene}}]{Mandal-2019}%
  \BibitemOpen
  \bibfield  {author} {\bibinfo {author} {\bibfnamefont {P.~R.}\ \bibnamefont
  {Mandal}}, \bibinfo {author} {\bibfnamefont {T.}~\bibnamefont {Sarkar}},\
  and\ \bibinfo {author} {\bibfnamefont {R.~L.}\ \bibnamefont {Greene}},\
  }\bibfield  {title} {\bibinfo {title} {Anomalous quantum criticality in the
  electron-doped cuprates},\ }\href {https://doi.org/10.1073/pnas.1817653116}
  {\bibfield  {journal} {\bibinfo  {journal} {Proc. Natl. Acad. Sci. U.S.A.}\
  }\textbf {\bibinfo {volume} {116}},\ \bibinfo {pages} {5991} (\bibinfo {year}
  {2019})}\BibitemShut {NoStop}%
\bibitem [{\citenamefont {Michon}\ \emph {et~al.}(2019)\citenamefont {Michon},
  \citenamefont {Girod}, \citenamefont {Badoux}, \citenamefont
  {Ka{\v{c}}mar{\v{c}}{\'\i}k}, \citenamefont {Ma}, \citenamefont {Dragomir},
  \citenamefont {Dabkowska}, \citenamefont {Gaulin}, \citenamefont {Zhou},
  \citenamefont {Pyon}, \citenamefont {Takayama}, \citenamefont {Takagi},
  \citenamefont {Verret}, \citenamefont {Doiron-Leyraud}, \citenamefont
  {Marcenat}, \citenamefont {Taillefer},\ and\ \citenamefont
  {Klein}}]{Michon-2019}%
  \BibitemOpen
  \bibfield  {author} {\bibinfo {author} {\bibfnamefont {B.}~\bibnamefont
  {Michon}}, \bibinfo {author} {\bibfnamefont {C.}~\bibnamefont {Girod}},
  \bibinfo {author} {\bibfnamefont {S.}~\bibnamefont {Badoux}}, \bibinfo
  {author} {\bibfnamefont {J.}~\bibnamefont {Ka{\v{c}}mar{\v{c}}{\'\i}k}},
  \bibinfo {author} {\bibfnamefont {Q.}~\bibnamefont {Ma}}, \bibinfo {author}
  {\bibfnamefont {M.}~\bibnamefont {Dragomir}}, \bibinfo {author}
  {\bibfnamefont {H.}~\bibnamefont {Dabkowska}}, \bibinfo {author}
  {\bibfnamefont {B.}~\bibnamefont {Gaulin}}, \bibinfo {author} {\bibfnamefont
  {J.-S.}\ \bibnamefont {Zhou}}, \bibinfo {author} {\bibfnamefont
  {S.}~\bibnamefont {Pyon}}, \bibinfo {author} {\bibfnamefont {T.}~\bibnamefont
  {Takayama}}, \bibinfo {author} {\bibfnamefont {H.}~\bibnamefont {Takagi}},
  \bibinfo {author} {\bibfnamefont {S.}~\bibnamefont {Verret}}, \bibinfo
  {author} {\bibfnamefont {N.}~\bibnamefont {Doiron-Leyraud}}, \bibinfo
  {author} {\bibfnamefont {C.}~\bibnamefont {Marcenat}}, \bibinfo {author}
  {\bibfnamefont {L.}~\bibnamefont {Taillefer}},\ and\ \bibinfo {author}
  {\bibfnamefont {T.}~\bibnamefont {Klein}},\ }\bibfield  {title} {\bibinfo
  {title} {Thermodynamic signatures of quantum criticality in cuprate
  superconductors},\ }\href {https://www.nature.com/articles/s41586-019-0932-x}
  {\bibfield  {journal} {\bibinfo  {journal} {Nature}\ }\textbf {\bibinfo
  {volume} {567}},\ \bibinfo {pages} {218} (\bibinfo {year}
  {2019})}\BibitemShut {NoStop}%
\bibitem [{\citenamefont {Legros}\ \emph {et~al.}(2019)\citenamefont {Legros},
  \citenamefont {Benhabib}, \citenamefont {Tabis}, \citenamefont
  {Lalibert{\'e}}, \citenamefont {Dion}, \citenamefont {Lizaire}, \citenamefont
  {Vignolle}, \citenamefont {Vignolles}, \citenamefont {Raffy}, \citenamefont
  {Li}, \citenamefont {Auban-Senzier}, \citenamefont {Doiron-Leyraud},
  \citenamefont {Fournier}, \citenamefont {Colson}, \citenamefont {Taillefer},\
  and\ \citenamefont {Proust}}]{Legros-2019}%
  \BibitemOpen
  \bibfield  {author} {\bibinfo {author} {\bibfnamefont {A.}~\bibnamefont
  {Legros}}, \bibinfo {author} {\bibfnamefont {S.}~\bibnamefont {Benhabib}},
  \bibinfo {author} {\bibfnamefont {W.}~\bibnamefont {Tabis}}, \bibinfo
  {author} {\bibfnamefont {F.}~\bibnamefont {Lalibert{\'e}}}, \bibinfo {author}
  {\bibfnamefont {M.}~\bibnamefont {Dion}}, \bibinfo {author} {\bibfnamefont
  {M.}~\bibnamefont {Lizaire}}, \bibinfo {author} {\bibfnamefont
  {B.}~\bibnamefont {Vignolle}}, \bibinfo {author} {\bibfnamefont
  {D.}~\bibnamefont {Vignolles}}, \bibinfo {author} {\bibfnamefont
  {H.}~\bibnamefont {Raffy}}, \bibinfo {author} {\bibfnamefont
  {Z.}~\bibnamefont {Li}}, \bibinfo {author} {\bibfnamefont {P.}~\bibnamefont
  {Auban-Senzier}}, \bibinfo {author} {\bibfnamefont {N.}~\bibnamefont
  {Doiron-Leyraud}}, \bibinfo {author} {\bibfnamefont {P.}~\bibnamefont
  {Fournier}}, \bibinfo {author} {\bibfnamefont {D.}~\bibnamefont {Colson}},
  \bibinfo {author} {\bibfnamefont {L.}~\bibnamefont {Taillefer}},\ and\
  \bibinfo {author} {\bibfnamefont {C.}~\bibnamefont {Proust}},\ }\bibfield
  {title} {\bibinfo {title} {Universal {T-linear} resistivity and {Planckian}
  dissipation in overdoped cuprates},\ }\href
  {https://www.nature.com/articles/s41567-018-0334-2} {\bibfield  {journal}
  {\bibinfo  {journal} {Nat. Phys.}\ }\textbf {\bibinfo {volume} {15}},\
  \bibinfo {pages} {142} (\bibinfo {year} {2019})}\BibitemShut {NoStop}%
\bibitem [{\citenamefont {Grissonnanche}\ \emph {et~al.}(2021)\citenamefont
  {Grissonnanche}, \citenamefont {Fang}, \citenamefont {Legros}, \citenamefont
  {Verret}, \citenamefont {Lalibert{\'e}}, \citenamefont {Collignon},
  \citenamefont {Zhou}, \citenamefont {Graf}, \citenamefont {Goddard},
  \citenamefont {Taillefer},\ and\ \citenamefont
  {Ramshaw}}]{Grissonnanche-2021}%
  \BibitemOpen
  \bibfield  {author} {\bibinfo {author} {\bibfnamefont {G.}~\bibnamefont
  {Grissonnanche}}, \bibinfo {author} {\bibfnamefont {Y.}~\bibnamefont {Fang}},
  \bibinfo {author} {\bibfnamefont {A.}~\bibnamefont {Legros}}, \bibinfo
  {author} {\bibfnamefont {S.}~\bibnamefont {Verret}}, \bibinfo {author}
  {\bibfnamefont {F.}~\bibnamefont {Lalibert{\'e}}}, \bibinfo {author}
  {\bibfnamefont {C.}~\bibnamefont {Collignon}}, \bibinfo {author}
  {\bibfnamefont {J.}~\bibnamefont {Zhou}}, \bibinfo {author} {\bibfnamefont
  {D.}~\bibnamefont {Graf}}, \bibinfo {author} {\bibfnamefont {P.~A.}\
  \bibnamefont {Goddard}}, \bibinfo {author} {\bibfnamefont {L.}~\bibnamefont
  {Taillefer}},\ and\ \bibinfo {author} {\bibfnamefont {B.~J.}\ \bibnamefont
  {Ramshaw}},\ }\bibfield  {title} {\bibinfo {title} {Linear-in temperature
  resistivity from an isotropic {Planckian} scattering rate},\ }\href
  {https://doi.org/10.1038/s41586-021-03697-8} {\bibfield  {journal} {\bibinfo
  {journal} {Nature}\ }\textbf {\bibinfo {volume} {595}},\ \bibinfo {pages}
  {667} (\bibinfo {year} {2021})}\BibitemShut {NoStop}%
\bibitem [{\citenamefont {Keimer}\ \emph {et~al.}(2015)\citenamefont {Keimer},
  \citenamefont {Kivelson}, \citenamefont {Norman}, \citenamefont {Uchida},\
  and\ \citenamefont {Zaanen}}]{Keimer-2015}%
  \BibitemOpen
  \bibfield  {author} {\bibinfo {author} {\bibfnamefont {B.}~\bibnamefont
  {Keimer}}, \bibinfo {author} {\bibfnamefont {S.~A.}\ \bibnamefont
  {Kivelson}}, \bibinfo {author} {\bibfnamefont {M.~R.}\ \bibnamefont
  {Norman}}, \bibinfo {author} {\bibfnamefont {S.}~\bibnamefont {Uchida}},\
  and\ \bibinfo {author} {\bibfnamefont {J.}~\bibnamefont {Zaanen}},\
  }\bibfield  {title} {\bibinfo {title} {From quantum matter to
  high-temperature superconductivity in copper oxides},\ }\href
  {https://www.nature.com/articles/nature14165} {\bibfield  {journal} {\bibinfo
   {journal} {Nature}\ }\textbf {\bibinfo {volume} {518}},\ \bibinfo {pages}
  {179} (\bibinfo {year} {2015})}\BibitemShut {NoStop}%
\bibitem [{\citenamefont {Dessau}\ \emph {et~al.}(1991)\citenamefont {Dessau},
  \citenamefont {Wells}, \citenamefont {Shen}, \citenamefont {Spicer},
  \citenamefont {Arko}, \citenamefont {List}, \citenamefont {Mitzi},\ and\
  \citenamefont {Kapitulnik}}]{Dessau-1991}%
  \BibitemOpen
  \bibfield  {author} {\bibinfo {author} {\bibfnamefont {D.~S.}\ \bibnamefont
  {Dessau}}, \bibinfo {author} {\bibfnamefont {B.~O.}\ \bibnamefont {Wells}},
  \bibinfo {author} {\bibfnamefont {Z.-X.}\ \bibnamefont {Shen}}, \bibinfo
  {author} {\bibfnamefont {W.~E.}\ \bibnamefont {Spicer}}, \bibinfo {author}
  {\bibfnamefont {A.~J.}\ \bibnamefont {Arko}}, \bibinfo {author}
  {\bibfnamefont {R.~S.}\ \bibnamefont {List}}, \bibinfo {author}
  {\bibfnamefont {D.~B.}\ \bibnamefont {Mitzi}},\ and\ \bibinfo {author}
  {\bibfnamefont {A.}~\bibnamefont {Kapitulnik}},\ }\bibfield  {title}
  {\bibinfo {title} {Anomalous spectral weight transfer at the superconducting
  transition of
  {${\mathrm{Bi}}_{2}$${\mathrm{Sr}}_{2}$${\mathrm{CaCu}}_{2}$${\mathrm{O}}_{8+\mathrm{\ensuremath{\delta}}}$}},\
  }\href {https://doi.org/10.1103/PhysRevLett.66.2160} {\bibfield  {journal}
  {\bibinfo  {journal} {Phys. Rev. Lett.}\ }\textbf {\bibinfo {volume} {66}},\
  \bibinfo {pages} {2160} (\bibinfo {year} {1991})}\BibitemShut {NoStop}%
\bibitem [{\citenamefont {Campuzano}\ \emph {et~al.}(1996)\citenamefont
  {Campuzano}, \citenamefont {Ding}, \citenamefont {Norman}, \citenamefont
  {Randeira}, \citenamefont {Bellman}, \citenamefont {Yokoya}, \citenamefont
  {Takahashi}, \citenamefont {Katayama-Yoshida}, \citenamefont {Mochiku},\ and\
  \citenamefont {Kadowaki}}]{Campuzano-1996}%
  \BibitemOpen
  \bibfield  {author} {\bibinfo {author} {\bibfnamefont {J.~C.}\ \bibnamefont
  {Campuzano}}, \bibinfo {author} {\bibfnamefont {H.}~\bibnamefont {Ding}},
  \bibinfo {author} {\bibfnamefont {M.~R.}\ \bibnamefont {Norman}}, \bibinfo
  {author} {\bibfnamefont {M.}~\bibnamefont {Randeira}}, \bibinfo {author}
  {\bibfnamefont {A.~F.}\ \bibnamefont {Bellman}}, \bibinfo {author}
  {\bibfnamefont {T.}~\bibnamefont {Yokoya}}, \bibinfo {author} {\bibfnamefont
  {T.}~\bibnamefont {Takahashi}}, \bibinfo {author} {\bibfnamefont
  {H.}~\bibnamefont {Katayama-Yoshida}}, \bibinfo {author} {\bibfnamefont
  {T.}~\bibnamefont {Mochiku}},\ and\ \bibinfo {author} {\bibfnamefont
  {K.}~\bibnamefont {Kadowaki}},\ }\bibfield  {title} {\bibinfo {title} {Direct
  observation of particle-hole mixing in the superconducting state by
  angle-resolved photoemission},\ }\href
  {https://doi.org/10.1103/PhysRevB.53.R14737} {\bibfield  {journal} {\bibinfo
  {journal} {Phys. Rev. B}\ }\textbf {\bibinfo {volume} {53}},\ \bibinfo
  {pages} {R14737} (\bibinfo {year} {1996})}\BibitemShut {NoStop}%
\bibitem [{\citenamefont {Loeser}\ \emph {et~al.}(1997)\citenamefont {Loeser},
  \citenamefont {Shen}, \citenamefont {Schabel}, \citenamefont {Kim},
  \citenamefont {Zhang}, \citenamefont {Kapitulnik},\ and\ \citenamefont
  {Fournier}}]{Loeser-1997}%
  \BibitemOpen
  \bibfield  {author} {\bibinfo {author} {\bibfnamefont {A.~G.}\ \bibnamefont
  {Loeser}}, \bibinfo {author} {\bibfnamefont {Z.-X.}\ \bibnamefont {Shen}},
  \bibinfo {author} {\bibfnamefont {M.~C.}\ \bibnamefont {Schabel}}, \bibinfo
  {author} {\bibfnamefont {C.}~\bibnamefont {Kim}}, \bibinfo {author}
  {\bibfnamefont {M.}~\bibnamefont {Zhang}}, \bibinfo {author} {\bibfnamefont
  {A.}~\bibnamefont {Kapitulnik}},\ and\ \bibinfo {author} {\bibfnamefont
  {P.}~\bibnamefont {Fournier}},\ }\bibfield  {title} {\bibinfo {title}
  {Temperature and doping dependence of the {Bi-Sr-Ca-Cu-O} electronic
  structure and fluctuation effects},\ }\href
  {https://doi.org/10.1103/PhysRevB.56.14185} {\bibfield  {journal} {\bibinfo
  {journal} {Phys. Rev. B}\ }\textbf {\bibinfo {volume} {56}},\ \bibinfo
  {pages} {14185} (\bibinfo {year} {1997})}\BibitemShut {NoStop}%
\bibitem [{\citenamefont {Shen}\ and\ \citenamefont
  {Schrieffer}(1997)}]{Shen-1997}%
  \BibitemOpen
  \bibfield  {author} {\bibinfo {author} {\bibfnamefont {Z.-X.}\ \bibnamefont
  {Shen}}\ and\ \bibinfo {author} {\bibfnamefont {J.~R.}\ \bibnamefont
  {Schrieffer}},\ }\bibfield  {title} {\bibinfo {title} {Momentum, temperature,
  and doping dependence of photoemission lineshape and implications for the
  nature of the pairing potential in high- ${T}_{c}$ superconducting
  materials},\ }\href {https://doi.org/10.1103/PhysRevLett.78.1771} {\bibfield
  {journal} {\bibinfo  {journal} {Phys. Rev. Lett.}\ }\textbf {\bibinfo
  {volume} {78}},\ \bibinfo {pages} {1771} (\bibinfo {year}
  {1997})}\BibitemShut {NoStop}%
\bibitem [{\citenamefont {Fedorov}\ \emph {et~al.}(1999)\citenamefont
  {Fedorov}, \citenamefont {Valla}, \citenamefont {Johnson}, \citenamefont
  {Li}, \citenamefont {Gu},\ and\ \citenamefont {Koshizuka}}]{Fedorov-1999}%
  \BibitemOpen
  \bibfield  {author} {\bibinfo {author} {\bibfnamefont {A.~V.}\ \bibnamefont
  {Fedorov}}, \bibinfo {author} {\bibfnamefont {T.}~\bibnamefont {Valla}},
  \bibinfo {author} {\bibfnamefont {P.~D.}\ \bibnamefont {Johnson}}, \bibinfo
  {author} {\bibfnamefont {Q.}~\bibnamefont {Li}}, \bibinfo {author}
  {\bibfnamefont {G.~D.}\ \bibnamefont {Gu}},\ and\ \bibinfo {author}
  {\bibfnamefont {N.}~\bibnamefont {Koshizuka}},\ }\bibfield  {title} {\bibinfo
  {title} {Temperature dependent photoemission studies of optimally doped
  {${\mathrm{Bi}}_{2}{\mathrm{Sr}}_{2}{\mathrm{CaCu}}_{2}{\mathrm{O}}_{8}$}},\
  }\href {https://doi.org/10.1103/PhysRevLett.82.2179} {\bibfield  {journal}
  {\bibinfo  {journal} {Phys. Rev. Lett.}\ }\textbf {\bibinfo {volume} {82}},\
  \bibinfo {pages} {2179} (\bibinfo {year} {1999})}\BibitemShut {NoStop}%
\bibitem [{\citenamefont {Kaminski}\ \emph {et~al.}(2016)\citenamefont
  {Kaminski}, \citenamefont {Rosenkranz}, \citenamefont {Norman}, \citenamefont
  {Randeria}, \citenamefont {Li}, \citenamefont {Raffy},\ and\ \citenamefont
  {Campuzano}}]{Kaminski-2016}%
  \BibitemOpen
  \bibfield  {author} {\bibinfo {author} {\bibfnamefont {A.}~\bibnamefont
  {Kaminski}}, \bibinfo {author} {\bibfnamefont {S.}~\bibnamefont
  {Rosenkranz}}, \bibinfo {author} {\bibfnamefont {M.~R.}\ \bibnamefont
  {Norman}}, \bibinfo {author} {\bibfnamefont {M.}~\bibnamefont {Randeria}},
  \bibinfo {author} {\bibfnamefont {Z.~Z.}\ \bibnamefont {Li}}, \bibinfo
  {author} {\bibfnamefont {H.}~\bibnamefont {Raffy}},\ and\ \bibinfo {author}
  {\bibfnamefont {J.~C.}\ \bibnamefont {Campuzano}},\ }\bibfield  {title}
  {\bibinfo {title} {Destroying coherence in high-temperature superconductors
  with current flow},\ }\href {https://doi.org/10.1103/PhysRevX.6.031040}
  {\bibfield  {journal} {\bibinfo  {journal} {Phys. Rev. X}\ }\textbf {\bibinfo
  {volume} {6}},\ \bibinfo {pages} {031040} (\bibinfo {year}
  {2016})}\BibitemShut {NoStop}%
\bibitem [{\citenamefont {Orenstein}\ and\ \citenamefont
  {Millis}(2000)}]{Orenstein-2000}%
  \BibitemOpen
  \bibfield  {author} {\bibinfo {author} {\bibfnamefont {J.}~\bibnamefont
  {Orenstein}}\ and\ \bibinfo {author} {\bibfnamefont {A.}~\bibnamefont
  {Millis}},\ }\bibfield  {title} {\bibinfo {title} {Advances in the physics of
  high-temperature superconductivity},\ }\href
  {https://doi.org/10.1126/science.288.5465.468} {\bibfield  {journal}
  {\bibinfo  {journal} {Science}\ }\textbf {\bibinfo {volume} {288}},\ \bibinfo
  {pages} {468} (\bibinfo {year} {2000})}\BibitemShut {NoStop}%
\bibitem [{\citenamefont {Hashimoto}\ \emph {et~al.}(2014)\citenamefont
  {Hashimoto}, \citenamefont {Vishik}, \citenamefont {He}, \citenamefont
  {Devereaux},\ and\ \citenamefont {Shen}}]{Hashimoto-2014}%
  \BibitemOpen
  \bibfield  {author} {\bibinfo {author} {\bibfnamefont {M.}~\bibnamefont
  {Hashimoto}}, \bibinfo {author} {\bibfnamefont {I.~M.}\ \bibnamefont
  {Vishik}}, \bibinfo {author} {\bibfnamefont {R.-H.}\ \bibnamefont {He}},
  \bibinfo {author} {\bibfnamefont {T.~P.}\ \bibnamefont {Devereaux}},\ and\
  \bibinfo {author} {\bibfnamefont {Z.-X.}\ \bibnamefont {Shen}},\ }\bibfield
  {title} {\bibinfo {title} {Energy gaps in high-transition-temperature cuprate
  superconductors},\ }\href {https://www.nature.com/articles/nphys3009}
  {\bibfield  {journal} {\bibinfo  {journal} {Nat. Phys.}\ }\textbf {\bibinfo
  {volume} {10}},\ \bibinfo {pages} {483} (\bibinfo {year} {2014})}\BibitemShut
  {NoStop}%
\bibitem [{\citenamefont {Emery}\ and\ \citenamefont
  {Kivelson}(1995{\natexlab{a}})}]{Emery-1994}%
  \BibitemOpen
  \bibfield  {author} {\bibinfo {author} {\bibfnamefont {V.~J.}\ \bibnamefont
  {Emery}}\ and\ \bibinfo {author} {\bibfnamefont {S.~A.}\ \bibnamefont
  {Kivelson}},\ }\bibfield  {title} {\bibinfo {title} {Superconductivity in bad
  metals},\ }\href {https://doi.org/10.1103/PhysRevLett.74.3253} {\bibfield
  {journal} {\bibinfo  {journal} {Phys. Rev. Lett.}\ }\textbf {\bibinfo
  {volume} {74}},\ \bibinfo {pages} {3253} (\bibinfo {year}
  {1995}{\natexlab{a}})}\BibitemShut {NoStop}%
\bibitem [{\citenamefont {Hartnoll}(2014)}]{Hartnoll-2014b}%
  \BibitemOpen
  \bibfield  {author} {\bibinfo {author} {\bibfnamefont {S.~A.}\ \bibnamefont
  {Hartnoll}},\ }\bibfield  {title} {\bibinfo {title} {Theory of universal
  incoherent metallic transport},\ }\href
  {https://www.nature.com/articles/nphys3174?} {\bibfield  {journal} {\bibinfo
  {journal} {Nat. Phys.}\ }\textbf {\bibinfo {volume} {11}},\ \bibinfo {pages}
  {54} (\bibinfo {year} {2014})}\BibitemShut {NoStop}%
\bibitem [{\citenamefont {Zaanen}(2019)}]{Zaanen-2019}%
  \BibitemOpen
  \bibfield  {author} {\bibinfo {author} {\bibfnamefont {J.}~\bibnamefont
  {Zaanen}},\ }\bibfield  {title} {\bibinfo {title} {{Planckian dissipation,
  minimal viscosity and the transport in cuprate strange metals}},\ }\href
  {https://doi.org/10.21468/SciPostPhys.6.5.061} {\bibfield  {journal}
  {\bibinfo  {journal} {SciPost Phys.}\ }\textbf {\bibinfo {volume} {6}},\
  \bibinfo {pages} {61} (\bibinfo {year} {2019})}\BibitemShut {NoStop}%
\bibitem [{\citenamefont {Feng}\ \emph {et~al.}(2000)\citenamefont {Feng},
  \citenamefont {Lu}, \citenamefont {Shen}, \citenamefont {Kim}, \citenamefont
  {Eisaki}, \citenamefont {Damascelli}, \citenamefont {Yoshizaki},
  \citenamefont {Shimoyama}, \citenamefont {Kishio}, \citenamefont {Gu},
  \citenamefont {Oh}, \citenamefont {Andrus}, \citenamefont {J.}, \citenamefont
  {N.},\ and\ \citenamefont {Shen}}]{Feng-2000}%
  \BibitemOpen
  \bibfield  {author} {\bibinfo {author} {\bibfnamefont {D.}~\bibnamefont
  {Feng}}, \bibinfo {author} {\bibfnamefont {D.}~\bibnamefont {Lu}}, \bibinfo
  {author} {\bibfnamefont {K.}~\bibnamefont {Shen}}, \bibinfo {author}
  {\bibfnamefont {C.}~\bibnamefont {Kim}}, \bibinfo {author} {\bibfnamefont
  {H.}~\bibnamefont {Eisaki}}, \bibinfo {author} {\bibfnamefont
  {A.}~\bibnamefont {Damascelli}}, \bibinfo {author} {\bibfnamefont
  {R.}~\bibnamefont {Yoshizaki}}, \bibinfo {author} {\bibfnamefont {J.-i.}\
  \bibnamefont {Shimoyama}}, \bibinfo {author} {\bibfnamefont {K.}~\bibnamefont
  {Kishio}}, \bibinfo {author} {\bibfnamefont {G.}~\bibnamefont {Gu}}, \bibinfo
  {author} {\bibfnamefont {S.}~\bibnamefont {Oh}}, \bibinfo {author}
  {\bibfnamefont {A.}~\bibnamefont {Andrus}}, \bibinfo {author} {\bibfnamefont
  {O.}~\bibnamefont {J.}}, \bibinfo {author} {\bibfnamefont {E.~J.}\
  \bibnamefont {N.}},\ and\ \bibinfo {author} {\bibfnamefont {Z.-X.}\
  \bibnamefont {Shen}},\ }\bibfield  {title} {\bibinfo {title} {Signature of
  superfluid density in the single-particle excitation spectrum of
  {Bi$_2$Sr$_2$CaCu$_2$O$_{8+\delta}$}},\ }\href
  {https://www.science.org/doi/full/10.1126/science.289.5477.277} {\bibfield
  {journal} {\bibinfo  {journal} {Science}\ }\textbf {\bibinfo {volume}
  {289}},\ \bibinfo {pages} {277} (\bibinfo {year} {2000})}\BibitemShut
  {NoStop}%
\bibitem [{\citenamefont {Zaanen}(2021)}]{Zaanen-2021_preprint}%
  \BibitemOpen
  \bibfield  {author} {\bibinfo {author} {\bibfnamefont {J.}~\bibnamefont
  {Zaanen}},\ }\bibfield  {title} {\bibinfo {title} {Lectures on quantum
  supreme matter},\ }\href {https://arxiv.org/abs/2110.00961} {\bibfield
  {journal} {\bibinfo  {journal} {arXiv:2110.00961}\ } (\bibinfo {year}
  {2021})}\BibitemShut {NoStop}%
\bibitem [{\citenamefont {Sachdev}(2000{\natexlab{a}})}]{Sachdev-2000}%
  \BibitemOpen
  \bibfield  {author} {\bibinfo {author} {\bibfnamefont {S.}~\bibnamefont
  {Sachdev}},\ }\bibfield  {title} {\bibinfo {title} {Quantum criticality:
  competing ground states in low dimensions},\ }\href
  {https://doi.org/10.1126/science.288.5465.475} {\bibfield  {journal}
  {\bibinfo  {journal} {Science}\ }\textbf {\bibinfo {volume} {288}},\ \bibinfo
  {pages} {475} (\bibinfo {year} {2000}{\natexlab{a}})}\BibitemShut {NoStop}%
\bibitem [{\citenamefont {Sachdev}(2000{\natexlab{b}})}]{Sachdev-2000QPT}%
  \BibitemOpen
  \bibfield  {author} {\bibinfo {author} {\bibfnamefont {S.}~\bibnamefont
  {Sachdev}},\ }\href {https://doi.org/10.1017/CBO9780511622540} {\emph
  {\bibinfo {title} {Quantum Phase Transitions}}}\ (\bibinfo  {publisher}
  {Cambridge University Press},\ \bibinfo {address} {Cambridge},\ \bibinfo
  {year} {2000})\BibitemShut {NoStop}%
\bibitem [{\citenamefont {Coleman}\ and\ \citenamefont
  {Schofield}(2005)}]{Coleman-2005}%
  \BibitemOpen
  \bibfield  {author} {\bibinfo {author} {\bibfnamefont {P.}~\bibnamefont
  {Coleman}}\ and\ \bibinfo {author} {\bibfnamefont {A.~J.}\ \bibnamefont
  {Schofield}},\ }\bibfield  {title} {\bibinfo {title} {Quantum criticality},\
  }\href
  {http://www.nature.com/nature/journal/v433/n7023/full/nature03279.html?foxtrotcallback=true}
  {\bibfield  {journal} {\bibinfo  {journal} {Nature}\ }\textbf {\bibinfo
  {volume} {433}},\ \bibinfo {pages} {226} (\bibinfo {year}
  {2005})}\BibitemShut {NoStop}%
\bibitem [{\citenamefont {Hartnoll}\ \emph {et~al.}(2018)\citenamefont
  {Hartnoll}, \citenamefont {Lucas},\ and\ \citenamefont
  {Sachdev}}]{Hartnoll-2018holo}%
  \BibitemOpen
  \bibfield  {author} {\bibinfo {author} {\bibfnamefont {S.~A.}\ \bibnamefont
  {Hartnoll}}, \bibinfo {author} {\bibfnamefont {A.}~\bibnamefont {Lucas}},\
  and\ \bibinfo {author} {\bibfnamefont {S.}~\bibnamefont {Sachdev}},\
  }\href@noop {} {\emph {\bibinfo {title} {Holographic quantum matter}}}\
  (\bibinfo  {publisher} {MIT press},\ \bibinfo {address} {Cambridge,
  Massachussets},\ \bibinfo {year} {2018})\BibitemShut {NoStop}%
\bibitem [{\citenamefont {Hartnoll}\ and\ \citenamefont
  {Mackenzie}(2022)}]{Hartnoll-2022}%
  \BibitemOpen
  \bibfield  {author} {\bibinfo {author} {\bibfnamefont {S.~A.}\ \bibnamefont
  {Hartnoll}}\ and\ \bibinfo {author} {\bibfnamefont {A.~P.}\ \bibnamefont
  {Mackenzie}},\ }\bibfield  {title} {\bibinfo {title} {Colloquium: {Planckian}
  dissipation in metals},\ }\href
  {https://doi.org/10.1103/RevModPhys.94.041002} {\bibfield  {journal}
  {\bibinfo  {journal} {Rev. Mod. Phys.}\ }\textbf {\bibinfo {volume} {94}},\
  \bibinfo {pages} {041002} (\bibinfo {year} {2022})}\BibitemShut {NoStop}%
\bibitem [{\citenamefont {Damascelli}\ \emph {et~al.}(2003)\citenamefont
  {Damascelli}, \citenamefont {Hussain},\ and\ \citenamefont
  {Shen}}]{Damascelli-2003}%
  \BibitemOpen
  \bibfield  {author} {\bibinfo {author} {\bibfnamefont {A.}~\bibnamefont
  {Damascelli}}, \bibinfo {author} {\bibfnamefont {Z.}~\bibnamefont
  {Hussain}},\ and\ \bibinfo {author} {\bibfnamefont {Z.-X.}\ \bibnamefont
  {Shen}},\ }\bibfield  {title} {\bibinfo {title} {Angle-resolved photoemission
  studies of the cuprate superconductors},\ }\href
  {https://doi.org/10.1103/RevModPhys.75.473} {\bibfield  {journal} {\bibinfo
  {journal} {Rev. Mod. Phys.}\ }\textbf {\bibinfo {volume} {75}},\ \bibinfo
  {pages} {473} (\bibinfo {year} {2003})}\BibitemShut {NoStop}%
\bibitem [{\citenamefont {Wang}\ \emph {et~al.}(2004)\citenamefont {Wang},
  \citenamefont {Yang}, \citenamefont {Sekharan}, \citenamefont {Ding},
  \citenamefont {Engelbrecht}, \citenamefont {Dai}, \citenamefont {Wang},
  \citenamefont {Kaminski}, \citenamefont {Valla}, \citenamefont {Kidd},
  \citenamefont {Fedorov},\ and\ \citenamefont {Johnson}}]{Wang-2004}%
  \BibitemOpen
  \bibfield  {author} {\bibinfo {author} {\bibfnamefont {S.-C.}\ \bibnamefont
  {Wang}}, \bibinfo {author} {\bibfnamefont {H.-B.}\ \bibnamefont {Yang}},
  \bibinfo {author} {\bibfnamefont {A.~K.~P.}\ \bibnamefont {Sekharan}},
  \bibinfo {author} {\bibfnamefont {H.}~\bibnamefont {Ding}}, \bibinfo {author}
  {\bibfnamefont {J.~R.}\ \bibnamefont {Engelbrecht}}, \bibinfo {author}
  {\bibfnamefont {X.}~\bibnamefont {Dai}}, \bibinfo {author} {\bibfnamefont
  {Z.}~\bibnamefont {Wang}}, \bibinfo {author} {\bibfnamefont {A.}~\bibnamefont
  {Kaminski}}, \bibinfo {author} {\bibfnamefont {T.}~\bibnamefont {Valla}},
  \bibinfo {author} {\bibfnamefont {T.}~\bibnamefont {Kidd}}, \bibinfo {author}
  {\bibfnamefont {A.~V.}\ \bibnamefont {Fedorov}},\ and\ \bibinfo {author}
  {\bibfnamefont {P.~D.}\ \bibnamefont {Johnson}},\ }\bibfield  {title}
  {\bibinfo {title} {Quasiparticle line shape of
  {${\mathrm{Sr}}_{2}{\mathrm{RuO}}_{4}$} and its relation to anisotropic
  transport},\ }\href {https://doi.org/10.1103/PhysRevLett.92.137002}
  {\bibfield  {journal} {\bibinfo  {journal} {Phys. Rev. Lett.}\ }\textbf
  {\bibinfo {volume} {92}},\ \bibinfo {pages} {137002} (\bibinfo {year}
  {2004})}\BibitemShut {NoStop}%
\bibitem [{\citenamefont {Sheehy}\ and\ \citenamefont
  {Schmalian}(2007)}]{Sheehy-2007}%
  \BibitemOpen
  \bibfield  {author} {\bibinfo {author} {\bibfnamefont {D.~E.}\ \bibnamefont
  {Sheehy}}\ and\ \bibinfo {author} {\bibfnamefont {J.}~\bibnamefont
  {Schmalian}},\ }\bibfield  {title} {\bibinfo {title} {Quantum critical
  scaling in graphene},\ }\href {https://doi.org/10.1103/PhysRevLett.99.226803}
  {\bibfield  {journal} {\bibinfo  {journal} {Phys. Rev. Lett.}\ }\textbf
  {\bibinfo {volume} {99}},\ \bibinfo {pages} {226803} (\bibinfo {year}
  {2007})}\BibitemShut {NoStop}%
\bibitem [{\citenamefont {Martin}\ \emph {et~al.}(2008)\citenamefont {Martin},
  \citenamefont {Akerman}, \citenamefont {Ulbricht}, \citenamefont {Lohmann},
  \citenamefont {Smet}, \citenamefont {Von~Klitzing},\ and\ \citenamefont
  {Yacoby}}]{Martin-2008}%
  \BibitemOpen
  \bibfield  {author} {\bibinfo {author} {\bibfnamefont {J.}~\bibnamefont
  {Martin}}, \bibinfo {author} {\bibfnamefont {N.}~\bibnamefont {Akerman}},
  \bibinfo {author} {\bibfnamefont {G.}~\bibnamefont {Ulbricht}}, \bibinfo
  {author} {\bibfnamefont {T.}~\bibnamefont {Lohmann}}, \bibinfo {author}
  {\bibfnamefont {J.~v.}\ \bibnamefont {Smet}}, \bibinfo {author}
  {\bibfnamefont {K.}~\bibnamefont {Von~Klitzing}},\ and\ \bibinfo {author}
  {\bibfnamefont {A.}~\bibnamefont {Yacoby}},\ }\bibfield  {title} {\bibinfo
  {title} {Observation of electron--hole puddles in graphene using a scanning
  single-electron transistor},\ }\href {https://doi.org/10.1038/nphys781}
  {\bibfield  {journal} {\bibinfo  {journal} {Nat. Phys.}\ }\textbf {\bibinfo
  {volume} {4}},\ \bibinfo {pages} {144} (\bibinfo {year} {2008})}\BibitemShut
  {NoStop}%
\bibitem [{\citenamefont {Crossno}\ \emph {et~al.}(2016)\citenamefont
  {Crossno}, \citenamefont {Shi}, \citenamefont {Wang}, \citenamefont {Liu},
  \citenamefont {Harzheim}, \citenamefont {Lucas}, \citenamefont {Sachdev},
  \citenamefont {Kim}, \citenamefont {Taniguchi}, \citenamefont {Watanabe},
  \citenamefont {Ohki},\ and\ \citenamefont {Fong}}]{Crossno-2016}%
  \BibitemOpen
  \bibfield  {author} {\bibinfo {author} {\bibfnamefont {J.}~\bibnamefont
  {Crossno}}, \bibinfo {author} {\bibfnamefont {J.~K.}\ \bibnamefont {Shi}},
  \bibinfo {author} {\bibfnamefont {K.}~\bibnamefont {Wang}}, \bibinfo {author}
  {\bibfnamefont {X.}~\bibnamefont {Liu}}, \bibinfo {author} {\bibfnamefont
  {A.}~\bibnamefont {Harzheim}}, \bibinfo {author} {\bibfnamefont
  {A.}~\bibnamefont {Lucas}}, \bibinfo {author} {\bibfnamefont
  {S.}~\bibnamefont {Sachdev}}, \bibinfo {author} {\bibfnamefont
  {P.}~\bibnamefont {Kim}}, \bibinfo {author} {\bibfnamefont {T.}~\bibnamefont
  {Taniguchi}}, \bibinfo {author} {\bibfnamefont {K.}~\bibnamefont {Watanabe}},
  \bibinfo {author} {\bibfnamefont {T.~A.}\ \bibnamefont {Ohki}},\ and\
  \bibinfo {author} {\bibfnamefont {K.~C.}\ \bibnamefont {Fong}},\ }\bibfield
  {title} {\bibinfo {title} {Observation of the {Dirac} fluid and the breakdown
  of the {Wiedemann}-{Franz} law in graphene},\ }\href
  {http://science.sciencemag.org/content/351/6277/1058/tab-pdf} {\bibfield
  {journal} {\bibinfo  {journal} {Science}\ }\textbf {\bibinfo {volume}
  {351}},\ \bibinfo {pages} {1058} (\bibinfo {year} {2016})}\BibitemShut
  {NoStop}%
\bibitem [{\citenamefont {Gallagher}\ \emph {et~al.}(2019)\citenamefont
  {Gallagher}, \citenamefont {Yang}, \citenamefont {Lyu}, \citenamefont {Tian},
  \citenamefont {Kou}, \citenamefont {Zhang}, \citenamefont {Watanabe},
  \citenamefont {Taniguchi},\ and\ \citenamefont {Wang}}]{Gallagher-2019}%
  \BibitemOpen
  \bibfield  {author} {\bibinfo {author} {\bibfnamefont {P.}~\bibnamefont
  {Gallagher}}, \bibinfo {author} {\bibfnamefont {C.-S.}\ \bibnamefont {Yang}},
  \bibinfo {author} {\bibfnamefont {T.}~\bibnamefont {Lyu}}, \bibinfo {author}
  {\bibfnamefont {F.}~\bibnamefont {Tian}}, \bibinfo {author} {\bibfnamefont
  {R.}~\bibnamefont {Kou}}, \bibinfo {author} {\bibfnamefont {H.}~\bibnamefont
  {Zhang}}, \bibinfo {author} {\bibfnamefont {K.}~\bibnamefont {Watanabe}},
  \bibinfo {author} {\bibfnamefont {T.}~\bibnamefont {Taniguchi}},\ and\
  \bibinfo {author} {\bibfnamefont {F.}~\bibnamefont {Wang}},\ }\bibfield
  {title} {\bibinfo {title} {Quantum-critical conductivity of the {Dirac} fluid
  in graphene},\ }\href {https://science.sciencemag.org/content/364/6436/158}
  {\bibfield  {journal} {\bibinfo  {journal} {Science}\ }\textbf {\bibinfo
  {volume} {364}},\ \bibinfo {pages} {158} (\bibinfo {year}
  {2019})}\BibitemShut {NoStop}%
\bibitem [{\citenamefont {Moll}\ \emph {et~al.}(2016)\citenamefont {Moll},
  \citenamefont {Kushwaha}, \citenamefont {Nandi}, \citenamefont {Schmidt},\
  and\ \citenamefont {Mackenzie}}]{Moll-2016}%
  \BibitemOpen
  \bibfield  {author} {\bibinfo {author} {\bibfnamefont {P.~J.}\ \bibnamefont
  {Moll}}, \bibinfo {author} {\bibfnamefont {P.}~\bibnamefont {Kushwaha}},
  \bibinfo {author} {\bibfnamefont {N.}~\bibnamefont {Nandi}}, \bibinfo
  {author} {\bibfnamefont {B.}~\bibnamefont {Schmidt}},\ and\ \bibinfo {author}
  {\bibfnamefont {A.~P.}\ \bibnamefont {Mackenzie}},\ }\bibfield  {title}
  {\bibinfo {title} {Evidence for hydrodynamic electron flow in {PdCoO$_2$}},\
  }\href {http://science.sciencemag.org/content/351/6277/1061/tab-pdf}
  {\bibfield  {journal} {\bibinfo  {journal} {Science}\ }\textbf {\bibinfo
  {volume} {351}},\ \bibinfo {pages} {1061} (\bibinfo {year}
  {2016})}\BibitemShut {NoStop}%
\bibitem [{\citenamefont {Mackenzie}(2017)}]{Mackenzie-2017}%
  \BibitemOpen
  \bibfield  {author} {\bibinfo {author} {\bibfnamefont {A.~P.}\ \bibnamefont
  {Mackenzie}},\ }\bibfield  {title} {\bibinfo {title} {The properties of
  ultrapure delafossite metals},\ }\href
  {https://doi.org/10.1088/1361-6633/aa50e5} {\bibfield  {journal} {\bibinfo
  {journal} {Rep. Prog. Phys.}\ }\textbf {\bibinfo {volume} {80}},\ \bibinfo
  {pages} {032501} (\bibinfo {year} {2017})}\BibitemShut {NoStop}%
\bibitem [{\citenamefont {Nandi}\ \emph {et~al.}(2018)\citenamefont {Nandi},
  \citenamefont {Scaffidi}, \citenamefont {Kushwaha}, \citenamefont {Khim},
  \citenamefont {Barber}, \citenamefont {Sunko}, \citenamefont {Mazzola},
  \citenamefont {King}, \citenamefont {Rosner}, \citenamefont {Moll},
  \citenamefont {K\"{o}nig}, \citenamefont {Moore}, \citenamefont {Hartnoll},\
  and\ \citenamefont {Mackenzie}}]{Nandi-2018}%
  \BibitemOpen
  \bibfield  {author} {\bibinfo {author} {\bibfnamefont {N.}~\bibnamefont
  {Nandi}}, \bibinfo {author} {\bibfnamefont {T.}~\bibnamefont {Scaffidi}},
  \bibinfo {author} {\bibfnamefont {P.}~\bibnamefont {Kushwaha}}, \bibinfo
  {author} {\bibfnamefont {S.}~\bibnamefont {Khim}}, \bibinfo {author}
  {\bibfnamefont {M.~E.}\ \bibnamefont {Barber}}, \bibinfo {author}
  {\bibfnamefont {V.}~\bibnamefont {Sunko}}, \bibinfo {author} {\bibfnamefont
  {F.}~\bibnamefont {Mazzola}}, \bibinfo {author} {\bibfnamefont {P.~D.}\
  \bibnamefont {King}}, \bibinfo {author} {\bibfnamefont {H.}~\bibnamefont
  {Rosner}}, \bibinfo {author} {\bibfnamefont {P.~J.}\ \bibnamefont {Moll}},
  \bibinfo {author} {\bibfnamefont {M.}~\bibnamefont {K\"{o}nig}}, \bibinfo
  {author} {\bibfnamefont {J.}~\bibnamefont {Moore}}, \bibinfo {author}
  {\bibfnamefont {S.}~\bibnamefont {Hartnoll}},\ and\ \bibinfo {author}
  {\bibfnamefont {A.}~\bibnamefont {Mackenzie}},\ }\bibfield  {title} {\bibinfo
  {title} {Unconventional magneto-transport in ultrapure {PdCoO$_2$} and
  {PtCoO$_2$}},\ }\href {https://www.nature.com/articles/s41535-018-0138-8}
  {\bibfield  {journal} {\bibinfo  {journal} {npj Quantum Mater.}\ }\textbf
  {\bibinfo {volume} {3}},\ \bibinfo {pages} {66} (\bibinfo {year}
  {2018})}\BibitemShut {NoStop}%
\bibitem [{\citenamefont {L\"ohneysen}\ \emph {et~al.}(1994)\citenamefont
  {L\"ohneysen}, \citenamefont {Pietrus}, \citenamefont {Portisch},
  \citenamefont {Schlager}, \citenamefont {Schr\"oder}, \citenamefont {Sieck},\
  and\ \citenamefont {Trappmann}}]{Lohneysen-1994}%
  \BibitemOpen
  \bibfield  {author} {\bibinfo {author} {\bibfnamefont {H.~v.}\ \bibnamefont
  {L\"ohneysen}}, \bibinfo {author} {\bibfnamefont {T.}~\bibnamefont
  {Pietrus}}, \bibinfo {author} {\bibfnamefont {G.}~\bibnamefont {Portisch}},
  \bibinfo {author} {\bibfnamefont {H.~G.}\ \bibnamefont {Schlager}}, \bibinfo
  {author} {\bibfnamefont {A.}~\bibnamefont {Schr\"oder}}, \bibinfo {author}
  {\bibfnamefont {M.}~\bibnamefont {Sieck}},\ and\ \bibinfo {author}
  {\bibfnamefont {T.}~\bibnamefont {Trappmann}},\ }\bibfield  {title} {\bibinfo
  {title} {Non-{Fermi}-liquid behavior in a heavy-fermion alloy at a magnetic
  instability},\ }\href {https://doi.org/10.1103/PhysRevLett.72.3262}
  {\bibfield  {journal} {\bibinfo  {journal} {Phys. Rev. Lett.}\ }\textbf
  {\bibinfo {volume} {72}},\ \bibinfo {pages} {3262} (\bibinfo {year}
  {1994})}\BibitemShut {NoStop}%
\bibitem [{\citenamefont {Trovarelli}\ \emph {et~al.}(2000)\citenamefont
  {Trovarelli}, \citenamefont {Geibel}, \citenamefont {Mederle}, \citenamefont
  {Langhammer}, \citenamefont {Grosche}, \citenamefont {Gegenwart},
  \citenamefont {Lang}, \citenamefont {Sparn},\ and\ \citenamefont
  {Steglich}}]{Trovarelli-2000}%
  \BibitemOpen
  \bibfield  {author} {\bibinfo {author} {\bibfnamefont {O.}~\bibnamefont
  {Trovarelli}}, \bibinfo {author} {\bibfnamefont {C.}~\bibnamefont {Geibel}},
  \bibinfo {author} {\bibfnamefont {S.}~\bibnamefont {Mederle}}, \bibinfo
  {author} {\bibfnamefont {C.}~\bibnamefont {Langhammer}}, \bibinfo {author}
  {\bibfnamefont {F.~M.}\ \bibnamefont {Grosche}}, \bibinfo {author}
  {\bibfnamefont {P.}~\bibnamefont {Gegenwart}}, \bibinfo {author}
  {\bibfnamefont {M.}~\bibnamefont {Lang}}, \bibinfo {author} {\bibfnamefont
  {G.}~\bibnamefont {Sparn}},\ and\ \bibinfo {author} {\bibfnamefont
  {F.}~\bibnamefont {Steglich}},\ }\bibfield  {title} {\bibinfo {title}
  {{${\mathrm{YbRh}}_{2}{\mathrm{Si}}_{2}$}: Pronounced non-{Fermi}-liquid
  effects above a low-lying magnetic phase transition},\ }\href
  {https://doi.org/10.1103/PhysRevLett.85.626} {\bibfield  {journal} {\bibinfo
  {journal} {Phys. Rev. Lett.}\ }\textbf {\bibinfo {volume} {85}},\ \bibinfo
  {pages} {626} (\bibinfo {year} {2000})}\BibitemShut {NoStop}%
\bibitem [{\citenamefont {Schr{\"o}der}\ \emph {et~al.}(2000)\citenamefont
  {Schr{\"o}der}, \citenamefont {Aeppli}, \citenamefont {Coldea}, \citenamefont
  {Adams}, \citenamefont {Stockert}, \citenamefont {L{\"o}hneysen},
  \citenamefont {Bucher}, \citenamefont {Ramazashvili},\ and\ \citenamefont
  {Coleman}}]{Schroder-2000}%
  \BibitemOpen
  \bibfield  {author} {\bibinfo {author} {\bibfnamefont {A.}~\bibnamefont
  {Schr{\"o}der}}, \bibinfo {author} {\bibfnamefont {G.}~\bibnamefont
  {Aeppli}}, \bibinfo {author} {\bibfnamefont {R.}~\bibnamefont {Coldea}},
  \bibinfo {author} {\bibfnamefont {M.}~\bibnamefont {Adams}}, \bibinfo
  {author} {\bibfnamefont {O.}~\bibnamefont {Stockert}}, \bibinfo {author}
  {\bibfnamefont {H.}~\bibnamefont {L{\"o}hneysen}}, \bibinfo {author}
  {\bibfnamefont {E.}~\bibnamefont {Bucher}}, \bibinfo {author} {\bibfnamefont
  {R.}~\bibnamefont {Ramazashvili}},\ and\ \bibinfo {author} {\bibfnamefont
  {P.}~\bibnamefont {Coleman}},\ }\bibfield  {title} {\bibinfo {title} {Onset
  of antiferromagnetism in heavy-fermion metals},\ }\href
  {https://doi.org/10.1038/35030039} {\bibfield  {journal} {\bibinfo  {journal}
  {Nature}\ }\textbf {\bibinfo {volume} {407}},\ \bibinfo {pages} {351}
  (\bibinfo {year} {2000})}\BibitemShut {NoStop}%
\bibitem [{\citenamefont {Si}\ \emph {et~al.}(2001)\citenamefont {Si},
  \citenamefont {Rabello}, \citenamefont {Ingersent},\ and\ \citenamefont
  {Smith}}]{Si-2001}%
  \BibitemOpen
  \bibfield  {author} {\bibinfo {author} {\bibfnamefont {Q.}~\bibnamefont
  {Si}}, \bibinfo {author} {\bibfnamefont {S.}~\bibnamefont {Rabello}},
  \bibinfo {author} {\bibfnamefont {K.}~\bibnamefont {Ingersent}},\ and\
  \bibinfo {author} {\bibfnamefont {J.~L.}\ \bibnamefont {Smith}},\ }\bibfield
  {title} {\bibinfo {title} {Locally critical quantum phase transitions in
  strongly correlated metals},\ }\href {https://doi.org/10.1038/35101507}
  {\bibfield  {journal} {\bibinfo  {journal} {Nature}\ }\textbf {\bibinfo
  {volume} {413}},\ \bibinfo {pages} {804} (\bibinfo {year}
  {2001})}\BibitemShut {NoStop}%
\bibitem [{\citenamefont {Stewart}(2001)}]{Stewart-2001}%
  \BibitemOpen
  \bibfield  {author} {\bibinfo {author} {\bibfnamefont {G.~R.}\ \bibnamefont
  {Stewart}},\ }\bibfield  {title} {\bibinfo {title} {Non-{Fermi}-liquid
  behavior in $d$- and $f$-electron metals},\ }\href
  {https://doi.org/10.1103/RevModPhys.73.797} {\bibfield  {journal} {\bibinfo
  {journal} {Rev. Mod. Phys.}\ }\textbf {\bibinfo {volume} {73}},\ \bibinfo
  {pages} {797} (\bibinfo {year} {2001})}\BibitemShut {NoStop}%
\bibitem [{\citenamefont {Paglione}\ \emph {et~al.}(2003)\citenamefont
  {Paglione}, \citenamefont {Tanatar}, \citenamefont {Hawthorn}, \citenamefont
  {Boaknin}, \citenamefont {Hill}, \citenamefont {Ronning}, \citenamefont
  {Sutherland}, \citenamefont {Taillefer}, \citenamefont {Petrovic},\ and\
  \citenamefont {Canfield}}]{Paglione-2003}%
  \BibitemOpen
  \bibfield  {author} {\bibinfo {author} {\bibfnamefont {J.}~\bibnamefont
  {Paglione}}, \bibinfo {author} {\bibfnamefont {M.~A.}\ \bibnamefont
  {Tanatar}}, \bibinfo {author} {\bibfnamefont {D.~G.}\ \bibnamefont
  {Hawthorn}}, \bibinfo {author} {\bibfnamefont {E.}~\bibnamefont {Boaknin}},
  \bibinfo {author} {\bibfnamefont {R.~W.}\ \bibnamefont {Hill}}, \bibinfo
  {author} {\bibfnamefont {F.}~\bibnamefont {Ronning}}, \bibinfo {author}
  {\bibfnamefont {M.}~\bibnamefont {Sutherland}}, \bibinfo {author}
  {\bibfnamefont {L.}~\bibnamefont {Taillefer}}, \bibinfo {author}
  {\bibfnamefont {C.}~\bibnamefont {Petrovic}},\ and\ \bibinfo {author}
  {\bibfnamefont {P.~C.}\ \bibnamefont {Canfield}},\ }\bibfield  {title}
  {\bibinfo {title} {Field-induced quantum critical point in
  {${\mathrm{C}\mathrm{e}\mathrm{C}\mathrm{o}\mathrm{I}\mathrm{n}}_{5}$}},\
  }\href {https://doi.org/10.1103/PhysRevLett.91.246405} {\bibfield  {journal}
  {\bibinfo  {journal} {Phys. Rev. Lett.}\ }\textbf {\bibinfo {volume} {91}},\
  \bibinfo {pages} {246405} (\bibinfo {year} {2003})}\BibitemShut {NoStop}%
\bibitem [{\citenamefont {Tokiwa}\ \emph {et~al.}(2013)\citenamefont {Tokiwa},
  \citenamefont {Bauer},\ and\ \citenamefont {Gegenwart}}]{Tokiwa-2013}%
  \BibitemOpen
  \bibfield  {author} {\bibinfo {author} {\bibfnamefont {Y.}~\bibnamefont
  {Tokiwa}}, \bibinfo {author} {\bibfnamefont {E.~D.}\ \bibnamefont {Bauer}},\
  and\ \bibinfo {author} {\bibfnamefont {P.}~\bibnamefont {Gegenwart}},\
  }\bibfield  {title} {\bibinfo {title} {Zero-field quantum critical point in
  {${\mathrm{CeCoIn}}_{5}$}},\ }\href
  {https://doi.org/10.1103/PhysRevLett.111.107003} {\bibfield  {journal}
  {\bibinfo  {journal} {Phys. Rev. Lett.}\ }\textbf {\bibinfo {volume} {111}},\
  \bibinfo {pages} {107003} (\bibinfo {year} {2013})}\BibitemShut {NoStop}%
\bibitem [{\citenamefont {Grbi{\'c}}\ \emph {et~al.}(2022)\citenamefont
  {Grbi{\'c}}, \citenamefont {O’Farrell}, \citenamefont {Matsumoto},
  \citenamefont {Kuga}, \citenamefont {Brando}, \citenamefont {K{\"u}chler},
  \citenamefont {Nevidomskyy}, \citenamefont {Yoshida}, \citenamefont
  {Sakakibara}, \citenamefont {Kono}, \citenamefont {Shimura}, \citenamefont
  {Sutherland}, \citenamefont {Takigawa},\ and\ \citenamefont
  {Nakatsuji}}]{Grbic-2022}%
  \BibitemOpen
  \bibfield  {author} {\bibinfo {author} {\bibfnamefont {M.~S.}\ \bibnamefont
  {Grbi{\'c}}}, \bibinfo {author} {\bibfnamefont {E.~C.}\ \bibnamefont
  {O’Farrell}}, \bibinfo {author} {\bibfnamefont {Y.}~\bibnamefont
  {Matsumoto}}, \bibinfo {author} {\bibfnamefont {K.}~\bibnamefont {Kuga}},
  \bibinfo {author} {\bibfnamefont {M.}~\bibnamefont {Brando}}, \bibinfo
  {author} {\bibfnamefont {R.}~\bibnamefont {K{\"u}chler}}, \bibinfo {author}
  {\bibfnamefont {A.~H.}\ \bibnamefont {Nevidomskyy}}, \bibinfo {author}
  {\bibfnamefont {M.}~\bibnamefont {Yoshida}}, \bibinfo {author} {\bibfnamefont
  {T.}~\bibnamefont {Sakakibara}}, \bibinfo {author} {\bibfnamefont
  {Y.}~\bibnamefont {Kono}}, \bibinfo {author} {\bibfnamefont {Y.}~\bibnamefont
  {Shimura}}, \bibinfo {author} {\bibfnamefont {M.~L.}\ \bibnamefont
  {Sutherland}}, \bibinfo {author} {\bibfnamefont {M.}~\bibnamefont
  {Takigawa}},\ and\ \bibinfo {author} {\bibfnamefont {S.}~\bibnamefont
  {Nakatsuji}},\ }\bibfield  {title} {\bibinfo {title} {Anisotropy-driven
  quantum criticality in an intermediate valence system},\ }\href
  {https://doi.org/10.1038/s41467-022-29757-9} {\bibfield  {journal} {\bibinfo
  {journal} {Nat. Comm.}\ }\textbf {\bibinfo {volume} {13}},\ \bibinfo {pages}
  {1} (\bibinfo {year} {2022})}\BibitemShut {NoStop}%
\bibitem [{\citenamefont {Gooth}\ \emph {et~al.}(2018)\citenamefont {Gooth},
  \citenamefont {Menges}, \citenamefont {Kumar}, \citenamefont {S{\"u}$\beta$},
  \citenamefont {Shekhar}, \citenamefont {Sun}, \citenamefont {Drechsler},
  \citenamefont {Zierold}, \citenamefont {Felser},\ and\ \citenamefont
  {Gotsmann}}]{Gooth-2018}%
  \BibitemOpen
  \bibfield  {author} {\bibinfo {author} {\bibfnamefont {J.}~\bibnamefont
  {Gooth}}, \bibinfo {author} {\bibfnamefont {F.}~\bibnamefont {Menges}},
  \bibinfo {author} {\bibfnamefont {N.}~\bibnamefont {Kumar}}, \bibinfo
  {author} {\bibfnamefont {V.}~\bibnamefont {S{\"u}$\beta$}}, \bibinfo {author}
  {\bibfnamefont {C.}~\bibnamefont {Shekhar}}, \bibinfo {author} {\bibfnamefont
  {Y.}~\bibnamefont {Sun}}, \bibinfo {author} {\bibfnamefont {U.}~\bibnamefont
  {Drechsler}}, \bibinfo {author} {\bibfnamefont {R.}~\bibnamefont {Zierold}},
  \bibinfo {author} {\bibfnamefont {C.}~\bibnamefont {Felser}},\ and\ \bibinfo
  {author} {\bibfnamefont {B.}~\bibnamefont {Gotsmann}},\ }\bibfield  {title}
  {\bibinfo {title} {Thermal and electrical signatures of a hydrodynamic
  electron fluid in tungsten diphosphide},\ }\href
  {https://doi.org/10.1038/s41467-018-06688-y} {\bibfield  {journal} {\bibinfo
  {journal} {Nature Comm.}\ }\textbf {\bibinfo {volume} {9}},\ \bibinfo {pages}
  {1} (\bibinfo {year} {2018})}\BibitemShut {NoStop}%
\bibitem [{\citenamefont {Jaoui}\ \emph {et~al.}(2018)\citenamefont {Jaoui},
  \citenamefont {Fauqu{\'e}}, \citenamefont {Rischau}, \citenamefont {Subedi},
  \citenamefont {Fu}, \citenamefont {Gooth}, \citenamefont {Kumar},
  \citenamefont {S{\"u}{\ss}}, \citenamefont {Maslov}, \citenamefont {Felser},\
  and\ \citenamefont {Behnia}}]{Jaoui-2018}%
  \BibitemOpen
  \bibfield  {author} {\bibinfo {author} {\bibfnamefont {A.}~\bibnamefont
  {Jaoui}}, \bibinfo {author} {\bibfnamefont {B.}~\bibnamefont {Fauqu{\'e}}},
  \bibinfo {author} {\bibfnamefont {C.~W.}\ \bibnamefont {Rischau}}, \bibinfo
  {author} {\bibfnamefont {A.}~\bibnamefont {Subedi}}, \bibinfo {author}
  {\bibfnamefont {C.}~\bibnamefont {Fu}}, \bibinfo {author} {\bibfnamefont
  {J.}~\bibnamefont {Gooth}}, \bibinfo {author} {\bibfnamefont
  {N.}~\bibnamefont {Kumar}}, \bibinfo {author} {\bibfnamefont
  {V.}~\bibnamefont {S{\"u}{\ss}}}, \bibinfo {author} {\bibfnamefont {D.~L.}\
  \bibnamefont {Maslov}}, \bibinfo {author} {\bibfnamefont {C.}~\bibnamefont
  {Felser}},\ and\ \bibinfo {author} {\bibfnamefont {K.}~\bibnamefont
  {Behnia}},\ }\bibfield  {title} {\bibinfo {title} {Departure from the
  {Wiedemann}--{Franz} law in {WP}$_2$ driven by mismatch in {T}-square
  resistivity prefactors},\ }\href
  {https://www.nature.com/articles/s41535-018-0136-x} {\bibfield  {journal}
  {\bibinfo  {journal} {npj Quantum Mater.}\ }\textbf {\bibinfo {volume} {3}},\
  \bibinfo {pages} {64} (\bibinfo {year} {2018})}\BibitemShut {NoStop}%
\bibitem [{Note1()}]{Note1}%
  \BibitemOpen
  \bibinfo {note} {$\Theta (x)$ is the Heaviside step function}\BibitemShut
  {NoStop}%
\bibitem [{\citenamefont {Balatsky}(1993)}]{Balatsky-1993}%
  \BibitemOpen
  \bibfield  {author} {\bibinfo {author} {\bibfnamefont {A.~V.}\ \bibnamefont
  {Balatsky}},\ }\bibfield  {title} {\bibinfo {title} {Superconducting
  instability in a non-{Fermi} liquid: scaling approach},\ }\href
  {https://doi.org/10.1080/09500839308242421} {\bibfield  {journal} {\bibinfo
  {journal} {Philos. Mag. Lett.}\ }\textbf {\bibinfo {volume} {68}},\ \bibinfo
  {pages} {251} (\bibinfo {year} {1993})}\BibitemShut {NoStop}%
\bibitem [{\citenamefont {Sudb\o{}}(1995)}]{Sudbo-1995}%
  \BibitemOpen
  \bibfield  {author} {\bibinfo {author} {\bibfnamefont {A.}~\bibnamefont
  {Sudb\o{}}},\ }\bibfield  {title} {\bibinfo {title} {Pair susceptibilities
  and gap equations in non-{Fermi} liquids},\ }\href
  {https://doi.org/10.1103/PhysRevLett.74.2575} {\bibfield  {journal} {\bibinfo
   {journal} {Phys. Rev. Lett.}\ }\textbf {\bibinfo {volume} {74}},\ \bibinfo
  {pages} {2575} (\bibinfo {year} {1995})}\BibitemShut {NoStop}%
\bibitem [{\citenamefont {Muthukumar}\ \emph {et~al.}(1995)\citenamefont
  {Muthukumar}, \citenamefont {Sa},\ and\ \citenamefont
  {Sardar}}]{Muthukumar-1995}%
  \BibitemOpen
  \bibfield  {author} {\bibinfo {author} {\bibfnamefont {V.~N.}\ \bibnamefont
  {Muthukumar}}, \bibinfo {author} {\bibfnamefont {D.}~\bibnamefont {Sa}},\
  and\ \bibinfo {author} {\bibfnamefont {M.}~\bibnamefont {Sardar}},\
  }\bibfield  {title} {\bibinfo {title} {Superconductivity from a
  non-{Fermi}-liquid: A {Ginzburg-Landau} approach},\ }\href
  {https://doi.org/10.1103/PhysRevB.52.9647} {\bibfield  {journal} {\bibinfo
  {journal} {Phys. Rev. B}\ }\textbf {\bibinfo {volume} {52}},\ \bibinfo
  {pages} {9647} (\bibinfo {year} {1995})}\BibitemShut {NoStop}%
\bibitem [{\citenamefont {Yin}\ and\ \citenamefont
  {Chakravarty}(1996)}]{Yin-1996}%
  \BibitemOpen
  \bibfield  {author} {\bibinfo {author} {\bibfnamefont {L.}~\bibnamefont
  {Yin}}\ and\ \bibinfo {author} {\bibfnamefont {S.}~\bibnamefont
  {Chakravarty}},\ }\bibfield  {title} {\bibinfo {title} {Spectral anomaly and
  high temperature superconductors},\ }\href
  {https://doi.org/10.1142/S0217979296000349} {\bibfield  {journal} {\bibinfo
  {journal} {Int. J. Mod. Phys. B}\ }\textbf {\bibinfo {volume} {10}},\
  \bibinfo {pages} {805} (\bibinfo {year} {1996})}\BibitemShut {NoStop}%
\bibitem [{\citenamefont {Mathur}\ \emph {et~al.}(1998)\citenamefont {Mathur},
  \citenamefont {Grosche}, \citenamefont {Julian}, \citenamefont {Walker},
  \citenamefont {Freye}, \citenamefont {Haselwimmer},\ and\ \citenamefont
  {Lonzarich}}]{Mathur-1998}%
  \BibitemOpen
  \bibfield  {author} {\bibinfo {author} {\bibfnamefont {N.}~\bibnamefont
  {Mathur}}, \bibinfo {author} {\bibfnamefont {F.}~\bibnamefont {Grosche}},
  \bibinfo {author} {\bibfnamefont {S.}~\bibnamefont {Julian}}, \bibinfo
  {author} {\bibfnamefont {I.}~\bibnamefont {Walker}}, \bibinfo {author}
  {\bibfnamefont {D.}~\bibnamefont {Freye}}, \bibinfo {author} {\bibfnamefont
  {R.}~\bibnamefont {Haselwimmer}},\ and\ \bibinfo {author} {\bibfnamefont
  {G.}~\bibnamefont {Lonzarich}},\ }\bibfield  {title} {\bibinfo {title}
  {Magnetically mediated superconductivity in heavy fermion compounds},\ }\href
  {https://www.nature.com/articles/27838} {\bibfield  {journal} {\bibinfo
  {journal} {Nature}\ }\textbf {\bibinfo {volume} {394}},\ \bibinfo {pages}
  {39} (\bibinfo {year} {1998})}\BibitemShut {NoStop}%
\bibitem [{\citenamefont {Petrovic}\ \emph {et~al.}(2001)\citenamefont
  {Petrovic}, \citenamefont {Pagliuso}, \citenamefont {Hundley}, \citenamefont
  {Movshovich}, \citenamefont {Sarrao}, \citenamefont {Thompson}, \citenamefont
  {Fisk},\ and\ \citenamefont {Monthoux}}]{Petrovic-2001}%
  \BibitemOpen
  \bibfield  {author} {\bibinfo {author} {\bibfnamefont {C.}~\bibnamefont
  {Petrovic}}, \bibinfo {author} {\bibfnamefont {P.~G.}\ \bibnamefont
  {Pagliuso}}, \bibinfo {author} {\bibfnamefont {M.~F.}\ \bibnamefont
  {Hundley}}, \bibinfo {author} {\bibfnamefont {R.}~\bibnamefont {Movshovich}},
  \bibinfo {author} {\bibfnamefont {J.~L.}\ \bibnamefont {Sarrao}}, \bibinfo
  {author} {\bibfnamefont {J.~D.}\ \bibnamefont {Thompson}}, \bibinfo {author}
  {\bibfnamefont {Z.}~\bibnamefont {Fisk}},\ and\ \bibinfo {author}
  {\bibfnamefont {P.}~\bibnamefont {Monthoux}},\ }\bibfield  {title} {\bibinfo
  {title} {Heavy-fermion superconductivity in {CeCoIn$_5$} at 2.3 {K}},\ }\href
  {https://doi.org/10.1088/0953-8984/13/17/103} {\bibfield  {journal} {\bibinfo
   {journal} {J. Phys. Cond. Mat.}\ }\textbf {\bibinfo {volume} {13}},\
  \bibinfo {pages} {L337} (\bibinfo {year} {2001})}\BibitemShut {NoStop}%
\bibitem [{\citenamefont {Sidorov}\ \emph {et~al.}(2002)\citenamefont
  {Sidorov}, \citenamefont {Nicklas}, \citenamefont {Pagliuso}, \citenamefont
  {Sarrao}, \citenamefont {Bang}, \citenamefont {Balatsky},\ and\ \citenamefont
  {Thompson}}]{Sidorov-2002}%
  \BibitemOpen
  \bibfield  {author} {\bibinfo {author} {\bibfnamefont {V.~A.}\ \bibnamefont
  {Sidorov}}, \bibinfo {author} {\bibfnamefont {M.}~\bibnamefont {Nicklas}},
  \bibinfo {author} {\bibfnamefont {P.~G.}\ \bibnamefont {Pagliuso}}, \bibinfo
  {author} {\bibfnamefont {J.~L.}\ \bibnamefont {Sarrao}}, \bibinfo {author}
  {\bibfnamefont {Y.}~\bibnamefont {Bang}}, \bibinfo {author} {\bibfnamefont
  {A.~V.}\ \bibnamefont {Balatsky}},\ and\ \bibinfo {author} {\bibfnamefont
  {J.~D.}\ \bibnamefont {Thompson}},\ }\bibfield  {title} {\bibinfo {title}
  {Superconductivity and quantum criticality in
  $\mathrm{C}\mathrm{e}\mathrm{C}\mathrm{o}\mathrm{I}{\mathrm{n}}_{\mathrm{5}}$},\
  }\href {https://doi.org/10.1103/PhysRevLett.89.157004} {\bibfield  {journal}
  {\bibinfo  {journal} {Phys. Rev. Lett.}\ }\textbf {\bibinfo {volume} {89}},\
  \bibinfo {pages} {157004} (\bibinfo {year} {2002})}\BibitemShut {NoStop}%
\bibitem [{\citenamefont {Nakatsuji}\ \emph {et~al.}(2008)\citenamefont
  {Nakatsuji}, \citenamefont {Kuga}, \citenamefont {Machida}, \citenamefont
  {Tayama}, \citenamefont {Sakakibara}, \citenamefont {Karaki}, \citenamefont
  {Ishimoto}, \citenamefont {Yonezawa}, \citenamefont {Maeno}, \citenamefont
  {Pearson}, \citenamefont {Lonzarich}, \citenamefont {Balicas}, \citenamefont
  {Lee},\ and\ \citenamefont {Z.}}]{Nakatsuji-2008}%
  \BibitemOpen
  \bibfield  {author} {\bibinfo {author} {\bibfnamefont {S.}~\bibnamefont
  {Nakatsuji}}, \bibinfo {author} {\bibfnamefont {K.}~\bibnamefont {Kuga}},
  \bibinfo {author} {\bibfnamefont {Y.}~\bibnamefont {Machida}}, \bibinfo
  {author} {\bibfnamefont {T.}~\bibnamefont {Tayama}}, \bibinfo {author}
  {\bibfnamefont {T.}~\bibnamefont {Sakakibara}}, \bibinfo {author}
  {\bibfnamefont {Y.}~\bibnamefont {Karaki}}, \bibinfo {author} {\bibfnamefont
  {H.}~\bibnamefont {Ishimoto}}, \bibinfo {author} {\bibfnamefont
  {S.}~\bibnamefont {Yonezawa}}, \bibinfo {author} {\bibfnamefont
  {Y.}~\bibnamefont {Maeno}}, \bibinfo {author} {\bibfnamefont
  {E.}~\bibnamefont {Pearson}}, \bibinfo {author} {\bibfnamefont {G.~G.}\
  \bibnamefont {Lonzarich}}, \bibinfo {author} {\bibfnamefont {L.}~\bibnamefont
  {Balicas}}, \bibinfo {author} {\bibfnamefont {H.}~\bibnamefont {Lee}},\ and\
  \bibinfo {author} {\bibfnamefont {F.}~\bibnamefont {Z.}},\ }\bibfield
  {title} {\bibinfo {title} {Superconductivity and quantum criticality in the
  heavy-fermion system {$\beta$-YbAlB$_4$}},\ }\href
  {https://doi.org/10.1038/nphys1002} {\bibfield  {journal} {\bibinfo
  {journal} {Nat. phys.}\ }\textbf {\bibinfo {volume} {4}},\ \bibinfo {pages}
  {603} (\bibinfo {year} {2008})}\BibitemShut {NoStop}%
\bibitem [{\citenamefont {Knebel}\ \emph {et~al.}(2011)\citenamefont {Knebel},
  \citenamefont {Aoki},\ and\ \citenamefont {Flouquet}}]{Knebel-2011}%
  \BibitemOpen
  \bibfield  {author} {\bibinfo {author} {\bibfnamefont {G.}~\bibnamefont
  {Knebel}}, \bibinfo {author} {\bibfnamefont {D.}~\bibnamefont {Aoki}},\ and\
  \bibinfo {author} {\bibfnamefont {J.}~\bibnamefont {Flouquet}},\ }\bibfield
  {title} {\bibinfo {title} {Antiferromagnetism and superconductivity in cerium
  based heavy-fermion compounds},\ }\href
  {https://doi.org/https://doi.org/10.1016/j.crhy.2011.05.002} {\bibfield
  {journal} {\bibinfo  {journal} {C. R. Phys.}\ }\textbf {\bibinfo {volume}
  {12}},\ \bibinfo {pages} {542} (\bibinfo {year} {2011})}\BibitemShut
  {NoStop}%
\bibitem [{\citenamefont {Kasahara}\ \emph {et~al.}(2010)\citenamefont
  {Kasahara}, \citenamefont {Shibauchi}, \citenamefont {Hashimoto},
  \citenamefont {Ikada}, \citenamefont {Tonegawa}, \citenamefont {Okazaki},
  \citenamefont {Shishido}, \citenamefont {Ikeda}, \citenamefont {Takeya},
  \citenamefont {Hirata}, \citenamefont {Terashima},\ and\ \citenamefont
  {Matsuda}}]{Kasahara-2010}%
  \BibitemOpen
  \bibfield  {author} {\bibinfo {author} {\bibfnamefont {S.}~\bibnamefont
  {Kasahara}}, \bibinfo {author} {\bibfnamefont {T.}~\bibnamefont {Shibauchi}},
  \bibinfo {author} {\bibfnamefont {K.}~\bibnamefont {Hashimoto}}, \bibinfo
  {author} {\bibfnamefont {K.}~\bibnamefont {Ikada}}, \bibinfo {author}
  {\bibfnamefont {S.}~\bibnamefont {Tonegawa}}, \bibinfo {author}
  {\bibfnamefont {R.}~\bibnamefont {Okazaki}}, \bibinfo {author} {\bibfnamefont
  {H.}~\bibnamefont {Shishido}}, \bibinfo {author} {\bibfnamefont
  {H.}~\bibnamefont {Ikeda}}, \bibinfo {author} {\bibfnamefont
  {H.}~\bibnamefont {Takeya}}, \bibinfo {author} {\bibfnamefont
  {K.}~\bibnamefont {Hirata}}, \bibinfo {author} {\bibfnamefont
  {T.}~\bibnamefont {Terashima}},\ and\ \bibinfo {author} {\bibfnamefont
  {Y.}~\bibnamefont {Matsuda}},\ }\bibfield  {title} {\bibinfo {title}
  {Evolution from non-{Fermi}- to {Fermi}-liquid transport via isovalent doping
  in {${\text{BaFe}}_{2}{({\text{As}}_{1\ensuremath{-}x}{\text{P}}_{x})}_{2}$}
  superconductors},\ }\href {https://doi.org/10.1103/PhysRevB.81.184519}
  {\bibfield  {journal} {\bibinfo  {journal} {Phys. Rev. B}\ }\textbf {\bibinfo
  {volume} {81}},\ \bibinfo {pages} {184519} (\bibinfo {year}
  {2010})}\BibitemShut {NoStop}%
\bibitem [{\citenamefont {B\"ohmer}\ \emph {et~al.}(2014)\citenamefont
  {B\"ohmer}, \citenamefont {Burger}, \citenamefont {Hardy}, \citenamefont
  {Wolf}, \citenamefont {Schweiss}, \citenamefont {Fromknecht}, \citenamefont
  {Reinecker}, \citenamefont {Schranz},\ and\ \citenamefont
  {Meingast}}]{Bohmer-2014}%
  \BibitemOpen
  \bibfield  {author} {\bibinfo {author} {\bibfnamefont {A.~E.}\ \bibnamefont
  {B\"ohmer}}, \bibinfo {author} {\bibfnamefont {P.}~\bibnamefont {Burger}},
  \bibinfo {author} {\bibfnamefont {F.}~\bibnamefont {Hardy}}, \bibinfo
  {author} {\bibfnamefont {T.}~\bibnamefont {Wolf}}, \bibinfo {author}
  {\bibfnamefont {P.}~\bibnamefont {Schweiss}}, \bibinfo {author}
  {\bibfnamefont {R.}~\bibnamefont {Fromknecht}}, \bibinfo {author}
  {\bibfnamefont {M.}~\bibnamefont {Reinecker}}, \bibinfo {author}
  {\bibfnamefont {W.}~\bibnamefont {Schranz}},\ and\ \bibinfo {author}
  {\bibfnamefont {C.}~\bibnamefont {Meingast}},\ }\bibfield  {title} {\bibinfo
  {title} {Nematic susceptibility of hole-doped and electron-doped
  {${\mathrm{Ba}\mathrm{F}\mathrm{e}}_{2}{\mathrm{As}}_{2}$} {Iron}-based
  superconductors from shear modulus measurements},\ }\href
  {https://doi.org/10.1103/PhysRevLett.112.047001} {\bibfield  {journal}
  {\bibinfo  {journal} {Phys. Rev. Lett.}\ }\textbf {\bibinfo {volume} {112}},\
  \bibinfo {pages} {047001} (\bibinfo {year} {2014})}\BibitemShut {NoStop}%
\bibitem [{\citenamefont {Shibauchi}\ \emph {et~al.}(2014)\citenamefont
  {Shibauchi}, \citenamefont {Carrington},\ and\ \citenamefont
  {Matsuda}}]{Shibauchi-2014}%
  \BibitemOpen
  \bibfield  {author} {\bibinfo {author} {\bibfnamefont {T.}~\bibnamefont
  {Shibauchi}}, \bibinfo {author} {\bibfnamefont {A.}~\bibnamefont
  {Carrington}},\ and\ \bibinfo {author} {\bibfnamefont {Y.}~\bibnamefont
  {Matsuda}},\ }\bibfield  {title} {\bibinfo {title} {A quantum critical point
  lying beneath the superconducting dome in {Iron} pnictides},\ }\href
  {https://doi.org/10.1146/annurev-conmatphys-031113-133921} {\bibfield
  {journal} {\bibinfo  {journal} {Annu. Rev. Condens. Matter Phys.}\ }\textbf
  {\bibinfo {volume} {5}},\ \bibinfo {pages} {113} (\bibinfo {year}
  {2014})}\BibitemShut {NoStop}%
\bibitem [{\citenamefont {Kuo}\ \emph {et~al.}(2016)\citenamefont {Kuo},
  \citenamefont {Chu}, \citenamefont {Palmstrom}, \citenamefont {Kivelson},\
  and\ \citenamefont {Fisher}}]{Kuo-2016}%
  \BibitemOpen
  \bibfield  {author} {\bibinfo {author} {\bibfnamefont {H.-H.}\ \bibnamefont
  {Kuo}}, \bibinfo {author} {\bibfnamefont {J.-H.}\ \bibnamefont {Chu}},
  \bibinfo {author} {\bibfnamefont {J.~C.}\ \bibnamefont {Palmstrom}}, \bibinfo
  {author} {\bibfnamefont {S.~A.}\ \bibnamefont {Kivelson}},\ and\ \bibinfo
  {author} {\bibfnamefont {I.~R.}\ \bibnamefont {Fisher}},\ }\bibfield  {title}
  {\bibinfo {title} {Ubiquitous signatures of nematic quantum criticality in
  optimally doped {Fe}-based superconductors},\ }\href
  {https://doi.org/10.1126/science.abb9280} {\bibfield  {journal} {\bibinfo
  {journal} {Science}\ }\textbf {\bibinfo {volume} {352}},\ \bibinfo {pages}
  {958} (\bibinfo {year} {2016})}\BibitemShut {NoStop}%
\bibitem [{\citenamefont {Bonesteel}\ \emph {et~al.}(1996)\citenamefont
  {Bonesteel}, \citenamefont {McDonald},\ and\ \citenamefont
  {Nayak}}]{Bonesteel-1996}%
  \BibitemOpen
  \bibfield  {author} {\bibinfo {author} {\bibfnamefont {N.~E.}\ \bibnamefont
  {Bonesteel}}, \bibinfo {author} {\bibfnamefont {I.~A.}\ \bibnamefont
  {McDonald}},\ and\ \bibinfo {author} {\bibfnamefont {C.}~\bibnamefont
  {Nayak}},\ }\bibfield  {title} {\bibinfo {title} {Gauge fields and pairing in
  double-layer composite fermion metals},\ }\href
  {https://doi.org/10.1103/PhysRevLett.77.3009} {\bibfield  {journal} {\bibinfo
   {journal} {Phys. Rev. Lett.}\ }\textbf {\bibinfo {volume} {77}},\ \bibinfo
  {pages} {3009} (\bibinfo {year} {1996})}\BibitemShut {NoStop}%
\bibitem [{\citenamefont {Son}(1999)}]{Son-1999}%
  \BibitemOpen
  \bibfield  {author} {\bibinfo {author} {\bibfnamefont {D.~T.}\ \bibnamefont
  {Son}},\ }\bibfield  {title} {\bibinfo {title} {Superconductivity by
  long-range color magnetic interaction in high-density quark matter},\ }\href
  {https://doi.org/10.1103/PhysRevD.59.094019} {\bibfield  {journal} {\bibinfo
  {journal} {Phys. Rev. D}\ }\textbf {\bibinfo {volume} {59}},\ \bibinfo
  {pages} {094019} (\bibinfo {year} {1999})}\BibitemShut {NoStop}%
\bibitem [{\citenamefont {Abanov}\ \emph
  {et~al.}(2001{\natexlab{a}})\citenamefont {Abanov}, \citenamefont
  {Chubukov},\ and\ \citenamefont {Finkel'stein}}]{Abanov-2001a}%
  \BibitemOpen
  \bibfield  {author} {\bibinfo {author} {\bibfnamefont {A.}~\bibnamefont
  {Abanov}}, \bibinfo {author} {\bibfnamefont {A.~V.}\ \bibnamefont
  {Chubukov}},\ and\ \bibinfo {author} {\bibfnamefont {A.~M.}\ \bibnamefont
  {Finkel'stein}},\ }\bibfield  {title} {\bibinfo {title} {Coherent
  vs.incoherent pairing in {2D} systems near magnetic instability},\ }\href
  {https://doi.org/10.1209/epl/i2001-00266-0} {\bibfield  {journal} {\bibinfo
  {journal} {EPL}\ }\textbf {\bibinfo {volume} {54}},\ \bibinfo {pages} {488}
  (\bibinfo {year} {2001}{\natexlab{a}})}\BibitemShut {NoStop}%
\bibitem [{\citenamefont {Abanov}\ \emph
  {et~al.}(2001{\natexlab{b}})\citenamefont {Abanov}, \citenamefont
  {Chubukov},\ and\ \citenamefont {Schmalian}}]{Abanov-2001b}%
  \BibitemOpen
  \bibfield  {author} {\bibinfo {author} {\bibfnamefont {A.}~\bibnamefont
  {Abanov}}, \bibinfo {author} {\bibfnamefont {A.~V.}\ \bibnamefont
  {Chubukov}},\ and\ \bibinfo {author} {\bibfnamefont {J.}~\bibnamefont
  {Schmalian}},\ }\bibfield  {title} {\bibinfo {title} {Quantum-critical
  superconductivity in underdoped cuprates},\ }\href
  {https://doi.org/10.1209/epl/i2001-00425-9} {\bibfield  {journal} {\bibinfo
  {journal} {EPL}\ }\textbf {\bibinfo {volume} {55}},\ \bibinfo {pages} {369}
  (\bibinfo {year} {2001}{\natexlab{b}})}\BibitemShut {NoStop}%
\bibitem [{\citenamefont {Chubukov}\ and\ \citenamefont
  {Schmalian}(2005)}]{Chubukov-2005}%
  \BibitemOpen
  \bibfield  {author} {\bibinfo {author} {\bibfnamefont {A.~V.}\ \bibnamefont
  {Chubukov}}\ and\ \bibinfo {author} {\bibfnamefont {J.}~\bibnamefont
  {Schmalian}},\ }\bibfield  {title} {\bibinfo {title} {Superconductivity due
  to massless boson exchange in the strong-coupling limit},\ }\href
  {https://doi.org/10.1103/PhysRevB.72.174520} {\bibfield  {journal} {\bibinfo
  {journal} {Phys. Rev. B}\ }\textbf {\bibinfo {volume} {72}},\ \bibinfo
  {pages} {174520} (\bibinfo {year} {2005})}\BibitemShut {NoStop}%
\bibitem [{\citenamefont {She}\ and\ \citenamefont {Zaanen}(2009)}]{She-2009}%
  \BibitemOpen
  \bibfield  {author} {\bibinfo {author} {\bibfnamefont {J.-H.}\ \bibnamefont
  {She}}\ and\ \bibinfo {author} {\bibfnamefont {J.}~\bibnamefont {Zaanen}},\
  }\bibfield  {title} {\bibinfo {title} {{BCS} superconductivity in quantum
  critical metals},\ }\href {https://doi.org/10.1103/PhysRevB.80.184518}
  {\bibfield  {journal} {\bibinfo  {journal} {Phys. Rev. B}\ }\textbf {\bibinfo
  {volume} {80}},\ \bibinfo {pages} {184518} (\bibinfo {year}
  {2009})}\BibitemShut {NoStop}%
\bibitem [{\citenamefont {Moon}\ and\ \citenamefont
  {Chubukov}(2010)}]{Moon-2010}%
  \BibitemOpen
  \bibfield  {author} {\bibinfo {author} {\bibfnamefont {E.-G.}\ \bibnamefont
  {Moon}}\ and\ \bibinfo {author} {\bibfnamefont {A.}~\bibnamefont
  {Chubukov}},\ }\bibfield  {title} {\bibinfo {title} {Quantum-critical pairing
  with varying exponents},\ }\href
  {https://link.springer.com/article/10.1007/s10909-010-0199-y} {\bibfield
  {journal} {\bibinfo  {journal} {J. Low Temp. Phys.}\ }\textbf {\bibinfo
  {volume} {161}},\ \bibinfo {pages} {263} (\bibinfo {year}
  {2010})}\BibitemShut {NoStop}%
\bibitem [{\citenamefont {Metlitski}\ \emph {et~al.}(2015)\citenamefont
  {Metlitski}, \citenamefont {Mross}, \citenamefont {Sachdev},\ and\
  \citenamefont {Senthil}}]{Metlitski-2015}%
  \BibitemOpen
  \bibfield  {author} {\bibinfo {author} {\bibfnamefont {M.~A.}\ \bibnamefont
  {Metlitski}}, \bibinfo {author} {\bibfnamefont {D.~F.}\ \bibnamefont
  {Mross}}, \bibinfo {author} {\bibfnamefont {S.}~\bibnamefont {Sachdev}},\
  and\ \bibinfo {author} {\bibfnamefont {T.}~\bibnamefont {Senthil}},\
  }\bibfield  {title} {\bibinfo {title} {Cooper pairing in non-{Fermi}
  liquids},\ }\href {https://doi.org/10.1103/PhysRevB.91.115111} {\bibfield
  {journal} {\bibinfo  {journal} {Phys. Rev. B}\ }\textbf {\bibinfo {volume}
  {91}},\ \bibinfo {pages} {115111} (\bibinfo {year} {2015})}\BibitemShut
  {NoStop}%
\bibitem [{\citenamefont {Roussev}\ and\ \citenamefont
  {Millis}(2001)}]{Roussev-2001}%
  \BibitemOpen
  \bibfield  {author} {\bibinfo {author} {\bibfnamefont {R.}~\bibnamefont
  {Roussev}}\ and\ \bibinfo {author} {\bibfnamefont {A.~J.}\ \bibnamefont
  {Millis}},\ }\bibfield  {title} {\bibinfo {title} {Quantum critical effects
  on transition temperature of magnetically mediated p-wave
  superconductivity},\ }\href {https://doi.org/10.1103/PhysRevB.63.140504}
  {\bibfield  {journal} {\bibinfo  {journal} {Phys. Rev. B}\ }\textbf {\bibinfo
  {volume} {63}},\ \bibinfo {pages} {140504} (\bibinfo {year}
  {2001})}\BibitemShut {NoStop}%
\bibitem [{\citenamefont {Raghu}\ \emph {et~al.}(2015)\citenamefont {Raghu},
  \citenamefont {Torroba},\ and\ \citenamefont {Wang}}]{Raghu-2015}%
  \BibitemOpen
  \bibfield  {author} {\bibinfo {author} {\bibfnamefont {S.}~\bibnamefont
  {Raghu}}, \bibinfo {author} {\bibfnamefont {G.}~\bibnamefont {Torroba}},\
  and\ \bibinfo {author} {\bibfnamefont {H.}~\bibnamefont {Wang}},\ }\bibfield
  {title} {\bibinfo {title} {Metallic quantum critical points with finite {BCS}
  couplings},\ }\href {https://doi.org/10.1103/PhysRevB.92.205104} {\bibfield
  {journal} {\bibinfo  {journal} {Phys. Rev. B}\ }\textbf {\bibinfo {volume}
  {92}},\ \bibinfo {pages} {205104} (\bibinfo {year} {2015})}\BibitemShut
  {NoStop}%
\bibitem [{\citenamefont {Berg}\ \emph {et~al.}(2012)\citenamefont {Berg},
  \citenamefont {Metlitski},\ and\ \citenamefont {Sachdev}}]{Berg-2012}%
  \BibitemOpen
  \bibfield  {author} {\bibinfo {author} {\bibfnamefont {E.}~\bibnamefont
  {Berg}}, \bibinfo {author} {\bibfnamefont {M.~A.}\ \bibnamefont
  {Metlitski}},\ and\ \bibinfo {author} {\bibfnamefont {S.}~\bibnamefont
  {Sachdev}},\ }\bibfield  {title} {\bibinfo {title} {Sign-problem--free
  quantum {Monte} carlo of the onset of antiferromagnetism in metals},\ }\href
  {https://doi.org/10.1126/science.1227769} {\bibfield  {journal} {\bibinfo
  {journal} {Science}\ }\textbf {\bibinfo {volume} {338}},\ \bibinfo {pages}
  {1606} (\bibinfo {year} {2012})}\BibitemShut {NoStop}%
\bibitem [{\citenamefont {Schattner}\ \emph
  {et~al.}(2016{\natexlab{a}})\citenamefont {Schattner}, \citenamefont
  {Gerlach}, \citenamefont {Trebst},\ and\ \citenamefont
  {Berg}}]{Schattner-2016a}%
  \BibitemOpen
  \bibfield  {author} {\bibinfo {author} {\bibfnamefont {Y.}~\bibnamefont
  {Schattner}}, \bibinfo {author} {\bibfnamefont {M.~H.}\ \bibnamefont
  {Gerlach}}, \bibinfo {author} {\bibfnamefont {S.}~\bibnamefont {Trebst}},\
  and\ \bibinfo {author} {\bibfnamefont {E.}~\bibnamefont {Berg}},\ }\bibfield
  {title} {\bibinfo {title} {Competing orders in a nearly antiferromagnetic
  metal},\ }\href {https://doi.org/10.1103/PhysRevLett.117.097002} {\bibfield
  {journal} {\bibinfo  {journal} {Phys. Rev. Lett.}\ }\textbf {\bibinfo
  {volume} {117}},\ \bibinfo {pages} {097002} (\bibinfo {year}
  {2016}{\natexlab{a}})}\BibitemShut {NoStop}%
\bibitem [{\citenamefont {Schattner}\ \emph
  {et~al.}(2016{\natexlab{b}})\citenamefont {Schattner}, \citenamefont
  {Lederer}, \citenamefont {Kivelson},\ and\ \citenamefont
  {Berg}}]{Schattner-2016b}%
  \BibitemOpen
  \bibfield  {author} {\bibinfo {author} {\bibfnamefont {Y.}~\bibnamefont
  {Schattner}}, \bibinfo {author} {\bibfnamefont {S.}~\bibnamefont {Lederer}},
  \bibinfo {author} {\bibfnamefont {S.~A.}\ \bibnamefont {Kivelson}},\ and\
  \bibinfo {author} {\bibfnamefont {E.}~\bibnamefont {Berg}},\ }\bibfield
  {title} {\bibinfo {title} {{Ising} nematic quantum critical point in a metal:
  A {Monte Carlo} study},\ }\href {https://doi.org/10.1103/PhysRevX.6.031028}
  {\bibfield  {journal} {\bibinfo  {journal} {Phys. Rev. X}\ }\textbf {\bibinfo
  {volume} {6}},\ \bibinfo {pages} {031028} (\bibinfo {year}
  {2016}{\natexlab{b}})}\BibitemShut {NoStop}%
\bibitem [{\citenamefont {Dumitrescu}\ \emph {et~al.}(2016)\citenamefont
  {Dumitrescu}, \citenamefont {Serbyn}, \citenamefont {Scalettar},\ and\
  \citenamefont {Vishwanath}}]{Dumitrescu-2016}%
  \BibitemOpen
  \bibfield  {author} {\bibinfo {author} {\bibfnamefont {P.~T.}\ \bibnamefont
  {Dumitrescu}}, \bibinfo {author} {\bibfnamefont {M.}~\bibnamefont {Serbyn}},
  \bibinfo {author} {\bibfnamefont {R.~T.}\ \bibnamefont {Scalettar}},\ and\
  \bibinfo {author} {\bibfnamefont {A.}~\bibnamefont {Vishwanath}},\ }\bibfield
   {title} {\bibinfo {title} {Superconductivity and nematic fluctuations in a
  model of doped {FeSe} monolayers: Determinant quantum {Monte Carlo} study},\
  }\href {https://doi.org/10.1103/PhysRevB.94.155127} {\bibfield  {journal}
  {\bibinfo  {journal} {Phys. Rev. B}\ }\textbf {\bibinfo {volume} {94}},\
  \bibinfo {pages} {155127} (\bibinfo {year} {2016})}\BibitemShut {NoStop}%
\bibitem [{\citenamefont {Lederer}\ \emph {et~al.}(2017)\citenamefont
  {Lederer}, \citenamefont {Yoni}, \citenamefont {Berg},\ and\ \citenamefont
  {Kivelson}}]{Lederer-2017}%
  \BibitemOpen
  \bibfield  {author} {\bibinfo {author} {\bibfnamefont {S.}~\bibnamefont
  {Lederer}}, \bibinfo {author} {\bibfnamefont {S.}~\bibnamefont {Yoni}},
  \bibinfo {author} {\bibfnamefont {E.}~\bibnamefont {Berg}},\ and\ \bibinfo
  {author} {\bibfnamefont {S.~A.}\ \bibnamefont {Kivelson}},\ }\bibfield
  {title} {\bibinfo {title} {Superconductivity and non-{Fermi} liquid behavior
  near a nematic quantum critical point},\ }\href
  {https://doi.org/10.1073/pnas.1620651114} {\bibfield  {journal} {\bibinfo
  {journal} {Proc. Natl. Acad. Sci. U. S. A.}\ }\textbf {\bibinfo {volume}
  {114}},\ \bibinfo {pages} {4905} (\bibinfo {year} {2017})}\BibitemShut
  {NoStop}%
\bibitem [{\citenamefont {Li}\ \emph {et~al.}(2017)\citenamefont {Li},
  \citenamefont {Wang}, \citenamefont {Yao},\ and\ \citenamefont
  {Lee}}]{Li-2017}%
  \BibitemOpen
  \bibfield  {author} {\bibinfo {author} {\bibfnamefont {Z.-X.}\ \bibnamefont
  {Li}}, \bibinfo {author} {\bibfnamefont {F.}~\bibnamefont {Wang}}, \bibinfo
  {author} {\bibfnamefont {H.}~\bibnamefont {Yao}},\ and\ \bibinfo {author}
  {\bibfnamefont {D.-H.}\ \bibnamefont {Lee}},\ }\bibfield  {title} {\bibinfo
  {title} {Nature of the effective interaction in electron-doped cuprate
  superconductors: A sign-problem-free quantum {Monte Carlo} study},\ }\href
  {https://doi.org/10.1103/PhysRevB.95.214505} {\bibfield  {journal} {\bibinfo
  {journal} {Phys. Rev. B}\ }\textbf {\bibinfo {volume} {95}},\ \bibinfo
  {pages} {214505} (\bibinfo {year} {2017})}\BibitemShut {NoStop}%
\bibitem [{\citenamefont {Wang}\ \emph {et~al.}(2017)\citenamefont {Wang},
  \citenamefont {Schattner}, \citenamefont {Berg},\ and\ \citenamefont
  {Fernandes}}]{Wang-2017}%
  \BibitemOpen
  \bibfield  {author} {\bibinfo {author} {\bibfnamefont {X.}~\bibnamefont
  {Wang}}, \bibinfo {author} {\bibfnamefont {Y.}~\bibnamefont {Schattner}},
  \bibinfo {author} {\bibfnamefont {E.}~\bibnamefont {Berg}},\ and\ \bibinfo
  {author} {\bibfnamefont {R.~M.}\ \bibnamefont {Fernandes}},\ }\bibfield
  {title} {\bibinfo {title} {Superconductivity mediated by quantum critical
  antiferromagnetic fluctuations: The rise and fall of hot spots},\ }\href
  {https://doi.org/10.1103/PhysRevB.95.174520} {\bibfield  {journal} {\bibinfo
  {journal} {Phys. Rev. B}\ }\textbf {\bibinfo {volume} {95}},\ \bibinfo
  {pages} {174520} (\bibinfo {year} {2017})}\BibitemShut {NoStop}%
\bibitem [{\citenamefont {Esterlis}\ \emph
  {et~al.}(2018{\natexlab{a}})\citenamefont {Esterlis}, \citenamefont
  {Nosarzewski}, \citenamefont {Huang}, \citenamefont {Moritz}, \citenamefont
  {Devereaux}, \citenamefont {Scalapino},\ and\ \citenamefont
  {Kivelson}}]{Esterlis-2018b}%
  \BibitemOpen
  \bibfield  {author} {\bibinfo {author} {\bibfnamefont {I.}~\bibnamefont
  {Esterlis}}, \bibinfo {author} {\bibfnamefont {B.}~\bibnamefont
  {Nosarzewski}}, \bibinfo {author} {\bibfnamefont {E.~W.}\ \bibnamefont
  {Huang}}, \bibinfo {author} {\bibfnamefont {B.}~\bibnamefont {Moritz}},
  \bibinfo {author} {\bibfnamefont {T.~P.}\ \bibnamefont {Devereaux}}, \bibinfo
  {author} {\bibfnamefont {D.~J.}\ \bibnamefont {Scalapino}},\ and\ \bibinfo
  {author} {\bibfnamefont {S.~A.}\ \bibnamefont {Kivelson}},\ }\bibfield
  {title} {\bibinfo {title} {Breakdown of the {Migdal}-{Eliashberg} theory: A
  determinant quantum {Monte} {Carlo} study},\ }\href
  {https://doi.org/10.1103/PhysRevB.97.140501} {\bibfield  {journal} {\bibinfo
  {journal} {Phys. Rev. B}\ }\textbf {\bibinfo {volume} {97}},\ \bibinfo
  {pages} {140501} (\bibinfo {year} {2018}{\natexlab{a}})}\BibitemShut
  {NoStop}%
\bibitem [{\citenamefont {Berg}\ \emph {et~al.}(2019)\citenamefont {Berg},
  \citenamefont {Lederer}, \citenamefont {Schattner},\ and\ \citenamefont
  {Trebst}}]{Berg-2019}%
  \BibitemOpen
  \bibfield  {author} {\bibinfo {author} {\bibfnamefont {E.}~\bibnamefont
  {Berg}}, \bibinfo {author} {\bibfnamefont {S.}~\bibnamefont {Lederer}},
  \bibinfo {author} {\bibfnamefont {Y.}~\bibnamefont {Schattner}},\ and\
  \bibinfo {author} {\bibfnamefont {S.}~\bibnamefont {Trebst}},\ }\bibfield
  {title} {\bibinfo {title} {{Monte Carlo} studies of quantum critical
  metals},\ }\href {https://doi.org/10.1146/annurev-conmatphys-031218-013339}
  {\bibfield  {journal} {\bibinfo  {journal} {Annu. Rev. Condens. Matter
  Phys.}\ }\textbf {\bibinfo {volume} {10}},\ \bibinfo {pages} {63} (\bibinfo
  {year} {2019})}\BibitemShut {NoStop}%
\bibitem [{\citenamefont {Maldacena}\ and\ \citenamefont
  {Stanford}(2016)}]{Maldacena-2016a}%
  \BibitemOpen
  \bibfield  {author} {\bibinfo {author} {\bibfnamefont {J.}~\bibnamefont
  {Maldacena}}\ and\ \bibinfo {author} {\bibfnamefont {D.}~\bibnamefont
  {Stanford}},\ }\bibfield  {title} {\bibinfo {title} {Remarks on the
  {Sachdev-Ye-Kitaev} model},\ }\href
  {https://doi.org/10.1103/PhysRevD.94.106002} {\bibfield  {journal} {\bibinfo
  {journal} {Phys. Rev. D}\ }\textbf {\bibinfo {volume} {94}},\ \bibinfo
  {pages} {106002} (\bibinfo {year} {2016})}\BibitemShut {NoStop}%
\bibitem [{\citenamefont {Maldacena}\ \emph
  {et~al.}(2016{\natexlab{a}})\citenamefont {Maldacena}, \citenamefont
  {Stanford},\ and\ \citenamefont {Yang}}]{Maldacena-2016b}%
  \BibitemOpen
  \bibfield  {author} {\bibinfo {author} {\bibfnamefont {J.}~\bibnamefont
  {Maldacena}}, \bibinfo {author} {\bibfnamefont {D.}~\bibnamefont
  {Stanford}},\ and\ \bibinfo {author} {\bibfnamefont {Z.}~\bibnamefont
  {Yang}},\ }\bibfield  {title} {\bibinfo {title} {Conformal symmetry and its
  breaking in two-dimensional nearly anti-de {Sitter} space},\ }\href
  {https://doi.org/10.1093/ptep/ptw124} {\bibfield  {journal} {\bibinfo
  {journal} {Prog. Theor. Exp. Phys.}\ }\textbf {\bibinfo {volume} {2016}},\
  \bibinfo {pages} {12C104} (\bibinfo {year} {2016}{\natexlab{a}})}\BibitemShut
  {NoStop}%
\bibitem [{\citenamefont {Kitaev}(2015{\natexlab{a}})}]{Kitaev2015a}%
  \BibitemOpen
  \bibfield  {author} {\bibinfo {author} {\bibfnamefont {A.}~\bibnamefont
  {Kitaev}},\ }\bibfield  {title} {\bibinfo {title} {A simple model of quantum
  holography - part 1},\ }\href
  {http://online.kitp.ucsb.edu/online/entangled15/kitaev/} {\bibfield
  {journal} {\bibinfo  {journal} {Talk at Kavli Institute for Theoretical
  Physics}\ } (\bibinfo {year} {2015}{\natexlab{a}})}\BibitemShut {NoStop}%
\bibitem [{\citenamefont {Kitaev}(2015{\natexlab{b}})}]{Kitaev2015b}%
  \BibitemOpen
  \bibfield  {author} {\bibinfo {author} {\bibfnamefont {A.}~\bibnamefont
  {Kitaev}},\ }\bibfield  {title} {\bibinfo {title} {A simple model of quantum
  holography - part 2},\ }\href
  {http://online.kitp.ucsb.edu/online/entangled15/kitaev2/} {\bibfield
  {journal} {\bibinfo  {journal} {Talk at Kavli Institute for Theoretical
  Physics}\ } (\bibinfo {year} {2015}{\natexlab{b}})}\BibitemShut {NoStop}%
\bibitem [{\citenamefont {Bagrets}\ \emph {et~al.}(2017)\citenamefont
  {Bagrets}, \citenamefont {Altland},\ and\ \citenamefont
  {Kamenev}}]{Bagrets-2017}%
  \BibitemOpen
  \bibfield  {author} {\bibinfo {author} {\bibfnamefont {D.}~\bibnamefont
  {Bagrets}}, \bibinfo {author} {\bibfnamefont {A.}~\bibnamefont {Altland}},\
  and\ \bibinfo {author} {\bibfnamefont {A.}~\bibnamefont {Kamenev}},\
  }\bibfield  {title} {\bibinfo {title} {Power-law out of time order
  correlation functions in the {SYK} model},\ }\href
  {https://doi.org/https://doi.org/10.1016/j.nuclphysb.2017.06.012} {\bibfield
  {journal} {\bibinfo  {journal} {Nucl. Phys. B}\ }\textbf {\bibinfo {volume}
  {921}},\ \bibinfo {pages} {727 } (\bibinfo {year} {2017})}\BibitemShut
  {NoStop}%
\bibitem [{\citenamefont {Garc\'{\i}a-Garc\'{\i}a}\ \emph
  {et~al.}(2018)\citenamefont {Garc\'{\i}a-Garc\'{\i}a}, \citenamefont
  {Loureiro}, \citenamefont {Romero-Berm\'udez},\ and\ \citenamefont
  {Tezuka}}]{Garcia-Garcia-2018}%
  \BibitemOpen
  \bibfield  {author} {\bibinfo {author} {\bibfnamefont {A.~M.}\ \bibnamefont
  {Garc\'{\i}a-Garc\'{\i}a}}, \bibinfo {author} {\bibfnamefont
  {B.}~\bibnamefont {Loureiro}}, \bibinfo {author} {\bibfnamefont
  {A.}~\bibnamefont {Romero-Berm\'udez}},\ and\ \bibinfo {author}
  {\bibfnamefont {M.}~\bibnamefont {Tezuka}},\ }\bibfield  {title} {\bibinfo
  {title} {Chaotic-integrable transition in the {Sachdev-Ye-Kitaev} model},\
  }\href {https://doi.org/10.1103/PhysRevLett.120.241603} {\bibfield  {journal}
  {\bibinfo  {journal} {Phys. Rev. Lett.}\ }\textbf {\bibinfo {volume} {120}},\
  \bibinfo {pages} {241603} (\bibinfo {year} {2018})}\BibitemShut {NoStop}%
\bibitem [{\citenamefont {Chowdhury}\ \emph {et~al.}(2022)\citenamefont
  {Chowdhury}, \citenamefont {Georges}, \citenamefont {Parcollet},\ and\
  \citenamefont {Sachdev}}]{Chowdhury-2022}%
  \BibitemOpen
  \bibfield  {author} {\bibinfo {author} {\bibfnamefont {D.}~\bibnamefont
  {Chowdhury}}, \bibinfo {author} {\bibfnamefont {A.}~\bibnamefont {Georges}},
  \bibinfo {author} {\bibfnamefont {O.}~\bibnamefont {Parcollet}},\ and\
  \bibinfo {author} {\bibfnamefont {S.}~\bibnamefont {Sachdev}},\ }\bibfield
  {title} {\bibinfo {title} {{Sachdev-Ye-Kitaev} models and beyond: Window into
  non-{Fermi} liquids},\ }\href {https://doi.org/10.1103/RevModPhys.94.035004}
  {\bibfield  {journal} {\bibinfo  {journal} {Rev. Mod. Phys.}\ }\textbf
  {\bibinfo {volume} {94}},\ \bibinfo {pages} {035004} (\bibinfo {year}
  {2022})}\BibitemShut {NoStop}%
\bibitem [{\citenamefont {Sachdev}\ and\ \citenamefont
  {Ye}(1993)}]{Sachdev-1993}%
  \BibitemOpen
  \bibfield  {author} {\bibinfo {author} {\bibfnamefont {S.}~\bibnamefont
  {Sachdev}}\ and\ \bibinfo {author} {\bibfnamefont {J.}~\bibnamefont {Ye}},\
  }\bibfield  {title} {\bibinfo {title} {Gapless spin-fluid ground state in a
  random quantum {Heisenberg} magnet},\ }\href
  {https://doi.org/10.1103/PhysRevLett.70.3339} {\bibfield  {journal} {\bibinfo
   {journal} {Phys. Rev. Lett.}\ }\textbf {\bibinfo {volume} {70}},\ \bibinfo
  {pages} {3339} (\bibinfo {year} {1993})}\BibitemShut {NoStop}%
\bibitem [{\citenamefont {Schmalian}\ and\ \citenamefont
  {Wolynes}(2000)}]{Schmalian-2000}%
  \BibitemOpen
  \bibfield  {author} {\bibinfo {author} {\bibfnamefont {J.}~\bibnamefont
  {Schmalian}}\ and\ \bibinfo {author} {\bibfnamefont {P.~G.}\ \bibnamefont
  {Wolynes}},\ }\bibfield  {title} {\bibinfo {title} {Stripe glasses:
  Self-generated randomness in a uniformly frustrated system},\ }\href
  {https://doi.org/10.1103/PhysRevLett.85.836} {\bibfield  {journal} {\bibinfo
  {journal} {Phys. Rev. Lett.}\ }\textbf {\bibinfo {volume} {85}},\ \bibinfo
  {pages} {836} (\bibinfo {year} {2000})}\BibitemShut {NoStop}%
\bibitem [{\citenamefont {Georges}\ \emph {et~al.}(2000)\citenamefont
  {Georges}, \citenamefont {Parcollet},\ and\ \citenamefont
  {Sachdev}}]{Georges-2000}%
  \BibitemOpen
  \bibfield  {author} {\bibinfo {author} {\bibfnamefont {A.}~\bibnamefont
  {Georges}}, \bibinfo {author} {\bibfnamefont {O.}~\bibnamefont {Parcollet}},\
  and\ \bibinfo {author} {\bibfnamefont {S.}~\bibnamefont {Sachdev}},\
  }\bibfield  {title} {\bibinfo {title} {Mean field theory of a quantum
  {Heisenberg} spin glass},\ }\href
  {https://doi.org/10.1103/PhysRevLett.85.840} {\bibfield  {journal} {\bibinfo
  {journal} {Phys. Rev. Lett.}\ }\textbf {\bibinfo {volume} {85}},\ \bibinfo
  {pages} {840} (\bibinfo {year} {2000})}\BibitemShut {NoStop}%
\bibitem [{\citenamefont {Georges}\ \emph {et~al.}(2001)\citenamefont
  {Georges}, \citenamefont {Parcollet},\ and\ \citenamefont
  {Sachdev}}]{Georges-2001}%
  \BibitemOpen
  \bibfield  {author} {\bibinfo {author} {\bibfnamefont {A.}~\bibnamefont
  {Georges}}, \bibinfo {author} {\bibfnamefont {O.}~\bibnamefont {Parcollet}},\
  and\ \bibinfo {author} {\bibfnamefont {S.}~\bibnamefont {Sachdev}},\
  }\bibfield  {title} {\bibinfo {title} {Quantum fluctuations of a nearly
  critical {Heisenberg} spin glass},\ }\href
  {https://doi.org/10.1103/PhysRevB.63.134406} {\bibfield  {journal} {\bibinfo
  {journal} {Phys. Rev. B}\ }\textbf {\bibinfo {volume} {63}},\ \bibinfo
  {pages} {134406} (\bibinfo {year} {2001})}\BibitemShut {NoStop}%
\bibitem [{\citenamefont {Westfahl}\ \emph {et~al.}(2003)\citenamefont
  {Westfahl}, \citenamefont {Schmalian},\ and\ \citenamefont
  {Wolynes}}]{Westfahl-2003}%
  \BibitemOpen
  \bibfield  {author} {\bibinfo {author} {\bibfnamefont {H.}~\bibnamefont
  {Westfahl}}, \bibinfo {author} {\bibfnamefont {J.}~\bibnamefont
  {Schmalian}},\ and\ \bibinfo {author} {\bibfnamefont {P.~G.}\ \bibnamefont
  {Wolynes}},\ }\bibfield  {title} {\bibinfo {title} {Dynamical mean-field
  theory of quantum stripe glasses},\ }\href
  {https://doi.org/10.1103/PhysRevB.68.134203} {\bibfield  {journal} {\bibinfo
  {journal} {Phys. Rev. B}\ }\textbf {\bibinfo {volume} {68}},\ \bibinfo
  {pages} {134203} (\bibinfo {year} {2003})}\BibitemShut {NoStop}%
\bibitem [{\citenamefont {Parcollet}\ and\ \citenamefont
  {Georges}(1999)}]{Parcollet-1999}%
  \BibitemOpen
  \bibfield  {author} {\bibinfo {author} {\bibfnamefont {O.}~\bibnamefont
  {Parcollet}}\ and\ \bibinfo {author} {\bibfnamefont {A.}~\bibnamefont
  {Georges}},\ }\bibfield  {title} {\bibinfo {title} {Non-{Fermi}-liquid regime
  of a doped {Mott} insulator},\ }\href
  {https://doi.org/10.1103/PhysRevB.59.5341} {\bibfield  {journal} {\bibinfo
  {journal} {Phys. Rev. B}\ }\textbf {\bibinfo {volume} {59}},\ \bibinfo
  {pages} {5341} (\bibinfo {year} {1999})}\BibitemShut {NoStop}%
\bibitem [{\citenamefont {Song}\ \emph {et~al.}(2017)\citenamefont {Song},
  \citenamefont {Jian},\ and\ \citenamefont {Balents}}]{Song-2017}%
  \BibitemOpen
  \bibfield  {author} {\bibinfo {author} {\bibfnamefont {X.-Y.}\ \bibnamefont
  {Song}}, \bibinfo {author} {\bibfnamefont {C.-M.}\ \bibnamefont {Jian}},\
  and\ \bibinfo {author} {\bibfnamefont {L.}~\bibnamefont {Balents}},\
  }\bibfield  {title} {\bibinfo {title} {Strongly correlated metal built from
  {Sachdev}-{Ye}-{Kitaev} models},\ }\href
  {https://doi.org/10.1103/PhysRevLett.119.216601} {\bibfield  {journal}
  {\bibinfo  {journal} {Phys. Rev. Lett.}\ }\textbf {\bibinfo {volume} {119}},\
  \bibinfo {pages} {216601} (\bibinfo {year} {2017})}\BibitemShut {NoStop}%
\bibitem [{\citenamefont {Gu}\ \emph {et~al.}(2020)\citenamefont {Gu},
  \citenamefont {Kitaev}, \citenamefont {Sachdev},\ and\ \citenamefont
  {Tarnopolsky}}]{Gu-2020}%
  \BibitemOpen
  \bibfield  {author} {\bibinfo {author} {\bibfnamefont {Y.}~\bibnamefont
  {Gu}}, \bibinfo {author} {\bibfnamefont {A.}~\bibnamefont {Kitaev}}, \bibinfo
  {author} {\bibfnamefont {S.}~\bibnamefont {Sachdev}},\ and\ \bibinfo {author}
  {\bibfnamefont {G.}~\bibnamefont {Tarnopolsky}},\ }\bibfield  {title}
  {\bibinfo {title} {Notes on the complex {Sachdev-Ye-Kitaev} model},\ }\href
  {https://link.springer.com/article/10.1007%2FJHEP02%282020%29157} {\bibfield
  {journal} {\bibinfo  {journal} {J. High En. Phys.}\ }\textbf {\bibinfo
  {volume} {2020}},\ \bibinfo {pages} {1} (\bibinfo {year} {2020})}\BibitemShut
  {NoStop}%
\bibitem [{\citenamefont {Kim}\ \emph {et~al.}(2020)\citenamefont {Kim},
  \citenamefont {Cao},\ and\ \citenamefont {Altman}}]{Kim-2020}%
  \BibitemOpen
  \bibfield  {author} {\bibinfo {author} {\bibfnamefont {J.}~\bibnamefont
  {Kim}}, \bibinfo {author} {\bibfnamefont {X.}~\bibnamefont {Cao}},\ and\
  \bibinfo {author} {\bibfnamefont {E.}~\bibnamefont {Altman}},\ }\bibfield
  {title} {\bibinfo {title} {Low-rank {Sachdev-Ye-Kitaev} models},\ }\href
  {https://doi.org/10.1103/PhysRevB.101.125112} {\bibfield  {journal} {\bibinfo
   {journal} {Phys. Rev. B}\ }\textbf {\bibinfo {volume} {101}},\ \bibinfo
  {pages} {125112} (\bibinfo {year} {2020})}\BibitemShut {NoStop}%
\bibitem [{\citenamefont {Esterlis}\ \emph {et~al.}(2021)\citenamefont
  {Esterlis}, \citenamefont {Guo}, \citenamefont {Patel},\ and\ \citenamefont
  {Sachdev}}]{Esterlis-2021}%
  \BibitemOpen
  \bibfield  {author} {\bibinfo {author} {\bibfnamefont {I.}~\bibnamefont
  {Esterlis}}, \bibinfo {author} {\bibfnamefont {H.}~\bibnamefont {Guo}},
  \bibinfo {author} {\bibfnamefont {A.~A.}\ \bibnamefont {Patel}},\ and\
  \bibinfo {author} {\bibfnamefont {S.}~\bibnamefont {Sachdev}},\ }\bibfield
  {title} {\bibinfo {title} {{Large-$N$} theory of critical {Fermi} surfaces},\
  }\href {https://doi.org/10.1103/PhysRevB.103.235129} {\bibfield  {journal}
  {\bibinfo  {journal} {Phys. Rev. B}\ }\textbf {\bibinfo {volume} {103}},\
  \bibinfo {pages} {235129} (\bibinfo {year} {2021})}\BibitemShut {NoStop}%
\bibitem [{\citenamefont {Guo}\ \emph {et~al.}(2022)\citenamefont {Guo},
  \citenamefont {Patel}, \citenamefont {Esterlis},\ and\ \citenamefont
  {Sachdev}}]{Guo-2022}%
  \BibitemOpen
  \bibfield  {author} {\bibinfo {author} {\bibfnamefont {H.}~\bibnamefont
  {Guo}}, \bibinfo {author} {\bibfnamefont {A.~A.}\ \bibnamefont {Patel}},
  \bibinfo {author} {\bibfnamefont {I.}~\bibnamefont {Esterlis}},\ and\
  \bibinfo {author} {\bibfnamefont {S.}~\bibnamefont {Sachdev}},\ }\bibfield
  {title} {\bibinfo {title} {{Large-$N$} theory of critical {Ferm}i surfaces.
  {II.} {Conductivity}},\ }\href {https://doi.org/10.1103/PhysRevB.106.115151}
  {\bibfield  {journal} {\bibinfo  {journal} {Phys. Rev. B}\ }\textbf {\bibinfo
  {volume} {106}},\ \bibinfo {pages} {115151} (\bibinfo {year}
  {2022})}\BibitemShut {NoStop}%
\bibitem [{\citenamefont {{Chowdhury}}\ and\ \citenamefont
  {{Berg}}(2020)}]{Chowdhury-2020a}%
  \BibitemOpen
  \bibfield  {author} {\bibinfo {author} {\bibfnamefont {D.}~\bibnamefont
  {{Chowdhury}}}\ and\ \bibinfo {author} {\bibfnamefont {E.}~\bibnamefont
  {{Berg}}},\ }\bibfield  {title} {\bibinfo {title} {Intrinsic superconducting
  instabilities of a solvable model for an incoherent metal},\ }\href
  {https://doi.org/10.1103/PhysRevResearch.2.013301} {\bibfield  {journal}
  {\bibinfo  {journal} {Phys. Rev. Research}\ }\textbf {\bibinfo {volume}
  {2}},\ \bibinfo {pages} {013301} (\bibinfo {year} {2020})}\BibitemShut
  {NoStop}%
\bibitem [{\citenamefont {Salvati}\ and\ \citenamefont
  {Tagliacozzo}(2021)}]{Salvati-2021}%
  \BibitemOpen
  \bibfield  {author} {\bibinfo {author} {\bibfnamefont {F.}~\bibnamefont
  {Salvati}}\ and\ \bibinfo {author} {\bibfnamefont {A.}~\bibnamefont
  {Tagliacozzo}},\ }\bibfield  {title} {\bibinfo {title} {Superconducting
  critical temperature in the extended diffusive {Sachdev-Ye-Kitaev} model},\
  }\href {https://doi.org/10.1103/PhysRevResearch.3.033117} {\bibfield
  {journal} {\bibinfo  {journal} {Phys. Rev. Research}\ }\textbf {\bibinfo
  {volume} {3}},\ \bibinfo {pages} {033117} (\bibinfo {year}
  {2021})}\BibitemShut {NoStop}%
\bibitem [{\citenamefont {Lantagne-Hurtubise}\ \emph
  {et~al.}(2021)\citenamefont {Lantagne-Hurtubise}, \citenamefont {Pathak},
  \citenamefont {Sahoo},\ and\ \citenamefont
  {Franz}}]{Lantagne-Hurtubise-2021}%
  \BibitemOpen
  \bibfield  {author} {\bibinfo {author} {\bibfnamefont {E.}~\bibnamefont
  {Lantagne-Hurtubise}}, \bibinfo {author} {\bibfnamefont {V.}~\bibnamefont
  {Pathak}}, \bibinfo {author} {\bibfnamefont {S.}~\bibnamefont {Sahoo}},\ and\
  \bibinfo {author} {\bibfnamefont {M.}~\bibnamefont {Franz}},\ }\bibfield
  {title} {\bibinfo {title} {Superconducting instabilities in a spinful
  {Sachdev-Ye-Kitaev} model},\ }\href
  {https://doi.org/10.1103/PhysRevB.104.L020509} {\bibfield  {journal}
  {\bibinfo  {journal} {Phys. Rev. B}\ }\textbf {\bibinfo {volume} {104}},\
  \bibinfo {pages} {L020509} (\bibinfo {year} {2021})}\BibitemShut {NoStop}%
\bibitem [{\citenamefont {Patel}\ \emph {et~al.}(2018)\citenamefont {Patel},
  \citenamefont {Lawler},\ and\ \citenamefont {Kim}}]{Patel-2018}%
  \BibitemOpen
  \bibfield  {author} {\bibinfo {author} {\bibfnamefont {A.~A.}\ \bibnamefont
  {Patel}}, \bibinfo {author} {\bibfnamefont {M.~J.}\ \bibnamefont {Lawler}},\
  and\ \bibinfo {author} {\bibfnamefont {E.-A.}\ \bibnamefont {Kim}},\
  }\bibfield  {title} {\bibinfo {title} {Coherent superconductivity with a
  large gap ratio from incoherent metals},\ }\href
  {https://doi.org/10.1103/PhysRevLett.121.187001} {\bibfield  {journal}
  {\bibinfo  {journal} {Phys. Rev. Lett.}\ }\textbf {\bibinfo {volume} {121}},\
  \bibinfo {pages} {187001} (\bibinfo {year} {2018})}\BibitemShut {NoStop}%
\bibitem [{\citenamefont {Wang}\ and\ \citenamefont
  {Chubukov}(2020)}]{Wang2020b}%
  \BibitemOpen
  \bibfield  {author} {\bibinfo {author} {\bibfnamefont {Y.}~\bibnamefont
  {Wang}}\ and\ \bibinfo {author} {\bibfnamefont {A.~V.}\ \bibnamefont
  {Chubukov}},\ }\bibfield  {title} {\bibinfo {title} {Quantum phase transition
  in the {Yukawa-SYK} model},\ }\href
  {https://doi.org/10.1103/PhysRevResearch.2.033084} {\bibfield  {journal}
  {\bibinfo  {journal} {Phys. Rev. Research}\ }\textbf {\bibinfo {volume}
  {2}},\ \bibinfo {pages} {033084} (\bibinfo {year} {2020})}\BibitemShut
  {NoStop}%
\bibitem [{\citenamefont {Wang}\ \emph {et~al.}(2021)\citenamefont {Wang},
  \citenamefont {Davis}, \citenamefont {Pan}, \citenamefont {Wang},\ and\
  \citenamefont {Meng}}]{Wang-2021}%
  \BibitemOpen
  \bibfield  {author} {\bibinfo {author} {\bibfnamefont {W.}~\bibnamefont
  {Wang}}, \bibinfo {author} {\bibfnamefont {A.}~\bibnamefont {Davis}},
  \bibinfo {author} {\bibfnamefont {G.}~\bibnamefont {Pan}}, \bibinfo {author}
  {\bibfnamefont {Y.}~\bibnamefont {Wang}},\ and\ \bibinfo {author}
  {\bibfnamefont {Z.~Y.}\ \bibnamefont {Meng}},\ }\bibfield  {title} {\bibinfo
  {title} {Phase diagram of the spin-$\frac{1}{2}$ {Yukawa--Sachdev-Ye-Kitaev}
  model: Non-{Fermi} liquid, insulator, and superconductor},\ }\href
  {https://doi.org/10.1103/PhysRevB.103.195108} {\bibfield  {journal} {\bibinfo
   {journal} {Phys. Rev. B}\ }\textbf {\bibinfo {volume} {103}},\ \bibinfo
  {pages} {195108} (\bibinfo {year} {2021})}\BibitemShut {NoStop}%
\bibitem [{\citenamefont {Esterlis}\ and\ \citenamefont
  {Schmalian}(2019)}]{Esterlis-2019}%
  \BibitemOpen
  \bibfield  {author} {\bibinfo {author} {\bibfnamefont {I.}~\bibnamefont
  {Esterlis}}\ and\ \bibinfo {author} {\bibfnamefont {J.}~\bibnamefont
  {Schmalian}},\ }\bibfield  {title} {\bibinfo {title} {Cooper pairing of
  incoherent electrons: An electron-phonon version of the {Sachdev-Ye-Kitaev}
  model},\ }\href {https://doi.org/10.1103/PhysRevB.100.115132} {\bibfield
  {journal} {\bibinfo  {journal} {Phys. Rev. B}\ }\textbf {\bibinfo {volume}
  {100}},\ \bibinfo {pages} {115132} (\bibinfo {year} {2019})}\BibitemShut
  {NoStop}%
\bibitem [{\citenamefont {Hauck}\ \emph {et~al.}(2020)\citenamefont {Hauck},
  \citenamefont {Klug}, \citenamefont {Esterlis},\ and\ \citenamefont
  {Schmalian}}]{Hauck-2020}%
  \BibitemOpen
  \bibfield  {author} {\bibinfo {author} {\bibfnamefont {D.}~\bibnamefont
  {Hauck}}, \bibinfo {author} {\bibfnamefont {M.~J.}\ \bibnamefont {Klug}},
  \bibinfo {author} {\bibfnamefont {I.}~\bibnamefont {Esterlis}},\ and\
  \bibinfo {author} {\bibfnamefont {J.}~\bibnamefont {Schmalian}},\ }\bibfield
  {title} {\bibinfo {title} {{Eliashberg} equations for an electron–phonon
  version of the {Sachdev}–{Ye}–{Kitaev} model: Pair breaking in
  non-{Fermi} liquid superconductors},\ }\href
  {http://www.sciencedirect.com/science/article/pii/S0003491620300531}
  {\bibfield  {journal} {\bibinfo  {journal} {Ann. Phys. (N. Y.)}\ }\textbf
  {\bibinfo {volume} {417}},\ \bibinfo {pages} {168120} (\bibinfo {year}
  {2020})},\ \bibinfo {note} {in "{Eliashberg} theory at 60: Strong-coupling
  superconductivity and beyond"}\BibitemShut {NoStop}%
\bibitem [{\citenamefont {Classen}\ and\ \citenamefont
  {Chubukov}(2021)}]{Classen-2022}%
  \BibitemOpen
  \bibfield  {author} {\bibinfo {author} {\bibfnamefont {L.}~\bibnamefont
  {Classen}}\ and\ \bibinfo {author} {\bibfnamefont {A.}~\bibnamefont
  {Chubukov}},\ }\bibfield  {title} {\bibinfo {title} {Superconductivity of
  incoherent electrons in the {Yukawa Sachdev-Ye-Kitaev} model},\ }\href
  {https://doi.org/10.1103/PhysRevB.104.125120} {\bibfield  {journal} {\bibinfo
   {journal} {Phys. Rev. B}\ }\textbf {\bibinfo {volume} {104}},\ \bibinfo
  {pages} {125120} (\bibinfo {year} {2021})}\BibitemShut {NoStop}%
\bibitem [{\citenamefont {Choi}\ \emph {et~al.}(2022)\citenamefont {Choi},
  \citenamefont {Tavakol},\ and\ \citenamefont {Kim}}]{Choi-2022}%
  \BibitemOpen
  \bibfield  {author} {\bibinfo {author} {\bibfnamefont {W.}~\bibnamefont
  {Choi}}, \bibinfo {author} {\bibfnamefont {O.}~\bibnamefont {Tavakol}},\ and\
  \bibinfo {author} {\bibfnamefont {Y.~B.}\ \bibnamefont {Kim}},\ }\bibfield
  {title} {\bibinfo {title} {{Pairing instabilities of the {Yukawa-SYK} models
  with controlled fermion incoherence}},\ }\href
  {https://doi.org/10.21468/SciPostPhys.12.5.151} {\bibfield  {journal}
  {\bibinfo  {journal} {SciPost Phys.}\ }\textbf {\bibinfo {volume} {12}},\
  \bibinfo {pages} {151} (\bibinfo {year} {2022})}\BibitemShut {NoStop}%
\bibitem [{\citenamefont {Li}\ \emph {et~al.}(2023)\citenamefont {Li},
  \citenamefont {Sachdev},\ and\ \citenamefont {Joshi}}]{Li-2023}%
  \BibitemOpen
  \bibfield  {author} {\bibinfo {author} {\bibfnamefont {C.}~\bibnamefont
  {Li}}, \bibinfo {author} {\bibfnamefont {S.}~\bibnamefont {Sachdev}},\ and\
  \bibinfo {author} {\bibfnamefont {D.~G.}\ \bibnamefont {Joshi}},\ }\bibfield
  {title} {\bibinfo {title} {Superconductivity of non-{Fermi} liquids described
  by {Sachdev-Ye-Kitaev} models},\ }\href
  {https://doi.org/10.1103/PhysRevResearch.5.013045} {\bibfield  {journal}
  {\bibinfo  {journal} {Phys. Rev. Res.}\ }\textbf {\bibinfo {volume} {5}},\
  \bibinfo {pages} {013045} (\bibinfo {year} {2023})}\BibitemShut {NoStop}%
\bibitem [{\citenamefont {Patel}\ and\ \citenamefont
  {Sachdev}(2017)}]{Patel-2017}%
  \BibitemOpen
  \bibfield  {author} {\bibinfo {author} {\bibfnamefont {A.~A.}\ \bibnamefont
  {Patel}}\ and\ \bibinfo {author} {\bibfnamefont {S.}~\bibnamefont
  {Sachdev}},\ }\bibfield  {title} {\bibinfo {title} {Quantum chaos on a
  critical {Fermi} surface},\ }\href {https://doi.org/10.1073/pnas.1618185114}
  {\bibfield  {journal} {\bibinfo  {journal} {Proc. Natl. Acad. Sci. U.S.A.}\
  }\textbf {\bibinfo {volume} {114}},\ \bibinfo {pages} {1844} (\bibinfo {year}
  {2017})}\BibitemShut {NoStop}%
\bibitem [{\citenamefont {Banerjee}\ and\ \citenamefont
  {Altman}(2017)}]{Banerjee-2017}%
  \BibitemOpen
  \bibfield  {author} {\bibinfo {author} {\bibfnamefont {S.}~\bibnamefont
  {Banerjee}}\ and\ \bibinfo {author} {\bibfnamefont {E.}~\bibnamefont
  {Altman}},\ }\bibfield  {title} {\bibinfo {title} {Solvable model for a
  dynamical quantum phase transition from fast to slow scrambling},\ }\href
  {https://doi.org/10.1103/PhysRevB.95.134302} {\bibfield  {journal} {\bibinfo
  {journal} {Phys. Rev. B}\ }\textbf {\bibinfo {volume} {95}},\ \bibinfo
  {pages} {134302} (\bibinfo {year} {2017})}\BibitemShut {NoStop}%
\bibitem [{\citenamefont {Kobrin}\ \emph {et~al.}(2021)\citenamefont {Kobrin},
  \citenamefont {Yang}, \citenamefont {Kahanamoku-Meyer}, \citenamefont
  {Olund}, \citenamefont {Moore}, \citenamefont {Stanford},\ and\ \citenamefont
  {Yao}}]{Kobrin-2021}%
  \BibitemOpen
  \bibfield  {author} {\bibinfo {author} {\bibfnamefont {B.}~\bibnamefont
  {Kobrin}}, \bibinfo {author} {\bibfnamefont {Z.}~\bibnamefont {Yang}},
  \bibinfo {author} {\bibfnamefont {G.~D.}\ \bibnamefont {Kahanamoku-Meyer}},
  \bibinfo {author} {\bibfnamefont {C.~T.}\ \bibnamefont {Olund}}, \bibinfo
  {author} {\bibfnamefont {J.~E.}\ \bibnamefont {Moore}}, \bibinfo {author}
  {\bibfnamefont {D.}~\bibnamefont {Stanford}},\ and\ \bibinfo {author}
  {\bibfnamefont {N.~Y.}\ \bibnamefont {Yao}},\ }\bibfield  {title} {\bibinfo
  {title} {Many-body chaos in the {Sachdev-Ye-Kitaev} model},\ }\href
  {https://doi.org/10.1103/PhysRevLett.126.030602} {\bibfield  {journal}
  {\bibinfo  {journal} {Phys. Rev. Lett.}\ }\textbf {\bibinfo {volume} {126}},\
  \bibinfo {pages} {030602} (\bibinfo {year} {2021})}\BibitemShut {NoStop}%
\bibitem [{\citenamefont {Zaanen}\ \emph {et~al.}(2015)\citenamefont {Zaanen},
  \citenamefont {Liu}, \citenamefont {Sun},\ and\ \citenamefont
  {Schalm}}]{Zaanen-2015holo}%
  \BibitemOpen
  \bibfield  {author} {\bibinfo {author} {\bibfnamefont {J.}~\bibnamefont
  {Zaanen}}, \bibinfo {author} {\bibfnamefont {Y.}~\bibnamefont {Liu}},
  \bibinfo {author} {\bibfnamefont {Y.-W.}\ \bibnamefont {Sun}},\ and\ \bibinfo
  {author} {\bibfnamefont {K.}~\bibnamefont {Schalm}},\ }\href
  {https://doi.org/10.1017/CBO9781139942492} {\emph {\bibinfo {title}
  {Holographic duality in condensed matter physics}}}\ (\bibinfo  {publisher}
  {Cambridge University Press},\ \bibinfo {address} {Cambridge},\ \bibinfo
  {year} {2015})\BibitemShut {NoStop}%
\bibitem [{\citenamefont {Kitaev}\ and\ \citenamefont
  {Suh}(2018)}]{Kitaev-2018}%
  \BibitemOpen
  \bibfield  {author} {\bibinfo {author} {\bibfnamefont {A.}~\bibnamefont
  {Kitaev}}\ and\ \bibinfo {author} {\bibfnamefont {S.~J.}\ \bibnamefont
  {Suh}},\ }\bibfield  {title} {\bibinfo {title} {The soft mode in the
  {Sachdev-Ye-Kitaev} model and its gravity dual},\ }\href
  {https://doi.org/10.1007/JHEP05(2018)183} {\bibfield  {journal} {\bibinfo
  {journal} {J. High En. Phys.}\ }\textbf {\bibinfo {volume} {2018}},\ \bibinfo
  {pages} {1} (\bibinfo {year} {2018})}\BibitemShut {NoStop}%
\bibitem [{\citenamefont {S{\'{a}}rosi}(2017)}]{Sarosi2017a}%
  \BibitemOpen
  \bibfield  {author} {\bibinfo {author} {\bibfnamefont {G.}~\bibnamefont
  {S{\'{a}}rosi}},\ }\bibfield  {title} {\bibinfo {title} {{AdS$_2$ holography
  and the SYK model}},\ }\href {https://doi.org/10.22323/1.323.0001} {\bibfield
   {journal} {\bibinfo  {journal} {PoS Proc. Sci.}\ }\textbf {\bibinfo {volume}
  {323}},\ \bibinfo {pages} {001} (\bibinfo {year} {2017})}\BibitemShut
  {NoStop}%
\bibitem [{\citenamefont {Nayak}\ \emph {et~al.}(2018)\citenamefont {Nayak},
  \citenamefont {Shukla}, \citenamefont {Soni}, \citenamefont {Trivedi},\ and\
  \citenamefont {Vishal}}]{Nayak2018}%
  \BibitemOpen
  \bibfield  {author} {\bibinfo {author} {\bibfnamefont {P.}~\bibnamefont
  {Nayak}}, \bibinfo {author} {\bibfnamefont {A.}~\bibnamefont {Shukla}},
  \bibinfo {author} {\bibfnamefont {R.~M.}\ \bibnamefont {Soni}}, \bibinfo
  {author} {\bibfnamefont {S.~P.}\ \bibnamefont {Trivedi}},\ and\ \bibinfo
  {author} {\bibfnamefont {V.}~\bibnamefont {Vishal}},\ }\bibfield  {title}
  {\bibinfo {title} {{On the dynamics of near-extremal black holes}},\ }\href
  {https://doi.org/10.1007/JHEP09(2018)048} {\bibfield  {journal} {\bibinfo
  {journal} {J. High En. Phys.}\ }\textbf {\bibinfo {volume} {2018}},\ \bibinfo
  {pages} {1} (\bibinfo {year} {2018})}\BibitemShut {NoStop}%
\bibitem [{\citenamefont {Moitra}\ \emph {et~al.}(2019)\citenamefont {Moitra},
  \citenamefont {Trivedi},\ and\ \citenamefont {Vishal}}]{Moitra2019}%
  \BibitemOpen
  \bibfield  {author} {\bibinfo {author} {\bibfnamefont {U.}~\bibnamefont
  {Moitra}}, \bibinfo {author} {\bibfnamefont {S.~P.}\ \bibnamefont
  {Trivedi}},\ and\ \bibinfo {author} {\bibfnamefont {V.}~\bibnamefont
  {Vishal}},\ }\bibfield  {title} {\bibinfo {title} {{Extremal and
  near-extremal black holes and near-CFT1}},\ }\href
  {https://doi.org/10.1007/JHEP07(2019)055} {\bibfield  {journal} {\bibinfo
  {journal} {J. High En. Phys.}\ }\textbf {\bibinfo {volume} {2019}},\ \bibinfo
  {pages} {1} (\bibinfo {year} {2019})}\BibitemShut {NoStop}%
\bibitem [{\citenamefont {Sachdev}(2019)}]{Sachdev2019}%
  \BibitemOpen
  \bibfield  {author} {\bibinfo {author} {\bibfnamefont {S.}~\bibnamefont
  {Sachdev}},\ }\bibfield  {title} {\bibinfo {title} {{Universal low
  temperature theory of charged black holes with AdS$_2$ horizons}},\ }\href
  {https://doi.org/10.1063/1.5092726} {\bibfield  {journal} {\bibinfo
  {journal} {J. Math. Phys.}\ }\textbf {\bibinfo {volume} {60}},\ \bibinfo
  {pages} {0} (\bibinfo {year} {2019})}\BibitemShut {NoStop}%
\bibitem [{\citenamefont {Inkof}\ \emph {et~al.}(2022)\citenamefont {Inkof},
  \citenamefont {Schalm},\ and\ \citenamefont {Schmalian}}]{Inkof-2022}%
  \BibitemOpen
  \bibfield  {author} {\bibinfo {author} {\bibfnamefont {G.-A.}\ \bibnamefont
  {Inkof}}, \bibinfo {author} {\bibfnamefont {K.}~\bibnamefont {Schalm}},\ and\
  \bibinfo {author} {\bibfnamefont {J.}~\bibnamefont {Schmalian}},\ }\bibfield
  {title} {\bibinfo {title} {Quantum critical {Eliashberg} theory, the
  {Sachdev-Ye-Kitaev} superconductor and their holographic duals},\ }\href
  {https://doi.org/10.1038/s41535-022-00460-8} {\bibfield  {journal} {\bibinfo
  {journal} {npj Quantum Mater.}\ }\textbf {\bibinfo {volume} {7}},\ \bibinfo
  {pages} {1} (\bibinfo {year} {2022})}\BibitemShut {NoStop}%
\bibitem [{\citenamefont {Schmalian}(2022)}]{Schmalian2022}%
  \BibitemOpen
  \bibfield  {author} {\bibinfo {author} {\bibfnamefont {J.}~\bibnamefont
  {Schmalian}},\ }\bibfield  {title} {\bibinfo {title} {Holographic
  superconductivity of a critical {Fermi} surface},\ }\href
  {https://arxiv.org/abs/2209.00474} {\bibfield  {journal} {\bibinfo  {journal}
  {arXiv:2209.00474}\ } (\bibinfo {year} {2022})}\BibitemShut {NoStop}%
\bibitem [{\citenamefont {Maldacena}(1999)}]{Maldacena-1999}%
  \BibitemOpen
  \bibfield  {author} {\bibinfo {author} {\bibfnamefont {J.}~\bibnamefont
  {Maldacena}},\ }\bibfield  {title} {\bibinfo {title} {The large-{N} limit of
  superconformal field theories and supergravity},\ }\href
  {https://link.springer.com/article/10.1023\%2FA\%3A1026654312961} {\bibfield
  {journal} {\bibinfo  {journal} {Int. J. Theor. Phys.}\ }\textbf {\bibinfo
  {volume} {38}},\ \bibinfo {pages} {1113} (\bibinfo {year}
  {1999})}\BibitemShut {NoStop}%
\bibitem [{\citenamefont {Franz}\ and\ \citenamefont
  {Rozali}(2018)}]{Franz-2018}%
  \BibitemOpen
  \bibfield  {author} {\bibinfo {author} {\bibfnamefont {M.}~\bibnamefont
  {Franz}}\ and\ \bibinfo {author} {\bibfnamefont {M.}~\bibnamefont {Rozali}},\
  }\bibfield  {title} {\bibinfo {title} {Mimicking black hole event horizons in
  atomic and solid-state systems},\ }\href
  {https://doi.org/10.1038/s41578-018-0058-z} {\bibfield  {journal} {\bibinfo
  {journal} {Nat. Rev. Mater.}\ }\textbf {\bibinfo {volume} {3}},\ \bibinfo
  {pages} {491} (\bibinfo {year} {2018})}\BibitemShut {NoStop}%
\bibitem [{\citenamefont {Kruchkov}\ \emph {et~al.}(2020)\citenamefont
  {Kruchkov}, \citenamefont {Patel}, \citenamefont {Kim},\ and\ \citenamefont
  {Sachdev}}]{Kruchkov-2020}%
  \BibitemOpen
  \bibfield  {author} {\bibinfo {author} {\bibfnamefont {A.}~\bibnamefont
  {Kruchkov}}, \bibinfo {author} {\bibfnamefont {A.~A.}\ \bibnamefont {Patel}},
  \bibinfo {author} {\bibfnamefont {P.}~\bibnamefont {Kim}},\ and\ \bibinfo
  {author} {\bibfnamefont {S.}~\bibnamefont {Sachdev}},\ }\bibfield  {title}
  {\bibinfo {title} {Thermoelectric power of {Sachdev-Ye-Kitaev} islands:
  Probing {Bekenstein-Hawking} entropy in quantum matter experiments},\ }\href
  {https://doi.org/10.1103/PhysRevB.101.205148} {\bibfield  {journal} {\bibinfo
   {journal} {Phys. Rev. B}\ }\textbf {\bibinfo {volume} {101}},\ \bibinfo
  {pages} {205148} (\bibinfo {year} {2020})}\BibitemShut {NoStop}%
\bibitem [{\citenamefont {Balm}\ \emph {et~al.}(2022)\citenamefont {Balm},
  \citenamefont {Chagnet}, \citenamefont {Arend}, \citenamefont {Aretz},
  \citenamefont {Grosvenor}, \citenamefont {Janse}, \citenamefont {Moors},
  \citenamefont {Post}, \citenamefont {Ohanesjan}, \citenamefont
  {Rodriguez-Fernandez}, \citenamefont {Schalm},\ and\ \citenamefont
  {Zaanen}}]{Balm-2022_preprint}%
  \BibitemOpen
  \bibfield  {author} {\bibinfo {author} {\bibfnamefont {F.}~\bibnamefont
  {Balm}}, \bibinfo {author} {\bibfnamefont {N.}~\bibnamefont {Chagnet}},
  \bibinfo {author} {\bibfnamefont {S.}~\bibnamefont {Arend}}, \bibinfo
  {author} {\bibfnamefont {J.}~\bibnamefont {Aretz}}, \bibinfo {author}
  {\bibfnamefont {K.}~\bibnamefont {Grosvenor}}, \bibinfo {author}
  {\bibfnamefont {M.}~\bibnamefont {Janse}}, \bibinfo {author} {\bibfnamefont
  {O.}~\bibnamefont {Moors}}, \bibinfo {author} {\bibfnamefont
  {J.}~\bibnamefont {Post}}, \bibinfo {author} {\bibfnamefont {V.}~\bibnamefont
  {Ohanesjan}}, \bibinfo {author} {\bibfnamefont {D.}~\bibnamefont
  {Rodriguez-Fernandez}}, \bibinfo {author} {\bibfnamefont {K.}~\bibnamefont
  {Schalm}},\ and\ \bibinfo {author} {\bibfnamefont {J.}~\bibnamefont
  {Zaanen}},\ }\bibfield  {title} {\bibinfo {title} {{T}-linear resistivity,
  optical conductivity and {Planckian} transport for a holographic local
  quantum critical metal in a periodic potential},\ }\href
  {https://arxiv.org/abs/2211.05492} {\bibfield  {journal} {\bibinfo  {journal}
  {arXiv:2211.05492}\ } (\bibinfo {year} {2022})}\BibitemShut {NoStop}%
\bibitem [{\citenamefont {Georges}\ \emph {et~al.}(1996)\citenamefont
  {Georges}, \citenamefont {Kotliar}, \citenamefont {Krauth},\ and\
  \citenamefont {Rozenberg}}]{Georges-1996}%
  \BibitemOpen
  \bibfield  {author} {\bibinfo {author} {\bibfnamefont {A.}~\bibnamefont
  {Georges}}, \bibinfo {author} {\bibfnamefont {G.}~\bibnamefont {Kotliar}},
  \bibinfo {author} {\bibfnamefont {W.}~\bibnamefont {Krauth}},\ and\ \bibinfo
  {author} {\bibfnamefont {M.~J.}\ \bibnamefont {Rozenberg}},\ }\bibfield
  {title} {\bibinfo {title} {Dynamical mean-field theory of strongly correlated
  fermion systems and the limit of infinite dimensions},\ }\href
  {https://doi.org/10.1103/RevModPhys.68.13} {\bibfield  {journal} {\bibinfo
  {journal} {Rev. Mod. Phys.}\ }\textbf {\bibinfo {volume} {68}},\ \bibinfo
  {pages} {13} (\bibinfo {year} {1996})}\BibitemShut {NoStop}%
\bibitem [{\citenamefont {Kotliar}\ and\ \citenamefont
  {Vollhardt}(2004)}]{Kotliar-2004}%
  \BibitemOpen
  \bibfield  {author} {\bibinfo {author} {\bibfnamefont {G.}~\bibnamefont
  {Kotliar}}\ and\ \bibinfo {author} {\bibfnamefont {D.}~\bibnamefont
  {Vollhardt}},\ }\bibfield  {title} {\bibinfo {title} {Strongly correlated
  materials: Insights from dynamical mean-field theory},\ }\href
  {https://doi.org/10.1063/1.1712502} {\bibfield  {journal} {\bibinfo
  {journal} {Phys. Today}\ }\textbf {\bibinfo {volume} {57}},\ \bibinfo {pages}
  {53} (\bibinfo {year} {2004})}\BibitemShut {NoStop}%
\bibitem [{\citenamefont {Patel}\ \emph {et~al.}(2022)\citenamefont {Patel},
  \citenamefont {Guo}, \citenamefont {Esterlis},\ and\ \citenamefont
  {Sachdev}}]{Patel-2023_preprint}%
  \BibitemOpen
  \bibfield  {author} {\bibinfo {author} {\bibfnamefont {A.~A.}\ \bibnamefont
  {Patel}}, \bibinfo {author} {\bibfnamefont {H.}~\bibnamefont {Guo}}, \bibinfo
  {author} {\bibfnamefont {I.}~\bibnamefont {Esterlis}},\ and\ \bibinfo
  {author} {\bibfnamefont {S.}~\bibnamefont {Sachdev}},\ }\bibfield  {title}
  {\bibinfo {title} {Universal theory of strange metals from spatially random
  interactions},\ }\href {https://doi.org/10.48550/arXiv.2203.04990} {\bibfield
   {journal} {\bibinfo  {journal} {arXiv:2203.04990}\ } (\bibinfo {year}
  {2022})}\BibitemShut {NoStop}%
\bibitem [{\citenamefont {Fu}\ and\ \citenamefont {Sachdev}(2016)}]{Fu-2016}%
  \BibitemOpen
  \bibfield  {author} {\bibinfo {author} {\bibfnamefont {W.}~\bibnamefont
  {Fu}}\ and\ \bibinfo {author} {\bibfnamefont {S.}~\bibnamefont {Sachdev}},\
  }\bibfield  {title} {\bibinfo {title} {Numerical study of fermion and boson
  models with infinite-range random interactions},\ }\href
  {https://doi.org/10.1103/PhysRevB.94.035135} {\bibfield  {journal} {\bibinfo
  {journal} {Phys. Rev. B}\ }\textbf {\bibinfo {volume} {94}},\ \bibinfo
  {pages} {035135} (\bibinfo {year} {2016})}\BibitemShut {NoStop}%
\bibitem [{\citenamefont {Lunkin}\ \emph {et~al.}(2018)\citenamefont {Lunkin},
  \citenamefont {Tikhonov},\ and\ \citenamefont {Feigel'man}}]{Lunkin-2018}%
  \BibitemOpen
  \bibfield  {author} {\bibinfo {author} {\bibfnamefont {A.~V.}\ \bibnamefont
  {Lunkin}}, \bibinfo {author} {\bibfnamefont {K.~S.}\ \bibnamefont
  {Tikhonov}},\ and\ \bibinfo {author} {\bibfnamefont {M.~V.}\ \bibnamefont
  {Feigel'man}},\ }\bibfield  {title} {\bibinfo {title} {{Sachdev-Ye-Kitaev}
  model with quadratic perturbations: The route to a non-{Fermi} liquid},\
  }\href {https://doi.org/10.1103/PhysRevLett.121.236601} {\bibfield  {journal}
  {\bibinfo  {journal} {Phys. Rev. Lett.}\ }\textbf {\bibinfo {volume} {121}},\
  \bibinfo {pages} {236601} (\bibinfo {year} {2018})}\BibitemShut {NoStop}%
\bibitem [{\citenamefont {Sun}\ and\ \citenamefont {Ye}(2020)}]{Sun-2018}%
  \BibitemOpen
  \bibfield  {author} {\bibinfo {author} {\bibfnamefont {F.}~\bibnamefont
  {Sun}}\ and\ \bibinfo {author} {\bibfnamefont {J.}~\bibnamefont {Ye}},\
  }\bibfield  {title} {\bibinfo {title} {Periodic table of the ordinary and
  supersymmetric {Sachdev-Ye-Kitaev} models},\ }\href
  {https://doi.org/10.1103/PhysRevLett.124.244101} {\bibfield  {journal}
  {\bibinfo  {journal} {Phys. Rev. Lett.}\ }\textbf {\bibinfo {volume} {124}},\
  \bibinfo {pages} {244101} (\bibinfo {year} {2020})}\BibitemShut {NoStop}%
\bibitem [{\citenamefont {Dai}\ \emph {et~al.}(2019)\citenamefont {Dai},
  \citenamefont {Jian},\ and\ \citenamefont {Yao}}]{Dai-2019}%
  \BibitemOpen
  \bibfield  {author} {\bibinfo {author} {\bibfnamefont {X.}~\bibnamefont
  {Dai}}, \bibinfo {author} {\bibfnamefont {S.-K.}\ \bibnamefont {Jian}},\ and\
  \bibinfo {author} {\bibfnamefont {H.}~\bibnamefont {Yao}},\ }\bibfield
  {title} {\bibinfo {title} {Global phase diagram of the one-dimensional
  {Sachdev-Ye-Kitaev} model at finite {$N$}},\ }\href
  {https://doi.org/10.1103/PhysRevB.100.235144} {\bibfield  {journal} {\bibinfo
   {journal} {Phys. Rev. B}\ }\textbf {\bibinfo {volume} {100}},\ \bibinfo
  {pages} {235144} (\bibinfo {year} {2019})}\BibitemShut {NoStop}%
\bibitem [{\citenamefont {Wang}\ \emph {et~al.}(2020)\citenamefont {Wang},
  \citenamefont {Chudnovskiy}, \citenamefont {Gorsky},\ and\ \citenamefont
  {Kamenev}}]{Wang-2020b}%
  \BibitemOpen
  \bibfield  {author} {\bibinfo {author} {\bibfnamefont {H.}~\bibnamefont
  {Wang}}, \bibinfo {author} {\bibfnamefont {A.~L.}\ \bibnamefont
  {Chudnovskiy}}, \bibinfo {author} {\bibfnamefont {A.}~\bibnamefont
  {Gorsky}},\ and\ \bibinfo {author} {\bibfnamefont {A.}~\bibnamefont
  {Kamenev}},\ }\bibfield  {title} {\bibinfo {title} {{Sachdev}-{Ye}-{Kitaev}
  superconductivity: Quantum {Kuramoto} and generalized {Richardson} models},\
  }\href {https://doi.org/10.1103/PhysRevResearch.2.033025} {\bibfield
  {journal} {\bibinfo  {journal} {Phys. Rev. Res.}\ }\textbf {\bibinfo {volume}
  {2}},\ \bibinfo {pages} {033025} (\bibinfo {year} {2020})}\BibitemShut
  {NoStop}%
\bibitem [{\citenamefont {Davison}\ \emph {et~al.}(2017)\citenamefont
  {Davison}, \citenamefont {Fu}, \citenamefont {Georges}, \citenamefont {Gu},
  \citenamefont {Jensen},\ and\ \citenamefont {Sachdev}}]{Davison-2017}%
  \BibitemOpen
  \bibfield  {author} {\bibinfo {author} {\bibfnamefont {R.~A.}\ \bibnamefont
  {Davison}}, \bibinfo {author} {\bibfnamefont {W.}~\bibnamefont {Fu}},
  \bibinfo {author} {\bibfnamefont {A.}~\bibnamefont {Georges}}, \bibinfo
  {author} {\bibfnamefont {Y.}~\bibnamefont {Gu}}, \bibinfo {author}
  {\bibfnamefont {K.}~\bibnamefont {Jensen}},\ and\ \bibinfo {author}
  {\bibfnamefont {S.}~\bibnamefont {Sachdev}},\ }\bibfield  {title} {\bibinfo
  {title} {Thermoelectric transport in disordered metals without
  quasiparticles: The {Sachdev}-{Ye}-{Kitaev} models and holography},\ }\href
  {https://doi.org/10.1103/PhysRevB.95.155131} {\bibfield  {journal} {\bibinfo
  {journal} {Phys. Rev. B}\ }\textbf {\bibinfo {volume} {95}},\ \bibinfo
  {pages} {155131} (\bibinfo {year} {2017})}\BibitemShut {NoStop}%
\bibitem [{\citenamefont {Grunwald}(2022)}]{Grunwald-thesis-2022}%
  \BibitemOpen
  \bibfield  {author} {\bibinfo {author} {\bibfnamefont {L.}~\bibnamefont
  {Grunwald}},\ }\emph {\bibinfo {title} {{SYK} Approach to non-{Fermi} Liquids
  and Incoherent Superconductivity}},\ \href@noop {} {Ph.D. thesis},\ \bibinfo
  {school} {University of Aachen} (\bibinfo {year} {2022})\BibitemShut
  {NoStop}%
\bibitem [{\citenamefont {Chen}\ \emph {et~al.}(2018)\citenamefont {Chen},
  \citenamefont {Ilan}, \citenamefont {de~Juan}, \citenamefont {Pikulin},\ and\
  \citenamefont {Franz}}]{Chen-2018}%
  \BibitemOpen
  \bibfield  {author} {\bibinfo {author} {\bibfnamefont {A.}~\bibnamefont
  {Chen}}, \bibinfo {author} {\bibfnamefont {R.}~\bibnamefont {Ilan}}, \bibinfo
  {author} {\bibfnamefont {F.}~\bibnamefont {de~Juan}}, \bibinfo {author}
  {\bibfnamefont {D.~I.}\ \bibnamefont {Pikulin}},\ and\ \bibinfo {author}
  {\bibfnamefont {M.}~\bibnamefont {Franz}},\ }\bibfield  {title} {\bibinfo
  {title} {Quantum holography in a {Graphene} flake with an irregular
  boundary},\ }\href {https://doi.org/10.1103/PhysRevLett.121.036403}
  {\bibfield  {journal} {\bibinfo  {journal} {Phys. Rev. Lett.}\ }\textbf
  {\bibinfo {volume} {121}},\ \bibinfo {pages} {036403} (\bibinfo {year}
  {2018})}\BibitemShut {NoStop}%
\bibitem [{\citenamefont {Altland}\ \emph
  {et~al.}(2019{\natexlab{a}})\citenamefont {Altland}, \citenamefont
  {Bagrets},\ and\ \citenamefont {Kamenev}}]{Altland-2019a}%
  \BibitemOpen
  \bibfield  {author} {\bibinfo {author} {\bibfnamefont {A.}~\bibnamefont
  {Altland}}, \bibinfo {author} {\bibfnamefont {D.}~\bibnamefont {Bagrets}},\
  and\ \bibinfo {author} {\bibfnamefont {A.}~\bibnamefont {Kamenev}},\
  }\bibfield  {title} {\bibinfo {title} {Quantum criticality of granular
  {Sachdev-Ye-Kitaev} matter},\ }\href
  {https://doi.org/10.1103/PhysRevLett.123.106601} {\bibfield  {journal}
  {\bibinfo  {journal} {Phys. Rev. Lett.}\ }\textbf {\bibinfo {volume} {123}},\
  \bibinfo {pages} {106601} (\bibinfo {year} {2019}{\natexlab{a}})}\BibitemShut
  {NoStop}%
\bibitem [{\citenamefont {Altland}\ \emph
  {et~al.}(2019{\natexlab{b}})\citenamefont {Altland}, \citenamefont
  {Bagrets},\ and\ \citenamefont {Kamenev}}]{Altland-2019b}%
  \BibitemOpen
  \bibfield  {author} {\bibinfo {author} {\bibfnamefont {A.}~\bibnamefont
  {Altland}}, \bibinfo {author} {\bibfnamefont {D.}~\bibnamefont {Bagrets}},\
  and\ \bibinfo {author} {\bibfnamefont {A.}~\bibnamefont {Kamenev}},\
  }\bibfield  {title} {\bibinfo {title} {{Sachdev-Ye-Kitaev} non-{Fermi}-liquid
  correlations in nanoscopic quantum transport},\ }\href
  {https://doi.org/10.1103/PhysRevLett.123.226801} {\bibfield  {journal}
  {\bibinfo  {journal} {Phys. Rev. Lett.}\ }\textbf {\bibinfo {volume} {123}},\
  \bibinfo {pages} {226801} (\bibinfo {year} {2019}{\natexlab{b}})}\BibitemShut
  {NoStop}%
\bibitem [{\citenamefont {Can}\ \emph {et~al.}(2019)\citenamefont {Can},
  \citenamefont {Nica},\ and\ \citenamefont {Franz}}]{Can-2019}%
  \BibitemOpen
  \bibfield  {author} {\bibinfo {author} {\bibfnamefont {O.}~\bibnamefont
  {Can}}, \bibinfo {author} {\bibfnamefont {E.~M.}\ \bibnamefont {Nica}},\ and\
  \bibinfo {author} {\bibfnamefont {M.}~\bibnamefont {Franz}},\ }\bibfield
  {title} {\bibinfo {title} {Charge transport in graphene-based mesoscopic
  realizations of {Sachdev-Ye-Kitaev} models},\ }\href
  {https://doi.org/10.1103/PhysRevB.99.045419} {\bibfield  {journal} {\bibinfo
  {journal} {Phys. Rev. B}\ }\textbf {\bibinfo {volume} {99}},\ \bibinfo
  {pages} {045419} (\bibinfo {year} {2019})}\BibitemShut {NoStop}%
\bibitem [{\citenamefont {Sahoo}\ \emph {et~al.}(2020)\citenamefont {Sahoo},
  \citenamefont {Lantagne-Hurtubise}, \citenamefont {Plugge},\ and\
  \citenamefont {Franz}}]{Sahoo-2020}%
  \BibitemOpen
  \bibfield  {author} {\bibinfo {author} {\bibfnamefont {S.}~\bibnamefont
  {Sahoo}}, \bibinfo {author} {\bibfnamefont {E.}~\bibnamefont
  {Lantagne-Hurtubise}}, \bibinfo {author} {\bibfnamefont {S.}~\bibnamefont
  {Plugge}},\ and\ \bibinfo {author} {\bibfnamefont {M.}~\bibnamefont
  {Franz}},\ }\bibfield  {title} {\bibinfo {title} {Traversable wormhole and
  {Hawking-Page} transition in coupled complex {SYK} models},\ }\href
  {https://doi.org/10.1103/PhysRevResearch.2.043049} {\bibfield  {journal}
  {\bibinfo  {journal} {Phys. Rev. Research}\ }\textbf {\bibinfo {volume}
  {2}},\ \bibinfo {pages} {043049} (\bibinfo {year} {2020})}\BibitemShut
  {NoStop}%
\bibitem [{\citenamefont {Florens}\ \emph {et~al.}(2013)\citenamefont
  {Florens}, \citenamefont {Mohan}, \citenamefont {Janani}, \citenamefont
  {Gupta},\ and\ \citenamefont {Narayanan}}]{Florens-2013}%
  \BibitemOpen
  \bibfield  {author} {\bibinfo {author} {\bibfnamefont {S.}~\bibnamefont
  {Florens}}, \bibinfo {author} {\bibfnamefont {P.}~\bibnamefont {Mohan}},
  \bibinfo {author} {\bibfnamefont {C.}~\bibnamefont {Janani}}, \bibinfo
  {author} {\bibfnamefont {T.}~\bibnamefont {Gupta}},\ and\ \bibinfo {author}
  {\bibfnamefont {R.}~\bibnamefont {Narayanan}},\ }\bibfield  {title} {\bibinfo
  {title} {Magnetic fluctuations near the {Mott} transition towards a spin
  liquid state},\ }\href {https://doi.org/10.1209\%2F0295-5075\%2F103\%2F17002}
  {\bibfield  {journal} {\bibinfo  {journal} {EPL}\ }\textbf {\bibinfo {volume}
  {103}},\ \bibinfo {pages} {17002} (\bibinfo {year} {2013})}\BibitemShut
  {NoStop}%
\bibitem [{\citenamefont {Werner}\ \emph {et~al.}(2008)\citenamefont {Werner},
  \citenamefont {Gull}, \citenamefont {Troyer},\ and\ \citenamefont
  {Millis}}]{Werner-2008}%
  \BibitemOpen
  \bibfield  {author} {\bibinfo {author} {\bibfnamefont {P.}~\bibnamefont
  {Werner}}, \bibinfo {author} {\bibfnamefont {E.}~\bibnamefont {Gull}},
  \bibinfo {author} {\bibfnamefont {M.}~\bibnamefont {Troyer}},\ and\ \bibinfo
  {author} {\bibfnamefont {A.~J.}\ \bibnamefont {Millis}},\ }\bibfield  {title}
  {\bibinfo {title} {Spin freezing transition and non-{Fermi}-liquid
  self-energy in a three-orbital model},\ }\href
  {https://doi.org/10.1103/PhysRevLett.101.166405} {\bibfield  {journal}
  {\bibinfo  {journal} {Phys. Rev. Lett.}\ }\textbf {\bibinfo {volume} {101}},\
  \bibinfo {pages} {166405} (\bibinfo {year} {2008})}\BibitemShut {NoStop}%
\bibitem [{\citenamefont {Werner}\ \emph {et~al.}(2018)\citenamefont {Werner},
  \citenamefont {Kim},\ and\ \citenamefont {Hoshino}}]{Werner-2018}%
  \BibitemOpen
  \bibfield  {author} {\bibinfo {author} {\bibfnamefont {P.}~\bibnamefont
  {Werner}}, \bibinfo {author} {\bibfnamefont {A.~J.}\ \bibnamefont {Kim}},\
  and\ \bibinfo {author} {\bibfnamefont {S.}~\bibnamefont {Hoshino}},\
  }\bibfield  {title} {\bibinfo {title} {Spin-freezing and the {Sachdev}-{Ye}
  model},\ }\href {https://doi.org/10.1209\%2F0295-5075\%2F124\%2F57002}
  {\bibfield  {journal} {\bibinfo  {journal} {EPL}\ }\textbf {\bibinfo {volume}
  {124}},\ \bibinfo {pages} {57002} (\bibinfo {year} {2018})}\BibitemShut
  {NoStop}%
\bibitem [{\citenamefont {Tsuji}\ and\ \citenamefont
  {Werner}(2019)}]{Tsuji-2019}%
  \BibitemOpen
  \bibfield  {author} {\bibinfo {author} {\bibfnamefont {N.}~\bibnamefont
  {Tsuji}}\ and\ \bibinfo {author} {\bibfnamefont {P.}~\bibnamefont {Werner}},\
  }\bibfield  {title} {\bibinfo {title} {Out-of-time-ordered correlators of the
  {Hubbard} model: {Sachdev}-{Ye}-{Kitaev} strange metal in the spin-freezing
  crossover region},\ }\href {https://doi.org/10.1103/PhysRevB.99.115132}
  {\bibfield  {journal} {\bibinfo  {journal} {Phys. Rev. B}\ }\textbf {\bibinfo
  {volume} {99}},\ \bibinfo {pages} {115132} (\bibinfo {year}
  {2019})}\BibitemShut {NoStop}%
\bibitem [{\citenamefont {Beenakker}(1997)}]{Beenakker-1997}%
  \BibitemOpen
  \bibfield  {author} {\bibinfo {author} {\bibfnamefont {C.~W.~J.}\
  \bibnamefont {Beenakker}},\ }\bibfield  {title} {\bibinfo {title}
  {Random-matrix theory of quantum transport},\ }\href
  {https://doi.org/10.1103/RevModPhys.69.731} {\bibfield  {journal} {\bibinfo
  {journal} {Rev. Mod. Phys.}\ }\textbf {\bibinfo {volume} {69}},\ \bibinfo
  {pages} {731} (\bibinfo {year} {1997})}\BibitemShut {NoStop}%
\bibitem [{\citenamefont {Gu}\ \emph {et~al.}(2017)\citenamefont {Gu},
  \citenamefont {Qi},\ and\ \citenamefont {Stanford}}]{Gu-2017}%
  \BibitemOpen
  \bibfield  {author} {\bibinfo {author} {\bibfnamefont {Y.}~\bibnamefont
  {Gu}}, \bibinfo {author} {\bibfnamefont {X.-L.}\ \bibnamefont {Qi}},\ and\
  \bibinfo {author} {\bibfnamefont {D.}~\bibnamefont {Stanford}},\ }\bibfield
  {title} {\bibinfo {title} {Local criticality, diffusion and chaos in
  generalized {Sachdev-Ye-Kitaev} models},\ }\href
  {https://doi.org/10.1007/JHEP05(2017)125} {\bibfield  {journal} {\bibinfo
  {journal} {J. High En. Phys.}\ }\textbf {\bibinfo {volume} {2017}},\ \bibinfo
  {pages} {125} (\bibinfo {year} {2017})}\BibitemShut {NoStop}%
\bibitem [{\citenamefont {Berkooz}\ \emph {et~al.}(2017)\citenamefont
  {Berkooz}, \citenamefont {Narayan}, \citenamefont {Rozali},\ and\
  \citenamefont {Sim{\'o}n}}]{Berkooz-2017}%
  \BibitemOpen
  \bibfield  {author} {\bibinfo {author} {\bibfnamefont {M.}~\bibnamefont
  {Berkooz}}, \bibinfo {author} {\bibfnamefont {P.}~\bibnamefont {Narayan}},
  \bibinfo {author} {\bibfnamefont {M.}~\bibnamefont {Rozali}},\ and\ \bibinfo
  {author} {\bibfnamefont {J.}~\bibnamefont {Sim{\'o}n}},\ }\bibfield  {title}
  {\bibinfo {title} {Higher dimensional generalizations of the {SYK} model},\
  }\href {https://link.springer.com/article/10.1007%2FJHEP01%282017%29138}
  {\bibfield  {journal} {\bibinfo  {journal} {J. High En. Phys.}\ }\textbf
  {\bibinfo {volume} {2017}},\ \bibinfo {pages} {138} (\bibinfo {year}
  {2017})}\BibitemShut {NoStop}%
\bibitem [{\citenamefont {Bi}\ \emph {et~al.}(2017)\citenamefont {Bi},
  \citenamefont {Jian}, \citenamefont {You}, \citenamefont {Pawlak},\ and\
  \citenamefont {Xu}}]{Bi-2017}%
  \BibitemOpen
  \bibfield  {author} {\bibinfo {author} {\bibfnamefont {Z.}~\bibnamefont
  {Bi}}, \bibinfo {author} {\bibfnamefont {C.-M.}\ \bibnamefont {Jian}},
  \bibinfo {author} {\bibfnamefont {Y.-Z.}\ \bibnamefont {You}}, \bibinfo
  {author} {\bibfnamefont {K.~A.}\ \bibnamefont {Pawlak}},\ and\ \bibinfo
  {author} {\bibfnamefont {C.}~\bibnamefont {Xu}},\ }\bibfield  {title}
  {\bibinfo {title} {Instability of the non-{Fermi}-liquid state of the
  {Sachdev-Ye-Kitaev} model},\ }\href
  {https://doi.org/10.1103/PhysRevB.95.205105} {\bibfield  {journal} {\bibinfo
  {journal} {Phys. Rev. B}\ }\textbf {\bibinfo {volume} {95}},\ \bibinfo
  {pages} {205105} (\bibinfo {year} {2017})}\BibitemShut {NoStop}%
\bibitem [{\citenamefont {Chowdhury}\ \emph {et~al.}(2018)\citenamefont
  {Chowdhury}, \citenamefont {Werman}, \citenamefont {Berg},\ and\
  \citenamefont {Senthil}}]{Chowdhury-2018}%
  \BibitemOpen
  \bibfield  {author} {\bibinfo {author} {\bibfnamefont {D.}~\bibnamefont
  {Chowdhury}}, \bibinfo {author} {\bibfnamefont {Y.}~\bibnamefont {Werman}},
  \bibinfo {author} {\bibfnamefont {E.}~\bibnamefont {Berg}},\ and\ \bibinfo
  {author} {\bibfnamefont {T.}~\bibnamefont {Senthil}},\ }\bibfield  {title}
  {\bibinfo {title} {Translationally invariant non-{Fermi}-liquid metals with
  critical {Fermi} surfaces: Solvable models},\ }\href
  {https://doi.org/10.1103/PhysRevX.8.031024} {\bibfield  {journal} {\bibinfo
  {journal} {Phys. Rev. X}\ }\textbf {\bibinfo {volume} {8}},\ \bibinfo {pages}
  {031024} (\bibinfo {year} {2018})}\BibitemShut {NoStop}%
\bibitem [{\citenamefont {Zhang}(2017)}]{Zhang-2017}%
  \BibitemOpen
  \bibfield  {author} {\bibinfo {author} {\bibfnamefont {P.}~\bibnamefont
  {Zhang}},\ }\bibfield  {title} {\bibinfo {title} {Dispersive
  {Sachdev-Ye-Kitaev} model: Band structure and quantum chaos},\ }\href
  {https://doi.org/10.1103/PhysRevB.96.205138} {\bibfield  {journal} {\bibinfo
  {journal} {Phys. Rev. B}\ }\textbf {\bibinfo {volume} {96}},\ \bibinfo
  {pages} {205138} (\bibinfo {year} {2017})}\BibitemShut {NoStop}%
\bibitem [{\citenamefont {Haldar}\ \emph {et~al.}(2018)\citenamefont {Haldar},
  \citenamefont {Banerjee},\ and\ \citenamefont {Shenoy}}]{Haldar-2018}%
  \BibitemOpen
  \bibfield  {author} {\bibinfo {author} {\bibfnamefont {A.}~\bibnamefont
  {Haldar}}, \bibinfo {author} {\bibfnamefont {S.}~\bibnamefont {Banerjee}},\
  and\ \bibinfo {author} {\bibfnamefont {V.~B.}\ \bibnamefont {Shenoy}},\
  }\bibfield  {title} {\bibinfo {title} {Higher-dimensional {Sachdev-Ye-Kitaev}
  non-{Fermi} liquids at {Lifshitz} transitions},\ }\href
  {https://doi.org/10.1103/PhysRevB.97.241106} {\bibfield  {journal} {\bibinfo
  {journal} {Phys. Rev. B}\ }\textbf {\bibinfo {volume} {97}},\ \bibinfo
  {pages} {241106} (\bibinfo {year} {2018})}\BibitemShut {NoStop}%
\bibitem [{\citenamefont {Garc\'{\i}a-Garc\'{\i}a}\ \emph
  {et~al.}(2021)\citenamefont {Garc\'{\i}a-Garc\'{\i}a}, \citenamefont
  {Zheng},\ and\ \citenamefont {Ziogas}}]{Garcia-Garcia-2021}%
  \BibitemOpen
  \bibfield  {author} {\bibinfo {author} {\bibfnamefont {A.~M.}\ \bibnamefont
  {Garc\'{\i}a-Garc\'{\i}a}}, \bibinfo {author} {\bibfnamefont {J.~P.}\
  \bibnamefont {Zheng}},\ and\ \bibinfo {author} {\bibfnamefont
  {V.}~\bibnamefont {Ziogas}},\ }\bibfield  {title} {\bibinfo {title} {Phase
  diagram of a two-site coupled complex {SYK} model},\ }\href
  {https://doi.org/10.1103/PhysRevD.103.106023} {\bibfield  {journal} {\bibinfo
   {journal} {Phys. Rev. D}\ }\textbf {\bibinfo {volume} {103}},\ \bibinfo
  {pages} {106023} (\bibinfo {year} {2021})}\BibitemShut {NoStop}%
\bibitem [{\citenamefont {Cha}\ \emph {et~al.}(2020)\citenamefont {Cha},
  \citenamefont {Wentzell}, \citenamefont {Parcollet}, \citenamefont
  {Georges},\ and\ \citenamefont {Kim}}]{Cha-2020}%
  \BibitemOpen
  \bibfield  {author} {\bibinfo {author} {\bibfnamefont {P.}~\bibnamefont
  {Cha}}, \bibinfo {author} {\bibfnamefont {N.}~\bibnamefont {Wentzell}},
  \bibinfo {author} {\bibfnamefont {O.}~\bibnamefont {Parcollet}}, \bibinfo
  {author} {\bibfnamefont {A.}~\bibnamefont {Georges}},\ and\ \bibinfo {author}
  {\bibfnamefont {E.-A.}\ \bibnamefont {Kim}},\ }\bibfield  {title} {\bibinfo
  {title} {Linear resistivity and {Sachdev-Ye-Kitaev (SYK)} spin liquid
  behavior in a quantum critical metal with spin-1/2 fermions},\ }\href
  {https://doi.org/10.1073/pnas.2003179117} {\bibfield  {journal} {\bibinfo
  {journal} {PNAS}\ }\textbf {\bibinfo {volume} {117}},\ \bibinfo {pages}
  {18341} (\bibinfo {year} {2020})}\BibitemShut {NoStop}%
\bibitem [{\citenamefont {Wang}(2020)}]{Wang-2020a}%
  \BibitemOpen
  \bibfield  {author} {\bibinfo {author} {\bibfnamefont {Y.}~\bibnamefont
  {Wang}},\ }\bibfield  {title} {\bibinfo {title} {Solvable strong-coupling
  quantum-dot model with a non-{Fermi}-liquid pairing transition},\ }\href
  {https://doi.org/10.1103/PhysRevLett.124.017002} {\bibfield  {journal}
  {\bibinfo  {journal} {Phys. Rev. Lett.}\ }\textbf {\bibinfo {volume} {124}},\
  \bibinfo {pages} {017002} (\bibinfo {year} {2020})}\BibitemShut {NoStop}%
\bibitem [{\citenamefont {Tinkham}(1996)}]{Tinkham-1996int}%
  \BibitemOpen
  \bibfield  {author} {\bibinfo {author} {\bibfnamefont {M.}~\bibnamefont
  {Tinkham}},\ }\href@noop {} {\emph {\bibinfo {title} {Introduction to
  superconductivity}}}\ (\bibinfo  {publisher} {Courier Corporation},\ \bibinfo
  {address} {North Chelmsford},\ \bibinfo {year} {1996})\BibitemShut {NoStop}%
\bibitem [{\citenamefont {Tinkham}\ and\ \citenamefont
  {Ferrell}(1959)}]{Tinkham-1959}%
  \BibitemOpen
  \bibfield  {author} {\bibinfo {author} {\bibfnamefont {M.}~\bibnamefont
  {Tinkham}}\ and\ \bibinfo {author} {\bibfnamefont {R.~A.}\ \bibnamefont
  {Ferrell}},\ }\bibfield  {title} {\bibinfo {title} {Determination of the
  superconducting skin depth from the energy gap and sum rule},\ }\href
  {https://doi.org/10.1103/PhysRevLett.2.331} {\bibfield  {journal} {\bibinfo
  {journal} {Phys. Rev. Lett.}\ }\textbf {\bibinfo {volume} {2}},\ \bibinfo
  {pages} {331} (\bibinfo {year} {1959})}\BibitemShut {NoStop}%
\bibitem [{\citenamefont {Uemura}\ \emph {et~al.}(1991)\citenamefont {Uemura},
  \citenamefont {Le}, \citenamefont {Luke}, \citenamefont {Sternlieb},
  \citenamefont {Wu}, \citenamefont {Brewer}, \citenamefont {Riseman},
  \citenamefont {Seaman}, \citenamefont {Maple}, \citenamefont {Ishikawa},
  \citenamefont {Hinks}, \citenamefont {Jorgensen}, \citenamefont {Saito},\
  and\ \citenamefont {Yamochi}}]{Uemura-1991}%
  \BibitemOpen
  \bibfield  {author} {\bibinfo {author} {\bibfnamefont {Y.~J.}\ \bibnamefont
  {Uemura}}, \bibinfo {author} {\bibfnamefont {L.~P.}\ \bibnamefont {Le}},
  \bibinfo {author} {\bibfnamefont {G.~M.}\ \bibnamefont {Luke}}, \bibinfo
  {author} {\bibfnamefont {B.~J.}\ \bibnamefont {Sternlieb}}, \bibinfo {author}
  {\bibfnamefont {W.~D.}\ \bibnamefont {Wu}}, \bibinfo {author} {\bibfnamefont
  {J.~H.}\ \bibnamefont {Brewer}}, \bibinfo {author} {\bibfnamefont {T.~M.}\
  \bibnamefont {Riseman}}, \bibinfo {author} {\bibfnamefont {C.~L.}\
  \bibnamefont {Seaman}}, \bibinfo {author} {\bibfnamefont {M.~B.}\
  \bibnamefont {Maple}}, \bibinfo {author} {\bibfnamefont {M.}~\bibnamefont
  {Ishikawa}}, \bibinfo {author} {\bibfnamefont {D.~G.}\ \bibnamefont {Hinks}},
  \bibinfo {author} {\bibfnamefont {J.~D.}\ \bibnamefont {Jorgensen}}, \bibinfo
  {author} {\bibfnamefont {G.}~\bibnamefont {Saito}},\ and\ \bibinfo {author}
  {\bibfnamefont {H.}~\bibnamefont {Yamochi}},\ }\bibfield  {title} {\bibinfo
  {title} {Basic similarities among cuprate, bismuthate, organic,
  {Chevrel}-phase, and heavy-fermion superconductors shown by penetration-depth
  measurements},\ }\href {https://doi.org/10.1103/PhysRevLett.66.2665}
  {\bibfield  {journal} {\bibinfo  {journal} {Phys. Rev. Lett.}\ }\textbf
  {\bibinfo {volume} {66}},\ \bibinfo {pages} {2665} (\bibinfo {year}
  {1991})}\BibitemShut {NoStop}%
\bibitem [{\citenamefont {Prozorov}\ and\ \citenamefont
  {Giannetta}(2006)}]{Prozorov-2006}%
  \BibitemOpen
  \bibfield  {author} {\bibinfo {author} {\bibfnamefont {R.}~\bibnamefont
  {Prozorov}}\ and\ \bibinfo {author} {\bibfnamefont {R.~W.}\ \bibnamefont
  {Giannetta}},\ }\bibfield  {title} {\bibinfo {title} {Magnetic penetration
  depth in unconventional superconductors},\ }\href
  {https://doi.org/10.1088/0953-2048/19/8/R01} {\bibfield  {journal} {\bibinfo
  {journal} {Supercond. Sci. Technol.}\ }\textbf {\bibinfo {volume} {19}},\
  \bibinfo {pages} {R41} (\bibinfo {year} {2006})}\BibitemShut {NoStop}%
\bibitem [{\citenamefont {Altland}\ and\ \citenamefont
  {Simons}(2010)}]{Altland-cm2010}%
  \BibitemOpen
  \bibfield  {author} {\bibinfo {author} {\bibfnamefont {A.}~\bibnamefont
  {Altland}}\ and\ \bibinfo {author} {\bibfnamefont {B.~D.}\ \bibnamefont
  {Simons}},\ }\href {https://doi.org/10.1017/CBO9780511789984} {\emph
  {\bibinfo {title} {Condensed Matter Field Theory}}},\ \bibinfo {edition}
  {2nd}\ ed.\ (\bibinfo  {publisher} {Cambridge University Press},\ \bibinfo
  {year} {2010})\BibitemShut {NoStop}%
\bibitem [{\citenamefont {Valentinis}\ \emph {et~al.}(2023)\citenamefont
  {Valentinis}, \citenamefont {Inkof},\ and\ \citenamefont
  {Schmalian}}]{short-paper}%
  \BibitemOpen
  \bibfield  {author} {\bibinfo {author} {\bibfnamefont {D.}~\bibnamefont
  {Valentinis}}, \bibinfo {author} {\bibfnamefont {G.~A.}\ \bibnamefont
  {Inkof}},\ and\ \bibinfo {author} {\bibfnamefont {J.}~\bibnamefont
  {Schmalian}},\ }\bibfield  {title} {\bibinfo {title} {{BCS} to incoherent
  superconductivity crossovers in the {Yukawa-SYK} model on a lattice},\ }\href
  {https://doi.org/10.48550/arXiv.2302.13134} {\bibfield  {journal} {\bibinfo
  {journal} {arXiv:2302.13134}\ } (\bibinfo {year} {2023})}\BibitemShut
  {NoStop}%
\bibitem [{\citenamefont {Combescot}(1995)}]{Combescot-1994}%
  \BibitemOpen
  \bibfield  {author} {\bibinfo {author} {\bibfnamefont {R.}~\bibnamefont
  {Combescot}},\ }\bibfield  {title} {\bibinfo {title} {Strong-coupling limit
  of {Eliashberg} theory},\ }\href {https://doi.org/10.1103/PhysRevB.51.11625}
  {\bibfield  {journal} {\bibinfo  {journal} {Phys. Rev. B}\ }\textbf {\bibinfo
  {volume} {51}},\ \bibinfo {pages} {11625} (\bibinfo {year}
  {1995})}\BibitemShut {NoStop}%
\bibitem [{\citenamefont {Berthod}(2018)}]{Berthod-2018}%
  \BibitemOpen
  \bibfield  {author} {\bibinfo {author} {\bibfnamefont {C.}~\bibnamefont
  {Berthod}},\ }\href {https://doi.org/10.1088/978-0-7503-1741-2} {\emph
  {\bibinfo {title} {Spectroscopic Probes of Quantum Matter}}},\ 2053-2563\
  (\bibinfo  {publisher} {IOP Publishing},\ \bibinfo {year} {2018})\BibitemShut
  {NoStop}%
\bibitem [{\citenamefont {Marsiglio}\ and\ \citenamefont
  {Carbotte}(2008)}]{Marsiglio-2008}%
  \BibitemOpen
  \bibfield  {author} {\bibinfo {author} {\bibfnamefont {F.}~\bibnamefont
  {Marsiglio}}\ and\ \bibinfo {author} {\bibfnamefont {J.~P.}\ \bibnamefont
  {Carbotte}},\ }\bibinfo {title} {Electron-phonon superconductivity},\ in\
  \href {https://doi.org/10.1007/978-3-540-73253-2_3} {\emph {\bibinfo
  {booktitle} {Superconductivity: Conventional and Unconventional
  Superconductors}}},\ \bibinfo {editor} {edited by\ \bibinfo {editor}
  {\bibfnamefont {K.~H.}\ \bibnamefont {Bennemann}}\ and\ \bibinfo {editor}
  {\bibfnamefont {J.~B.}\ \bibnamefont {Ketterson}}}\ (\bibinfo  {publisher}
  {Springer Berlin Heidelberg},\ \bibinfo {address} {Berlin, Heidelberg},\
  \bibinfo {year} {2008})\ pp.\ \bibinfo {pages} {73--162}\BibitemShut
  {NoStop}%
\bibitem [{\citenamefont {Marsiglio}(2020)}]{Marsiglio-2020}%
  \BibitemOpen
  \bibfield  {author} {\bibinfo {author} {\bibfnamefont {F.}~\bibnamefont
  {Marsiglio}},\ }\bibfield  {title} {\bibinfo {title} {Eliashberg theory: A
  short review},\ }\href
  {http://www.sciencedirect.com/science/article/pii/S000349162030035X}
  {\bibfield  {journal} {\bibinfo  {journal} {Ann. Phys. (N. Y.)}\ }\textbf
  {\bibinfo {volume} {417}},\ \bibinfo {pages} {168102} (\bibinfo {year}
  {2020})}\BibitemShut {NoStop}%
\bibitem [{Note2()}]{Note2}%
  \BibitemOpen
  \bibinfo {note} {Furthermore, since the fermions are dispersionless,
  $A(\omega )$ from Eq.\spacefactor \@m {} (\ref {eq:A_def}) also counts how
  many states are available at the energy $\omega $, i.e.\spacefactor \@m {},
  it is equivalent to the single-particle density of states of the
  system}\BibitemShut {NoStop}%
\bibitem [{\citenamefont {Anderson}\ \emph {et~al.}(2010)\citenamefont
  {Anderson}, \citenamefont {Guionnet},\ and\ \citenamefont
  {Zeitouni}}]{Anderson-2010int}%
  \BibitemOpen
  \bibfield  {author} {\bibinfo {author} {\bibfnamefont {G.~W.}\ \bibnamefont
  {Anderson}}, \bibinfo {author} {\bibfnamefont {A.}~\bibnamefont {Guionnet}},\
  and\ \bibinfo {author} {\bibfnamefont {O.}~\bibnamefont {Zeitouni}},\ }\href
  {https://www.cambridge.org/core/books/an-introduction-to-random-matrices/8992DA8EB0386651E8DA8214A1FC7241}
  {\emph {\bibinfo {title} {An introduction to random matrices}}},\ Vol.\
  \bibinfo {volume} {118}\ (\bibinfo  {publisher} {Cambridge university
  press},\ \bibinfo {address} {Cambridge},\ \bibinfo {year} {2010})\BibitemShut
  {NoStop}%
\bibitem [{\citenamefont {Maldacena}\ \emph
  {et~al.}(2016{\natexlab{b}})\citenamefont {Maldacena}, \citenamefont
  {Shenker},\ and\ \citenamefont {Stanford}}]{Maldacena-2016c}%
  \BibitemOpen
  \bibfield  {author} {\bibinfo {author} {\bibfnamefont {J.}~\bibnamefont
  {Maldacena}}, \bibinfo {author} {\bibfnamefont {S.~H.}\ \bibnamefont
  {Shenker}},\ and\ \bibinfo {author} {\bibfnamefont {D.}~\bibnamefont
  {Stanford}},\ }\bibfield  {title} {\bibinfo {title} {A bound on chaos},\
  }\href {https://doi.org/10.1007/JHEP08(2016)106} {\bibfield  {journal}
  {\bibinfo  {journal} {J. High En. Phys.}\ }\textbf {\bibinfo {volume}
  {2016}},\ \bibinfo {pages} {1} (\bibinfo {year}
  {2016}{\natexlab{b}})}\BibitemShut {NoStop}%
\bibitem [{\citenamefont {Schmalian}\ \emph {et~al.}(1998)\citenamefont
  {Schmalian}, \citenamefont {Pines},\ and\ \citenamefont
  {Stojkovi\ifmmode~\acute{c}\else \'{c}\fi{}}}]{Schmalian-1998}%
  \BibitemOpen
  \bibfield  {author} {\bibinfo {author} {\bibfnamefont {J.}~\bibnamefont
  {Schmalian}}, \bibinfo {author} {\bibfnamefont {D.}~\bibnamefont {Pines}},\
  and\ \bibinfo {author} {\bibfnamefont {B.}~\bibnamefont
  {Stojkovi\ifmmode~\acute{c}\else \'{c}\fi{}}},\ }\bibfield  {title} {\bibinfo
  {title} {Weak pseudogap behavior in the underdoped cuprate superconductors},\
  }\href {https://doi.org/10.1103/PhysRevLett.80.3839} {\bibfield  {journal}
  {\bibinfo  {journal} {Phys. Rev. Lett.}\ }\textbf {\bibinfo {volume} {80}},\
  \bibinfo {pages} {3839} (\bibinfo {year} {1998})}\BibitemShut {NoStop}%
\bibitem [{\citenamefont {Schmalian}\ \emph {et~al.}(1999)\citenamefont
  {Schmalian}, \citenamefont {Pines},\ and\ \citenamefont
  {Stojkovi\ifmmode~\acute{c}\else \'{c}\fi{}}}]{Schmalian-1999}%
  \BibitemOpen
  \bibfield  {author} {\bibinfo {author} {\bibfnamefont {J.}~\bibnamefont
  {Schmalian}}, \bibinfo {author} {\bibfnamefont {D.}~\bibnamefont {Pines}},\
  and\ \bibinfo {author} {\bibfnamefont {B.}~\bibnamefont
  {Stojkovi\ifmmode~\acute{c}\else \'{c}\fi{}}},\ }\bibfield  {title} {\bibinfo
  {title} {Microscopic theory of weak pseudogap behavior in the underdoped
  cuprate superconductors: General theory and quasiparticle properties},\
  }\href {https://doi.org/10.1103/PhysRevB.60.667} {\bibfield  {journal}
  {\bibinfo  {journal} {Phys. Rev. B}\ }\textbf {\bibinfo {volume} {60}},\
  \bibinfo {pages} {667} (\bibinfo {year} {1999})}\BibitemShut {NoStop}%
\bibitem [{\citenamefont {Sachdev}(2015)}]{Sachdev-2015}%
  \BibitemOpen
  \bibfield  {author} {\bibinfo {author} {\bibfnamefont {S.}~\bibnamefont
  {Sachdev}},\ }\bibfield  {title} {\bibinfo {title} {{Bekenstein-Hawking}
  entropy and strange metals},\ }\href
  {https://doi.org/10.1103/PhysRevX.5.041025} {\bibfield  {journal} {\bibinfo
  {journal} {Phys. Rev. X}\ }\textbf {\bibinfo {volume} {5}},\ \bibinfo {pages}
  {041025} (\bibinfo {year} {2015})}\BibitemShut {NoStop}%
\bibitem [{\citenamefont {Milekhin}(2022)}]{Milekhin-2022_preprint}%
  \BibitemOpen
  \bibfield  {author} {\bibinfo {author} {\bibfnamefont {A.}~\bibnamefont
  {Milekhin}},\ }\bibfield  {title} {\bibinfo {title} {On minimal residual
  entropy in non-{Fermi} liquids},\ }\href {https://arxiv.org/abs/2207.01588}
  {\bibfield  {journal} {\bibinfo  {journal} {arXiv:2207.01588}\ } (\bibinfo
  {year} {2022})}\BibitemShut {NoStop}%
\bibitem [{\citenamefont {Emery}\ and\ \citenamefont
  {Kivelson}(1995{\natexlab{b}})}]{Emery-1995}%
  \BibitemOpen
  \bibfield  {author} {\bibinfo {author} {\bibfnamefont {V.}~\bibnamefont
  {Emery}}\ and\ \bibinfo {author} {\bibfnamefont {S.}~\bibnamefont
  {Kivelson}},\ }\bibfield  {title} {\bibinfo {title} {Importance of phase
  fluctuations in superconductors with small superfluid density},\ }\href
  {https://doi.org/10.1038/374434a0} {\bibfield  {journal} {\bibinfo  {journal}
  {Nature}\ }\textbf {\bibinfo {volume} {374}},\ \bibinfo {pages} {434}
  (\bibinfo {year} {1995}{\natexlab{b}})}\BibitemShut {NoStop}%
\bibitem [{\citenamefont {Bardeen}\ \emph
  {et~al.}(1957{\natexlab{a}})\citenamefont {Bardeen}, \citenamefont {Cooper},\
  and\ \citenamefont {Schrieffer}}]{Bardeen-1957a}%
  \BibitemOpen
  \bibfield  {author} {\bibinfo {author} {\bibfnamefont {J.}~\bibnamefont
  {Bardeen}}, \bibinfo {author} {\bibfnamefont {L.~N.}\ \bibnamefont
  {Cooper}},\ and\ \bibinfo {author} {\bibfnamefont {J.~R.}\ \bibnamefont
  {Schrieffer}},\ }\bibfield  {title} {\bibinfo {title} {Microscopic theory of
  superconductivity},\ }\href {https://doi.org/10.1103/PhysRev.106.162}
  {\bibfield  {journal} {\bibinfo  {journal} {Phys. Rev.}\ }\textbf {\bibinfo
  {volume} {106}},\ \bibinfo {pages} {162} (\bibinfo {year}
  {1957}{\natexlab{a}})}\BibitemShut {NoStop}%
\bibitem [{\citenamefont {Bardeen}\ \emph
  {et~al.}(1957{\natexlab{b}})\citenamefont {Bardeen}, \citenamefont {Cooper},\
  and\ \citenamefont {Schrieffer}}]{Bardeen-1957b}%
  \BibitemOpen
  \bibfield  {author} {\bibinfo {author} {\bibfnamefont {J.}~\bibnamefont
  {Bardeen}}, \bibinfo {author} {\bibfnamefont {L.~N.}\ \bibnamefont
  {Cooper}},\ and\ \bibinfo {author} {\bibfnamefont {J.~R.}\ \bibnamefont
  {Schrieffer}},\ }\bibfield  {title} {\bibinfo {title} {Theory of
  superconductivity},\ }\href {https://doi.org/10.1103/PhysRev.108.1175}
  {\bibfield  {journal} {\bibinfo  {journal} {Phys. Rev.}\ }\textbf {\bibinfo
  {volume} {108}},\ \bibinfo {pages} {1175} (\bibinfo {year}
  {1957}{\natexlab{b}})}\BibitemShut {NoStop}%
\bibitem [{\citenamefont {Schrieffer}(2018)}]{Schrieffer-1963th}%
  \BibitemOpen
  \bibfield  {author} {\bibinfo {author} {\bibfnamefont {J.~R.}\ \bibnamefont
  {Schrieffer}},\ }\href@noop {} {\emph {\bibinfo {title} {Theory of
  superconductivity}}}\ (\bibinfo  {publisher} {CRC Press},\ \bibinfo {address}
  {Cleveland},\ \bibinfo {year} {2018})\BibitemShut {NoStop}%
\bibitem [{\citenamefont {Anderson}(1959)}]{Anderson-1959}%
  \BibitemOpen
  \bibfield  {author} {\bibinfo {author} {\bibfnamefont {P.}~\bibnamefont
  {Anderson}},\ }\bibfield  {title} {\bibinfo {title} {Theory of dirty
  superconductors},\ }\href
  {https://doi.org/https://doi.org/10.1016/0022-3697(59)90036-8} {\bibfield
  {journal} {\bibinfo  {journal} {J. Phys. Chem. Solids}\ }\textbf {\bibinfo
  {volume} {11}},\ \bibinfo {pages} {26} (\bibinfo {year} {1959})}\BibitemShut
  {NoStop}%
\bibitem [{\citenamefont {Abrikosov}\ and\ \citenamefont
  {Gorkov}(1959{\natexlab{a}})}]{Abrikosov-1959a}%
  \BibitemOpen
  \bibfield  {author} {\bibinfo {author} {\bibfnamefont {A.}~\bibnamefont
  {Abrikosov}}\ and\ \bibinfo {author} {\bibfnamefont {L.}~\bibnamefont
  {Gorkov}},\ }\bibfield  {title} {\bibinfo {title} {On the theory of
  superconducting alloys. 1. the electrodynamics of alloys at absolute zero},\
  }\href@noop {} {\bibfield  {journal} {\bibinfo  {journal} {Zh. Eksp. Teor.
  Fiz.}\ }\textbf {\bibinfo {volume} {35}},\ \bibinfo {pages} {1558} (\bibinfo
  {year} {1959}{\natexlab{a}})},\ \bibinfo {note}
  {\href{http://83.149.229.155/cgi-bin/dn/e_008_06_1090.pdf}{[Sov. Phys. JETP
  \textbf{8}, 1090 (1959)]}}\BibitemShut {NoStop}%
\bibitem [{\citenamefont {Abrikosov}\ and\ \citenamefont
  {Gorkov}(1959{\natexlab{b}})}]{Abrikosov-1959b}%
  \BibitemOpen
  \bibfield  {author} {\bibinfo {author} {\bibfnamefont {A.}~\bibnamefont
  {Abrikosov}}\ and\ \bibinfo {author} {\bibfnamefont {L.}~\bibnamefont
  {Gorkov}},\ }\bibfield  {title} {\bibinfo {title} {Superconducting alloys at
  finite temperatures},\ }\href@noop {} {\bibfield  {journal} {\bibinfo
  {journal} {Zh. Eksp. Teor. Fiz.}\ }\textbf {\bibinfo {volume} {36}},\
  \bibinfo {pages} {319} (\bibinfo {year} {1959}{\natexlab{b}})},\ \bibinfo
  {note} {\href{http://83.149.229.155/cgi-bin/dn/e_009_01_0220.pdf}{[Sov. Phys.
  JETP \textbf{9}, 220 (1959)]}}\BibitemShut {NoStop}%
\bibitem [{\citenamefont {Abrikosov}\ and\ \citenamefont
  {Gorkov}(1961)}]{Abrikosov-1961}%
  \BibitemOpen
  \bibfield  {author} {\bibinfo {author} {\bibfnamefont {A.}~\bibnamefont
  {Abrikosov}}\ and\ \bibinfo {author} {\bibfnamefont {L.}~\bibnamefont
  {Gorkov}},\ }\bibfield  {title} {\bibinfo {title} {Contribution to the theory
  of superconducting alloys with paramagnetic impurities},\ }\href@noop {}
  {\bibfield  {journal} {\bibinfo  {journal} {Zh. Eksp. Teor. Fiz.}\ }\textbf
  {\bibinfo {volume} {39}},\ \bibinfo {pages} {1781} (\bibinfo {year}
  {1961})},\ \bibinfo {note}
  {\href{http://83.149.229.155/cgi-bin/dn/e_009_01_0220.pdf}{[Sov. Phys. JETP
  \textbf{12}, 1243 (1961)]}}\BibitemShut {NoStop}%
\bibitem [{\citenamefont {Potter}\ and\ \citenamefont
  {Lee}(2011)}]{Potter-2011}%
  \BibitemOpen
  \bibfield  {author} {\bibinfo {author} {\bibfnamefont {A.~C.}\ \bibnamefont
  {Potter}}\ and\ \bibinfo {author} {\bibfnamefont {P.~A.}\ \bibnamefont
  {Lee}},\ }\bibfield  {title} {\bibinfo {title} {Engineering a $p+\mathit{ip}$
  superconductor: Comparison of topological insulator and {Rashba}
  spin-orbit-coupled materials},\ }\href
  {https://doi.org/10.1103/PhysRevB.83.184520} {\bibfield  {journal} {\bibinfo
  {journal} {Phys. Rev. B}\ }\textbf {\bibinfo {volume} {83}},\ \bibinfo
  {pages} {184520} (\bibinfo {year} {2011})}\BibitemShut {NoStop}%
\bibitem [{\citenamefont {Abrikosov}\ \emph {et~al.}(2012)\citenamefont
  {Abrikosov}, \citenamefont {Gorkov},\ and\ \citenamefont
  {Dzyaloshinski}}]{Abrikosov-2012meth}%
  \BibitemOpen
  \bibfield  {author} {\bibinfo {author} {\bibfnamefont {A.~A.}\ \bibnamefont
  {Abrikosov}}, \bibinfo {author} {\bibfnamefont {L.~P.}\ \bibnamefont
  {Gorkov}},\ and\ \bibinfo {author} {\bibfnamefont {I.~E.}\ \bibnamefont
  {Dzyaloshinski}},\ }\href@noop {} {\emph {\bibinfo {title} {Methods of
  quantum field theory in statistical physics}}}\ (\bibinfo  {publisher}
  {Courier Corporation},\ \bibinfo {address} {North Chelmsford},\ \bibinfo
  {year} {2012})\BibitemShut {NoStop}%
\bibitem [{\citenamefont {Kang}\ and\ \citenamefont
  {Fernandes}(2016)}]{Kang-2016}%
  \BibitemOpen
  \bibfield  {author} {\bibinfo {author} {\bibfnamefont {J.}~\bibnamefont
  {Kang}}\ and\ \bibinfo {author} {\bibfnamefont {R.~M.}\ \bibnamefont
  {Fernandes}},\ }\bibfield  {title} {\bibinfo {title} {Robustness of quantum
  critical pairing against disorder},\ }\href
  {https://doi.org/10.1103/PhysRevB.93.224514} {\bibfield  {journal} {\bibinfo
  {journal} {Phys. Rev. B}\ }\textbf {\bibinfo {volume} {93}},\ \bibinfo
  {pages} {224514} (\bibinfo {year} {2016})}\BibitemShut {NoStop}%
\bibitem [{\citenamefont {Millis}\ \emph {et~al.}(1988)\citenamefont {Millis},
  \citenamefont {Sachdev},\ and\ \citenamefont {Varma}}]{Millis-1988}%
  \BibitemOpen
  \bibfield  {author} {\bibinfo {author} {\bibfnamefont {A.~J.}\ \bibnamefont
  {Millis}}, \bibinfo {author} {\bibfnamefont {S.}~\bibnamefont {Sachdev}},\
  and\ \bibinfo {author} {\bibfnamefont {C.~M.}\ \bibnamefont {Varma}},\
  }\bibfield  {title} {\bibinfo {title} {Inelastic scattering and pair breaking
  in anisotropic and isotropic superconductors},\ }\href
  {https://doi.org/10.1103/PhysRevB.37.4975} {\bibfield  {journal} {\bibinfo
  {journal} {Phys. Rev. B}\ }\textbf {\bibinfo {volume} {37}},\ \bibinfo
  {pages} {4975} (\bibinfo {year} {1988})}\BibitemShut {NoStop}%
\bibitem [{\citenamefont {Abanov}\ \emph {et~al.}(2008)\citenamefont {Abanov},
  \citenamefont {Chubukov},\ and\ \citenamefont {Norman}}]{Abanov-2008}%
  \BibitemOpen
  \bibfield  {author} {\bibinfo {author} {\bibfnamefont {A.}~\bibnamefont
  {Abanov}}, \bibinfo {author} {\bibfnamefont {A.~V.}\ \bibnamefont
  {Chubukov}},\ and\ \bibinfo {author} {\bibfnamefont {M.~R.}\ \bibnamefont
  {Norman}},\ }\bibfield  {title} {\bibinfo {title} {Gap anisotropy and
  universal pairing scale in a spin-fluctuation model of cuprate
  superconductors},\ }\href {https://doi.org/10.1103/PhysRevB.78.220507}
  {\bibfield  {journal} {\bibinfo  {journal} {Phys. Rev. B}\ }\textbf {\bibinfo
  {volume} {78}},\ \bibinfo {pages} {220507} (\bibinfo {year}
  {2008})}\BibitemShut {NoStop}%
\bibitem [{\citenamefont {Allen}\ and\ \citenamefont
  {Dynes}(1975)}]{Allen-1975}%
  \BibitemOpen
  \bibfield  {author} {\bibinfo {author} {\bibfnamefont {P.~B.}\ \bibnamefont
  {Allen}}\ and\ \bibinfo {author} {\bibfnamefont {R.~C.}\ \bibnamefont
  {Dynes}},\ }\bibfield  {title} {\bibinfo {title} {Transition temperature of
  strong-coupled superconductors reanalyzed},\ }\href
  {https://doi.org/10.1103/PhysRevB.12.905} {\bibfield  {journal} {\bibinfo
  {journal} {Phys. Rev. B}\ }\textbf {\bibinfo {volume} {12}},\ \bibinfo
  {pages} {905} (\bibinfo {year} {1975})}\BibitemShut {NoStop}%
\bibitem [{\citenamefont {Wang}\ \emph {et~al.}(2016)\citenamefont {Wang},
  \citenamefont {Abanov}, \citenamefont {Altshuler}, \citenamefont
  {Yuzbashyan},\ and\ \citenamefont {Chubukov}}]{Wang-2016}%
  \BibitemOpen
  \bibfield  {author} {\bibinfo {author} {\bibfnamefont {Y.}~\bibnamefont
  {Wang}}, \bibinfo {author} {\bibfnamefont {A.}~\bibnamefont {Abanov}},
  \bibinfo {author} {\bibfnamefont {B.~L.}\ \bibnamefont {Altshuler}}, \bibinfo
  {author} {\bibfnamefont {E.~A.}\ \bibnamefont {Yuzbashyan}},\ and\ \bibinfo
  {author} {\bibfnamefont {A.~V.}\ \bibnamefont {Chubukov}},\ }\bibfield
  {title} {\bibinfo {title} {Superconductivity near a quantum-critical point:
  The special role of the first {Matsubara} frequency},\ }\href
  {https://doi.org/10.1103/PhysRevLett.117.157001} {\bibfield  {journal}
  {\bibinfo  {journal} {Phys. Rev. Lett.}\ }\textbf {\bibinfo {volume} {117}},\
  \bibinfo {pages} {157001} (\bibinfo {year} {2016})}\BibitemShut {NoStop}%
\bibitem [{\citenamefont {Wu}\ \emph {et~al.}(2019{\natexlab{a}})\citenamefont
  {Wu}, \citenamefont {Abanov}, \citenamefont {Wang},\ and\ \citenamefont
  {Chubukov}}]{Wu-2019}%
  \BibitemOpen
  \bibfield  {author} {\bibinfo {author} {\bibfnamefont {Y.-M.}\ \bibnamefont
  {Wu}}, \bibinfo {author} {\bibfnamefont {A.}~\bibnamefont {Abanov}}, \bibinfo
  {author} {\bibfnamefont {Y.}~\bibnamefont {Wang}},\ and\ \bibinfo {author}
  {\bibfnamefont {A.~V.}\ \bibnamefont {Chubukov}},\ }\bibfield  {title}
  {\bibinfo {title} {Special role of the first {Matsubara} frequency for
  superconductivity near a quantum critical point: Nonlinear gap equation below
  ${T}_{c}$ and spectral properties in real frequencies},\ }\href
  {https://doi.org/10.1103/PhysRevB.99.144512} {\bibfield  {journal} {\bibinfo
  {journal} {Phys. Rev. B}\ }\textbf {\bibinfo {volume} {99}},\ \bibinfo
  {pages} {144512} (\bibinfo {year} {2019}{\natexlab{a}})}\BibitemShut
  {NoStop}%
\bibitem [{\citenamefont {Carbotte}(1990)}]{Carbotte-1990}%
  \BibitemOpen
  \bibfield  {author} {\bibinfo {author} {\bibfnamefont {J.~P.}\ \bibnamefont
  {Carbotte}},\ }\bibfield  {title} {\bibinfo {title} {Properties of
  boson-exchange superconductors},\ }\href
  {https://doi.org/10.1103/RevModPhys.62.1027} {\bibfield  {journal} {\bibinfo
  {journal} {Rev. Mod. Phys.}\ }\textbf {\bibinfo {volume} {62}},\ \bibinfo
  {pages} {1027} (\bibinfo {year} {1990})}\BibitemShut {NoStop}%
\bibitem [{\citenamefont {Wu}\ \emph {et~al.}(2019{\natexlab{b}})\citenamefont
  {Wu}, \citenamefont {Abanov},\ and\ \citenamefont {Chubukov}}]{Wu-2019b}%
  \BibitemOpen
  \bibfield  {author} {\bibinfo {author} {\bibfnamefont {Y.-M.}\ \bibnamefont
  {Wu}}, \bibinfo {author} {\bibfnamefont {A.}~\bibnamefont {Abanov}},\ and\
  \bibinfo {author} {\bibfnamefont {A.~V.}\ \bibnamefont {Chubukov}},\
  }\bibfield  {title} {\bibinfo {title} {Pairing in quantum critical systems:
  Transition temperature, pairing gap, and their ratio},\ }\href
  {https://doi.org/10.1103/PhysRevB.99.014502} {\bibfield  {journal} {\bibinfo
  {journal} {Phys. Rev. B}\ }\textbf {\bibinfo {volume} {99}},\ \bibinfo
  {pages} {014502} (\bibinfo {year} {2019}{\natexlab{b}})}\BibitemShut
  {NoStop}%
\bibitem [{\citenamefont {Marsiglio}\ and\ \citenamefont
  {Carbotte}(1991)}]{Marsiglio-1991}%
  \BibitemOpen
  \bibfield  {author} {\bibinfo {author} {\bibfnamefont {F.}~\bibnamefont
  {Marsiglio}}\ and\ \bibinfo {author} {\bibfnamefont {J.~P.}\ \bibnamefont
  {Carbotte}},\ }\bibfield  {title} {\bibinfo {title} {Gap function and density
  of states in the strong-coupling limit for an electron-boson system},\ }\href
  {https://doi.org/10.1103/PhysRevB.43.5355} {\bibfield  {journal} {\bibinfo
  {journal} {Phys. Rev. B}\ }\textbf {\bibinfo {volume} {43}},\ \bibinfo
  {pages} {5355} (\bibinfo {year} {1991})}\BibitemShut {NoStop}%
\bibitem [{\citenamefont {Karakozov}\ \emph {et~al.}(1991)\citenamefont
  {Karakozov}, \citenamefont {Maksimov},\ and\ \citenamefont
  {Mikhailovsky}}]{Karakozov-1991}%
  \BibitemOpen
  \bibfield  {author} {\bibinfo {author} {\bibfnamefont {A.}~\bibnamefont
  {Karakozov}}, \bibinfo {author} {\bibfnamefont {E.}~\bibnamefont
  {Maksimov}},\ and\ \bibinfo {author} {\bibfnamefont {A.}~\bibnamefont
  {Mikhailovsky}},\ }\bibfield  {title} {\bibinfo {title} {The investigation of
  {Eliashberg} equations for superconductors with strong electron-phonon
  interaction},\ }\href
  {https://doi.org/https://doi.org/10.1016/0038-1098(91)90556-B} {\bibfield
  {journal} {\bibinfo  {journal} {Solid State Commun.}\ }\textbf {\bibinfo
  {volume} {79}},\ \bibinfo {pages} {329} (\bibinfo {year} {1991})}\BibitemShut
  {NoStop}%
\bibitem [{\citenamefont {Fisher}\ \emph {et~al.}(1973)\citenamefont {Fisher},
  \citenamefont {Barber},\ and\ \citenamefont {Jasnow}}]{Fisher-1973}%
  \BibitemOpen
  \bibfield  {author} {\bibinfo {author} {\bibfnamefont {M.~E.}\ \bibnamefont
  {Fisher}}, \bibinfo {author} {\bibfnamefont {M.~N.}\ \bibnamefont {Barber}},\
  and\ \bibinfo {author} {\bibfnamefont {D.}~\bibnamefont {Jasnow}},\
  }\bibfield  {title} {\bibinfo {title} {Helicity modulus, superfluidity, and
  scaling in isotropic systems},\ }\href
  {https://doi.org/10.1103/PhysRevA.8.1111} {\bibfield  {journal} {\bibinfo
  {journal} {Phys. Rev. A}\ }\textbf {\bibinfo {volume} {8}},\ \bibinfo {pages}
  {1111} (\bibinfo {year} {1973})}\BibitemShut {NoStop}%
\bibitem [{\citenamefont {Taylor}\ \emph {et~al.}(2006)\citenamefont {Taylor},
  \citenamefont {Griffin}, \citenamefont {Fukushima},\ and\ \citenamefont
  {Ohashi}}]{Taylor-2006}%
  \BibitemOpen
  \bibfield  {author} {\bibinfo {author} {\bibfnamefont {E.}~\bibnamefont
  {Taylor}}, \bibinfo {author} {\bibfnamefont {A.}~\bibnamefont {Griffin}},
  \bibinfo {author} {\bibfnamefont {N.}~\bibnamefont {Fukushima}},\ and\
  \bibinfo {author} {\bibfnamefont {Y.}~\bibnamefont {Ohashi}},\ }\bibfield
  {title} {\bibinfo {title} {Pairing fluctuations and the superfluid density
  through the {BCS-BEC} crossover},\ }\href
  {https://doi.org/10.1103/PhysRevA.74.063626} {\bibfield  {journal} {\bibinfo
  {journal} {Phys. Rev. A}\ }\textbf {\bibinfo {volume} {74}},\ \bibinfo
  {pages} {063626} (\bibinfo {year} {2006})}\BibitemShut {NoStop}%
\bibitem [{\citenamefont {Taylor}\ \emph {et~al.}(2007)\citenamefont {Taylor},
  \citenamefont {Griffin}, \citenamefont {Fukushima},\ and\ \citenamefont
  {Ohashi}}]{Taylor-2007}%
  \BibitemOpen
  \bibfield  {author} {\bibinfo {author} {\bibfnamefont {E.}~\bibnamefont
  {Taylor}}, \bibinfo {author} {\bibfnamefont {A.}~\bibnamefont {Griffin}},
  \bibinfo {author} {\bibfnamefont {N.}~\bibnamefont {Fukushima}},\ and\
  \bibinfo {author} {\bibfnamefont {Y.}~\bibnamefont {Ohashi}},\ }\bibfield
  {title} {\bibinfo {title} {Publisher's note: Pairing fluctuations and the
  superfluid density through the {BCS-BEC} crossover {[Phys. Rev. A 74, 063626
  (2006)]}},\ }\href {https://doi.org/10.1103/PhysRevA.75.019902} {\bibfield
  {journal} {\bibinfo  {journal} {Phys. Rev. A}\ }\textbf {\bibinfo {volume}
  {75}},\ \bibinfo {pages} {019902} (\bibinfo {year} {2007})}\BibitemShut
  {NoStop}%
\bibitem [{\citenamefont {Dressel}\ and\ \citenamefont
  {Gr\"{u}ner}(2002)}]{Dressel-2001}%
  \BibitemOpen
  \bibfield  {author} {\bibinfo {author} {\bibfnamefont {M.}~\bibnamefont
  {Dressel}}\ and\ \bibinfo {author} {\bibfnamefont {G.}~\bibnamefont
  {Gr\"{u}ner}},\ }\href
  {http://www.cambridge.org/catalogue/catalogue.asp?isbn=051182422X} {\emph
  {\bibinfo {title} {Electrodynamics of Solids}}}\ (\bibinfo  {publisher}
  {Cambridge University Press},\ \bibinfo {address} {Cambridge},\ \bibinfo
  {year} {2002})\BibitemShut {NoStop}%
\bibitem [{\citenamefont {Robertson}(1940)}]{Robertson-1940}%
  \BibitemOpen
  \bibfield  {author} {\bibinfo {author} {\bibfnamefont {H.~P.}\ \bibnamefont
  {Robertson}},\ }\bibfield  {title} {\bibinfo {title} {The invariant theory of
  isotropic turbulence},\ }in\ \href
  {https://doi.org/10.1017/S0305004100017199} {\emph {\bibinfo {booktitle}
  {Math. Proc. Cambridge Philos. Soc.}}},\ Vol.~\bibinfo {volume} {36}\
  (\bibinfo {organization} {Cambridge University Press},\ \bibinfo {year}
  {1940})\ pp.\ \bibinfo {pages} {209--223}\BibitemShut {NoStop}%
\bibitem [{Note3()}]{Note3}%
  \BibitemOpen
  \bibinfo {note} {The connection between the kernel $\protect \qopname \relax
  m{lim}_{q \rightarrow 0}\protect \mathcal {K}(\protect \bm {q},0)$ and the
  macroscopic magnetic response (\ref {eq:Ampere_law_gen}) can be realized by
  utilizing either the longitudinal or the transverse part of the kernel (\ref
  {eq:Kernel_micro}): in fact, in the long-wavelength limit $q \rightarrow 0$,
  the distinction between transverse and longitudinal waves with respect to $q$
  disappears, as can also be explicitly checked from our expression (\ref
  {eq:Kernel_micro}) of the electromagnetic kernel in the vanishing-momentum
  limit.}\BibitemShut {Stop}%
\bibitem [{\citenamefont {Leggett}(1998)}]{Leggett-1998}%
  \BibitemOpen
  \bibfield  {author} {\bibinfo {author} {\bibfnamefont {A.}~\bibnamefont
  {Leggett}},\ }\bibfield  {title} {\bibinfo {title} {On the superfluid
  fraction of an arbitrary many-body system at {T= 0}},\ }\href
  {https://doi.org/10.1023/B:JOSS.0000033170.38619.6c} {\bibfield  {journal}
  {\bibinfo  {journal} {J. Stat. Phys.}\ }\textbf {\bibinfo {volume} {93}},\
  \bibinfo {pages} {927} (\bibinfo {year} {1998})}\BibitemShut {NoStop}%
\bibitem [{\citenamefont {Leggett}(2006)}]{Leggett-2006quantum}%
  \BibitemOpen
  \bibfield  {author} {\bibinfo {author} {\bibfnamefont {A.~J.}\ \bibnamefont
  {Leggett}},\ }\href {https://academic.oup.com/book/32814?login=false} {\emph
  {\bibinfo {title} {Quantum liquids: {Bose} condensation and {Cooper} pairing
  in condensed-matter systems}}}\ (\bibinfo  {publisher} {Oxford university
  press},\ \bibinfo {address} {Oxford},\ \bibinfo {year} {2006})\BibitemShut
  {NoStop}%
\bibitem [{\citenamefont {Fisher}\ \emph {et~al.}(1988)\citenamefont {Fisher},
  \citenamefont {Gordon},\ and\ \citenamefont {Phillips}}]{Fisher-1988}%
  \BibitemOpen
  \bibfield  {author} {\bibinfo {author} {\bibfnamefont {R.}~\bibnamefont
  {Fisher}}, \bibinfo {author} {\bibfnamefont {J.}~\bibnamefont {Gordon}},\
  and\ \bibinfo {author} {\bibfnamefont {N.}~\bibnamefont {Phillips}},\
  }\bibfield  {title} {\bibinfo {title} {Specific heat of the {high-$T_c$}
  oxide superconductors},\ }\href {https://doi.org/10.1007/BF00617717}
  {\bibfield  {journal} {\bibinfo  {journal} {J. Supercond.}\ }\textbf
  {\bibinfo {volume} {1}},\ \bibinfo {pages} {231} (\bibinfo {year}
  {1988})}\BibitemShut {NoStop}%
\bibitem [{\citenamefont {Loram}\ \emph {et~al.}(2001)\citenamefont {Loram},
  \citenamefont {Luo}, \citenamefont {Cooper}, \citenamefont {Liang},\ and\
  \citenamefont {Tallon}}]{Loram-2001}%
  \BibitemOpen
  \bibfield  {author} {\bibinfo {author} {\bibfnamefont {J.}~\bibnamefont
  {Loram}}, \bibinfo {author} {\bibfnamefont {J.}~\bibnamefont {Luo}}, \bibinfo
  {author} {\bibfnamefont {J.}~\bibnamefont {Cooper}}, \bibinfo {author}
  {\bibfnamefont {W.}~\bibnamefont {Liang}},\ and\ \bibinfo {author}
  {\bibfnamefont {J.}~\bibnamefont {Tallon}},\ }\bibfield  {title} {\bibinfo
  {title} {Evidence on the pseudogap and condensate from the electronic
  specific heat},\ }\href
  {https://doi.org/https://doi.org/10.1016/S0022-3697(00)00101-3} {\bibfield
  {journal} {\bibinfo  {journal} {J. Phys. Chem. Solids}\ }\textbf {\bibinfo
  {volume} {62}},\ \bibinfo {pages} {59} (\bibinfo {year} {2001})}\BibitemShut
  {NoStop}%
\bibitem [{\citenamefont {Movshovich}\ \emph {et~al.}(2001)\citenamefont
  {Movshovich}, \citenamefont {Jaime}, \citenamefont {Thompson}, \citenamefont
  {Petrovic}, \citenamefont {Fisk}, \citenamefont {Pagliuso},\ and\
  \citenamefont {Sarrao}}]{Movshovich-2001}%
  \BibitemOpen
  \bibfield  {author} {\bibinfo {author} {\bibfnamefont {R.}~\bibnamefont
  {Movshovich}}, \bibinfo {author} {\bibfnamefont {M.}~\bibnamefont {Jaime}},
  \bibinfo {author} {\bibfnamefont {J.~D.}\ \bibnamefont {Thompson}}, \bibinfo
  {author} {\bibfnamefont {C.}~\bibnamefont {Petrovic}}, \bibinfo {author}
  {\bibfnamefont {Z.}~\bibnamefont {Fisk}}, \bibinfo {author} {\bibfnamefont
  {P.~G.}\ \bibnamefont {Pagliuso}},\ and\ \bibinfo {author} {\bibfnamefont
  {J.~L.}\ \bibnamefont {Sarrao}},\ }\bibfield  {title} {\bibinfo {title}
  {Unconventional superconductivity in {$\mathrm{CeIrIn}_{5}$ and
  ${\mathrm{CeCoIn}}_{5}$}: Specific heat and thermal conductivity studies},\
  }\href {https://doi.org/10.1103/PhysRevLett.86.5152} {\bibfield  {journal}
  {\bibinfo  {journal} {Phys. Rev. Lett.}\ }\textbf {\bibinfo {volume} {86}},\
  \bibinfo {pages} {5152} (\bibinfo {year} {2001})}\BibitemShut {NoStop}%
\bibitem [{\citenamefont {Esterlis}\ \emph
  {et~al.}(2018{\natexlab{b}})\citenamefont {Esterlis}, \citenamefont
  {Kivelson},\ and\ \citenamefont {Scalapino}}]{Esterlis-2018a}%
  \BibitemOpen
  \bibfield  {author} {\bibinfo {author} {\bibfnamefont {I.}~\bibnamefont
  {Esterlis}}, \bibinfo {author} {\bibfnamefont {S.}~\bibnamefont {Kivelson}},\
  and\ \bibinfo {author} {\bibfnamefont {D.}~\bibnamefont {Scalapino}},\
  }\bibfield  {title} {\bibinfo {title} {A bound on the superconducting
  transition temperature},\ }\href {https://doi.org/10.1038/s41535-018-0133-0}
  {\bibfield  {journal} {\bibinfo  {journal} {npj Quantum Mater.}\ }\textbf
  {\bibinfo {volume} {3}},\ \bibinfo {pages} {1} (\bibinfo {year}
  {2018}{\natexlab{b}})}\BibitemShut {NoStop}%
\bibitem [{\citenamefont {Chowdhury}\ and\ \citenamefont
  {Berg}(2020)}]{Chowdhury-2020eff}%
  \BibitemOpen
  \bibfield  {author} {\bibinfo {author} {\bibfnamefont {D.}~\bibnamefont
  {Chowdhury}}\ and\ \bibinfo {author} {\bibfnamefont {E.}~\bibnamefont
  {Berg}},\ }\bibfield  {title} {\bibinfo {title} {The unreasonable
  effectiveness of {Eliashberg} theory for pairing of non-{Fermi} liquids},\
  }\href {https://doi.org/https://doi.org/10.1016/j.aop.2020.168125} {\bibfield
   {journal} {\bibinfo  {journal} {Ann. Phys. (N. Y.)}\ }\textbf {\bibinfo
  {volume} {417}},\ \bibinfo {pages} {168125} (\bibinfo {year} {2020})},\
  \bibinfo {note} {in "{Eliashberg} theory at 60: Strong-coupling
  superconductivity and beyond"}\BibitemShut {NoStop}%
\bibitem [{\citenamefont {Aji}\ and\ \citenamefont {Varma}(2007)}]{Aji-2007}%
  \BibitemOpen
  \bibfield  {author} {\bibinfo {author} {\bibfnamefont {V.}~\bibnamefont
  {Aji}}\ and\ \bibinfo {author} {\bibfnamefont {C.~M.}\ \bibnamefont
  {Varma}},\ }\bibfield  {title} {\bibinfo {title} {Theory of the quantum
  critical fluctuations in cuprate superconductors},\ }\href
  {https://doi.org/10.1103/PhysRevLett.99.067003} {\bibfield  {journal}
  {\bibinfo  {journal} {Phys. Rev. Lett.}\ }\textbf {\bibinfo {volume} {99}},\
  \bibinfo {pages} {067003} (\bibinfo {year} {2007})}\BibitemShut {NoStop}%
\bibitem [{\citenamefont {Lederer}\ \emph {et~al.}(2015)\citenamefont
  {Lederer}, \citenamefont {Schattner}, \citenamefont {Berg},\ and\
  \citenamefont {Kivelson}}]{Lederer-2015}%
  \BibitemOpen
  \bibfield  {author} {\bibinfo {author} {\bibfnamefont {S.}~\bibnamefont
  {Lederer}}, \bibinfo {author} {\bibfnamefont {Y.}~\bibnamefont {Schattner}},
  \bibinfo {author} {\bibfnamefont {E.}~\bibnamefont {Berg}},\ and\ \bibinfo
  {author} {\bibfnamefont {S.~A.}\ \bibnamefont {Kivelson}},\ }\bibfield
  {title} {\bibinfo {title} {Enhancement of superconductivity near a nematic
  quantum critical point},\ }\href
  {https://doi.org/10.1103/PhysRevLett.114.097001} {\bibfield  {journal}
  {\bibinfo  {journal} {Phys. Rev. Lett.}\ }\textbf {\bibinfo {volume} {114}},\
  \bibinfo {pages} {097001} (\bibinfo {year} {2015})}\BibitemShut {NoStop}%
\bibitem [{\citenamefont {Fernandes}\ and\ \citenamefont
  {Chubukov}(2016)}]{Fernandes-2016}%
  \BibitemOpen
  \bibfield  {author} {\bibinfo {author} {\bibfnamefont {R.~M.}\ \bibnamefont
  {Fernandes}}\ and\ \bibinfo {author} {\bibfnamefont {A.~V.}\ \bibnamefont
  {Chubukov}},\ }\bibfield  {title} {\bibinfo {title} {Low-energy microscopic
  models for {Iron}-based superconductors: a review},\ }\href
  {https://iopscience.iop.org/article/10.1088/1361-6633/80/1/014503} {\bibfield
   {journal} {\bibinfo  {journal} {Reports on Progress in Physics}\ }\textbf
  {\bibinfo {volume} {80}},\ \bibinfo {pages} {014503} (\bibinfo {year}
  {2016})}\BibitemShut {NoStop}%
\bibitem [{\citenamefont {Caprara}\ \emph {et~al.}(2017)\citenamefont
  {Caprara}, \citenamefont {Di~Castro}, \citenamefont {Seibold},\ and\
  \citenamefont {Grilli}}]{Caprara-2017}%
  \BibitemOpen
  \bibfield  {author} {\bibinfo {author} {\bibfnamefont {S.}~\bibnamefont
  {Caprara}}, \bibinfo {author} {\bibfnamefont {C.}~\bibnamefont {Di~Castro}},
  \bibinfo {author} {\bibfnamefont {G.}~\bibnamefont {Seibold}},\ and\ \bibinfo
  {author} {\bibfnamefont {M.}~\bibnamefont {Grilli}},\ }\bibfield  {title}
  {\bibinfo {title} {Dynamical charge density waves rule the phase diagram of
  cuprates},\ }\href {https://doi.org/10.1103/PhysRevB.95.224511} {\bibfield
  {journal} {\bibinfo  {journal} {Phys. Rev. B}\ }\textbf {\bibinfo {volume}
  {95}},\ \bibinfo {pages} {224511} (\bibinfo {year} {2017})}\BibitemShut
  {NoStop}%
\bibitem [{\citenamefont {Huang}\ \emph {et~al.}(2019)\citenamefont {Huang},
  \citenamefont {Sheppard}, \citenamefont {Moritz},\ and\ \citenamefont
  {Devereaux}}]{Huang-2019}%
  \BibitemOpen
  \bibfield  {author} {\bibinfo {author} {\bibfnamefont {E.~W.}\ \bibnamefont
  {Huang}}, \bibinfo {author} {\bibfnamefont {R.}~\bibnamefont {Sheppard}},
  \bibinfo {author} {\bibfnamefont {B.}~\bibnamefont {Moritz}},\ and\ \bibinfo
  {author} {\bibfnamefont {T.~P.}\ \bibnamefont {Devereaux}},\ }\bibfield
  {title} {\bibinfo {title} {Strange metallicity in the doped {Hubbard}
  model},\ }\href {https://science.sciencemag.org/content/366/6468/987}
  {\bibfield  {journal} {\bibinfo  {journal} {Science}\ }\textbf {\bibinfo
  {volume} {366}},\ \bibinfo {pages} {987} (\bibinfo {year}
  {2019})}\BibitemShut {NoStop}%
\bibitem [{\citenamefont {Chubukov}\ and\ \citenamefont
  {Schmalian}(2020)}]{Chubukov-2020}%
  \BibitemOpen
  \bibfield  {author} {\bibinfo {author} {\bibfnamefont {A.~V.}\ \bibnamefont
  {Chubukov}}\ and\ \bibinfo {author} {\bibfnamefont {J.}~\bibnamefont
  {Schmalian}},\ }\bibfield  {title} {\bibinfo {title} {Pairing glue in cuprate
  superconductors from the self-energy revealed via machine learning},\ }\href
  {https://doi.org/10.1103/PhysRevB.101.180510} {\bibfield  {journal} {\bibinfo
   {journal} {Phys. Rev. B}\ }\textbf {\bibinfo {volume} {101}},\ \bibinfo
  {pages} {180510} (\bibinfo {year} {2020})}\BibitemShut {NoStop}%
\bibitem [{\citenamefont {Phillips}\ \emph {et~al.}(2020)\citenamefont
  {Phillips}, \citenamefont {Yeo},\ and\ \citenamefont
  {Huang}}]{Phillips-2020}%
  \BibitemOpen
  \bibfield  {author} {\bibinfo {author} {\bibfnamefont {P.~W.}\ \bibnamefont
  {Phillips}}, \bibinfo {author} {\bibfnamefont {L.}~\bibnamefont {Yeo}},\ and\
  \bibinfo {author} {\bibfnamefont {E.~W.}\ \bibnamefont {Huang}},\ }\bibfield
  {title} {\bibinfo {title} {Exact theory for superconductivity in a doped
  {Mott} insulator},\ }\href {https://doi.org/10.1038/s41567-020-0988-4}
  {\bibfield  {journal} {\bibinfo  {journal} {Nat. Phys.}\ }\textbf {\bibinfo
  {volume} {16}},\ \bibinfo {pages} {1175} (\bibinfo {year}
  {2020})}\BibitemShut {NoStop}%
\bibitem [{\citenamefont {Caprara}\ \emph {et~al.}(2022)\citenamefont
  {Caprara}, \citenamefont {Castro}, \citenamefont {Mirarchi}, \citenamefont
  {Seibold},\ and\ \citenamefont {Grilli}}]{Caprara-2022}%
  \BibitemOpen
  \bibfield  {author} {\bibinfo {author} {\bibfnamefont {S.}~\bibnamefont
  {Caprara}}, \bibinfo {author} {\bibfnamefont {C.~D.}\ \bibnamefont {Castro}},
  \bibinfo {author} {\bibfnamefont {G.}~\bibnamefont {Mirarchi}}, \bibinfo
  {author} {\bibfnamefont {G.}~\bibnamefont {Seibold}},\ and\ \bibinfo {author}
  {\bibfnamefont {M.}~\bibnamefont {Grilli}},\ }\bibfield  {title} {\bibinfo
  {title} {Dissipation-driven strange metal behavior},\ }\href
  {https://doi.org/10.1038/s42005-021-00786-y} {\bibfield  {journal} {\bibinfo
  {journal} {Commun. Phys.}\ }\textbf {\bibinfo {volume} {5}},\ \bibinfo
  {pages} {1} (\bibinfo {year} {2022})}\BibitemShut {NoStop}%
\bibitem [{\citenamefont {Collins}\ \emph {et~al.}(1989)\citenamefont
  {Collins}, \citenamefont {Schlesinger}, \citenamefont {Holtzberg},
  \citenamefont {Chaudhari},\ and\ \citenamefont {Feild}}]{Collins-1989}%
  \BibitemOpen
  \bibfield  {author} {\bibinfo {author} {\bibfnamefont {R.~T.}\ \bibnamefont
  {Collins}}, \bibinfo {author} {\bibfnamefont {Z.}~\bibnamefont
  {Schlesinger}}, \bibinfo {author} {\bibfnamefont {F.}~\bibnamefont
  {Holtzberg}}, \bibinfo {author} {\bibfnamefont {P.}~\bibnamefont
  {Chaudhari}},\ and\ \bibinfo {author} {\bibfnamefont {C.}~\bibnamefont
  {Feild}},\ }\bibfield  {title} {\bibinfo {title} {Reflectivity and
  conductivity of
  {${\mathrm{YBa}}_{2}$${\mathrm{Cu}}_{3}$${\mathrm{O}}_{7}$}},\ }\href
  {https://doi.org/10.1103/PhysRevB.39.6571} {\bibfield  {journal} {\bibinfo
  {journal} {Phys. Rev. B}\ }\textbf {\bibinfo {volume} {39}},\ \bibinfo
  {pages} {6571} (\bibinfo {year} {1989})}\BibitemShut {NoStop}%
\bibitem [{\citenamefont {Orenstein}\ \emph {et~al.}(1990)\citenamefont
  {Orenstein}, \citenamefont {Thomas}, \citenamefont {Millis}, \citenamefont
  {Cooper}, \citenamefont {Rapkine}, \citenamefont {Timusk}, \citenamefont
  {Schneemeyer},\ and\ \citenamefont {Waszczak}}]{Orenstein-1990}%
  \BibitemOpen
  \bibfield  {author} {\bibinfo {author} {\bibfnamefont {J.}~\bibnamefont
  {Orenstein}}, \bibinfo {author} {\bibfnamefont {G.~A.}\ \bibnamefont
  {Thomas}}, \bibinfo {author} {\bibfnamefont {A.~J.}\ \bibnamefont {Millis}},
  \bibinfo {author} {\bibfnamefont {S.~L.}\ \bibnamefont {Cooper}}, \bibinfo
  {author} {\bibfnamefont {D.~H.}\ \bibnamefont {Rapkine}}, \bibinfo {author}
  {\bibfnamefont {T.}~\bibnamefont {Timusk}}, \bibinfo {author} {\bibfnamefont
  {L.~F.}\ \bibnamefont {Schneemeyer}},\ and\ \bibinfo {author} {\bibfnamefont
  {J.~V.}\ \bibnamefont {Waszczak}},\ }\bibfield  {title} {\bibinfo {title}
  {Frequency- and temperature-dependent conductivity in
  {${\mathrm{YBa}}_{2}$${\mathrm{Cu}}_{3}$${\mathrm{O}}_{6+\mathit{x}}$}
  crystals},\ }\href {https://doi.org/10.1103/PhysRevB.42.6342} {\bibfield
  {journal} {\bibinfo  {journal} {Phys. Rev. B}\ }\textbf {\bibinfo {volume}
  {42}},\ \bibinfo {pages} {6342} (\bibinfo {year} {1990})}\BibitemShut
  {NoStop}%
\bibitem [{\citenamefont {Marel}\ \emph {et~al.}(2003)\citenamefont {Marel},
  \citenamefont {Molegraaf}, \citenamefont {Zaanen}, \citenamefont {Nussinov},
  \citenamefont {Carbone}, \citenamefont {Damascelli}, \citenamefont {Eisaki},
  \citenamefont {Greven}, \citenamefont {Kes},\ and\ \citenamefont
  {Li}}]{vanderMarel-2003}%
  \BibitemOpen
  \bibfield  {author} {\bibinfo {author} {\bibfnamefont {D.~v.~d.}\
  \bibnamefont {Marel}}, \bibinfo {author} {\bibfnamefont {H.}~\bibnamefont
  {Molegraaf}}, \bibinfo {author} {\bibfnamefont {J.}~\bibnamefont {Zaanen}},
  \bibinfo {author} {\bibfnamefont {Z.}~\bibnamefont {Nussinov}}, \bibinfo
  {author} {\bibfnamefont {F.}~\bibnamefont {Carbone}}, \bibinfo {author}
  {\bibfnamefont {A.}~\bibnamefont {Damascelli}}, \bibinfo {author}
  {\bibfnamefont {H.}~\bibnamefont {Eisaki}}, \bibinfo {author} {\bibfnamefont
  {M.}~\bibnamefont {Greven}}, \bibinfo {author} {\bibfnamefont
  {P.}~\bibnamefont {Kes}},\ and\ \bibinfo {author} {\bibfnamefont
  {M.}~\bibnamefont {Li}},\ }\bibfield  {title} {\bibinfo {title} {Quantum
  critical behaviour in a high-{$T_c$} superconductor},\ }\href
  {https://doi.org/10.1038/nature01978} {\bibfield  {journal} {\bibinfo
  {journal} {Nature}\ }\textbf {\bibinfo {volume} {425}},\ \bibinfo {pages}
  {271} (\bibinfo {year} {2003})}\BibitemShut {NoStop}%
\bibitem [{\citenamefont {van Heumen}\ \emph {et~al.}(2022)\citenamefont {van
  Heumen}, \citenamefont {Feng}, \citenamefont {Cassanelli}, \citenamefont
  {Neubrand}, \citenamefont {de~Jager}, \citenamefont {Berben}, \citenamefont
  {Huang}, \citenamefont {Kondo}, \citenamefont {Takeuchi},\ and\ \citenamefont
  {Zaanen}}]{vanHeumen-2022}%
  \BibitemOpen
  \bibfield  {author} {\bibinfo {author} {\bibfnamefont {E.}~\bibnamefont {van
  Heumen}}, \bibinfo {author} {\bibfnamefont {X.}~\bibnamefont {Feng}},
  \bibinfo {author} {\bibfnamefont {S.}~\bibnamefont {Cassanelli}}, \bibinfo
  {author} {\bibfnamefont {L.}~\bibnamefont {Neubrand}}, \bibinfo {author}
  {\bibfnamefont {L.}~\bibnamefont {de~Jager}}, \bibinfo {author}
  {\bibfnamefont {M.}~\bibnamefont {Berben}}, \bibinfo {author} {\bibfnamefont
  {Y.}~\bibnamefont {Huang}}, \bibinfo {author} {\bibfnamefont
  {T.}~\bibnamefont {Kondo}}, \bibinfo {author} {\bibfnamefont
  {T.}~\bibnamefont {Takeuchi}},\ and\ \bibinfo {author} {\bibfnamefont
  {J.}~\bibnamefont {Zaanen}},\ }\bibfield  {title} {\bibinfo {title} {Strange
  metal electrodynamics across the phase diagram of
  {${\mathrm{Bi}}_{2\ensuremath{-}x}{\mathrm{Pb}}_{x}{\mathrm{Sr}}_{2\ensuremath{-}y}{\mathrm{La}}_{y}{\mathrm{CuO}}_{6+\ensuremath{\delta}}$}
  cuprates},\ }\href {https://doi.org/10.1103/PhysRevB.106.054515} {\bibfield
  {journal} {\bibinfo  {journal} {Phys. Rev. B}\ }\textbf {\bibinfo {volume}
  {106}},\ \bibinfo {pages} {054515} (\bibinfo {year} {2022})}\BibitemShut
  {NoStop}%
\bibitem [{\citenamefont {Michon}\ \emph {et~al.}(2023)\citenamefont {Michon},
  \citenamefont {Berthod}, \citenamefont {Rischau}, \citenamefont {Ataei},
  \citenamefont {Chen}, \citenamefont {Komiya}, \citenamefont {Ono},
  \citenamefont {Taillefer}, \citenamefont {van~der Marel},\ and\ \citenamefont
  {Georges}}]{Michon-2022_preprint}%
  \BibitemOpen
  \bibfield  {author} {\bibinfo {author} {\bibfnamefont {B.}~\bibnamefont
  {Michon}}, \bibinfo {author} {\bibfnamefont {C.}~\bibnamefont {Berthod}},
  \bibinfo {author} {\bibfnamefont {C.}~\bibnamefont {Rischau}}, \bibinfo
  {author} {\bibfnamefont {A.}~\bibnamefont {Ataei}}, \bibinfo {author}
  {\bibfnamefont {L.}~\bibnamefont {Chen}}, \bibinfo {author} {\bibfnamefont
  {S.}~\bibnamefont {Komiya}}, \bibinfo {author} {\bibfnamefont
  {S.}~\bibnamefont {Ono}}, \bibinfo {author} {\bibfnamefont {L.}~\bibnamefont
  {Taillefer}}, \bibinfo {author} {\bibfnamefont {D.}~\bibnamefont {van~der
  Marel}},\ and\ \bibinfo {author} {\bibfnamefont {A.}~\bibnamefont
  {Georges}},\ }\bibfield  {title} {\bibinfo {title} {Planckian behavior of
  cuprate superconductors: Reconciling the scaling of optical conductivity with
  resistivity and specific heat},\ }\href
  {https://doi.org/10.1038/s41467-023-38762-5} {\bibfield  {journal} {\bibinfo
  {journal} {Nat. Comm.}\ }\textbf {\bibinfo {volume} {14}},\ \bibinfo {pages}
  {3033} (\bibinfo {year} {2023})}\BibitemShut {NoStop}%
\bibitem [{\citenamefont {Lee}\ \emph {et~al.}(2006)\citenamefont {Lee},
  \citenamefont {Nagaosa},\ and\ \citenamefont {Wen}}]{Lee-2006}%
  \BibitemOpen
  \bibfield  {author} {\bibinfo {author} {\bibfnamefont {P.~A.}\ \bibnamefont
  {Lee}}, \bibinfo {author} {\bibfnamefont {N.}~\bibnamefont {Nagaosa}},\ and\
  \bibinfo {author} {\bibfnamefont {X.-G.}\ \bibnamefont {Wen}},\ }\bibfield
  {title} {\bibinfo {title} {Doping a {Mott} insulator: Physics of
  high-temperature superconductivity},\ }\href
  {https://doi.org/10.1103/RevModPhys.78.17} {\bibfield  {journal} {\bibinfo
  {journal} {Rev. Mod. Phys.}\ }\textbf {\bibinfo {volume} {78}},\ \bibinfo
  {pages} {17} (\bibinfo {year} {2006})}\BibitemShut {NoStop}%
\bibitem [{\citenamefont {Azeyanagi}\ \emph {et~al.}(2018)\citenamefont
  {Azeyanagi}, \citenamefont {Ferrari},\ and\ \citenamefont
  {Massolo}}]{Azeyanagi-2018}%
  \BibitemOpen
  \bibfield  {author} {\bibinfo {author} {\bibfnamefont {T.}~\bibnamefont
  {Azeyanagi}}, \bibinfo {author} {\bibfnamefont {F.}~\bibnamefont {Ferrari}},\
  and\ \bibinfo {author} {\bibfnamefont {F.~I.~S.}\ \bibnamefont {Massolo}},\
  }\bibfield  {title} {\bibinfo {title} {Phase diagram of planar matrix quantum
  mechanics, tensor, and {Sachdev-Ye-Kitaev} models},\ }\href
  {https://doi.org/10.1103/PhysRevLett.120.061602} {\bibfield  {journal}
  {\bibinfo  {journal} {Phys. Rev. Lett.}\ }\textbf {\bibinfo {volume} {120}},\
  \bibinfo {pages} {061602} (\bibinfo {year} {2018})}\BibitemShut {NoStop}%
\bibitem [{\citenamefont {Ferrari}\ and\ \citenamefont
  {Schaposnik~Massolo}(2019)}]{Ferrari-2019}%
  \BibitemOpen
  \bibfield  {author} {\bibinfo {author} {\bibfnamefont {F.}~\bibnamefont
  {Ferrari}}\ and\ \bibinfo {author} {\bibfnamefont {F.~I.}\ \bibnamefont
  {Schaposnik~Massolo}},\ }\bibfield  {title} {\bibinfo {title} {Phases of
  melonic quantum mechanics},\ }\href
  {https://doi.org/10.1103/PhysRevD.100.026007} {\bibfield  {journal} {\bibinfo
   {journal} {Phys. Rev. D}\ }\textbf {\bibinfo {volume} {100}},\ \bibinfo
  {pages} {026007} (\bibinfo {year} {2019})}\BibitemShut {NoStop}%
\bibitem [{\citenamefont {Sorokhaibam}(2020)}]{Sorokhaibam-2020}%
  \BibitemOpen
  \bibfield  {author} {\bibinfo {author} {\bibfnamefont {N.}~\bibnamefont
  {Sorokhaibam}},\ }\bibfield  {title} {\bibinfo {title} {Phase transition and
  chaos in charged {SYK} model},\ }\href
  {https://link.springer.com/article/10.1007/JHEP07(2020)055} {\bibfield
  {journal} {\bibinfo  {journal} {J. High Energy Phys.}\ }\textbf {\bibinfo
  {volume} {2020}}\bibinfo  {number} { (7)},\ \bibinfo {pages} {1}}\BibitemShut
  {NoStop}%
\bibitem [{\citenamefont {Smit}\ \emph {et~al.}(2021)\citenamefont {Smit},
  \citenamefont {Valentinis}, \citenamefont {Schmalian},\ and\ \citenamefont
  {Kopietz}}]{Smit-2021}%
  \BibitemOpen
\bibfield  {number} {  }\bibfield  {author} {\bibinfo {author} {\bibfnamefont
  {R.~L.}\ \bibnamefont {Smit}}, \bibinfo {author} {\bibfnamefont
  {D.}~\bibnamefont {Valentinis}}, \bibinfo {author} {\bibfnamefont
  {J.}~\bibnamefont {Schmalian}},\ and\ \bibinfo {author} {\bibfnamefont
  {P.}~\bibnamefont {Kopietz}},\ }\bibfield  {title} {\bibinfo {title} {Quantum
  discontinuity fixed point and renormalization group flow of the
  {Sachdev-Ye-Kitaev} model},\ }\href
  {https://doi.org/10.1103/PhysRevResearch.3.033089} {\bibfield  {journal}
  {\bibinfo  {journal} {Phys. Rev. Research}\ }\textbf {\bibinfo {volume}
  {3}},\ \bibinfo {pages} {033089} (\bibinfo {year} {2021})}\BibitemShut
  {NoStop}%
\bibitem [{\citenamefont {van~der Marel}(1990)}]{Vandermarel-1990}%
  \BibitemOpen
  \bibfield  {author} {\bibinfo {author} {\bibfnamefont {D.}~\bibnamefont
  {van~der Marel}},\ }\bibfield  {title} {\bibinfo {title} {Anomalous behaviour
  of the chemical potential in superconductors with a low density of charge
  carriers},\ }\href {https://doi.org/10.1016/0921-4534(90)90429-I} {\bibfield
  {journal} {\bibinfo  {journal} {Physica C}\ }\textbf {\bibinfo {volume}
  {165}},\ \bibinfo {pages} {35} (\bibinfo {year} {1990})}\BibitemShut
  {NoStop}%
\bibitem [{\citenamefont {Valentinis}\ \emph
  {et~al.}(2016{\natexlab{a}})\citenamefont {Valentinis}, \citenamefont
  {van~der Marel},\ and\ \citenamefont {Berthod}}]{Valentinis-2016a}%
  \BibitemOpen
  \bibfield  {author} {\bibinfo {author} {\bibfnamefont {D.}~\bibnamefont
  {Valentinis}}, \bibinfo {author} {\bibfnamefont {D.}~\bibnamefont {van~der
  Marel}},\ and\ \bibinfo {author} {\bibfnamefont {C.}~\bibnamefont
  {Berthod}},\ }\bibfield  {title} {\bibinfo {title} {{BCS} superconductivity
  near the band edge: Exact results for one and several bands},\ }\href
  {https://doi.org/10.1103/PhysRevB.94.024511} {\bibfield  {journal} {\bibinfo
  {journal} {Phys. Rev. B}\ }\textbf {\bibinfo {volume} {94}},\ \bibinfo
  {pages} {024511} (\bibinfo {year} {2016}{\natexlab{a}})}\BibitemShut
  {NoStop}%
\bibitem [{\citenamefont {Valentinis}\ \emph
  {et~al.}(2016{\natexlab{b}})\citenamefont {Valentinis}, \citenamefont
  {van~der Marel},\ and\ \citenamefont {Berthod}}]{Valentinis-2016b}%
  \BibitemOpen
  \bibfield  {author} {\bibinfo {author} {\bibfnamefont {D.}~\bibnamefont
  {Valentinis}}, \bibinfo {author} {\bibfnamefont {D.}~\bibnamefont {van~der
  Marel}},\ and\ \bibinfo {author} {\bibfnamefont {C.}~\bibnamefont
  {Berthod}},\ }\bibfield  {title} {\bibinfo {title} {Rise and fall of shape
  resonances in thin films of {BCS} superconductors},\ }\href
  {https://doi.org/10.1103/PhysRevB.94.054516} {\bibfield  {journal} {\bibinfo
  {journal} {Phys. Rev. B}\ }\textbf {\bibinfo {volume} {94}},\ \bibinfo
  {pages} {054516} (\bibinfo {year} {2016}{\natexlab{b}})}\BibitemShut
  {NoStop}%
\bibitem [{\citenamefont {Valentinis}\ \emph {et~al.}(2017)\citenamefont
  {Valentinis}, \citenamefont {Gariglio}, \citenamefont {F\^ete}, \citenamefont
  {Triscone}, \citenamefont {Berthod},\ and\ \citenamefont {van~der
  Marel}}]{Valentinis-2017}%
  \BibitemOpen
  \bibfield  {author} {\bibinfo {author} {\bibfnamefont {D.}~\bibnamefont
  {Valentinis}}, \bibinfo {author} {\bibfnamefont {S.}~\bibnamefont
  {Gariglio}}, \bibinfo {author} {\bibfnamefont {A.}~\bibnamefont {F\^ete}},
  \bibinfo {author} {\bibfnamefont {J.-M.}\ \bibnamefont {Triscone}}, \bibinfo
  {author} {\bibfnamefont {C.}~\bibnamefont {Berthod}},\ and\ \bibinfo {author}
  {\bibfnamefont {D.}~\bibnamefont {van~der Marel}},\ }\bibfield  {title}
  {\bibinfo {title} {Modulation of the superconducting critical temperature due
  to quantum confinement at the {${\mathrm{LaAlO}}_{3}/{\mathrm{SrTiO}}_{3}$}
  interface},\ }\href {https://doi.org/10.1103/PhysRevB.96.094518} {\bibfield
  {journal} {\bibinfo  {journal} {Phys. Rev. B}\ }\textbf {\bibinfo {volume}
  {96}},\ \bibinfo {pages} {094518} (\bibinfo {year} {2017})}\BibitemShut
  {NoStop}%
\bibitem [{\citenamefont {Inkof}(2021)}]{Inkof-thesis-2021}%
  \BibitemOpen
  \bibfield  {author} {\bibinfo {author} {\bibfnamefont {G.~A.}\ \bibnamefont
  {Inkof}},\ }\emph {\bibinfo {title} {The quantum critical {SYK}
  superconductor, {Lifshitz} points and their holographic duals}},\ \href@noop
  {} {Ph.D. thesis},\ \bibinfo  {school} {Karlsruhe Institute of Technology}
  (\bibinfo {year} {2021})\BibitemShut {NoStop}%
\bibitem [{\citenamefont {Cao}\ \emph {et~al.}(2018{\natexlab{a}})\citenamefont
  {Cao}, \citenamefont {Fatemi}, \citenamefont {Fang}, \citenamefont
  {Watanabe}, \citenamefont {Taniguchi}, \citenamefont {Kaxiras},\ and\
  \citenamefont {Jarillo-Herrero}}]{Cao-2018b}%
  \BibitemOpen
  \bibfield  {author} {\bibinfo {author} {\bibfnamefont {Y.}~\bibnamefont
  {Cao}}, \bibinfo {author} {\bibfnamefont {V.}~\bibnamefont {Fatemi}},
  \bibinfo {author} {\bibfnamefont {S.}~\bibnamefont {Fang}}, \bibinfo {author}
  {\bibfnamefont {K.}~\bibnamefont {Watanabe}}, \bibinfo {author}
  {\bibfnamefont {T.}~\bibnamefont {Taniguchi}}, \bibinfo {author}
  {\bibfnamefont {E.}~\bibnamefont {Kaxiras}},\ and\ \bibinfo {author}
  {\bibfnamefont {P.}~\bibnamefont {Jarillo-Herrero}},\ }\bibfield  {title}
  {\bibinfo {title} {Unconventional superconductivity in magic-angle graphene
  superlattices},\ }\href {https://doi.org/10.1038/nature26160} {\bibfield
  {journal} {\bibinfo  {journal} {Nature}\ }\textbf {\bibinfo {volume} {556}},\
  \bibinfo {pages} {43} (\bibinfo {year} {2018}{\natexlab{a}})}\BibitemShut
  {NoStop}%
\bibitem [{\citenamefont {Cao}\ \emph {et~al.}(2018{\natexlab{b}})\citenamefont
  {Cao}, \citenamefont {Fatemi}, \citenamefont {Demir}, \citenamefont {Fang},
  \citenamefont {Tomarken}, \citenamefont {Luo}, \citenamefont
  {Sanchez-Yamagishi}, \citenamefont {Watanabe}, \citenamefont {Taniguchi},
  \citenamefont {Kaxiras}, \citenamefont {Ashoori},\ and\ \citenamefont
  {Jarillo-Herrero}}]{Cao-2018a}%
  \BibitemOpen
  \bibfield  {author} {\bibinfo {author} {\bibfnamefont {Y.}~\bibnamefont
  {Cao}}, \bibinfo {author} {\bibfnamefont {V.}~\bibnamefont {Fatemi}},
  \bibinfo {author} {\bibfnamefont {A.}~\bibnamefont {Demir}}, \bibinfo
  {author} {\bibfnamefont {S.}~\bibnamefont {Fang}}, \bibinfo {author}
  {\bibfnamefont {S.~L.}\ \bibnamefont {Tomarken}}, \bibinfo {author}
  {\bibfnamefont {J.~Y.}\ \bibnamefont {Luo}}, \bibinfo {author} {\bibfnamefont
  {J.~D.}\ \bibnamefont {Sanchez-Yamagishi}}, \bibinfo {author} {\bibfnamefont
  {K.}~\bibnamefont {Watanabe}}, \bibinfo {author} {\bibfnamefont
  {T.}~\bibnamefont {Taniguchi}}, \bibinfo {author} {\bibfnamefont
  {E.}~\bibnamefont {Kaxiras}}, \bibinfo {author} {\bibfnamefont {R.~C.}\
  \bibnamefont {Ashoori}},\ and\ \bibinfo {author} {\bibfnamefont
  {P.}~\bibnamefont {Jarillo-Herrero}},\ }\bibfield  {title} {\bibinfo {title}
  {Correlated insulator behaviour at half-filling in magic-angle graphene
  superlattices},\ }\href {https://www.nature.com/articles/nature26154}
  {\bibfield  {journal} {\bibinfo  {journal} {Nature}\ }\textbf {\bibinfo
  {volume} {556}},\ \bibinfo {pages} {80} (\bibinfo {year}
  {2018}{\natexlab{b}})}\BibitemShut {NoStop}%
\bibitem [{\citenamefont {Yankowitz}\ \emph {et~al.}(2019)\citenamefont
  {Yankowitz}, \citenamefont {Chen}, \citenamefont {Polshyn}, \citenamefont
  {Zhang}, \citenamefont {Watanabe}, \citenamefont {Taniguchi}, \citenamefont
  {Graf}, \citenamefont {Young},\ and\ \citenamefont {Dean}}]{Yankowitz-2019}%
  \BibitemOpen
  \bibfield  {author} {\bibinfo {author} {\bibfnamefont {M.}~\bibnamefont
  {Yankowitz}}, \bibinfo {author} {\bibfnamefont {S.}~\bibnamefont {Chen}},
  \bibinfo {author} {\bibfnamefont {H.}~\bibnamefont {Polshyn}}, \bibinfo
  {author} {\bibfnamefont {Y.}~\bibnamefont {Zhang}}, \bibinfo {author}
  {\bibfnamefont {K.}~\bibnamefont {Watanabe}}, \bibinfo {author}
  {\bibfnamefont {T.}~\bibnamefont {Taniguchi}}, \bibinfo {author}
  {\bibfnamefont {D.}~\bibnamefont {Graf}}, \bibinfo {author} {\bibfnamefont
  {A.~F.}\ \bibnamefont {Young}},\ and\ \bibinfo {author} {\bibfnamefont
  {C.~R.}\ \bibnamefont {Dean}},\ }\bibfield  {title} {\bibinfo {title} {Tuning
  superconductivity in twisted bilayer graphene},\ }\href
  {https://doi.org/10.1126/science.aav1910} {\bibfield  {journal} {\bibinfo
  {journal} {Science}\ }\textbf {\bibinfo {volume} {363}},\ \bibinfo {pages}
  {1059} (\bibinfo {year} {2019})}\BibitemShut {NoStop}%
\bibitem [{\citenamefont {Balents}\ \emph {et~al.}(2020)\citenamefont
  {Balents}, \citenamefont {Dean}, \citenamefont {Efetov},\ and\ \citenamefont
  {Young}}]{Balents-2020}%
  \BibitemOpen
  \bibfield  {author} {\bibinfo {author} {\bibfnamefont {L.}~\bibnamefont
  {Balents}}, \bibinfo {author} {\bibfnamefont {C.~R.}\ \bibnamefont {Dean}},
  \bibinfo {author} {\bibfnamefont {D.~K.}\ \bibnamefont {Efetov}},\ and\
  \bibinfo {author} {\bibfnamefont {A.~F.}\ \bibnamefont {Young}},\ }\bibfield
  {title} {\bibinfo {title} {Superconductivity and strong correlations in
  moir{\'e} flat bands},\ }\href {https://doi.org/10.1038/s41567-020-0906-9}
  {\bibfield  {journal} {\bibinfo  {journal} {Nat. Phys.}\ }\textbf {\bibinfo
  {volume} {16}},\ \bibinfo {pages} {725} (\bibinfo {year} {2020})}\BibitemShut
  {NoStop}%
\bibitem [{\citenamefont {Peng}\ \emph {et~al.}(2022)\citenamefont {Peng},
  \citenamefont {Martinelli}, \citenamefont {Li}, \citenamefont {Rossi},
  \citenamefont {Mitrano}, \citenamefont {Arpaia}, \citenamefont {Sala},
  \citenamefont {Gao}, \citenamefont {Guo}, \citenamefont {De~Luca},
  \citenamefont {Walters}, \citenamefont {Nag}, \citenamefont {Barbour},
  \citenamefont {Gu}, \citenamefont {Pelliciari}, \citenamefont {Brookes},
  \citenamefont {Abbamonte}, \citenamefont {Salluzzo}, \citenamefont {Zhou},
  \citenamefont {Zhou}, \citenamefont {Bisogni}, \citenamefont {Braicovich},
  \citenamefont {Johnston},\ and\ \citenamefont {Ghiringhelli}}]{Peng-2022}%
  \BibitemOpen
  \bibfield  {author} {\bibinfo {author} {\bibfnamefont {Y.}~\bibnamefont
  {Peng}}, \bibinfo {author} {\bibfnamefont {L.}~\bibnamefont {Martinelli}},
  \bibinfo {author} {\bibfnamefont {Q.}~\bibnamefont {Li}}, \bibinfo {author}
  {\bibfnamefont {M.}~\bibnamefont {Rossi}}, \bibinfo {author} {\bibfnamefont
  {M.}~\bibnamefont {Mitrano}}, \bibinfo {author} {\bibfnamefont
  {R.}~\bibnamefont {Arpaia}}, \bibinfo {author} {\bibfnamefont {M.~M.}\
  \bibnamefont {Sala}}, \bibinfo {author} {\bibfnamefont {Q.}~\bibnamefont
  {Gao}}, \bibinfo {author} {\bibfnamefont {X.}~\bibnamefont {Guo}}, \bibinfo
  {author} {\bibfnamefont {G.~M.}\ \bibnamefont {De~Luca}}, \bibinfo {author}
  {\bibfnamefont {A.}~\bibnamefont {Walters}}, \bibinfo {author} {\bibfnamefont
  {A.}~\bibnamefont {Nag}}, \bibinfo {author} {\bibfnamefont {A.}~\bibnamefont
  {Barbour}}, \bibinfo {author} {\bibfnamefont {G.}~\bibnamefont {Gu}},
  \bibinfo {author} {\bibfnamefont {J.}~\bibnamefont {Pelliciari}}, \bibinfo
  {author} {\bibfnamefont {N.~B.}\ \bibnamefont {Brookes}}, \bibinfo {author}
  {\bibfnamefont {P.}~\bibnamefont {Abbamonte}}, \bibinfo {author}
  {\bibfnamefont {M.}~\bibnamefont {Salluzzo}}, \bibinfo {author}
  {\bibfnamefont {X.}~\bibnamefont {Zhou}}, \bibinfo {author} {\bibfnamefont
  {K.-J.}\ \bibnamefont {Zhou}}, \bibinfo {author} {\bibfnamefont
  {V.}~\bibnamefont {Bisogni}}, \bibinfo {author} {\bibfnamefont
  {L.}~\bibnamefont {Braicovich}}, \bibinfo {author} {\bibfnamefont
  {S.}~\bibnamefont {Johnston}},\ and\ \bibinfo {author} {\bibfnamefont
  {G.}~\bibnamefont {Ghiringhelli}},\ }\bibfield  {title} {\bibinfo {title}
  {Doping dependence of the electron-phonon coupling in two families of bilayer
  superconducting cuprates},\ }\href
  {https://doi.org/10.1103/PhysRevB.105.115105} {\bibfield  {journal} {\bibinfo
   {journal} {Phys. Rev. B}\ }\textbf {\bibinfo {volume} {105}},\ \bibinfo
  {pages} {115105} (\bibinfo {year} {2022})}\BibitemShut {NoStop}%
\bibitem [{\citenamefont {Wang}\ \emph {et~al.}(2019)\citenamefont {Wang},
  \citenamefont {Bagrets}, \citenamefont {Chudnovskiy},\ and\ \citenamefont
  {Kamenev}}]{Wang-2019}%
  \BibitemOpen
  \bibfield  {author} {\bibinfo {author} {\bibfnamefont {H.}~\bibnamefont
  {Wang}}, \bibinfo {author} {\bibfnamefont {D.}~\bibnamefont {Bagrets}},
  \bibinfo {author} {\bibfnamefont {A.}~\bibnamefont {Chudnovskiy}},\ and\
  \bibinfo {author} {\bibfnamefont {A.}~\bibnamefont {Kamenev}},\ }\bibfield
  {title} {\bibinfo {title} {On the replica structure of {Sachdev-Ye-Kitaev}
  model},\ }\href
  {https://link.springer.com/article/10.1007/JHEP09(2019)057#citeas} {\bibfield
   {journal} {\bibinfo  {journal} {J. High En. Phys.}\ }\textbf {\bibinfo
  {volume} {2019}},\ \bibinfo {pages} {57} (\bibinfo {year}
  {2019})}\BibitemShut {NoStop}%
\bibitem [{\citenamefont {Tulipman}\ and\ \citenamefont
  {Berg}(2020)}]{Tulipman-2020}%
  \BibitemOpen
  \bibfield  {author} {\bibinfo {author} {\bibfnamefont {E.}~\bibnamefont
  {Tulipman}}\ and\ \bibinfo {author} {\bibfnamefont {E.}~\bibnamefont
  {Berg}},\ }\bibfield  {title} {\bibinfo {title} {Strongly coupled quantum
  phonon fluid in a solvable model},\ }\href
  {https://doi.org/10.1103/PhysRevResearch.2.033431} {\bibfield  {journal}
  {\bibinfo  {journal} {Phys. Rev. Research}\ }\textbf {\bibinfo {volume}
  {2}},\ \bibinfo {pages} {033431} (\bibinfo {year} {2020})}\BibitemShut
  {NoStop}%
\bibitem [{\citenamefont {Tulipman}\ and\ \citenamefont
  {Berg}(2021)}]{Tulipman-2021}%
  \BibitemOpen
  \bibfield  {author} {\bibinfo {author} {\bibfnamefont {E.}~\bibnamefont
  {Tulipman}}\ and\ \bibinfo {author} {\bibfnamefont {E.}~\bibnamefont
  {Berg}},\ }\bibfield  {title} {\bibinfo {title} {Strongly coupled phonon
  fluid and {Goldstone} modes in an anharmonic quantum solid: Transport and
  chaos},\ }\href {https://doi.org/10.1103/PhysRevB.104.195113} {\bibfield
  {journal} {\bibinfo  {journal} {Phys. Rev. B}\ }\textbf {\bibinfo {volume}
  {104}},\ \bibinfo {pages} {195113} (\bibinfo {year} {2021})}\BibitemShut
  {NoStop}%
\bibitem [{\citenamefont {Kennes}\ \emph {et~al.}(2017)\citenamefont {Kennes},
  \citenamefont {Wilner}, \citenamefont {Reichman},\ and\ \citenamefont
  {Millis}}]{Kennes-2017}%
  \BibitemOpen
  \bibfield  {author} {\bibinfo {author} {\bibfnamefont {D.~M.}\ \bibnamefont
  {Kennes}}, \bibinfo {author} {\bibfnamefont {E.~Y.}\ \bibnamefont {Wilner}},
  \bibinfo {author} {\bibfnamefont {D.~R.}\ \bibnamefont {Reichman}},\ and\
  \bibinfo {author} {\bibfnamefont {A.~J.}\ \bibnamefont {Millis}},\ }\bibfield
   {title} {\bibinfo {title} {Transient superconductivity from electronic
  squeezing of optically pumped phonons},\ }\href
  {https://doi.org/10.1038/nphys4024} {\bibfield  {journal} {\bibinfo
  {journal} {Nat. Phys.}\ }\textbf {\bibinfo {volume} {13}},\ \bibinfo {pages}
  {479} (\bibinfo {year} {2017})}\BibitemShut {NoStop}%
\bibitem [{\citenamefont {Lunkin}\ and\ \citenamefont
  {Feigel’man}(2022)}]{Lunkin-2022a}%
  \BibitemOpen
  \bibfield  {author} {\bibinfo {author} {\bibfnamefont {A.~V.}\ \bibnamefont
  {Lunkin}}\ and\ \bibinfo {author} {\bibfnamefont {M.~V.}\ \bibnamefont
  {Feigel’man}},\ }\bibfield  {title} {\bibinfo {title} {{Non-equilibrium
  {Sachdev-Ye-Kitaev} model with quadratic perturbation}},\ }\href
  {https://doi.org/10.21468/SciPostPhys.12.1.031} {\bibfield  {journal}
  {\bibinfo  {journal} {SciPost Phys.}\ }\textbf {\bibinfo {volume} {12}},\
  \bibinfo {pages} {031} (\bibinfo {year} {2022})}\BibitemShut {NoStop}%
\bibitem [{\citenamefont {Lunkin}\ and\ \citenamefont
  {Feigel'man}(2022)}]{Lunkin-2022b}%
  \BibitemOpen
  \bibfield  {author} {\bibinfo {author} {\bibfnamefont {A.~V.}\ \bibnamefont
  {Lunkin}}\ and\ \bibinfo {author} {\bibfnamefont {M.~.~V.}\ \bibnamefont
  {Feigel'man}},\ }\bibfield  {title} {\bibinfo {title} {{High-frequency
  transport and zero-sound in an array of {SYK} quantum dots}},\ }\href
  {https://doi.org/10.21468/SciPostPhys.13.3.073} {\bibfield  {journal}
  {\bibinfo  {journal} {SciPost Phys.}\ }\textbf {\bibinfo {volume} {13}},\
  \bibinfo {pages} {073} (\bibinfo {year} {2022})}\BibitemShut {NoStop}%
\bibitem [{\citenamefont {Kennes}\ and\ \citenamefont
  {Rubio}(2022)}]{Kennes-2022_preprint}%
  \BibitemOpen
  \bibfield  {author} {\bibinfo {author} {\bibfnamefont {D.}~\bibnamefont
  {Kennes}}\ and\ \bibinfo {author} {\bibfnamefont {A.}~\bibnamefont {Rubio}},\
  }\bibfield  {title} {\bibinfo {title} {A new era of quantum materials mastery
  and quantum simulators in and out of equilibrium},\ }\href
  {https://arxiv.org/abs/2204.11928} {\bibfield  {journal} {\bibinfo  {journal}
  {arXiv:2204.11928}\ } (\bibinfo {year} {2022})}\BibitemShut {NoStop}%
\bibitem [{\citenamefont {Patel}\ and\ \citenamefont
  {Sachdev}(2019)}]{Patel-2019}%
  \BibitemOpen
  \bibfield  {author} {\bibinfo {author} {\bibfnamefont {A.~A.}\ \bibnamefont
  {Patel}}\ and\ \bibinfo {author} {\bibfnamefont {S.}~\bibnamefont
  {Sachdev}},\ }\bibfield  {title} {\bibinfo {title} {Theory of a {Planckian}
  metal},\ }\href {https://doi.org/10.1103/PhysRevLett.123.066601} {\bibfield
  {journal} {\bibinfo  {journal} {Phys. Rev. Lett.}\ }\textbf {\bibinfo
  {volume} {123}},\ \bibinfo {pages} {066601} (\bibinfo {year}
  {2019})}\BibitemShut {NoStop}%
\bibitem [{\citenamefont {Schmalian}\ \emph {et~al.}(1996)\citenamefont
  {Schmalian}, \citenamefont {Langer}, \citenamefont {Grabowski},\ and\
  \citenamefont {Bennemann}}]{Schmalian-1996}%
  \BibitemOpen
  \bibfield  {author} {\bibinfo {author} {\bibfnamefont {J.}~\bibnamefont
  {Schmalian}}, \bibinfo {author} {\bibfnamefont {M.}~\bibnamefont {Langer}},
  \bibinfo {author} {\bibfnamefont {S.}~\bibnamefont {Grabowski}},\ and\
  \bibinfo {author} {\bibfnamefont {K.~H.}\ \bibnamefont {Bennemann}},\
  }\bibfield  {title} {\bibinfo {title} {Self-consistent summation of
  many-particle diagrams on the real frequency axis and its application to the
  {FLEX} approximation},\ }\href
  {https://www.sciencedirect.com/science/article/pii/0010465595001344}
  {\bibfield  {journal} {\bibinfo  {journal} {Comput. Phys. Commun.}\ }\textbf
  {\bibinfo {volume} {93}},\ \bibinfo {pages} {141} (\bibinfo {year}
  {1996})}\BibitemShut {NoStop}%
\bibitem [{\citenamefont {Bruus}\ and\ \citenamefont
  {Flensberg}(2004)}]{Bruus-2004mb}%
  \BibitemOpen
  \bibfield  {author} {\bibinfo {author} {\bibfnamefont {H.}~\bibnamefont
  {Bruus}}\ and\ \bibinfo {author} {\bibfnamefont {K.}~\bibnamefont
  {Flensberg}},\ }\href
  {https://global.oup.com/academic/product/many-body-quantum-theory-in-condensed-matter-physics-9780198566335?cc=us&lang=en&}
  {\emph {\bibinfo {title} {Many-body quantum theory in condensed matter
  physics: an introduction}}}\ (\bibinfo  {publisher} {Oxford University
  Press},\ \bibinfo {address} {Oxford},\ \bibinfo {year} {2004})\BibitemShut
  {NoStop}%
\bibitem [{\citenamefont {Peierls}(1933)}]{Peierls-1933}%
  \BibitemOpen
  \bibfield  {author} {\bibinfo {author} {\bibfnamefont {R.}~\bibnamefont
  {Peierls}},\ }\bibfield  {title} {\bibinfo {title} {Zur theorie des
  diamagnetismus von leitungselektronen},\ }\href
  {https://doi.org/10.1007/bf01342591} {\bibfield  {journal} {\bibinfo
  {journal} {Z. Phys.}\ }\textbf {\bibinfo {volume} {80}},\ \bibinfo {pages}
  {763} (\bibinfo {year} {1933})}\BibitemShut {NoStop}%
\bibitem [{\citenamefont {Li}\ \emph {et~al.}(2020)\citenamefont {Li},
  \citenamefont {Golez}, \citenamefont {Mazza}, \citenamefont {Millis},
  \citenamefont {Georges},\ and\ \citenamefont {Eckstein}}]{Li-2020}%
  \BibitemOpen
  \bibfield  {author} {\bibinfo {author} {\bibfnamefont {J.}~\bibnamefont
  {Li}}, \bibinfo {author} {\bibfnamefont {D.}~\bibnamefont {Golez}}, \bibinfo
  {author} {\bibfnamefont {G.}~\bibnamefont {Mazza}}, \bibinfo {author}
  {\bibfnamefont {A.~J.}\ \bibnamefont {Millis}}, \bibinfo {author}
  {\bibfnamefont {A.}~\bibnamefont {Georges}},\ and\ \bibinfo {author}
  {\bibfnamefont {M.}~\bibnamefont {Eckstein}},\ }\bibfield  {title} {\bibinfo
  {title} {Electromagnetic coupling in tight-binding models for strongly
  correlated light and matter},\ }\href
  {https://doi.org/10.1103/PhysRevB.101.205140} {\bibfield  {journal} {\bibinfo
   {journal} {Phys. Rev. B}\ }\textbf {\bibinfo {volume} {101}},\ \bibinfo
  {pages} {205140} (\bibinfo {year} {2020})}\BibitemShut {NoStop}%
\end{thebibliography}

%

\end{document}